\documentclass[12pt]{article}

\usepackage[reqno]{amsmath}
\usepackage{epsfig}
\usepackage{array}
 \usepackage{bbm}
\usepackage{cite}

\def\gappeq{\mathrel{\rlap {\raise.5ex\hbox{$>$}}
{\lower.5ex\hbox{$\sim$}}}}
\def\lappeq{\mathrel{\rlap{\raise.5ex\hbox{$<$}}
{\lower.5ex\hbox{$\sim$}}}}
\def\CPV{CP \! \! \! \! \! \! /~~}
\def\beq{\begin{equation}} \def\eeq{\end{equation}}
\def\bea{\begin{eqnarray}} \def\eea{\end{eqnarray}}
\def\bq{\begin{quote}} \def\eq{\end{quote}}

\def\dd{\displaystyle}

\def\gs{\mathrel{
   \rlap{\raise 0.511ex \hbox{$>$}}{\lower 0.511ex \hbox{$\sim$}}}}
\def\ls{\mathrel{
   \rlap{\raise 0.511ex \hbox{$<$}}{\lower 0.511ex \hbox{$\sim$}}}}

\input epsf.tex
\def\DESepsf(#1 width #2){\epsfxsize=#2 \epsfbox{#1}}
\usepackage{graphicx}
\def \DESepsf(#1 width #2){\bf #1  here: just uncomment the macro.}

\begin{document}



\title{\bf Theory of Neutrinos: A White Paper\\} \date{}
\maketitle
\begin{center}
\author{  {\bf \normalsize R.N. Mohapatra$^{*}$}
({\it \normalsize Group Leader})\\
{\bf  \normalsize S. Antusch$^1$, K.S. Babu$^2$, G. Barenboim$^3$,
M.-C. Chen$^4$, S. Davidson$^5$,} \\
{\bf \normalsize A. de Gouv\^ea$^6$, P. de Holanda$^{7}$,
B. Dutta$^8$, Y. Grossman$^9$, A. Joshipura$^{10}$,} \\
{\bf \normalsize B. Kayser$^{11}$, J. Kersten$^{12}$, Y.Y. Keum$^{13}$,
S.F. King$^{14}$ P. Langacker$^{15}$,} \\
{\bf \normalsize M. Lindner$^{16}$, W. Loinaz$^{17}$, I. Masina$^{18}$,
I. Mocioiu$^{19}$, S. Mohanty$^{10}$,} \\
{\bf \normalsize H. Murayama$^{20}$, S. Pascoli$^{21}$, S.T. Petcov$^{22,23}$,
A. Pilaftsis$^{24}$, P. Ramond$^{25}$,} \\
{\bf \normalsize M.~Ratz$^{26}$, W.~Rodejohann$^{16}$, R. Shrock$^{27}$,
T. Takeuchi$^{28}$, T. Underwood$^{5}$,} \\
{\bf \normalsize L. Wolfenstein$^{29}$ }
\normalsize \\[0.3cm]
$^{*}$ University of Maryland, College Park, MD 20742, USA\\
 \normalsize $^1$ Universidad Aut\'onoma de Madrid, 28049 Madrid, Spain 
and University of Southampton, Southampton SO17 1BJ, United Kingdom\\
 \normalsize $^2$ Oklahoma State University, Stillwater, OK-74078, USA\\
 \normalsize $^3$ University of Valencia, Valencia, Spain\\
 \normalsize $^4$  Fermilab, Batavia, IL 60540\\
 \normalsize $^5$ IPPP, University of Durham, Durham, DH1 3LE, Great Britain\\
 \normalsize $^6$ Northwestern University, Evanston, Illinois 60208-3112, USA\\
 \normalsize $^7$ Instituto de F\'\i sica Gleb Wataghin, UNICAMP
 PO BOX 6165, \\ \normalsize  CEP 13083-970, Campinas - SP, Brazil\\
 \normalsize $^8$ University of Regina,  Regina, Saskatchewan, Canada\\
 \normalsize $^9$ SLAC, Stanford, CA-94305, USA\\
 \normalsize $^{10}$  Physical Research Laboratory, Ahmedabad 380009, India\\
 \normalsize $^{11}$ Fermilab, Batavia, Il-60617, USA\\
 \normalsize $^{12}$ Deutsches Elektronen-Synchrotron DESY, 22603
   Hamburg, Germany\\
 \normalsize $^{13}$ Institute of Physics, Academia Sinica, Taipei,
 Taiwan 115, Republic of China  \\

 \normalsize $^{14}$ University of Southampton, Southampton SO17 1BJ, 
United Kingdom\\
 \normalsize $^{15}$ University of Pennsylvania, Philadelphia, PA 19104-6396,
USA\\
 \normalsize $^{16}$ Technische Universit\"at M\"unchen,
 \normalsize James-Franck-Stra{\ss}e, 85748~Garching, Germany\\
 \normalsize $^{17}$ Amherst College, Amherst, MA 01002-5000, USA\\
 \normalsize $^{18}$  Fermi Center, Via Panisperna 89/A, I-00184 Roma, Italy\\
 \normalsize and INFN, Sezione di Roma, ``La Sapienza''
              Univ., P.le A. Moro 2, I-00185 Roma, Italy\\
  \normalsize $^{19}$ Pennsylvania State University, University Park, PA 
16802, USA\\
 \normalsize $^{20}$  School of Natural Sciences,
Institute for Advanced Study,
        Princeton, NJ 08, USA\footnote{ On leave of absence from Department of
Physics, University of California, Berkeley, CA 94720.}\\
 \normalsize $^{21}$ UCLA, Los Angeles, CA 90095-1547, USA
and Department of Physics, Theory Division, CERN, CH-1211 Geneva 23, Switzerland\\
 \normalsize $^{22}$  SISSA/INFN-sezione di Trieste, Trieste, Italy\\
 \normalsize $^{23}$  INRNE, Bulgarian Academy of Sciences, Sofia, Bulgaria\\
 \normalsize $^{24}$  School of Physics and Astronomy, 
  University of Manchester, Manchester M13 9PL, United Kingdom\\
 \normalsize $^{25}$ University of Florida, Gainesville, FL 32611, USA\\
 \normalsize $^{26}$ Physikalisches Institut der Universit\"at Bonn, Nussallee
 12, 53115 Bonn,
Germany\\
\normalsize $^{27}$ Department of Physics, Sloan Laboratory, Yale University,
New Haven, CT 06250, USA\\
 \normalsize $^{28}$ Virginia Tech, Blacksburg, VA 24061, USA\\
 \normalsize $^{29}$ Carnegie-Mellon University, Pittsburgh, PA 15213, 
USA
}
\date{}
\begin{abstract}
During 2004, four divisions of the American Physical Society commissioned
a study of neutrino physics to take stock of where the field is at the
moment and where it is going in the near and far future. Several working
groups looked at various aspects of this vast field. The summary was
published as a main report entitled ``The Neutrino Matrix'' accompanied by
short 50 page versions of the report of each working group. Theoretical
research in this field has been quite extensive and touches many areas
and the short 50 page report \cite{short_report} provided only a brief
summary and overview
of few of the important points. The theory discussion group felt that
it may be of value to the community to publish the entire study as a white
paper and the result is the current article. After a brief overview of the
present knowledge of neutrino masses and mixing and some popular ways to
probe the new physics implied by recent data,
the white paper summarizes what can be learned about physics beyond the
Standard Model from the various proposed neutrino experiments. It also
comments on the impact of the experiments on our understanding of the
origin of the matter-antimatter asymmetry of the Universe and
the basic nature of neutrino interactions as well as the
existence of possible additional neutrinos. Extensive references to
original literature are provided.
\end{abstract}

\end{center}
\newpage
\tableofcontents

\newpage

\section{Introduction}

Our understanding of neutrinos has changed dramatically in the past six
years. Thanks to many neutrino oscillation experiments involving solar,
atmospheric, accelerator and reactor (anti)-neutrinos \cite{expt,pdg}, we have
learned that neutrinos produced in a well defined flavor eigenstate can be
detected, after propagating a macroscopic distance, as a different flavor
eigenstate.  The simplest interpretation of this phenomenon is that, like all
charged fermions, the neutrinos have mass and that, similar to quarks, the
neutrino weak, or flavor, eigenstates are different from neutrino mass
eigenstates {\it i.e.}, neutrinos mix.  This new state of affairs has also
raised many other issues \cite{barger}
which did not exist for massless neutrinos: For
example, (i) massive Dirac neutrinos,
like charged leptons and quarks, can have nonzero magnetic dipole moments
and massive Dirac and Majorana neutrinos can
have nonzero transition dipole moments; (ii) the heavier neutrinos decay
into lighter ones, like charged leptons and quarks, and (iii) (most
importantly) the neutrinos can be either Majorana or Dirac fermions (see later
for details).

 Learning about all these possibilities can not only bring our
knowledge of neutrinos to the same level as that of charged
leptons and quarks, but may also lead to a plethora of laboratory as
well as astrophysical and cosmological
consequences with far-reaching implications.
Most importantly, knowing neutrino properties in detail may
also play a crucial role in clarifying the blueprint of
new physical laws beyond those embodied in the Standard Model.

One may also consider the possibility that there could be new
neutrino species
beyond the three known ones $(\nu_e,\nu_\mu, \nu_\tau)$. In addition to
being a question whose answer would be a revolutionary milestone pointing
to unexpected new physics, it may also become a necessity if
 the LSND results are confirmed by the MiniBooNE
 experiment, now in progress at Fermilab. This would, undoubtedly, be a
second revolution in our
thinking about neutrinos and the nature of unification.

The existence of neutrino masses qualifies as the first evidence of new physics
beyond the Standard Model. The answers to the neutrino-questions mentioned
above will add substantially to our knowledge about the precise nature of this
new physics, and in turn about the nature of new forces beyond the Standard
Model. They also have the potential to unravel some of the deepest and most
long-standing mysteries of cosmology and astrophysics, such as the origin of
matter, the origin of the heavy elements, and, perhaps, even the nature of dark
energy.

 Active endeavors are under way to launch the era of precision neutrino
measurement science, that will surely broaden the horizon of our knowledge
about neutrinos. We undertake this survey to pin down how different
experimental results expected in the coming decades can elucidate the nature of
neutrinos and our quest for new physics. In particular, we would like to know
(i) the implications of neutrinos for such long-standing ideas as grand
unification, supersymmetry, string theory,
extra dimensions, etc; (ii) the implications of the
possible existence of additional neutrino species for physics and cosmology,
and (iii) whether neutrinos have anything to do with the origin of the observed
matter-antimatter asymmetry in the universe and, if so, whether there is any
way to determine this via low-energy experiments. Once the answers to these
questions are at hand, we will have considerably narrowed the choices of new
physics, providing a giant leap in our understanding of the physical Universe.

This review grew out of a year long study of the future of neutrino
physics conducted by four divisions of the American Physical Society
and is meant to be an overview of where we stand in neutrino physics
today,\footnote{The bulk of this report was finalized at the end of 
January 2005. We have not  included the (sometimes substantial) progress 
that has been obtained in several areas of neutrino physics since then.} 
where we are going in the next decades and the implications of
this new knowledge for the nature of new physics and for the early
universe. We apologize for surely missing vast parts of the neutrino
literature in our references. We expect this overview to be supplemented
by other excellent existing reviews of the subject in the literature.
Regarding more references and the more experimental aspects of the topics
under study, we refer to the other working group reports,
the Solar and Atmospheric Experiments \cite{APSsol},
the Reactor \cite{APSrea}, the
Neutrino Factory and Beta Beam Experiments and Development
\cite{APSnufac}, the Neutrinoless Double Beta Decay and
Direct Searches for Neutrino Mass \cite{APSB} and the
Neutrino Astrophysics and Cosmology \cite{APSastro} WGs.
In particular, we have not discussed  theoretical 
models for neutrino masses except giving a broad outline of ideas and 
getting beyond it only when there is a need to make 
some phenomenological point. Nonetheless, we hope to have captured
in this study the essential issues in neutrino physics that will be 
relevant as we proceed to the next level in our exploration of this 
fascinating field.

\subsection{Our present knowledge about masses and mixings}

\subsubsection{Dirac versus Majorana Neutrinos}
The fact that the neutrino has no electric charge endows it with certain
properties not shared by the charged fermions of the Standard Model.  One can
write two kinds of Lorentz invariant mass terms for the neutrino, Dirac and
Majorana masses, whereas for the charged fermions, conservation of electric
charge allows only Dirac-type mass terms. In the four component notation for
describing fermions, commonly used for writing the Dirac equation for the
electron, the Dirac mass has the form $\bar{\psi}\psi$, connecting fields of
opposite chirality, whereas the Majorana mass is of the form $\psi^TC^{-1}\psi$
connecting fields of the same chirality, where $\psi$ is the four component
spinor and $C$ is the charge conjugation matrix. In the first case, the fermion
$\psi$ is different from its antiparticle, whereas in the latter case it is its
own antiparticle. A Majorana neutrino implies a whole new class of experimental
signatures, the most prominent among them being the process of neutrinoless
double beta decay of heavy nuclei, ($\beta\beta_{0\nu}$). Since
$\beta\beta_{0\nu}$ arises due to the presence of neutrino Majorana masses, a
measurement of its rate can provide very precise information about neutrino
masses and mixing, provided (i) one can satisfactorily eliminate other
contributions to this process that may arise from other interactions in a full
beyond-the-standard-model theory, as we discuss below, (ii) one can precisely
estimate the values of the nuclear matrix elements associated with the
$\beta\beta_{0\nu}$ in question.

The expressions for the Dirac and Majorana mass terms make it clear that a
theory forbids Majorana masses for a fermion only if there is an additional
global symmetry under which it has nonzero charge.  As noted above, for charged
fermions such as the electron and the muon, Majorana mass-terms are forbidden
by the fact that they have nonzero electric charge and the theory has
electromagnetic $U(1)$ invariance. Hence all charged fermions are Dirac
fermions. On the other hand, a Lagrangian with both Majorana and Dirac masses
describes, necessarily, a pair of Majorana fermions, irrespective of how small
the Majorana mass term is (although it may prove very difficult to address
whether the fermion is of the Dirac or the Majorana type when the Majorana
mass-term is significantly smaller than the Dirac mass term).  Hence, since the
neutrino has no electric charge, the ``simplest" theories predict that
the neutrino
is a Majorana fermion meaning that a Majorana neutrino is more natural (or at
least requires fewer assumptions) than a Dirac neutrino.  In most of the
discussions below we assume that the neutrino is a Majorana fermion, unless
otherwise noted.

\subsubsection{Neutrino mixings}

We will use a notation where the electroweak-doublet neutrino eigenstate
(defined as the neutrino that is produced in a charged-current weak interaction
process associated with a well-defined charged lepton) is denoted by
$\nu_{\alpha}$, with $\alpha = e, {\mu}, {\tau}$.  We will also consider
$\nu_\alpha$ to include a set of $n_s$ possible electroweak-singlet
(``sterile") neutrinos.  Corresponding to these $3+n_s$ neutrino interaction
eigenstates are $3+n_s$ mass eigenstates of neutrinos, $\nu_i$.  We will order
the basis of mass eigenstates so that $m_1^2<m^2_2$ and $\Delta
m^2_{12}<|\Delta m^2_{13}|$, where $\Delta m^2_{ij}\equiv m_j^2-m_i^2$. The
neutrino interaction eigenstates are expressed in terms of the mass eigenstates
as follows: $\nu_{\alpha}=\sum_i U_{\alpha i}\nu_i$, where $U$ is a $(3+n_s)
\times (3+n_s)$ dimensional unitary matrix.  For the active neutrinos, with
$\alpha=e,\mu,\tau$, the relevant submatrix is thus a rectangular matrix with
three rows and $3+n_s$ columns.  In seesaw models, the entries in the columns
$4,...3+n_s$ are very small, of order $m_D/m_R$, where $m_D$ is a typical Dirac
mass and $m_R$ is a large mass of a right-handed Majorana neutrino.  Motivated
by these models, one commonly assumes a decoupling, so that to good
approximation the electroweak-doublet neutrinos can be expressed as linear
combinations of just three mass eigenstates, and hence one deals with a $3
\times 3$ truncation of the full $(3+n_s) \times (3+n_s)$ neutrino mixing
matrix. Since only the three electroweak-doublet neutrinos couple to the $W$,
the actual observed lepton mixing matrix that appears in the charged weak
current involves the product of the $3 \times (3+n_s)$ rectangular submatrix of
the full lepton mixing matrix with the adjoint of the $3 \times 3$ unitary
transformation mapping the mass to weak eigenstates of the charged leptons.
Thus, the lepton mixing matrix occurring in the charge-lowering weak current
has three rows and $3+n_s$ columns, corresponding to the fact that, in general,
a charged lepton $\alpha$ couples to a $\nu_\alpha$ which is
a linear combination of $3+n_s$ mass eigenstates.
Henceforth, unless explicitly indicated, we shall
assume the above-mentioned decoupling, so that the neutrino mixing matrix is $3
\times 3$, and will use $U$ to refer to the observed lepton mixing matrix,
incorporating both the mixings in the neutrino and charged lepton
sector. Neutrino oscillations and the mixing of two mass eigenstates of
neutrinos, $\nu_1$ and $\nu_2$, to form the weak eigenstates $\nu_e$ and
$\nu_\mu$ were first discussed by Pontecorvo and by
Maki, Nakagawa, and Sakata \cite{BPont57}.
The $3 \times 3$ truncation of the full neutrino mixing matrix
is often called the MNS, MNSP, or PMNS matrix in honor of these pioneers.

For the case of three Majorana neutrinos, the lepton mixing matrix
$U$ can be written as $V K$, where $V$ will be parametrized as
\begin{equation}
V =
\begin{pmatrix}
c_{12}c_{13} & s_{12}c_{13} & s_{13} e^{-i\delta} \cr
-s_{12}c_{23}-c_{12}s_{23}s_{13} e^{i\delta} &
c_{12}c_{23}-s_{12}s_{23}s_{13} e^{i\delta}
& s_{23}c_{13} \cr
s_{12}s_{23}-c_{12}c_{23}s_{13} e^{i\delta} &
-c_{12}s_{23}-s_{12}c_{23}s_{13} e^{i\delta} & c_{23}c_{13}
\end{pmatrix},
\label{V}
\end{equation}
while
$K~=~\mathrm{diag}\,(1, e^{i\phi_1},e^{i(\phi_2 + \delta)})$
\cite{Valle,BHP80}.

Neutrino oscillation experiments have already provided measurements for  the
neutrino mass-squared differences, as well as the mixing angles.
At the 3$\sigma$ level, the allowed ranges are \cite{Maltoni:2004ei}
$\sin^2 2\theta_{23}\geq 0.87$;
$1.4\times 10^{-3}~{\rm eV}^2 \leq |\Delta m^2_{13}|
 \leq 3.3\times 10^{-3}~{\rm eV}^2$;
$0.70 \leq \sin^2 2\theta_{12} \leq 0.94$;
$7.1\times 10^{-5}~{\rm eV}^2 \leq \Delta m^2_{12}
\leq 8.9\times 10^{-5}~{\rm eV}^2$;
$\sin^2 \theta_{13}\leq 0.051$~\cite{CHOOZ}.
There is currently no constraint on any of the CP odd phases or on the sign of
$\Delta m^2_{13}$. Note that in contrast to the quark sector we have
two large angles (one possibly maximal) and one small (possibly zero)
angle.

\subsubsection{Matter effect on neutrino propagation}

A very important fact about neutrinos that we seem to have learned
from solar neutrino data is that neutrino propagation in matter is
substantially different from that in vacuum. This effect is known
as the MSW (Mikheev-Smirnov-Wolfenstein) effect \cite{msw} and
has been widely discussed in the literature \cite{othermsw}.
There is however an important aspect of the favored large mixing
angle (LMA) MSW solution which needs to be tested in future
experiments. The LMA solution predicts a rise in the survival
probability in the energy region of a few MeV as we move down
from higher to lower solar neutrino energies. Since the present
data do not cover this energy region, new solar neutrino data is
needed in order to conclusively establish the LMA solution
\cite{bahcall}.

\subsubsection{Neutrino masses}

Given the current precision of neutrino oscillation experiments and the
fact that  neutrino oscillations  are only sensitive to
mass-squared differences, three possible arrangements of the neutrino 
masses are allowed:
\renewcommand{\labelenumi}{(\roman{enumi})}
\begin{enumerate}
\item Normal hierarchy, i.e.\ $m_1 < m_2 \ll m_3$. In this case $\Delta 
m^2_{23}\equiv m^2_3-m^2_2 > 0$, and
$m_3 \simeq \sqrt{\Delta m^2_{23}}\simeq 0.03-0.07$~eV.
 The solar neutrino oscillation involves the two lighter levels. The mass
of the lightest neutrino is unconstrained. If $m_1\ll m_2$, then we find
the value of $m_2 \simeq 0.008$ eV.

\item Inverted hierarchy, i.e.\ $m_1 \simeq m_2 \gg m_3$ \cite{silk} with
$m_{1,2} \simeq \sqrt{\Delta m^2_{23}}\simeq 0.03-0.07$ eV. In this case,
solar neutrino oscillation takes place between the heavier levels and we
have $\Delta m^2_{23}\equiv m^2_3-m^2_2 < 0$. We have no information about
$m_3$ except that its value is much less than the other two masses.

\item Degenerate neutrinos \cite{cald1} i.e., $m_1\simeq m_2 \simeq m_3$.

\end{enumerate}
\renewcommand{\labelenumi}{\arabic{enumi}.}
The behaviors of masses for different mass patterns are shown in Fig.\ 1.
\begin{figure}[t]
\begin{center}
\hspace*{-7mm}
\epsfig{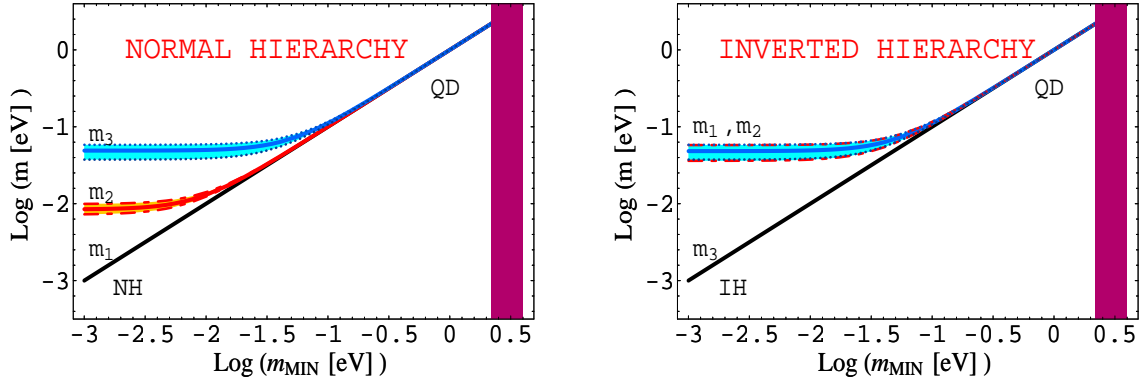}\\
\caption{{\small The three light neutrino masses as a function of the
lightest mass for the normal (left plot) and inverted (right plot) hierarchy.}}
\label{numass}
\end{center}
\end{figure}

\subsubsection{Overall scale for masses}
Oscillation experiments cannot tell us about the overall scale of masses.
It is therefore important to explore to what extent the absolute values
of the masses can be determined. While discussing the question of absolute
masses, it is good to keep in mind that none of the methods discussed
below can provide any information about the lightest neutrino mass in the cases of
a normal or inverted mass-hierarchy. They are most useful for determining
absolute masses in the case of degenerate neutrinos {\it i.e.,}\/ when all
$m_i\geq 0.1$ eV.

\noindent{\it Neutrino mass from beta decay}

  One can directly search for the kinematic effect of nonzero neutrino masses
in beta-decay by modifications of the Kurie plot.  This search is sensitive to
neutrino masses regardless of whether the neutrinos are Dirac or Majorana
particles.  These may be due to the emission, via mixing, of massive neutrinos
that cause kinks in this plot. If the masses are small, then the effects will
occur near to the end point of the electron energy spectrum and will be
sensitive to the quantity $m_{\beta}\equiv \sqrt{\sum_i |U_{ei}|^2m^2_i}$. The
Mainz \cite{Weinheimer:tn} and Troitsk \cite{Lobashev:tp} 
experiments place the 
present upper limit on $m_\beta \leq 2.3$~eV and 2.2 eV, respectively. 
The proposed KATRIN 
\cite{Osipowicz:2001sq} experiment is projected to be sensitive to
$m_{\beta}>0.2$~eV, which will have important implications for the theory of
neutrino masses. For instance, if the result is positive, it will imply a
degenerate spectrum; on the other hand a negative result will be a very useful
constraint.

\subsubsection{Neutrino masses and neutrinoless double beta decay}

Another sensitive probe for the absolute scale of the neutrino masses  is the
search for neutrinoless double beta decay, $\beta\beta_{0\nu}$, whose rate is
potentially measurable if the neutrinos are Majorana fermions and
$m_{ee}~=~ \sum U^2_{ei} m_i$
is large enough
\cite{BiPet87,ElliotVogel02},
or if there are new lepton number violating interactions \cite{moh1}.
In the absence of new lepton number violating interactions, a positive sign of
 $\beta\beta_{0\nu}$ would allow one to measure $m_{ee}$. Either way, we would
learn that the neutrinos are Majorana fermions
\cite{valle1}.
 However, if $m_{ee}$ is very small, and there are new lepton number
violating interactions,
neutrinoless double beta decay will measure the strength of the new
interactions (such as doubly charged Higgs fields or R-parity violating
interactions) rather than neutrino mass. There are many examples of models
where new interactions can lead to a  $\beta\beta_{0\nu}$ decay rate in the
observable range without at the same time yielding a significant Majorana mass
for the neutrinos. As a result, one must be careful in interpreting any nonzero
signal in $\beta\beta_{0\nu}$ experiments and not jump to the conclusion that a
direct measurement of neutrino mass has been made. The way to tell whether such
a nonzero signal is due to neutrino masses or is a reflection of new
interactions is to supplement  $\beta\beta_{0\nu}$ decay results with collider
searches for these new interactions. Thus collider experiments, such as those
at LHC, and double beta experiments play complementary roles.

The present best upper bounds on  $\beta\beta_{0\nu}$ decay
lifetimes come from the Heidelberg-Moscow
\cite{Klapdor-Kleingrothaus:2000sn}
and the IGEX \cite{Aalseth:2002rf}
experiments and  can be translated into an upper
limit on $m_{ee}\ls 0.9 $~eV \cite{Lisi}. There is a claim of discovery of
neutrinoless double beta decay of enriched $^{76}Ge$ experiment by
the Heidelberg-Moscow collaboration
\cite{klapdor04}.
Interpreted in terms of a Majorana mass of
the neutrino, this implies $m_{ee}$ between 0.12 eV to
0.90 eV. If confirmed, this result is of fundamental significance.
For a thorough discussions of this result
(see also \cite{klapdor}) \cite{antiklapdor}, we refer readers to the 
report of the double beta decay working group \cite{APSB}.

\subsubsection{Cosmology and neutrino masses}

 A very different way to get
information on the absolute scale of neutrino masses is from the study of the
cosmic microwave radiation spectrum as well as the study of the large scale
structure in the universe. A qualitative way of understanding why this 
is the case is that if
neutrinos
are present in abundance in the universe at the epoch of structure
formation and have a sizable mass the formation of structure is affected. For
instance, for a given neutrino mass $m$, all structure on a scale smaller
than a certain value given by the inverse of neutrino mass is washed away by
neutrino free-streaming. This implies a reduced power on smaller scales. Thus,
accurate measurements of the galaxy power spectrum for small scales can
help constrain or determine neutrino masses. Recent results from the WMAP
and surveys of large scale structure have set a limit
on the sum of neutrino masses $\sum m_i \leq 0.7-2$~eV
\cite{wmap,hannestad}. More recent results from the Sloan Digital Sky
Survey (SDSS) place
the limit of $\sum m_i \leq 1.6$ eV.  Hannestad \cite{hannestad} has
emphasized that these upper limits can change if there are more neutrino
species --- e.g. for 5 neutrinos, $\sum m_i \leq 2.12$~eV if they are in
equilibrium at the epoch of BBN.

A point worth emphasizing is that the above result is valid for both
Majorana and Dirac neutrinos as long as the ``right-handed'' neutrinos
decouple sufficiently earlier than the BBN epoch and are not regenerated
subsequently\footnote{ In the Dirac case the ``right-handed''
degrees of freedom are decoupled because of the smallness of the corresponding
Yukawa couplings. However, for very small temperatures,
i.e.\ long after BBN, it is no longer appropriate to describe neutrinos
in terms of chiral states. This means that strictly speaking there is a
regeneration, but this does not affect BBN (see,
e.g.,~\cite{Dolgov:1994vu}).}.

These limits already provide nontrivial information about neutrino masses:
the limit $\sum_i m_{i}=0.7$~eV, if  taken at face value,
implies that each individual neutrino mass is smaller than $0.23$~eV, which
is similar to the projected sensitivity of the
proposed KATRIN experiment.
PLANCK satellite observations are expected to be sensitive to even 
smaller values of
$\sum_i m_i$,  thereby providing a
completely independent source of information on neutrino masses.
These results may have implications for models of sterile
neutrinos that attempt to explain the LSND results.

\subsubsection{CP violation}
It is clear from Eq.~\eqref{V} that, for Majorana neutrinos, there are 
three CP-odd
phases that characterize neutrino mixings \cite{Valle,BHP80},
and our understanding
of the leptonic sector will remain incomplete without knowledge of these 
\cite{boris,GKM}.
There are two possible ways to explore CP phases: (i) one way is to perform
long-baseline oscillation experiments and look for differences between neutrino
and anti-neutrino survival probabilities \cite{minakata}; (ii) another way
is to use possible connections with cosmology. It has often been argued
that neutrinoless double beta decay may also provide an alternative way to
explore CP violation \cite{BGKP96}.
This is discussed in Sec.~\ref{sec:0nubbCP}.

In summary, the most important
goals of the next phase of neutrino oscillation experiments are:

(i) To determine the value of $\theta_{13}$ as precisely as possible;

(ii) To determine the sign of $\Delta m^2_{13}$, or the character of the 
neutrino mass hierarchy;

(iii) To improve the accuracy of the measurement of the other angles and 
the mass-squared differences;

(iv) To probe the existence of the three CP odd phases as best as possible.

The discussion above assumes a minimal picture for massive neutrinos where
 the most general Majorana mass for three neutrinos has been added.
While this may be the picture to the leading order, it is
quite conceivable that there are other interesting subdominant effects
that go beyond this.
It is of utmost interest to determine to what extent one can constrain (or
perhaps discover) these new nonstandard phenomena, since their absence
up to a certain level (or, of course, their presence) will provide
crucial insight into the detailed nature of the New Physics.

\subsubsection{Prospects for determining whether neutrinos are Majorana
or Dirac}
As an example of what we can learn from future experiments, we focus on
three experiments --- searches for neutrinoless double beta decay (down to the
level of $0.03$ eV level), studies to determine the sign of
$\Delta m^2_{23}\equiv (m^2_3-m^2_2)$, and the KATRIN experiment, which is sensitive
to the effects of a nonzero neutrino mass down to $0.2$ eV in tritium beta
decay. The interplay between the possible results of these three experiments is summarized in Table~\ref{tab:NatureOfNeutrinos}
 \begin{table}
\centering
\caption{Different possible conclusions regarding the
nature of the neutrinos and their mass hierarchy from the three
complementary experiments.}
\label{tab:NatureOfNeutrinos}
\begin{tabular}{|c||c||c||c|}
\hline
$\beta\beta_{0\nu}$ & $\Delta m^2_{13}$ & KATRIN & Conclusion \\ \hline
yes & $>0$ & yes & Degenerate Hierarchy, Majorana \\
yes & $>0$ & no & Degenerate Hierarchy, Majorana or\\
 & & & Normal Hierarchy, Majorana with heavy particle contribution\\
yes & $<0$ & no & Inverted Hierarchy, Majorana \\
yes & $<0$ & yes & Degenerate Hierarchy, Majorana\\
no & $>0$ & no & Normal Hierarchy, Dirac or Majorana\\
no & $<0$ & no & Dirac\\
no & $<0$ & yes & Dirac \\
no & $>0$ & yes & Dirac \\ \hline
\end{tabular}
\end{table}

We see that extremely valuable information will follow from the results of
these experiments.

\subsubsection{Sterile neutrinos}

A question of great importance in neutrino physics is the number of neutrino
species. Measurement of the invisible $Z$-width in LEP-SLC experiments tell us
that there are three types of light standard-model electroweak-doublet
neutrinos that couple to the $W$ and $Z$ boson. These are the three known
neutrinos $\nu_{e,\mu,\tau}$. This implies that if there are other
neutrino-like interaction eigenstates, then they must either be sufficiently
massive that they cannot occur in the decay of the $Z$
or they must be electroweak singlets with no coupling to the $W$
or $Z$.  In the latter case, the interaction eigenstates are called sterile
neutrinos.  In general, a neutrino mass eigenstate will be a linear combination
of the three electroweak-doublet neutrinos and some unknown number of
electroweak-singlet (= sterile) neutrinos.  In the presence of
electroweak-singlet neutrinos, the neutral weak current is not, in general,
diagonal \cite{LeeShrock:77,valle80}.  In common parlance, the word sterile
neutrino is often used to denote a light electroweak-singlet neutrino and hence
to exclude the heavy electroweak-singlet neutrino-like states that may well
play a role in the seesaw mechanism. So the question is: are there any (light)
sterile neutrinos and if so, how many are they and do they mix with the ordinary
neutrinos?

Light sterile neutrinos have been postulated in order to 
explain~\cite{caldwell}
the data from the Los Alamos Liquid Scintillation Detector (LSND)
experiment~\cite{lsnd}, where neutrino flavor conversion both from a stopped
muon (DAR) as well as the one accompanying the muon in pion decay have
apparently been observed. The evidence from the DAR is statistically more
significant and is interpreted as an oscillation from $\bar{\nu}_\mu$ to
$\bar{\nu}_e$. The mass and mixing parameter range that fits data is:
\begin{eqnarray}
 \Delta m^2 \simeq 0.2 - 2~\mathrm{eV}^2\;,
 \quad \sin^22\theta \simeq 0.003-0.03\;.
\end{eqnarray}
There are points at higher masses specifically at 6 eV$^2$ which are also
allowed by the present LSND data for small mixings. The KARMEN experiment at
the Rutherford laboratory has very strongly constrained the allowed parameter
range of the LSND data~\cite{karmen}. Currently the MiniBooNE experiment at
Fermilab is under way to probe the LSND parameter region~\cite{louis}.

Since this $\Delta m^2_\mathrm{LSND}$ is so different from that $\Delta
m^2_{\odot, A}$, the simplest way to explain these results is to add
one~\cite{caldwell,other} or two~\cite{sorel} sterile neutrinos. The sterile
neutrinos raise important issues of consistency with cosmology as well as
physics beyond the simple three neutrino picture and will be discussed in a
subsequent section.

\subsubsection{Neutrino electromagnetic dipole moments and neutrino decay}

 A massive Dirac neutrino can have a diagonal magnetic (and a CP-violating
electric) dipole moment.  Because a Majorana neutrino is the same as its
antiparticle, it has vanishing diagonal magnetic and electric dipole moments.
A massive Dirac or Majorana neutrino can have nondiagonal, i.e., transition,
magnetic and electric dipole moments.  Some discussions of diagonal and
transition neutrino electromagnetic moments in renormalizable electroweak gauge
theories (where these can be calculated) include \cite{mmom1,mmom2},
\cite{LeeShrock:77}, \cite{leftright}--\cite{mmom2004}.  In the standard model
extended to contain massive Dirac neutrinos, $\mu_{\nu_j} = 3 e G_F
m_{\nu_j}/(8 \pi^2 \sqrt{2}) = 1.6 \times 10^{-19} (m_{\nu_j}/(1 \ {\rm eV})) \
\mu_B$ for the neutrino mass eigenstate $\nu_j$ \cite{fs}, where
$\mu_B~=~\frac{e}{2m_e}$ is a Bohr magneton.  In left-right models and others
with new physics beyond the standard model, this may be larger (e.g.,
\cite{leftright,rs82,mmom6,mmom7,mmom2004}).  
In contrast to the magnetic dipole
moment, the neutrino electric dipole moment vanishes at one-loop order for a
massive Dirac neutrino in the extended standard model \cite{rs82}. However, in
a left-right model, a Dirac neutrino may acquire an electric dipole moment at
the one-loop level \cite{rs82}.  In the more generic case of a Majorana
neutrino, one's interest focuses on the neutrino transition magnetic (and
electric) dipole moments.  The presence of these diagonal or transition moments
allows for new electromagnetic interactions between neutrinos and other
fermions of the Standard Model. In particular in neutrino-electron scattering,
in addition to the usual weak interaction contribution, there will be a photon
exchange contribution to the scattering cross section. The existing neutrino
scattering measurements therefore provide an upper limit on the neutrino
magnetic moment: $\mu_{\nu_e}\leq (1-1.3)\times 10^{-10}\mu_B$.
As we will discuss in more detail later, the observation of nonzero 
neutrino magnetic
moment would be considered evidence of new physics at the TeV scale. The 
reason for that is
that if all new physics is parameterized by (Majorana or Dirac) neutrino 
masses, or, equivalently,
if all new physics effects are suppressed by the very large naive seesaw 
energy scale (close to the GUT scale)
the neutrino magnetic moments are expected to be of order
$10^{-19}\mu_B\left(\frac{m_\nu}{1~eV}\right)$. High-precision searches 
for a magnetic moment provide, therefore,
complementary tools to probe the physics that is expected to lie just 
beyond the electroweak symmetry breaking scale.

A neutrino magnetic or electric dipole moment leads to new processes that can
alter our understanding of energy balance in astrophysical systems such as in
stars and supernovae~\cite{raffelt}. It can also affect considerations
involving the neutrinos in the early universe such as the BBN. In
Sec.\ \ref{mm} we discuss more details on neutrino magnetic moments and what
one can learn from various proposed experiments.

The existence of a neutrino magnetic or electric transition moment is also
related to neutrino decays. For instance, it would allow heavier neutrinos to
decay radiatively to the lighter ones
\cite{mmom1,ms77,pw82,LeeShrock:77,STP77,Marciano77}. Such decays can be 
detectable in
astrophysical experiments. Present upper limits coupled with the general idea
about spectra of neutrinos from oscillation experiments, imply that lifetimes
of the primary mass eigenstates in electroweak-doublet neutrinos are larger
than $10^{20}$ sec., much longer than the age of the universe
\cite{mmom1,ms77,pw82,LeeShrock:77,STP77,Marciano77}. Such decays do
not therefore affect the evolution of the universe.

It is however possible that there are other scalar particles to which the
neutrinos decay; one such example is the majoron, which is a Goldstone boson
corresponding to the spontaneous breaking of a global $B-L$
symmetry~\cite{cmp}. The decay to these scalar bosons may occur at a faster
rate\cite{gelmini} than that to photons and may therefore have 
astrophysical and cosmological
implications~\cite{beacom}. This will be the subject of another
working group \cite{APSastro};
so we only focus on the implications of the magnetic moment in one of the
subsequent sections.

\subsection{Neutrino probes of other fundamental symmetries}

Neutrino experiments can also be used to probe the validity of other
fundamental symmetries, some of which are often commonly assumed in
theoretical discussions, as well as the basic assumptions of local quantum
field theories on which the Standard Model is based. Some examples of
these are:
\begin{itemize}
\item Violation of Lorentz invariance;

\item CPT violation;

\item Possible existence of new long range forces in nature associated
with lepton number;

\item Nonstandard interactions of neutrinos such as flavor changing
neutral currents involving neutrinos.

\end{itemize}

We will explore to what extent existing limits on these departures from
standard scenarios can be improved.

\subsection{Why neutrino mass necessarily means physics beyond the
Standard Model?}

Neutrino oscillations are, to date, the only evidence for the existence of
physics beyond the Standard Model  (in the domain of particle
physics). It is of utmost importance
to decipher the kind of new physics indicated by the existing data
and to anticipate the signals of new physics that might appear in future planned observations.
We must understand how and if
they fit into the different big pictures that have been advocated for
independent reasons, including the gauge hierarchy problem and gauge coupling unification.
To discuss this, we first introduce the
Standard Model and possible ways to extend it to accommodate the neutrino
observations.

In the Standard Model, which is based on the gauge group
$SU(3)_c\times SU(2)_L\times U(1)_Y$, the quarks and
leptons transform as  $Q_L(3,2, {1\over 3})$, $u_R(3, 1, {4\over 3})$,
  $ d_R(3, 1,-\frac{2}{3})$, $L (1, 2, -1)$,  $e_R(1,1,-2)$. The Higgs
boson $H$, responsible for electroweak symmetry breaking, transforms as $(1, 2, +1)$.
The electroweak symmetry $SU(2)_L\times U(1)_Y$ is broken by the vacuum
expectation of the Higgs doublet $\langle H^0\rangle=v_{\rm wk}\simeq 246$ GeV, which
 renders the $W^{\pm}$ and $Z^0$ gauge bosons and the electrically charged fermions massive.
The reason neutrinos do not get mass as a result of the Higgs
mechanism is that the right-handed
neutrino $N_R$ was not included in the list of fermions in the
Standard Model; as a result
 there is no coupling of the form $h_\nu\bar{L}HN_R$ that could have
given mass to the neutrinos after symmetry breaking.

 One seemingly straightforward way to understand the neutrino mass would be to
extend the Standard Model to include the $N_R$. This would also be desirable
from the point of view of making the model quark lepton symmetric. There are
two problems with this na\"{\i}vely trivial modification. One is that by quark
lepton symmetry one would expect the neutrino masses arising from the Yukawa
coupling $h_\nu\bar{L}HN_R$ to be of the same order as the quark and charged
leptons. Observations suggest that neutrino masses are at least $10^6$
times smaller than the smallest quark and lepton masses. Therefore, a
nonzero neutrino mass not only suggests the existence of right-handed neutrinos
(of which there would be three if they correspond to the usual generations),
but some new physics that will enable us to understand why $M_\nu \ll
m_{q,\ell}$.  The seesaw mechanism provides a plausible basis for this
understanding, since it makes use of the fact that, among the known fermions,
only neutrinos can have Majorana mass terms.  Thus, ironically, we may have a
better way to understand the lightness of the neutrinos than we do to
understand the generational hierarchy factor of $\sim 10^6$ between the masses
of the top quark and the electron, for which there is no accepted explanation
at present.

The other problem with introducing a set of right-handed neutrino fields
is the fact that they are Standard Model gauge singlets.
This means that, as far as the symmetries of the Standard Model are
concerned, a Majorana mass for the $N_R$ fields is allowed.
If such a mass term is present, however, the neutrino masses are not
simply given by the $h_{\nu}v_{\rm wk}$, but are determined by a more
complicated function of $h_{\nu}v_{\rm wk}$ and the Majorana masses of the
right-handed neutrinos. In order to avoid the presence of a Majorana mass
for the right-handed neutrinos one is required to impose an extra {\sl
symmetry} to the Standard Model Lagrangian (say, lepton number) ---
a very nontrivial modification of what is traditionally referred to as
the Standard Model of electroweak interactions.

\subsubsection{Seesaw mechanism for small neutrino masses}

A simple way to understand the smallness of neutrino mass within this minimally
extended Standard Model is to break lepton number symmetry (or $B-L$ symmetry)
and add a Majorana mass for the right-handed neutrino $M_RN^T_RC^{-1}N_R$. Thus
the two terms that give mass to the neutrinos have the form $h_\nu v_{\rm wk}
 \bar{\nu}_LN_R + M_R N^T_RC^{-1}N_R$ + h.c. Assuming $n$ ``left-handed'' neutrinos $\nu_L$
 and $m$ ``right-handed neutrinos $N_R$, the $(n+m)\times (n+m)$ Majorana neutrino mass matrix is
\begin{equation}
{\cal M}~=~\begin{pmatrix}0 &
h_\nu v_{\rm wk}\cr  h_\nu v_{\rm wk} & M_R\end{pmatrix}.
\label{eq:6by6}
\end{equation}
In the limit $M_R\gg hv_{\rm wk}$, the eigenvalues of
this matrix are given by $-\frac{(h_\nu v_{\rm wk})^2}{M_R}$ and $M_R$, with respective approximate eigenvectors
$\nu_L$ and $N_R$. The effective active neutrino masses are
clearly much smaller than typical charged fermion masses (which are of order
$h_\nu v_{\rm wk}$) as long $M_R \gg v_{\rm wk}$. This is the well-known seesaw mechanism
\cite{minkowski,
Yanagida:1980,Gell-Mann:1980vs,Glashow:1979vf,Mohapatra:1980ia}. If we
take as a guide a value for $h_\nu\leq 1$, then atmospheric neutrino data
requires that  $M_R\leq 10^{15}$ GeV. It should be emphasized that there is very little concrete information or
experimental guidance regarding the magnitude of $M_R$, which is virtually unconstrained \cite{LSND_seesaw}.
One question which arises is why this
value rather than $M_{\rm Pl}\simeq 10^{18}$ GeV, which, one may argue, would have been a
more natural value? Could this be an indication of a new symmetry? The answer
to this question is obviously of fundamental significance.

An example of such a symmetry is the $B-L$ symmetry embodied in the left-right
symmetric models based on the gauge group $SU(2)_L\times SU(2)_R\times
U(1)_{B-L}$~\cite{moh}. This gauge group is also a subgroup of $SO(10)$ grand
unification group. The above mentioned value of $M_R$ is rather close to the
conventional GUT scale of $10^{16}$ GeV. This makes the seesaw mechanism a very
attractive framework for discussing the neutrino mass.  We will discuss further
consequences of grand unification for neutrino masses in a subsequent
section. We will also explore in this review unification-model independent
consequences of the seesaw mechanism.

\subsubsection{Type I vs type II seesaw mechanism}

If there are indeed right-handed neutrinos, the most general Majorana mass matrix  that mixes active and
sterile neutrinos is given by Eq.~(\ref{eq:6by6}) with the $n\times n$ zero matrix in the upper left-hand corner replaced
by a generic (symmetric) matrix $M_L$. This phenomenon occurs, for example,  when
the theory containing the $N_R$ becomes parity symmetric as is the case for
$SU(2)_L\times SU(2)_R\times U(1)_{B-L}$ or $SO(10)$ based models.  In this case
the seesaw formula is modified to
\begin{equation} { M}_\nu^{\rm II} =
 M_L - M_\nu^D M_R^{-1} (M_\nu^D)^T\;,
\label{eq:typeII} \end{equation} 
where, in an  $SU(2)_L\times SU(2)_R\times U(1)_{B-L}$ symmetric model,
$M_L = f v_L$ and $M_R=f v_R$, where $v_{L,R}$ are the vacuum expectation
values of Higgs fields that couple to the right and left-handed neutrinos. Eq.~(\ref{eq:typeII})
is called the type II seesaw relation
\cite{seesaw2}.  It should be noted that in the absence of a discrete
left-right symmetry, $M_L$ is in general not related to $M_R$.

\subsubsection{Triplet seesaw}
An alternative way to understand the small neutrino mass without
introducing the right handed neutrino is the triplet seesaw mechanism. It
was pointed out in various papers \cite{triplet1} in early 1980 that
if the standard model is extended by the addition of a triplet Higgs
$\Delta_L$ with weak hypercharge $Y=2$, a
vev for it can lead to Majorana mass for the neutrinos from the
interaction $f_\nu \psi_L^TC^{-1}\tau_2\vec{\tau}\cdot
\vec{\Delta}\psi_L$ ($\psi_L$ being the lepton doublet
$(\nu_L,e_L)$). However, one has to tune the Yukawa coupling $f_\nu$ by
about $10^{-10}$ or so to get desirable neutrino masses. It has
subsequently been shown \cite{triplet2} that in the context of grand
unified theories, the triplet vev is given by the formula
$<\Delta^0_L>\sim \frac{v^2_{wk}}{M_U}$, where $M_U$ is close to the grand
unification scale and corresponds to the physical mass of the triplet
Higgs field. Since $M_U \gg v_{wk}$, this provides a natural suppression
of the triplet vev and the right order for the neutrino mass emerges.
Note also that this can also emerge from the type II seesaw formula in the
limit of $M_{N_R}\rightarrow \infty$. In this case, the neutrino mass
matrix is directly proportional to the coupling matrix $f_\nu$.

\subsubsection{Seesaw with triplet fermions}
Yet another possible extension of the standard model without right
 handed neutrinos which leads to small neutrino masses is to postulate 
the existence of  triplet
vectorlike fermions: $\vec{\Lambda}$ \cite{ma}. Since a vectorlike triplet
can have an arbitrary mass, it also leads to seesaw mass formula.

\subsubsection{Understanding large mixings within the seesaw mechanism}

 A major puzzle of quark--lepton physics is the fact that the quark
mixing matrix and the leptonic
one are qualitatively different.
In order to understand the mixing angles~\cite{barger,king}, we
have to study the
mass matrices for the charged leptons and neutrinos.

\noindent{\it A general approach}

To see the possible origin of neutrino mixings, one can start with the
following form for the mass part of the neutrino Lagrangian:
\begin{eqnarray}
{\cal L}_{mass}~=~\bar{\nu}_L M_{\nu}^D N_R + \bar{e}_L M_\ell e_R +
N^T_R M_R N_R ~~+ ~~ h.c.
\end{eqnarray}
Using the seesaw mechanism one can derive from this equation, the formula
for neutrino masses can be written as for the case of type I seesaw:
\begin{equation}
{ M}_\nu^{\rm I} = - M_\nu^D M_R^{-1} (M_\nu^D)^T,
\end{equation}
To obtain the lepton mixing matrix, one can diagonalize the charged lepton mass
matrix by $M_\ell ~=~ U_\ell M^d_\ell V^{\dagger}$ and
$m_{\nu}=(U^{*})_{\nu }m^d_\nu(U^{\dagger})_{\nu}$ and find that
$U=~U^{\dagger}_\ell U_\nu$.

With  this theoretical preamble, understanding of neutrino mixings can
proceed along two paths. In theories where quark and lepton mixings are
disconnected (such as many weak scale theories), one may like to pass to a
basis where the charged lepton masses are diagonal. In that case, all the
neutrino mixing information is in the effective neutrino mass matrix.
One can then look for the types of mass matrices
for neutrinos that can lead to bi-large mixings and try to understand
them in terms of new physics. Here we give a brief overview of some
generic structures for $M_\nu$ that do the job.

\noindent (i) {\it The case of normal hierarchy:}
In this case, one neutrino mass matrix that leads to ``bi-large'' mixing
has the
 form:
\begin{equation}
M_\nu~=~m_0\,\begin{pmatrix}\epsilon & \epsilon & \epsilon\cr \epsilon &
1+\epsilon &
1\cr \epsilon & 1 & 1\end{pmatrix}
\end{equation}
where $m_0$ is $\sqrt{\Delta m^2_{\rm A}}$. We have omitted order one
coefficients in front of the $\epsilon$'s. This matrix leads to $\tan
\theta_A \simeq 1$, $\Delta m^2_{\odot}/\Delta m^2_{\rm A} \simeq \epsilon^2$ and
also to a large solar angle. For the LMA solution, we find the interesting
result that $\epsilon \sim \lambda$ where $\lambda$ is the Cabibbo angle
($\simeq 0.22$). This  could be a signal of hidden quark lepton
connection\footnote{Alternatively,
the relation $\theta_{\odot} + \lambda = \pi/4$,
nowadays known as quark--lepton--complementarity,
can also be interpreted as such a connection \cite{cablampi4}.}.
In fact we will see below that in the context of a minimal $SO(10)$ model,
this connection is realized in a natural manner.

\noindent(ii) {\it The case of inverted hierarchy:}
The elements of the neutrino mass matrix in this case have a pattern
\begin{eqnarray}
{ m}_\nu=m_0~\left(\begin{array}{ccc} \epsilon &
c & s\\ c & \epsilon & \epsilon\\ s & \epsilon &
\epsilon\end{array}\right).
\end{eqnarray}
where $c=cos\theta$ and $s=\sin\theta$ and it denotes the atmospheric
neutrino mixing angle. An interesting point about this mass matrix is that
in the limit of $\epsilon\rightarrow 0$, it possesses an $L_e-L_\mu-L_\tau$
symmetry~\cite{inverted}. One therefore might hope that if inverted
hierarchy structure is
confirmed, it may provide evidence for this leptonic symmetry. This
can be an important clue to new physics beyond the Standard Model.
This issue of leptonic symmetries will be discussed in the main body of this report.

\noindent (iii) {\it Degenerate neutrinos:}
One may either add a unit matrix to the just mentioned mass matrices and look
for new physics models for them; alternatively, one may look for some dynamical
ways by which large mixings can arise. It turns out that if neutrinos are mass
degenerate, one can generate large mixings out of small
mixings~\cite{Babu:1993qv,Balaji:2000gd,Antusch:2002fr,Mohapatra:2003tw,Hagedorn:2004ba} purely
as a consequence of radiative corrections. We will call this possibility
radiative magnification and will discuss it in a future section.

In grand unified theories, quark and lepton mass matrices are connected. One
may therefore lose crucial information about symmetries if one works in a basis
where the charged leptons are diagonal. Furthermore, if either of the quark (up
or down) mass matrices are chosen diagonal, it may not even be possible to go
to the diagonal charged lepton basis. Thus in this case, we have $U~=~
U^{\dagger}_\ell U_\nu$. So one may seek an understanding of large mixings in
the charged lepton sector. For example in $SU(5)$ type theories, $U_\ell$ is
related to the mixings of right-handed quarks which are unobservable in low
energy weak interactions and can therefore be the source of large mixings.
Models of this type are called lopsided mixing models \cite{lop}.

The basic strategy then would be to look for clues for new symmetries in the
structure of the mass matrices, which could then provide information about the
nature of physics beyond the Standard Model. The symmetries of course may
become obscured by our choice of basis where the charged leptons are diagonal.
It is this which gives different possibilities for arriving at the bi-large
mixings and the hope is that different strategies will lead to different
predictions for observables, which can then be put to experimental test.

\subsubsection{Alternatives to high-scale seesaw}

While the high-scale seesaw mechanism is the simplest and perhaps the most
elegant way to understand the small neutrino masses and become couched in a
quark leptonic and parity symmetric framework leading to simple grand
unification theories, there are alternatives to seesaw which can also explain
the small neutrino 
masses~\cite{zee,li,babu2,dchang,valle2,gluza,nt,lrs,chacko,pl}. In
such a case the neutrinos can either be Dirac or Majorana fermions depending on
the theory.

 Unlike the non-supersymmetric seesaw models, alternatives such as the one
presented in \cite{valle2} often predict observable charged lepton
lepton-flavor violating signals \cite{Ilakovac}, e.g., $\mu \rightarrow e
\gamma$, $\mu, \tau \to eee$, etc.  More generically, searches for
charged-lepton flavor violation crucially help distinguish among the several
theoretical interpretations of the origin of neutrino masses.

\subsection{Summary of the Introduction}

Some of the questions that we would like to answers in the course of
this work are:

\begin{itemize}

\item Can we decide whether the neutrino is a Dirac or Majorana particle?

\item To what extent can the planned neutrino experiments pin down
the structure of the three neutrino mass matrix? This involves such questions
as determining the sign of $\Delta m^2_{23}$, higher precision measurement
of mixing parameters, etc.

\item What is the impact of a $\theta_{13}$ measurement (and the improved
determination of the other elements of the lepton mixing matrix) on the general
landscape of physics beyond the Standard Model? We find that
$\theta_{13}$ is a powerful discriminator of models.

\item Can we test the seesaw hypothesis and discriminate between different
types of seesaw using lepton flavor violation and other ``non-neutrino probes?''

\item How can one experimentally discover or limit physics
beyond the standard scenario? This will address such aspects as:

(1) Flavor changing neutral currents for neutrinos; present limits
and future prospects;

(2) Admixtures of sterile neutrinos, both heavy and light;

(3) magnetic moments of neutrinos.

\item What can we learn about CP violation in the lepton sector and how
can we connect it to the question of the origin of matter via
leptogenesis. Given what we know about the neutrino masses, assuming
thermal leptogenesis, do we have an explanation of the observed baryon to
photon ratio?

\end{itemize}

\newpage
\section{What can we learn about neutrino mass matrices from
experiments?} 

In this section we briefly review our ability to reconstruct the neutrino mass
matrix. We will also discuss (from ``the bottom up") what we hope to learn from
the neutrino mass matrix itself, instead of trying to quantify what different
models predict for the neutrino mass matrix. See, for example,
\cite{barger, king} for
reviews of a few different models. In a subsequent section, we will discuss
the connection of neutrino masses to GUTs, and will spend a little more time on
``top-down" predictions for neutrino masses and mixing angles.

As mentioned earlier, we will assume that the neutrinos are Majorana fermions.
While there is no experimental evidence that this is the case, the majority of
the theoretical HEP community considers it more likely that the neutrinos are
Majorana fermions, and a larger amount of phenomenological research effort has
gone into understanding and interpreting Majorana neutrino mass matrices than
Dirac mass matrices. For some discussions of Dirac neutrino mass matrices and
how they are related to the large mixing in the leptonic sector and the
neutrino mass-squared differences, see, for example,~\cite{Dirac}.

Below the electroweak phase transition, the Majorana neutrino mass matrix
$m_{\nu}$ is the coefficient of the operator (using four-component-spinor
notation)
\begin{equation}
\frac{1}{2}m_{\nu}^{\alpha\beta}\nu^{T}_{\alpha}C^{-1}\nu_{\beta}+H.c.,
\label{m_operator}
\end{equation}
where $\alpha,\beta=e,\mu,\tau,\ldots$ are flavor indices, and
$m_{\nu}^{\alpha\beta}$ are the components of the neutrino mass matrix
(note that $m_{\nu}$ is symmetric, {\it i.e.,}
$m_{\nu}^{\alpha\beta}=m_{\nu}^{\beta\alpha}$). In this section we will
concentrate on a purely active $3\times 3$ mass matrix. A detailed
discussion of $4\times 4$ (and larger) mass matrices, which also allow
for the existence of fourth generation and/or sterile neutrinos is the
subject of subsequent sections. Note that Eq.~(\ref{m_operator}) is not
sensitive to the mechanism that generates neutrino masses. These will be
discussed in more detail in a later section.

In general, one cannot work back from a knowledge of the observed lepton mixing
matrix to the individual nondiagonal mass matrices in the charged lepton and
neutrino sectors.  It is, indeed, the diagonalization of both of these mass
matrices that gives rise to the observed lepton mixing, and models exist where
the mixing in the charged lepton sector is large.
One can always choose to work on the weak basis where the charged lepton mass matrix is diagonal ---
the price one pays for doing this is that the flavor structure of the theory may not be manifest.
In this case, one can calculate the neutrino mass matrix in terms of the observed
lepton mixing matrix as
\begin{equation}
m^{\alpha\beta}_{\nu}=\sum_i(U^{*})_{\alpha i}m_i (U^{\dagger})_{i\beta},
\label{m_ab}
\end{equation}
We choose sign conventions such that the neutrino mass eigenvalues are real and
positive. By choosing to write $U=V K$, where $V$ and $K$ are given by Eq.~(1)
we have removed all of the redundancy contained in $m_{\nu}$ associated with
re-defining the neutrino fields by a complex phase. Hence, $m_{\nu}$ as defined
by Eq.~(\ref{m_ab}) is only a function of observable parameters. The phases in
$K$ are the so-called Majorana phases
\cite{Valle,BHP80}. They can be redefined away by allowing
the neutrino mass eigenvalues to be complex. In this case, $U=V$, $m_1$ is real
and positive, and $m_2=|m_2|e^{-2i\phi_1}$, $m_3=|m_3|e^{-2i\phi_2}$.

In the near future, we hope to significantly improve the determination of
the elements of the neutrino mass matrix, although some uncertainty will
still remain (for a detailed discussion, see, for example,
\cite{reconstruction}). Through neutrino oscillation experiments, all
three mixing angles $\theta_{12}, \theta_{23}$, and $\theta_{13}$ are
expected to be determined with good precision (this is one of the main
goals of next-generation neutrino oscillation experiments, discussed in
great lengths in this report), while there is hope that the ``Dirac
phase" $\delta$ can be probed via long-baseline $\nu_{\mu}\to\nu_e$
oscillation searches. Neutrino oscillation experiments will also
determine with good precision the neutrino mass-squared differences
($\Delta m^2_{12}$ at the 5\%--10\% level, $\Delta m^2_{13}$ [including
the sign] at the few percent level). In order to complete the picture,
three other quantities must also be measured, none of which is directly
related to neutrino oscillations.

One is the overall scale for neutrino masses. As already briefly
discussed, this will be probed, according to our current understanding,
by studies of the end-point spectrum of beta-decay, searches for
neutrinoless double beta decay, and cosmological observations (especially
studies of large-scale structure formation). Note that neutrinoless
double beta decay experiments are sensitive to $|m_{\nu}^{ee}|$, {\it
i.e.}, they directly measure the absolute value of an element of $m_{\nu}$.

The other two remaining observables are the
``Majorana" phases. Neutrinoless double beta decay experiments are
sensitive to a particular combination of these, the so-called
effective Majorana mass,
\begin{equation}
\left|m_{\nu}^{ee} \right| \equiv \langle m \rangle_{eff}
= \left|\cos^2\theta_{13}
\left(|m_1|\cos^2\theta_{12}+|m_2|e^{2i\phi_1}\sin^2\theta_{12}\right)
+\sin^2\theta_{13}|m_3|e^{2i\phi_2}\right|\;.
\end{equation}
With present uncertainties in the nuclear matrix elements, however,
it seems at least very challenging \cite{noMaj} to
obtain any information regarding Majorana phases from neutrinoless
double beta decay. For a detailed study, see, for example,
\cite{noMajresp}.

A few comments are in order. First, the relation between the rate for
neutrinoless double beta decay and the Majorana phases and neutrino
masses only holds under the assumption that the neutrino masses are the
only source of lepton-number violation (as far as neutrinoless
double beta decay is concerned). Second, only one or a combination of the two
independent Majorana phases can be determined in this way. It is fair to
say that there is no realistic measurement one can look forward to making
in the near future that will add any information and help us disentangle
the ``other" Majorana phase. Third, it is curious to note that the effect
the Majorana phases have on the rate for neutrinoless double beta decay
is CP even \cite{GKM}. While Majorana phases can mediate CP violating
phenomena \cite{GKM}, it seems unlikely that any of them can be
realistically studied experimentally in the foreseeable future. For
further discussion of CP violation among neutrinos see Ref.\ \cite{jenkins}.

In spite of all the uncertainty due to our inability to measure Majorana
phases, it is fair to say that we expect to correctly reconstruct several
features of the neutrino mass matrix \cite{reconstruction}, especially if
the overall mass scale and the neutrino mass hierarchy are determined
experimentally.
What do we hope to accomplish by reconstructing the neutrino mass
matrix? The answer is that we wish to uncover whether there are new
fundamental organizing principles responsible for explaining in a more
satisfying way the values of the neutrino masses and the leptonic mixing
angles. In other words, we would like to establish whether there is a
fundamental reason behind the fact that the $\nu_3$ state contains almost
the same amount of $\nu_{\mu}$
and $\nu_{\tau}$, while at the same time containing a relatively small
amount of $\nu_e$. Are there flavor (or family) symmetries, capable of
dynamically distinguishing the different generations of quarks and
leptons and, we hope, explaining why there are three quasi-identical
particles for each matter field?

In the neutrino sector, we are only getting started. We have, for
example, identified several textures for the neutrino mass matrix that
lead to the currently observed mass-squared differences and mixing
angles, and have identified some of the measurements that will allow us
to identify which textures best describe Nature. As has been already
pointed out, it is not clear whether this is the best avenue to pursue as
far as identifying whether there is a deep explanation for the patterns
we observe in experiments. For example, it may turn out that we have made
a weak-basis choice that renders the job more complicated (it is possible
that the mixing angles are ``determined by the charged lepton
sector" \cite{fpr,charged_lepton,Antusch:2004re}), or that all the structure contained
in the neutrino sector is obscured after heavy degrees of freedom are
integrated out (as may happen in type-I seesaw models). Nonetheless, we
will discuss a few of these textures in order to exemplify some of the
measurements (and how precise they should be) that will shed a
significant amount of light in the issue of interpreting neutrino masses
and mixing angles.

Arguably the simplest assumption one can make is that there is no
symmetry or dynamical principle that explains why leptonic mixing angles
are large \cite{anarchy}. This ``flavorless" neutrino flavor model is
often referred to as ``neutrino mass anarchy" and is, currently,
compatible with data \cite{anarchy,anarchy_stat}. Curiously, the
anarchical hypothesis is not without predictions: it requires that the
unobserved magnitude of the $U_{e3}$ element of the leptonic mixing
matrix is $|U_{e3}|^2>0.01$ at the 95\% confidence level (see
\cite{anarchy_stat} for details and a proper definition of this
bound). This means that after the next-generation of reactor and/or
long-baseline experiments analyze their data we will know whether we can
afford a ``random" leptonic mixing matrix or not. It should be noted that
this model applies only for the leptonic mixing matrix --- it has nothing
specific to say about the order of magnitude of neutral and charged
lepton masses, or their hierarchies.

If one assumes that there is a nontrivial texture to the neutrino
mass matrix, and that this texture ``explains" the observed values of the
mixing parameters, there are several completely different options. Some
are tabulated in Table~\ref{texturetable}, and will be discussed
briefly. Before proceeding, however, it is important to explain how
these textures should be interpreted. The hypothesis is that, at leading
order, the neutrino mass matrix can be parametrized by far fewer than
the usual six complex coefficients. These are chosen in such a way that
the dominant features of neutrino masses and mixings are explained. These
are: (i) $m^2_3$ is either much larger or much smaller than
$m_1^2,m_2^2$. This splitting determines the atmospheric mass-squared
difference. (ii) the $\nu_e$ content of the $\nu_3$ state is
zero. (iii) the $\nu_{\mu}$ and $\nu_{\tau}$ contents of the $\nu_3$
state are similar (or, perhaps, identical). In order to accommodate the
other observed features (like $|U_{e3}|$, the solar mass-squared difference and the
solar angle) one includes sub-leading effects that violate the
leading-order structure. The structure of these sub-leading effects
determines the ``predictions" for the observables that are not determined
by the leading order mass-texture. In Table~\ref{texturetable}, we list
the predictions obtained in the case of a structureless sub-leading
mass matrix, {\it i.e.,}\/ one proportional to the anarchical texture
\cite{theta_23_andre}. In the case of a structured sub-leading mass
matrix, expectations may vary significantly from these quoted in
Table~\ref{texturetable}.
\renewcommand\arraystretch{0.8}
\begin{table}
\setlength{\tabcolsep}{0.5pt}
\centering
\begin{tabular}{|c|c|c|c|c|c|c|} \hline
Case & Texture & $\,$Hierarchy$\,$ & $|U_{e3}|$ &
$\begin{array}{c} |\cos2\theta_{23}|\\ \hbox{(n.s.)} \end{array}$ &
$\,|\cos2\theta_{23}|$ & $\,$Solar Angle$\,$ \\ \hline
A & $\frac{\sqrt{\Delta m^2_{13}}}{2} \left(\begin{array}{ccc}0&0&0\\
0&1&1 \\ 0&1&1\end{array}\right)$& Normal &$\sqrt{\frac{\Delta
m^2_{12}}{\Delta m^2_{13}}}$ & ${\cal{O}}(1)$ & $\sqrt{\frac{\Delta
m^2_{12}}{\Delta m^2_{13}}}$ & ${\cal{O}}(1)$ \\ \hline
B &  $\sqrt{\Delta m^2_{13}}\left(\begin{array}{ccc}1&0&0\\
0&\frac{1}{2}&-\frac{1}{2} \\
0&-\frac{1}{2}&\frac{1}{2}\end{array}\right)$& Inverted & $\frac{\Delta
m^2_{12}}{|\Delta m^2_{13}|}$ & -- & $\frac{\Delta m^2_{12}}{|\Delta
m^2_{13}|}$ & ${\cal{O}}(1)$ \\ \hline
C &  $\frac{\sqrt{\Delta m^2_{13}}}{\sqrt{2}}
\left(\begin{array}{ccc}0&1&1\\ 1&0&0\\ 1&0&0\end{array}\right)$ &
Inverted & $\frac{\Delta m^2_{12}}{|\Delta m^2_{13}|}$ & ${\cal{O}}(1)$ &
$\frac{\Delta m^2_{12}}{|\Delta m^2_{13}|}$ &
$\begin{array}{c}
 |\cos2\theta_{12}| \\ \sim\frac{\Delta m^2_{12}}{|\Delta m^2_{13}|}
\end{array}$
\\ \hline
Anarchy & $\sqrt{\Delta m^2_{13}}\left(\begin{array}{ccc}1&1&1\\ 1&1&1 \\
1&1&1\end{array}\right)$ & Normal
& $>0.1$ & ${\cal{O}}(1)$ & -- & ${\cal{O}}(1)$ \\ \hline
\end{tabular}
\caption{Different leading-order neutrino mass textures and their
``predictions" for various observables. The fifth column indicates the
``prediction" for $|\cos2\theta_{23}|$ when there is no symmetry relating
the different order one entries of the leading-order texture (`n.s.'
stands for `no structure', meaning that the entries of the matrices in
the second column should all be multiplied by and order one coefficient),
while the sixth column indicates the ``prediction" for
$|\cos2\theta_{23}|$ when the coefficients of the leading order texture
are indeed related as prescribed by the matrix contained in the second
column. See text for details. One may argue that the
anarchical texture prefers but does not require a normal mass hierarchy.}
\label{texturetable}
\end{table}

Case A is characterized by large entries in the ``$\mu-\tau$''
sub-matrix, and small entries in the ``$e$'' column and row. The
determinant of the ``$\mu-\tau$'' sub-matrix is constrained to be small
in order to guarantee a hierarchy between the two independent
mass-squared differences. The hierarchy of the neutrino masses is
predicted to be normal ($m_3^2>m_2^2>m_1^2$). Maximal atmospheric mixing
can be imposed at the leading order by requiring that the
``$\mu-\tau$" sub-matrix is democratic. The introduction of sub-leading
effects leads to a ``large" $|U_{e3}|$ and $\cos2\theta_{23}$, of order
the square-root of the ratio of mass-squared differences, which is
${\cal{O}}(0.1)$. If this texture is indeed realized in nature, we expect to
observe a nonzero $|U_{e3}|$ and a deviation of the atmospheric mixing
from maximal at next-generation experiments. It may prove difficult to
distinguish between case A and the anarchical texture via neutrino
oscillation measurements alone. One potential discriminant seems to be
the expected rate for neutrinoless double beta decay.

Case B is characterized by small ``$e-\mu$" and ``$e-\tau$" entries, a
small determinant of the ``$\mu-\tau$''  submatrix, and the constraint
that the trace of $m_{\nu}$ is  close to $2m_{\nu}^{ee}$. In this case,
one predicts an inverted mass hierarchy ($m_2^2>m_1^2\gg m_3^2$), and
both $|U_{e3}|$ and $\cos2\theta_{23}$ are constrained to be of order the
ratio of the mass-squared differences $({\cal{O}}(0.01))$. The system is
constrained enough that it is hard to obtain a much larger deviation of
the atmospheric angle from maximal or a much larger $|U_{e3}|$, while
much smaller ones are, of course, obtainable if the sub-leading
contributions are structured. If case B is indeed realized in Nature,
there is a good chance that no $|U_{e3}|$ effects will be observed in
next-generation oscillation experiments, while precise measurements of
the atmospheric mixing angle will remain consistent with
$\theta_{23}=\pi/4$ (equivalently, if a large deviation of the
atmospheric angle is detected this texture will be ruled out). On a more
positive note, one should expect a ``large" rate for neutrinoless double
beta decay ($m_{\nu}^{ee}\sim\sqrt{\Delta m^2_{13}}$). A texture which is
na\"{\i}vely similar to case B is to change the sign of the
``$\mu-\tau$'' sub-matrix, such that the trace of the leading order mass
matrix is close to zero. This case, however, is disfavored by solar data,
as the solar angle is constrained to be too small (for a more detailed
discussion see, for example, \cite{theta_23_andre}).

Case C is characterized by ``$e-\mu$" and ``$e-\tau$" entries which are
much larger than all the other ones (set to zero at leading order)
and thus corresponds to the case of approximate
$L_e - L_{\mu} - L_{\tau}$ symmetry \cite{inverted}. It
leads to an inverted mass hierarchy, and a close to bi-maximal
\cite{bi-maximal} leading order mixing matrix. The solar angle is (at
leading order) exactly maximal, while the atmospheric angle is
generically large, becoming maximal in the limit when $m_{\nu}^{e\mu}=\pm
m_{\nu}^{e\tau}$ (for real $m_{\nu}^{e\alpha}$). Sub-leading corrections
to case C which are responsible for splitting the two heavy leading-order
mass eigenstates will induce a $|U_{e3}|$, $\cos2\theta_{23}$ and
$\cos2\theta_{12}$ of order the ratio of the mass-squared differences
$({\cal{O}}(0.01))$, or smaller.
Hence, similar to case B, it seems unlikely that
$U_{e3}$-effects will be measured at next-generation experiments. This
scenario is currently disfavored, as it also predicts a solar angle
$\theta_{12}$ very close to $\pi/4$ \cite{lms,fpr,theta_23_andre}. One should
not conclude, however, that scenarios based on ``perturbations" around
bi-maximal mixing are ruled out. A related issue has been discussed in
detail recently by the authors of \cite{fpr,charged_lepton}.

A realistic three generation extension of the mass matrix in case (A) that
leads to large but not maximal solar neutrino mixing is given by
\begin{eqnarray}
M_\nu~=~\frac{\sqrt{\Delta
m^2_A}}{2}\left(\begin{array}{ccc}c\epsilon^n
&b\epsilon &d\epsilon\\ b\epsilon & 1+c\epsilon & -1 \\
d\epsilon & -1 & 1+\epsilon\end{array}\right)
\end{eqnarray}
Note that if $b=d$ and $c=1$, the atmospheric neutrino mixing is maximal
and the mixing parameter $\theta_{13}=0$ \cite{mutau}. In this case 
the mass matrix 
has $\mu-\tau$ interchange symmetry. Depending on how this symmetry is
broken, the parameter $\theta_{13}$ is either of order $\sqrt{\frac{\Delta
m^2_{\odot}}{\Delta m^2_A}}$ (for $c\neq 1$) or $\frac{\Delta
m^2_{\odot}}{\Delta m^2_A}$ (for $b\neq d$) \cite{rabianjan}. Therefore
search for $\theta_{13}$ down to the level of $0.01$ will be a big help in
determining the structure of the neutrino mass matrix for the case of
normal hierarchy. In both these cases, there is an important correlation
between $\theta_{13}$ and $\theta_A-\frac{\pi}{4}$ \cite{rabianjan}.


\begin{figure}[h]
\begin{center}
\psfig{file=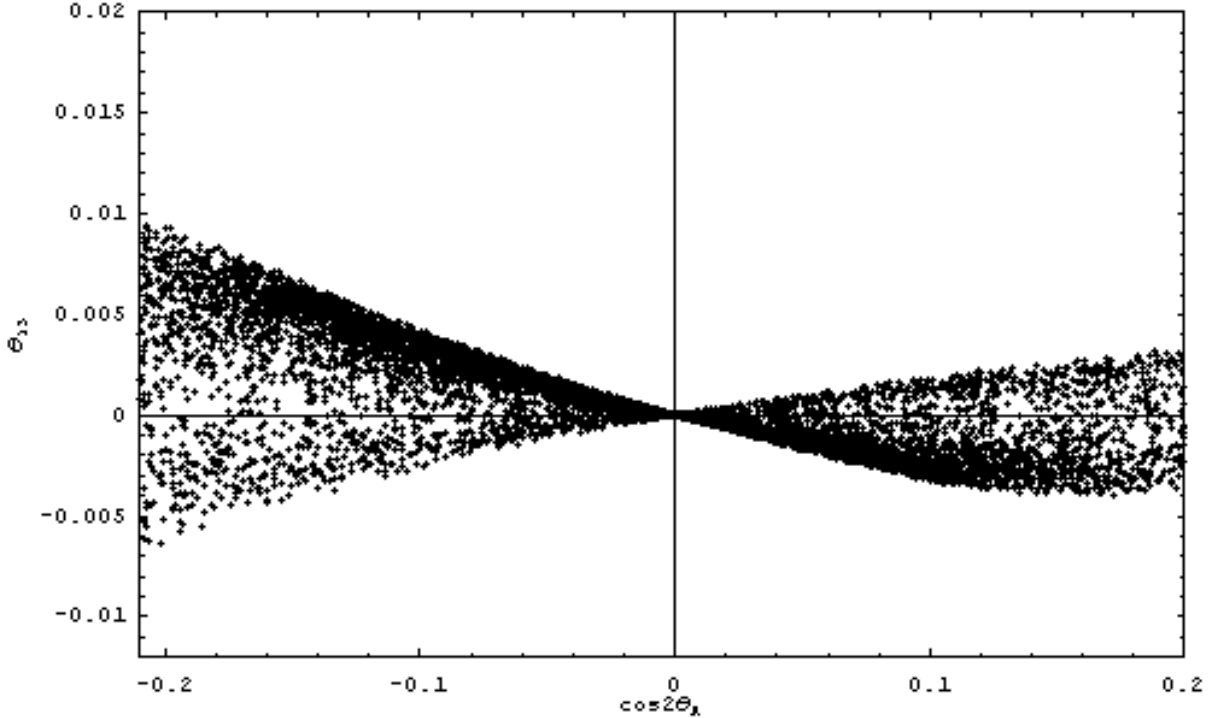,width=1\textwidth}
\caption{Departure from $\mu-\tau$ symmetry and correlation
between $\theta_{13}$ and $\theta_A$ (Case (i)): opposite to case
(ii) (figure by H. Yu) }
\end{center}
\end{figure}

In general, a neutrino mixing matrix originating from a $\mu-\tau$ 
symmetric mass matrix has the following structure 
(for simplicity, we did not include here the Majorana phases)
\begin{eqnarray} 
U = 
\left(\begin{array}{ccc}
cos \theta_{12} &  sin \theta_{12} &  0 \\[0.2cm]
-\frac{\sin \theta_{12}}{\sqrt{2}} &  \frac{\cos \theta_{12}}{\sqrt{2} }
& \frac{1}{\sqrt{2}} \\[0.2cm]
\frac{ \sin \theta_{12}}{\sqrt{2}} &  -\frac{ \cos \theta_{12}}{\sqrt{2}}
& \frac{1}{\sqrt{2}} \\[0.3cm]
\end{array}
\right)~. 
\end{eqnarray}
Note that the mass spectrum of the 
neutrinos is not predicted by the $\mu-\tau$ symmetry. Depending on the 
value of $\theta_{12}$, several interesting mixing schemes can arise: 
if $\theta_{12} = \pi/4$ then we have bi--maximal mixing. 
However, as mentioned, the observed deviation from $\pi/4$ is rather large. 

Much closer to current data is the so--called tri--bimaximal mixing scheme 
\cite{tbm0}, corresponding to 
$\sin^2 \theta_{12} = 1/3$ and leading to the following often--studied 
mixing matrix: 
\begin{eqnarray} 
U = \left(
\begin{array}{ccc}
\sqrt{\frac{2}{3}} & \sqrt{\frac{1}{3}} & 0 \\[0.2cm]
-\sqrt{\frac{1}{6}} & \sqrt{\frac{1}{3}} & \sqrt{\frac{1}{2}}  \\[0.2cm]
\sqrt{\frac{1}{6}} & -\sqrt{\frac{1}{3}} & \sqrt{\frac{1}{2}}  
\end{array} 
\right)~. 
\end{eqnarray}
Models which give rise to such a matrix (for some recent 
attempts see \cite{tbm1}) 
are typically quite intricate and not as straightforward to construct 
as models leading to bi--maximal mixing.

There are other viable neutrino mass textures, including some that lead
to degenerate neutrino masses. We refer readers to the literature for a
more thorough discussion (\cite{barger, king} and references therein).
The point
we wish to emphasize here is that the amount of information we have
concerning neutrino masses and leptonic mixing is still very
limited. This is reflected in the fact that too many different hypothesis
can be raised in order to ``explain" the same set of observables. The
situation is bound to change in the near future, and there is hope that
the data will ``select" one specific neutrino mass matrix. Our job will
then be to interpret what Nature is trying to ``say" through $m_{\nu}$. A
more accurate determination of a few observables will already shine a
significant amount of light in the currently obscure picture we are
trying to obtain: (i) what is the neutrino mass hierarchy? (ii) is
$|U_{e3}|$ larger than 0.1? (iii) is $|\cos2\theta_{23}|>0.1$? All three
of these can be answered in a next-generation $\nu_{\mu}\to\nu_{e}$
long-baseline accelerator experiment, while the second one can be
addressed by a next-generation reactor neutrino experiment (with a
baseline of ${\cal{O}}$(1~km)). Of course,
in order to be sure we are on the right
track we also need to (iv) determine that lepton number is not a conserved
symmetry.

\newpage
\section{$\boldsymbol{\beta\beta_{0\nu}}$-Decay and CP Violation}
\label{sec:0nubbCP}
\subsection[$\beta\beta_{0\nu}$ decay]{$\boldsymbol{\beta\beta_{0\nu}}$ decay}

In this section, we focus our attention on what we can learn from
neutrinoless double beta decay experiments. As already alluded to in the
introduction, in a given theory, neutrinoless double beta decay can arise
from two sources: (i) neutrino Majorana mass or/and (ii) lepton number
violating interactions. While the absence of a signal in a
$\beta\beta_{0\nu}$ experiment will constrain both sources (and associated
theories), a positive signal cannot necessarily be considered as evidence
for one or the other exclusively. For instance, one must supplement the
results from $\beta\beta_{0\nu}$ experiment with collider experiments such
as from LHC or a possible $e^+e^-$ type to get a definitive information
regarding the source. Alternatively, one may conduct the double beta decay
experiment with different nuclei and if the matrix elements happen to
differ substantially for the two sources, one may be able to disentangle
the different sources. Therefore as we interpret any positive signal for
$\beta\beta_{0\nu}$ decay one must keep this in mind. Below, we discuss
what we can learn about neutrino masses and mixings, once it is
established that the source of the positive signal is the Majorana mass
for the neutrino.
The experiments with solar and atmospheric neutrinos and with reactor
antineutrinos have provided data on $\theta_{12}$, $\theta_{23}$ and
$\theta_{13}$, and on the neutrino mass squared differences driving the solar
and atmospheric neutrino oscillations, $\Delta m^2_{12}$ and $\Delta m^2_{13}$.
Future oscillation experiments will improve considerably the precision on these
basic parameters.  However, these experiments are insensitive to the nature of
massive neutrinos $\nu_j$ which can be Dirac or Majorana particles
\cite{BHP80,Lang87} (see also, e.g., \cite{BiPet87}).
They cannot give information on the absolute scale of
neutrino masses (i.e., on the value of $m_1$), and on the two Majorana CP
violation phases --- the latter do not enter into the expressions for the
probabilities of flavor neutrino oscillations \cite{BHP80}.
Determining the nature of massive neutrinos
and obtaining information on the absolute neutrino
mass scale is one of the fundamental problems in the studies of neutrino
mixing.

   Neutrinos are predicted to be Majorana particles in the
seesaw model of neutrino mass generation.
This model gives a natural explanation of the smallness
of the neutrino masses and, through the leptogenesis theory,
provides an explanation of the observed baryon
asymmetry in the Universe, which thus is linked
to the existence of neutrino mixing.
The only experiments which have the potential of establishing the
Majorana nature of massive neutrinos
are the neutrinoless double beta decay experiments
searching for the process
$(A,Z) \rightarrow (A,Z+2) + e^- + e^-$
(see, e.g., \cite{BiPet87,ElliotVogel02}).
The observation of the $\beta\beta_{0\nu}$-decay
and the measurement of the corresponding $\beta\beta_{0\nu}$-decay
rate with a sufficient accuracy,
would not only be a proof that the total
lepton charge is not conserved
in nature, but might provide also a unique information
on i) the type of the neutrino mass spectrum,
ii) the absolute scale of neutrino masses, and
on iii) the values of the Majorana CP violation phases.
Let us add that in supersymmetric
theories with seesaw mechanism
of neutrino mass generation, the rates of
lepton flavor violating decays
$\mu \rightarrow e + \gamma$,
$\tau \rightarrow \mu + \gamma$ can be interestingly large (e.g., 
\cite{ihl}) and 
may depend on the Majorana CP violating phases in the lepton
mixing matrix (see, e.g., \cite{RaidalJE,PPY}). Furthermore, the values
of the Majorana phases can be important for the stability under
RGE running of the neutrino mass and mixing parameters,
see Sec.\ \ref{sec:RGE}.

   Let us recall that the SK atmospheric neutrino and K2K data
do not allow one to determine the sign of
$\Delta m^2_{\rm A}$. This implies that if we identify
$\Delta m^2_{\rm A}$ with $\Delta m^2_{13}$
in the case of 3-neutrino mixing,
one can have $\Delta m^2_{13} > 0$
or $\Delta m^2_{13} < 0$. The two
possibilities correspond to two different
types of neutrino mass spectrum:
with normal hierarchy, $m_1 < m_2 < m_3$, and
with inverted hierarchy,
$m_3 < m_1 < m_2$.
In the case of strong
inequalities between the masses,
the spectra are called {\it normal hierarchical} (NH)
and {\it inverted hierarchical} (IH).
The NH and IH spectra correspond to
 $m_1 \ll 0.02$ eV and $m_3 \ll 0.02$ eV, respectively.
If $m_1 \cong m_2 \cong m_3 \cong m_0$ and
$m_j^2 \gg |\Delta m^2_{\rm A}|,\Delta m^2_\odot$,
the spectrum is {\it quasi-degenerate} (QD).
The QD spectrum is realized if
$m_{1,2,3} > 0.20$ eV roughly requiring that the largest mass
difference is about 10\% of the common mass.

Under the assumptions of (1) 3-neutrino mixing,
for which we have compelling evidence
from the experiments with
solar and atmospheric neutrinos and from the KamLAND
experiment, (2) massive neutrinos $\nu_j$ being
Majorana particles, and
(3) $\beta\beta_{0\nu}$-decay generated
{\it only by the (V-A) charged current
weak interaction via the exchange of the three
Majorana neutrinos  $\nu_j$},
the effective Majorana mass
in $\beta\beta_{0\nu}$-decay
of interest is given by
(see, e.g., \cite{BiPet87,BPP1}):
\begin{equation}
\langle m \rangle_{eff}  = \left| m_1 |U_{e 1}|^2
+ m_2 |U_{e 2}|^2~e^{2i\phi_{1}}
 + m_3 |U_{e 3}|^2~e^{2i\phi_2} \right|~,
\label{effmass2}
\end{equation}
\noindent where
$U_{ej}$, $j=1,2,3$, are the elements of the
first row of the lepton mixing matrix $U$,
$m_j > 0$ is the mass of the Majorana neutrino $\nu_j$,
and $\phi_{1}$ and $\phi_{2}$
are the two Majorana CP violating phases
\cite{Valle,Lang87}.
In the case of CP conservation we have
$e^{2i\phi_{1,2}} \equiv \eta_{21(31)}= \pm 1$,
$\eta_{ij}$ being the relative CP parity
of the neutrinos $\nu_i$ and $\nu_j$.

\begin{figure}[!h]
\centerline{
\includegraphics*[height=12.8cm,width=8cm]{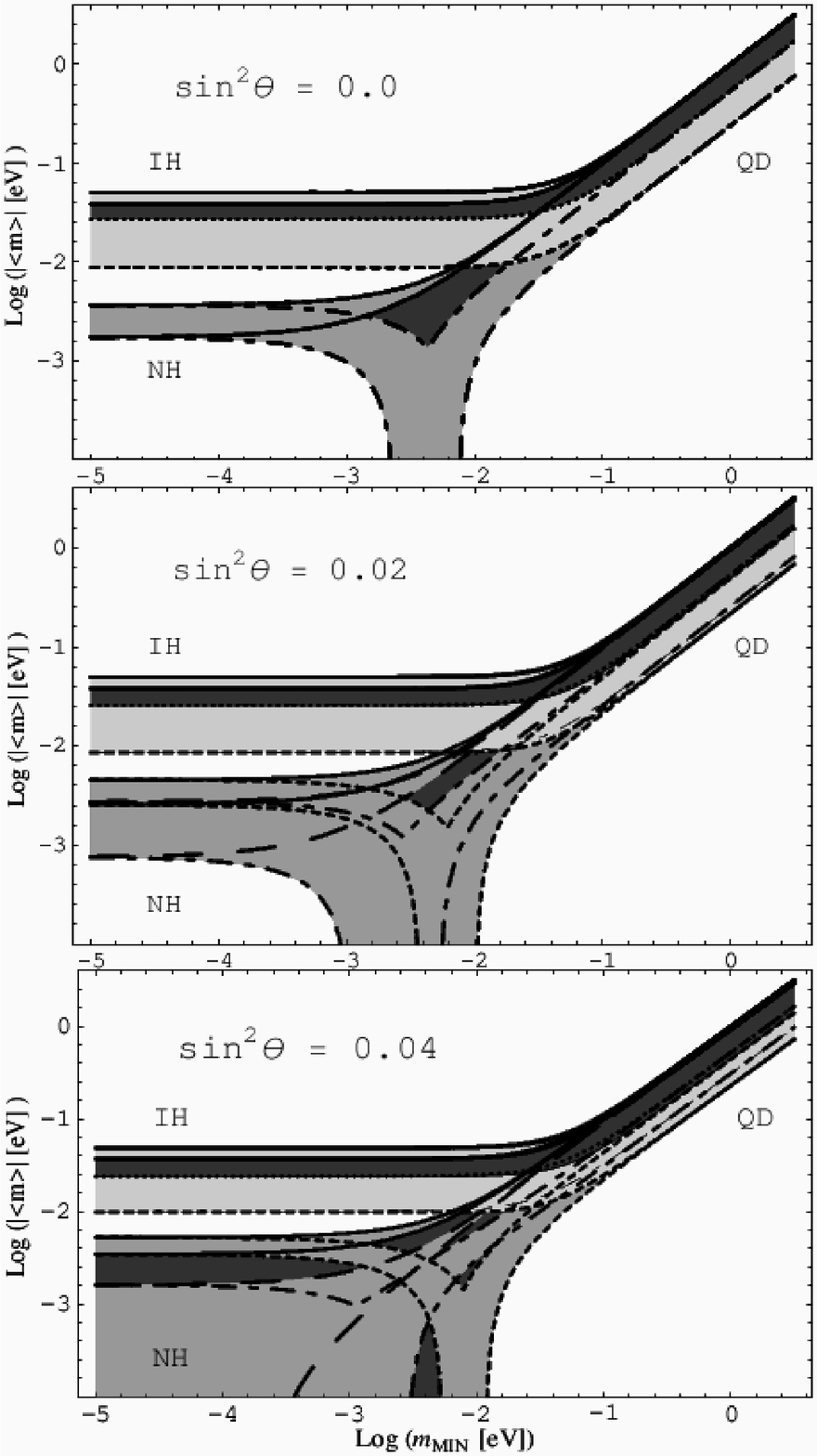}
}
\vspace{-3mm}
\caption{
The dependence of $\langle m\rangle_{eff}$ on $m_{min}$
in the case of the LMA-I solution,
for   normal and inverted hierarchy, and
for the $90 \%$~C.L. allowed regions
of $\Delta m^2_{\odot}$ and $\sin^2 \theta_\odot$ found in
Ref.~\protect\cite{SNO3BCGPR}
and of $\Delta m^2_{\rm A}$ in Ref.~\protect\cite{Fogliatm0308055},
and a fixed value of $\sin^2 \theta = 0.0 (0.02) [0.04]$
in the upper (middle) [lower] panel.
In the case of CP conservation,
the allowed values of $\langle m\rangle_{eff}$ are constrained
to lie: for i) normal hierarchy
and the middle and lower panels (upper panel) -
in the medium-gray and light-gray regions
{\it a)} between the two lower  thick solid lines
(between the two lower  thick solid lines) if
$\eta_{21} = \eta_{31} = 1$,
{\it b)}
between the two  long-dashed lines
(between the two lower  thick solid lines) if
$\eta_{21} = - \eta_{31} = 1$,
{\it c)}
between the three thick  dash-dotted lines and the axes
(between the dash-dotted lines and the axes)
if $\eta_{21} = - \eta_{31} = - 1$,
{\it d)}
between the three thick  short-dashed lines and the axes
(between the dash-dotted lines and the axes)
if $\eta_{21} = \eta_{31} = - 1$;
and for ii) inverted hierarchy
and the middle  and lower panels  (upper) -
in the  light-gray regions
{\it a)}
between the two upper  thick solid lines
(between the two upper  thick solid lines) if
$\eta_{32} = \eta_{31} = \pm 1$,
{\it b)}
between the dotted  and
the thin dash-dotted lines
(between the dotted  and
the thick short-dashed lines)
if $\eta_{32} = - \eta_{31} =  1$,
{\it c)}
between the dotted  and
the upper thick short-dashed lines
(between the dotted  and
the thick short-dashed lines)
if $\eta_{32} = - \eta_{31} = - 1$.
In the case of CP violation, the allowed regions
for $\langle m\rangle_{eff}$ cover all the gray regions.
Values of $\langle m\rangle_{eff}$ in the dark gray regions
signal CP violation.(From Ref.~\protect\cite{PPaddendum}).
}
\label{figmeff2}
\end{figure}

  One can express \cite{SPAS94} two of the three
neutrino masses, say, $m_{2,3}$, in terms of the third mass,
$m_1$, and of $\Delta m^2_\odot$ and $\Delta m^2_{\rm A}$, while
the elements $|U_{ej}|$ can be expressed in terms of
$\theta_\odot$ and $\theta$ (a concise discussion of the relevant
formalism can be found, e.g., in
Refs.~\cite{BPP1,fsv,PPVenice03}). Within the convention employed
in the present study in both cases of neutrino mass spectrum with
normal and inverted hierarchy one has: $\Delta m^2_\odot = \Delta
m_{12}^2 > 0$, and $m_2 = \sqrt{m_1^2 + \Delta m^2_\odot}$. In the
case of normal hierarchy, $\Delta m^2_{\rm A} = \Delta m_{13}^2 >
0$ and $m_3 = \sqrt{m_1^2 + \Delta m^2_{\rm A}}$, while if the
spectrum is with inverted hierarchy, $\Delta m^2_{\rm A} = \Delta
m_{23}^2 > 0$ and thus $m_1 = \sqrt{m_3^2 + \Delta m^2_{\rm A}-
\Delta m^2_\odot}$. For both types of hierarchy, the following
relations hold: $|U_{\mathrm{e} 1}|^2 = \cos^2\theta_{\odot} (1 -
|U_{\mathrm{e} 3}|^2)$, $|U_{\mathrm{e} 2}|^2 =
\sin^2\theta_{\odot} (1 - |U_{\mathrm{e} 3}|^2)$, and
$|U_{\mathrm{e} 3}|^2 \equiv \sin^2\theta$. We denote the smallest
neutrino mass as $m_{min}$ and we have $m_{min}=m_{1 \,(3)}$ for
the case of normal (inverted) hierarchy.
\begin{table}[ht]
\centering
\begin{tabular}{|c|c|c|c|c|}
\hline
\rule{0pt}{0.5cm} $\sin^2 \theta$
& ${\langle m\rangle_{eff}}_{\rm max}^{\rm NH}$ &
${\langle m\rangle_{eff}}_{\rm min}^{\rm IH}$ &
${\langle m\rangle_{eff}}_{\rm max}^{\rm IH}$ &
${\langle m\rangle_{eff}}_{\rm min}^{\rm QD} $
\\ \hline \hline
0.0   & 2.6 (2.6)   &  19.9 (17.3)    &  50.5 (44.2)   & 79.9 \\ \hline
 0.02 & 3.6 (3.5)   &  19.5 (17.0)    &  49.5 (43.3)   & 74.2 \\ \hline
 0.04 & 4.6 (4.3)   &  19.1 (16.6)    &  48.5 (42.4)   & 68.5 \\ \hline
\end{tabular}
\caption{ \label{tabmeff1}
The maximal values of
$\langle m\rangle_{eff}${} (in units of meV)
for the NH  and IH spectra, and the minimal values of
$\langle m\rangle_{eff}$ (in units of  meV) for the IH and QD spectra,
for the best fit values of the oscillation parameters and
$\sin^2\theta = 0.0$, $0.02$ and
$0.04$.
The results for the NH and IH spectra
are obtained for
$|\Delta m^2_{\rm A}| = 2.6 \times 10^{-3} \ \mathrm{eV}^2~
(2.0 \times 10^{-3} \ \mathrm{eV}^2 -~{\rm values~in~brackets})$ and
$m_1 = 10^{-4}$ eV, while
those for the QD spectrum correspond to
$m_0 = 0.2$ eV. (From Ref.~\protect\cite{PPaddendum}).
}
\end{table}
\begin{table}[ht]
\centering
\begin{tabular}{|c|c|c|c|c|}
\hline
\rule{0pt}{0.5cm} $\sin^2 \theta$
& ${\langle m\rangle_{eff}}_{\rm max}^{\rm NH}$ &
${\langle m\rangle_{eff}}_{\rm min}^{\rm IH}$ &
${\langle m\rangle_{eff}}_{\rm max}^{\rm IH}$ &
${\langle m\rangle_{eff}}_{\rm min}^{\rm QD} $
\\ \hline \hline
0.0  & 3.7 (3.7) & 10.1 (8.7) & 56.3 (50.6) & 47.9 \\ \hline
0.02 & 4.7 (4.6) &  9.9 (8.6) & 55.1 (49.6) & 42.8 \\ \hline
0.04 & 5.5 (5.3) & 11.4 (9.9) & 54.0 (48.6) & 45.4 \\ \hline
\end{tabular}
\caption{\label{tabmeff2}
The same as in Table~\ref{tabmeff1} but
for the 90\% C.L.\ allowed regions of
$\Delta m^2_\odot$ and $\theta_\odot$ obtained in \protect\cite{SNO3BCGPR},
and of $\Delta m^2_{\rm A}$ obtained in
Ref.\ \protect\cite{FogliatmKamL}
(Ref.~\protect\cite{Fogliatm0308055} --- results in brackets).
(From Ref.~\protect\cite{PPaddendum}).}
\end{table}
  The problem of obtaining the allowed values of $\langle m\rangle_{eff}$
given the constraints on the relevant parameters
following from the neutrino oscillation data
has been first studied in~Ref.~\cite{SPAS94}
and subsequently in a large number of papers, see, e.g.,
\cite{BGKP96,BGGKP99,BPP1,bbpapers1,bbpapers2,bbpapers3}.
Detailed analyses were performed, e.g.,
in Refs.~\cite{BPP1,noMajresp,PPVenice03,Lisi},
and in particular in Ref.~\cite{PPaddendum},
where the allowed values of $\Delta m^2_{\rm A}$,
$\Delta m^2_\odot$, $\theta_\odot$ and $\theta$, obtained
from the most recent neutrino oscillation
data, were used. The results are summarized
in Tabs.\ \ref{tabmeff1} and \ref{tabmeff2}, and
in 
Fig. \ref{figmeff2}.

In Fig.
 ~\ref{figmeff2}
(taken from Ref.~\cite{PPaddendum})
we show the allowed ranges of values of $\langle m\rangle_{eff}$
as a function of
$m_{min}$ for the cases of NH and IH spectrum
The predictions for $\langle m\rangle_{eff}$
are obtained  by using 
 the allowed
at 90\%~C.L. (Fig.~\ref{figmeff2}), values of
$\Delta m^2_\odot$, $\theta_\odot$ and $\Delta m^2_{\rm A}$
from Refs.~\cite{SNO3BCGPR} and
\cite{Fogliatm0308055}
and for three fixed values of $\sin^2 \theta$.

  The existence of significant lower bounds on $\langle m\rangle_{eff}$
in the cases of IH and QD spectra \cite{PPSNO2bb},
$\langle m\rangle_{eff}^{\rm{IH}} \geq 10 \,\mathrm{meV}$ and
$\langle m\rangle_{eff}^{\rm{IH}} \geq 43 \,\mathrm{meV}$, respectively
\cite{PPaddendum},
which lie either partially (IH spectrum) or completely
(QD spectrum) within the range of sensitivity of
the next generation of $\beta\beta_{0\nu}$-decay experiments,
is one of the most important features of
the predictions of $\langle m\rangle_{eff}$.
The indicated minimal values are
given, up to small corrections, by
$\Delta m^2_{\rm A} \cos2\theta_{\odot}$ and
$m_0 \cos2\theta_{\odot}$. According to the most
recent combined analyses of the solar
and reactor neutrino data,
including the latest SNO and KamLAND results
(see, e.g., \cite{BCGPRKL2})
i) the possibility of  $\cos2\theta_{\odot} = 0$
is excluded at more than 6 s.d.,
ii) the best fit value of
$\cos2\theta_{\odot}$ is
$\cos2\theta_{\odot} = 0.40$, and
iii) at 95\% C.L.\ one has
for $\sin^2\theta = 0~(0.02)$,
$\cos{2\theta_{\odot}} \geq 0.27~(0.32)$.
The quoted results on $\cos{2\theta_{\odot}}$
together with the range of possible values
of $\Delta m^2_{\rm A}$ and $m_0$, lead to the
conclusion about the existence
of significant and robust lower
bounds on $\langle m\rangle_{eff}$ in the cases of
IH and QD spectrum \cite{PPaddendum,Carlosbb03}.
At the same time
one can {\it always} have $\langle m\rangle_{eff} = 0$
in the case of neutrino
mass spectrum with normal hierarchy \cite{PPW}.
It follows from Tabs.\ \ref{tabmeff1} and \ref{tabmeff2}
that in this case
$\langle m\rangle_{eff}$ cannot exceed $5.5$ meV.
This implies that the maximal value
of $\langle m\rangle_{eff}$ in the case of neutrino mass
spectrum with normal hierarchy
is considerably smaller than
the minimal values of $\langle m\rangle_{eff}$
for the inverted hierarchy and
quasi-degenerate neutrino mass spectrum.
This opens the possibility of obtaining
information about the type of neutrino
mass spectrum from a measurement of
$\langle m\rangle_{eff} \neq 0$, or from obtaining a
sufficiently stringent upper bound on $\langle m\rangle_{eff}$.
In particular, a positive result in the future generation
of $\beta\beta_{0\nu}$-decay experiments with
$\langle m\rangle_{eff} > 10 \ $meV
would imply that the NH spectrum is excluded.
The uncertainty in the relevant nuclear matrix elements\footnote{Recently,
encouraging results
in what regards the problem of the
calculation of the nuclear matrix elements have been
obtained \cite{FesSimVogel03}.}
and prospective experimental errors
in  the values of the oscillation parameters,
in $\langle m\rangle_{eff}$, and for the case of QD spectrum --- in $m_0$,
weaken, but do not invalidate, the reported results
(see, e.g., Refs.~\cite{PPRSNO2bb}).
If the neutrino mass spectrum turned out
to be of the QD type, a measurement of
$\langle m\rangle_{eff}$ in $\beta\beta_{0\nu}$-decay experiment and
of $m_0$ in the KATRIN experiment \cite{Osipowicz:2001sq}
could be used, in particular, to check the validity of
the light Majorana neutrino exchange mechanism
for the $\beta\beta_{0\nu}$-decay and search for indications
for contributions from other types of mechanisms
(see, e.g., \cite{moh1,bb0nunmi}).

It follows from Fig.  \ref{figmeff2}
that a measurement
of $\langle m\rangle_{eff} \geq 10$ meV would either
i) determine a relatively narrow
interval of possible values of the lightest
neutrino mass $m_{min}$, or
ii) would establish an upper
limit on the possible values of $m_{min}$.
If a sufficiently stringent
upper limit on $\langle m\rangle_{eff}$ is experimentally
obtained below 100 meV,
this would lead to a significant upper limit
on the possible value of $m_{min}$.

  The possibility of establishing CP violation in the lepton sector
due to Majorana CP violating phases
has been studied in \cite{PPW,bbpapers3} and in
much greater detail in Ref.~\cite{noMajresp}.
It was found that it is very challenging\footnote{A very
pessimistic conclusion
about the prospects to establish  CP
violation in the lepton sector
due to Majorana CP violating phases
from a measurement of $\langle m\rangle_{eff}$ and, e.g., of
$m_0$, was reached in \cite{noMaj}.}:
it requires quite accurate measurements
of $\langle m\rangle_{eff}$ and of $m_1$,
and holds only for a limited range of
values of the relevant parameters.
For the IH and the QD spectra, which are
of interest, the ``just CP violation'' region~\cite{BPP1}
--- an experimental point in this region
would signal unambiguously CP violation
associated with Majorana neutrinos,
is larger for smaller values of
$\cos 2 \theta_\odot$.
More specifically,
proving that CP violation associated with
Majorana neutrinos takes place
requires, in particular, a relative
experimental error on the measured value of
$\langle m\rangle_{eff}$ not bigger than (15 -- 20)\%,
a ``theoretical uncertainty'' in the value of
$\langle m\rangle_{eff}$ due to an imprecise knowledge of the
corresponding nuclear matrix elements
smaller than a factor of 2, a value of
$\tan^2\theta_{\odot} \geq 0.55$,
and values of the relevant Majorana
CP violating phases ($2\phi_{1,2}$) typically
within the ranges of $\sim (\pi/2 - 3\pi/4)$ and
$\sim (5\pi/4 - 3\pi/2)$ \cite{noMajresp}.


\subsection{The MNSP Lepton Mixing Matrix and CP Violation in the Lepton 
Sector}


  It is well known that, in general, in gauge theories with massive neutrinos
the MNSP lepton mixing matrix results from a product of two matrices:
\begin{equation}
U = U_{\ell}^\dagger \, U_\nu~.
\label{UPMNSCPV}
\end{equation}
%
where $U_{\ell}$ and $U_\nu$ are two $3\times 3$
unitary matrices: $U_{\ell}$ arises from the
diagonalization of the
charged lepton mass matrix,
while $U_\nu$ diagonalizes the
neutrino Majorana mass term.
Any $3\times 3$  unitary matrix contains
3 moduli and 6 phases and
can be written as \cite{wir}
(see also \cite{SKing02}):
\begin{equation} \label{eq:unit}
U = e^{i \Phi} \, P \, \tilde{U} \, Q~,
\end{equation}
%
where $P \equiv {\rm diag} (1,e^{i \phi},e^{i \omega})$
and $Q \equiv {\rm diag} (1,e^{i \rho},e^{i \sigma}) $
are diagonal phase matrices having
2 phases each, and $\tilde{U}$ is a unitary
``CKM-like'' matrix containing 1 phase and 3 angles.
The charged lepton Dirac mass term,
$M_{\ell}$,
is diagonalized by a bi-unitary transformation:
\begin{equation}
M_{\ell} = U_L \, M_{\ell}^{\rm diag} \, U_R^\dagger ~,
\end{equation}
%
where $U_{L,R}$ are $3\times 3$ unitary matrices
and $M_{\ell}^{\rm diag}$ is the diagonal matrix
containing the masses of the charged leptons.
Casting $U_{L,R}$ in the form (\ref{eq:unit}), i.e.,
$U_{L,R} = e^{i \Phi_{L,R}} \, P_{L,R} \, \tilde{U}_{L,R} \, Q_{L,R}$,
we find
\begin{equation}
M_{\ell} = e^{i (\Phi_L - \Phi_R)} \, Q_L \,
\tilde{U}_L \, P_L \, M_{\ell}^{\rm diag} \,
Q_R^\dagger \, \tilde{U}_R^\dagger \, P_R^\dagger~.
\end{equation}
%
The term $P_L \, M_{\ell}^{\rm diag} \, Q_R^\dagger$ contains
only 2 relative phases, which can be
associated with the right-handed charged
lepton fields.
The three independent phases in
$e^{i (\Phi_L - \Phi_R)} \, Q_L$ can be
absorbed by a redefinition of the
left-handed charged lepton fields.
Therefore, $U_{\ell}$ is effectively given
by $\tilde{U}_L$
and contains three angles and one phase.

   The neutrino mass matrix $M_\nu$ is diagonalized via
\begin{equation}
M_\nu = U_\nu^\ast \, M_\nu^{\rm diag} \, U_\nu^\dagger ~.
\end{equation}
%
The unitary matrix $U_\nu$ can be written
in the form (\ref{eq:unit}).
It is not possible to
absorb phases in the neutrino
fields since the neutrino mass term is
of Majorana type \cite{Valle,Lang87}. Thus,
\begin{equation} \label{eq:pmnscpv}
U = U_{\ell}^\dagger \, U_\nu = e^{i \Phi \nu} \,
\tilde{U}_{\ell}^\dagger \, P_\nu \, \tilde{U}_\nu \, Q_\nu ~.
\end{equation}
%
The common phase $\Phi_\nu$ has no physical meaning and
we will ignore it.
Consequently, in the most general case,
the elements of
$U$ given by Eq.\ (\ref{eq:pmnscpv})
are expressed in terms of six real
parameters and six phases
in $\tilde{U}_{\ell}$ and $U_\nu$.
Only six combinations of those ---
the three angles and the three phases of
$U$, are observable,
in principle, at low energies.
Note that the two phases
in $Q_\nu$ are ``Majorana-like'',
since they will not appear in
the probabilities describing
the flavor neutrino oscillations
\cite{Valle,Lang87}. Note also that
if $U_{\ell} = {\mathbbm 1}$,
the phases in the matrix $P_\nu$ can be
eliminated by a redefinition of the
charged lepton fields.

  If one assumes that, e.g., $\tilde{U}_\nu$ is bimaximal,
\begin{equation}
\label{eq:Ubimax}
\tilde{U}_\nu \equiv U_{\rm bimax} =
\begin{pmatrix}
\frac{1}{\sqrt{2}} &  \frac{1}{\sqrt{2}} &  0 \cr
-\frac{1}{2} &  \frac{1}{2} &  \frac{1}{\sqrt{2}} \cr
\frac{1}{2} &  -\frac{1}{2} &  \frac{1}{\sqrt{2}}
\end{pmatrix}~,
\end{equation}
%
\noindent which permits a rather simple explanation of the
smallness of $\sin \theta_{13}$ and the
deviation of $\theta_{\odot}$
from $\pi/4$, then $\tilde{U}_\nu$ is real .
In this case the three angles
and the Dirac phase in the neutrino mixing matrix $U$
will depend in a complicated manner
on the three angles and the phase in
$\tilde{U}_{\ell}$ and on
the two phases in $P_\nu$.
The two Majorana phases will depend
in addition on the parameters in $Q_\nu$. See \cite{fpr} for details.

  It should be emphasized that the form of $U$
given in Eq.\ (\ref{eq:pmnscpv})
is the most general one.
A specific model in the framework of which
Eq.\ (\ref{eq:pmnscpv}) is obtained  might
imply symmetries or textures
both in  $m_{\ell}$ and $M_\nu$,
which will reduce the number
of independent parameters in
$U_{\ell}^\dagger$  and/or $U_\nu$.

  In the scheme with three massive
Majorana neutrinos under discussion there exist
three rephasing invariants related
to the three CP violating phases
in $U$, $\delta$ and $\phi_{1,2}$
\cite{CJ85,PKSP3nu88,JMaj87,BrancoLR86,ASBranco00}.
The first is the standard Dirac one $J_{CP}$ \cite{CJ85},
associated with the Dirac phase $\delta$:
\begin{equation}
\label{eq:JCP}
J_{CP} =
{\rm Im} \left\{ U_{e1} \, U_{\mu 2} \, U_{e 2}^\ast \, U_{\mu 1}^\ast
\right\}~.
\end{equation}
%
It determines the magnitude
of CP violation effects in neutrino oscillations
\cite{PKSP3nu88}. Let us note that if $U_{\ell} = {\mathbbm 1}$
and $\tilde{U_\nu}$ is a real matrix, one has $J_{CP} = 0$.

   The two additional
invariants, $S_1$ and $S_2$, whose existence is
related to the Majorana nature of massive
neutrinos, i.e., to the phases
$\alpha$ and $\beta$, can be chosen as \cite{JMaj87,ASBranco00}
(see also \cite{BPP1})%
\footnote{We assume that the fields
of massive Majorana neutrinos satisfy
Majorana conditions which
do not contain phase factors.}:
\begin{equation}
S_1 = {\rm Im}\left\{ U_{e1} \, U_{e3}^\ast \right\}~,~~~~
S_2 = {\rm Im}\left\{ U_{e2} \, U_{e3}^\ast \right\}~.
\label{eq:SCP}
\end{equation}
%
If $S_1 \neq 0$ and/or $S_2 \neq 0$, CP is not conserved
due to the Majorana phases $\phi_1$ and/or $\phi_2$.
The effective Majorana mass in
$\beta\beta_{0\nu}$-decay, $\langle m\rangle_{eff}$, depends, in general, on
$S_1$ and $S_2$ \cite{BPP1} and not on $J_{CP}$.
Let us note, however, even if $S_{1,2} = 0$
(which can take place if, e.g., $|U_{e3}| = 0$),
the two Majorana phases $\phi_{1,2} $
can still be a source of CP non-conservation
in the lepton sector provided
${\rm Im}\left\{ U_{e1} \, U_{e2}^\ast \right\} \neq 0$
and ${\rm Im}\left\{ U_{\mu 2} \, U_{\mu 3}^\ast \right\}\neq 0$
\cite{ASBranco00}.

  Let us denote the phase in
$\tilde{U}_{\ell}$ by $\psi$.
We will include it
in $\tilde{U}_{\ell}$
in the same way
this is done for
the phase $\delta$
in the standard parametrization
of the $U$ matrix.
If we write $P_\nu = {\rm diag} (1,e^{i \phi},e^{i \omega})$
and $Q_\nu \equiv {\rm diag} (1,e^{i \rho},e^{i \sigma})$,
the Dirac phase $\delta$ in $U$, which has observable
consequences in neutrino oscillation
experiments,
is determined {\it only by the ``Dirac phase'' in  $\tilde{U}_{\nu}$
and the phases $\psi$, $\phi$  and $\omega$}.
The Majorana phases in $U$ 
receive contributions also from the
two remaining phases $\rho$ and $\sigma$.
Allowing the phases $\delta$ and $\phi_{1,2}$
to vary between 0 and $2 \pi$, permits
to constrain (without loss of generality)
the mixing angles in
$\tilde{U}_{\ell}$,
$\theta_{ij}$, to lie between 0 and $\pi/2$.

  There are interesting specific cases in which
there are direct relations between
all 3 CP violating phases in the
$U$ matrix \cite{fpr}.

\newpage
\section{Testing Seesaw Models}

Although it is far from clear that the seesaw mechanism
\cite{Yanagida:1980,Gell-Mann:1980vs,Glashow:1979vf,Mohapatra:1980ia} is
responsible for neutrino masses, most physicists consider that it is by far the
most elegant mechanism. It fits very well into the big picture of other areas
of particle physics such as supersymmetry, grand unification etc. It is
therefore important to discuss how we can test the seesaw models.  Evidently,
such tests are indirect, since the right-handed electroweak-singlet neutrinos
are much too heavy to be produced at colliders.  In this section, we consider
two aspects of the seesaw mechanism: (i) indirect signals of seesaw mechanism
in lepton flavor violating processes, which can be measured in near future;
(ii) leptogenesis, which can give us a hint as to the CP violating phases in
the lepton sector as well as perhaps the spectrum of the RH neutrinos. The
presence of CP violating phases needed for leptogenesis (see Sec.\
\ref{sec:Leptogenesis}) in turn can lead to CP violating low energy observables
in the seesaw models. We explore these probes of the seesaw mechanism in this
section.

\subsection{\label{sec:LFV}Lepton Flavor Violation and Lepton Electric
Dipole Moments}

Neutrino oscillation experiments have revealed that the violation of flavor
symmetry is much greater in the lepton sector than in the quark sector.
We will discuss how this flavor violation manifests itself via the seesaw
mechanism in other observable quantities where lepton flavor and/or CP are
violated.  As we are going to discuss, among the laboratory observables
particularly interesting are Lepton Flavor Violating (LFV) decays, like $\mu
\rightarrow e \gamma$ and $\tau \rightarrow \mu \gamma$, and Lepton Electric
Dipole Moments (LEDM), like $d_e$ and $d_\mu$.

\begin{table}[ht]
\centering
\begin{tabular}{|c||c|c|c|c|}
\hline 
 & BR($\mu \rightarrow e \gamma$) & BR($\tau \rightarrow \mu \gamma$) &$d_e$[e cm] & $d_\mu$[e cm] \\
\hline \hline
present \cite{present} &  $< 1.2~ 10^{-11} $ & $<1.1~ 10^{-6} $ & $ < 1.5~ 10^{-27} $  & $< 10^{-18} $  \\
\hline
planned \cite{planned} & $< 10^{-14}$ & $ < 10^{-8}$  & $< 10^{-29(-32)}$  & $ < 10^{-24(-26)}$  \\
\hline
\end{tabular}
\caption{Present status and future prospects for LFV decays and LEDM.}
\label{experimental}
\end{table}

Searches for LFV decays and for LEDM are experimentally very promising, since
the present upper bounds could be strengthened by many orders of magnitude, as
summarized in Table \ref{experimental}.  Their impact on theory is also very
promising: finding LFV and LEDM means discovering new low energy physics beyond
the SM supplemented with the seesaw \cite{STP77,ChengLi80}, like e.g. supersymmetry.
This can be easily understood by identifying the SM with the operators of dimension $d \le
4$ of a low energy effective theory valid up to a cutoff $\Lambda$.
Flavor and CP are accidentally conserved in
the leptonic sector of the SM, hence there is
no room for neutrino oscillations nor for LFV decays and LEDM.
Their possible sources have to be found among the operators of $d>4$: neutrino masses arise
from the $d=5$ operator $ \nu^T C^{-1} \nu \langle H^0 \rangle^2 / \Lambda$,
while LFV decays and LEDM from the $d=6$ operator $ {\bar \ell} \sigma^{\mu
\nu} (1+\gamma_5) \ell F_{\mu \nu} \langle H^0 \rangle / \Lambda^2 $.
In the seesaw lepton flavour is no longer conserved but
$\Lambda \sim M_R \ls 10^{15}$ GeV and, as a consequence, the $d=6$
operator above is so strongly suppressed that its effects are negligibly small
\cite{STP77,ChengLi80}.
However, if additional physics is present at smaller mass scales and if this
additional physics violates LF and/or CP, the suppression is milder and LFV
decays and LEDM could be raised up to the experimentally interesting range.

This enhancement due to new low-energy physics is precisely what happens
in low-energy supersymmetry
\footnote{It can also happen in quite different theories,
such as extended technicolor \cite{dml,qdml}.}
where, due to loops with sleptons and gauginos, the $d=6$ operator is suppressed by powers of
$m_\mathrm{SUSY}$. The experimental limits on LFV decays and LEDM then imply
such severe constraints \cite{constraints} on the amount of LF and CP
violations in slepton masses (defined in the basis where charged fermions are
diagonal), that one would expect LF and CP to be exact symmetries of the
supersymmetry breaking mass terms defined at the appropriate cutoff scale (the
Planck scale for supergravity, the messenger mass for gauge mediation, etc).
It is important to stress that, if below this scale there are LF and CP
violating Yukawa interactions, in the running down to $m_\mathrm{SUSY}$ they
nevertheless induce a small amount of LF and CP violations in slepton masses.

It is well known that this is the case for the seesaw interactions of the
right-handed neutrinos \cite{bormas} and/or the GUT interactions of the heavy
triplets \cite{barhall}.  Remarkably enough, this radiative contribution to the
LFV decays and LEDM, which essentially depends on the supersymmetric spectrum
and on the pattern of the Yukawa interactions, might be close to or even exceed
the present or planned experimental limits.  Clearly, this has an impact on
seesaw models, possibly embedded also in a GUT framework.  In the following we
will discuss separately the case of type I and type II seesaw.

\subsubsection{Type I Seesaw: LFV}

\begin{figure}[!ht]
\centerline{\psfig{file=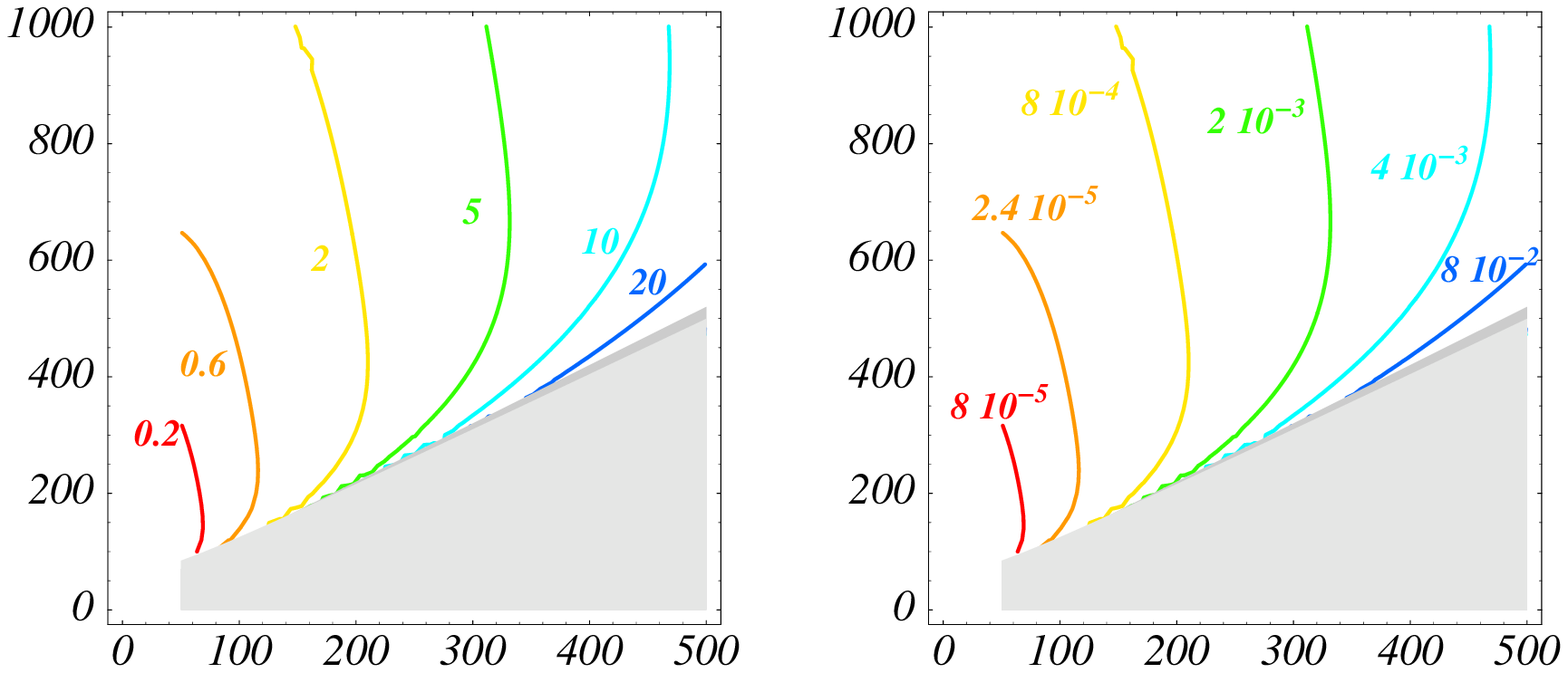,width=1\textwidth}
\put(-235,178){ $m_{\tilde{e}_R}$ }
\put(-460,178){ $m_{\tilde{e}_R}$ }
\put(-280,0){ $\tilde M_1$ }
\put(-50,0){ $\tilde M_1$ }
\put(-150,220){ Upper limit on $|C_{\mu e}|$ }
\put(-370,220){ Upper limit on $|C_{\tau \mu }|$ }
\put(-120,60){ $\times \frac{BR(\mu \rightarrow e \gamma)}{10^{-14}}
\frac{20}{\tan \beta}$ }
\put(-340,60){ $\times \frac{BR(\tau \rightarrow \mu \gamma)}{10^{-8}}
\frac{20}{\tan \beta}$ }}
 \caption{Upper bounds on $|C_{\tau \mu}|$, $|C_{\mu e}|$ in the plane
$(\tilde M_1, m_{\tilde e_R})$,
respectively the bino and right slepton masses expressed in GeV.
We choose as reference values the planned experimental sensitivities
$BR(\tau \rightarrow \mu \gamma)< 10^{-8}$, $BR(\mu \rightarrow e
\gamma) < 10^{-14}$,
with $\tan \beta = 20$, reminding how to adapt the numbers in the plots
for different values
of these parameters.
We also assume mSUGRA with $m_0=A_0$ and $\mu$ from radiative electroweak
symmetry breaking.
The plots are adapted from those in the last Ref.\ of \protect\cite{isaLFV}.}
\label{FR}
\end{figure}

For type I seesaw, in the low energy basis where charged leptons are diagonal,
the $ij$ mass term of $L$-sleptons, ${m^2}^{LL}_{ij}$, is the relevant one in
the decay $\ell_i \rightarrow \ell_j \gamma$.  Assuming for the sake of
simplicity the mSUGRA \cite{mSUGRA} spectrum at $M_{\rm Pl}$, one obtains at
the leading log \cite{bormas} (see also \cite{Hisano:1995cp}):
\begin{equation}
{m^2}^{LL}_{ij} = \frac{1}{8 \pi^2} (3 m_0^2 + 2 A_0^2) C_{ij} ~, ~~~
C_{ij} \equiv \sum_k ~ (Y_\nu)_{ik}~ ({Y_\nu})^*_{jk}
~ \dd\ln\dd\frac{M_{\rm Pl}}{M_k} ~,
\end{equation}
where $Y_\nu = M_\nu^D /v_{\rm wk}$
and $m_0$ and $A_0$ are the universal scalar masses and trilinear couplings at
$M_{\rm Pl}$, respectively, and we have chosen the basis where $M_R$ is
diagonal.  For the full RG results, see \cite{Petcov:2003zb}.  The seesaw model
dependence thus resides in $|C_{ij}|$, and an experimental limit on BR$(\ell_i
\rightarrow \ell_j \gamma)$ corresponds to an upper bound on $|C_{ij}|$.  For
$\mu \rightarrow e \gamma$ and $\tau \rightarrow \mu \gamma$ this bound
\cite{isaLFV} is shown in Fig.\ \ref{FR} as a function of the lightest charged
slepton and gaugino masses.

It has been shown that many seesaw models predict $|C_{\mu e}|$ and/or
$|C_{\tau \mu}|$ close to the experimentally accessible range \cite{modelsLFV,
casasibarra} and, in particular, this might be the case for models based on
$U(1)$ flavor symmetries \cite{U1}.  To reduce the uncertainty due to the
supersymmetric spectrum, it is interesting to exploit the correlation between
LFV decays and muon $g-2$ \cite{LFVeg-2} or neutralino dark matter
\cite{LFVeDM}.

Planned searches could also help in discriminating between categories of seesaw
models \cite{isaLFV}.  To give some hints on the latter issue consider,
e.g., hierarchical eigenvalues of $Y_\nu$.  The different $N^c$ thresholds can
then be neglected and in first approximation one has $|C_{ij}| \approx
|{V_L}_{i3}| |{V_L}_{j3}| y_3^2$ $ \log (M_{\rm Pl} /M_3)$, where $V_L$ is the
lepton analog of the CKM mixing matrix.  In $SO(10)$-inspired models $y_3 =y_t
\sim 1$, and the model dependence essentially resides in the magnitude of
$|{V_L}_{i3}| |{V_L}_{j3}|$, namely on the amount of LF violation present in
the left-mixings of $Y_\nu$.  Under the above assumptions $|C_{\tau \mu}| =
{\cal{O}}( 10 \times |{V_L}_{32}| )$.  If at high energy LF is strongly
violated in the $\tau -\mu$ sector
(models with 'lopsided' $y_\nu$ as the one studied in \cite{Blazek:2001zm})
planned searches for $\tau \rightarrow \mu \gamma$ could be successful for a
significant region of the supersymmetric parameter space.  If this violation is
on the contrary tiny like in the quark sector --- in which case the large LF
violation observed at low energy purely arises from a magnification effect of
the seesaw \cite{magnif} --- $\tau \rightarrow \mu \gamma$ would not be
observed.  Progress in the experimental sensitivity to the latter decay would
thus offer precious information.  The prediction for $\mu \rightarrow e
\gamma$, linked to the product $|{V_L}_{23}| |{V_L}_{13}|$, is more model
dependent but, on the other hand, the present experimental bound is already
very severe.  For instance, simple $U(1)$ flavor symmetries, those with all
lepton charges of the same sign, predict $|C_{\mu e}| ={\cal{O}}(10 \times
\Delta m^2_{\odot} / \Delta m^2_{\rm A} )$ and for LMA are already in crisis
\cite{U1, isaLFV}.  Since the present limit corresponds to a significantly
smaller degree of LFV at high energy, this means that a much richer flavor
symmetry has to be at work.  Notice also that in the future we could test
$|C_{\mu e}|$ up to the CKM-level \cite{mvv}; in fact, if $y_3 = {\cal{O}}(1)$
and $V_{L} \approx V_{CKM}$, then $C_{\mu e} = {\cal{O}} (10^{-3})$ which is
well inside Fig.\ \ref{FR}.

\subsubsection{Type I Seesaw: EDM}

Let us now discuss the consequences of type I seesaw models for lepton EDM.  It
is well known that in the simplest supersymmetric models (with or without
neutrino mass) the dipole moments of electrons and muons obey a simple scaling
law $d_e/d_\mu \approx m_e/m_\mu$. Given the present bound on $d_e$, this
implies $d_\mu < 10^{-25}$ e cm, which is at the level of the best experimental
prospects.

Things can change in seesaw models due to the fact that interactions involving
right-handed neutrinos via radiative corrections can affect the scaling law. In
type I seesaw with degenerate $N^c$, the radiative contributions to $d_e$ and
$d_\mu$ still preserve the scaling law.  However, with hierarchical $N^c$ this
proportionality is broken due to threshold effects arising from both the flavor
conserving $A$-term contribution \cite{ellisetal} and by the FV
double-insertion contributions \cite{isaEDM, FarPesk}, which dominate for $\tan
\beta > 10$.  Nevertheless, if only the type I seesaw radiative contributions
are taken into account, $d_e$ and $d_\mu$ turn out to be barely at hand of
future experimental searches (but can be for very particular textures
\cite{ellisetal}).

Discovering LEDM would then suggest the presence of additional particles and
interactions beyond those of the supersymmetric type I seesaw.  The heavy color
triplets of GUT theories are excellent candidates for this \cite{barhall,
romstr, isaEDM}.  In particular, it turns out that the limits on $d_e$ are
competitive with those on proton lifetime in constraining the pattern of GUT
theories where heavy triplets and right-handed neutrinos are simultaneously
present \cite{IeC}.

\subsubsection{Type II Seesaw}

We will now  consider a class of models where the right-handed neutrino mass
 arises from a renormalizable coupling of the
form $f N N \Delta_R$, where $N$ is a right-handed neutrino, $f$ is a
coupling
constant and $\Delta_R$ is a Higgs field whose vacuum expectation value (vev)
gives mass to the right-handed neutrino. This is a natural feature of
models with asymptotic parity conservation, such as those based on
$SU(2)_L\times SU(2)_R \times U(1)_{B-L}$ or any higher gauge group
such as $SO(10)$, where the $\Delta_R$ field is part of an
$SU(2)_R$ triplet field. Parity invariance then implies that
 we also have an $f \nu \nu \Delta_L$ coupling term as a parity partner
of the $NN\Delta_R$ coupling. In this class of
theories, whenever $\Delta_R$ acquires a vev, so does $\Delta_L$ and they
are related by the formula
$\langle\Delta_L \rangle \equiv v_L = \frac{v^2_{\rm wk}}{\gamma
v_R}$, where $v_{\rm wk}$ is the weak scale, $v_R$ is the $\Delta_R$ vev
and $\gamma$ is a coupling constant in the Higgs potential.
The $\Delta_L$ vev contributes a separate seesaw suppressed
Majorana mass to neutrinos leading to the type II seesaw formula (see
Eq.~(\ref{eq:typeII}))~\cite{seesaw2}.
In the case where right-handed Majorana masses are heavy enough,
the second term in the type II seesaw formula
can be negligible, and the first term, $M_L~=~f v_L$,
 is dominant. We will call this  type II seesaw. The type II seesaw gives
rise to the most simple
explanation of the neutrino sector and is phenomenologically very
successful  especially when we try to construct
realistic models.

The simplest model that can be imagined for type II seesaw has just MSSM and
right-handed neutrinos below the GUT scale and hence there is no new
symmetry breaking scale.  The right-handed neutrino masses can
have hierarchies and therefore get decoupled at different
scales below the GUT scale. Due to the radiative corrections from the
RGEs, the flavor-violating pieces
present
in $Y_{\nu}$  and $f$ get transmitted to the flavor universal scalar masses and
produce lepton flavor violation. The $f$ term contribution is the
additional contribution that is typical of type II seesaw
models~\cite{Babu:2002tb}:
  \begin{eqnarray}
dY_{e}/dt &=& {1 \over 16 \pi^2}(Y_\nu Y_\nu^{\dag}+\cdots)Y_e~,\\\nonumber
dY_{\nu}/dt &=& {1 \over 16 \pi^2}(f f^{\dag}+\cdots)Y_\nu~,\\\nonumber
dm^{2}_{LL}/dt &=& {1 \over 16 \pi^2}(Y_\nu Y_\nu^{\dag}m^2_{LL}+
m^2_{LL}Y_\nu Y_\nu^{\dag}+\cdots)~.
\end{eqnarray}
Here, $m^2_{LL}$ represents the soft left-handed scalar masses.
The flavor non-diagonal pieces generate the lepton flavor violation. In
the type II seesaw
the structure $f$ coupling generates the neutrino mixing parameters.
Both $f$ and $Y_{\nu}$ are determined by the particular model which
explains the quark and
lepton masses.

In order to calculate the BRs of $\mu\rightarrow e\gamma$, $\tau\rightarrow
\mu\gamma$ and the electric dipole moments for the electron and muon, we use
the minimal SUGRA universal boundary conditions at the GUT scale.  The unifying
framework of $SO(10)$ has been chosen and the values of quark masses and the
CKM CP violation are satisfied.  The values of the universal scalar mass $m_0$,
universal gaugino mass $m_{1/2}$, universal trilinear term $A_0$, $\tan\beta$
and the sign of $\mu$ as free parameters determine the final result. The
assumption of universality allows us to probe the flavor violation originating
from the neutrino sector. We also assume that there is no phase associated with
the SUSY breaking. The Yukawa and/or the Majorana couplings are responsible for
CP violation in these models.

\begin{figure}[tbp]
    \centering
    \includegraphics*[angle=0,width=8cm]{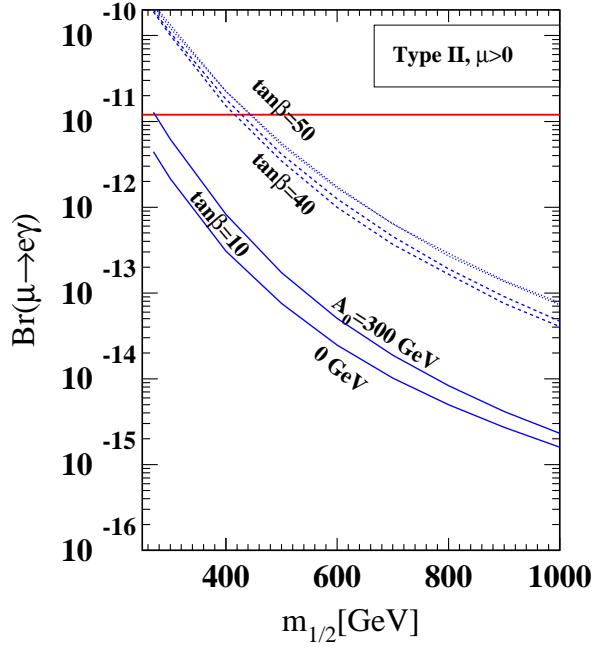}
    \vspace{0.1cm}
    \caption{\label{fig:fig1}  BR[$\mu\rightarrow
e\gamma$] is plotted as a function of $m_{1/2}$ for different values
$A_0$ and $\tan\beta=10$, 40 and 50 in pure type II seesaw.}
\end{figure}

\begin{figure}
    \centering
    \includegraphics*[angle=0,width=8cm]{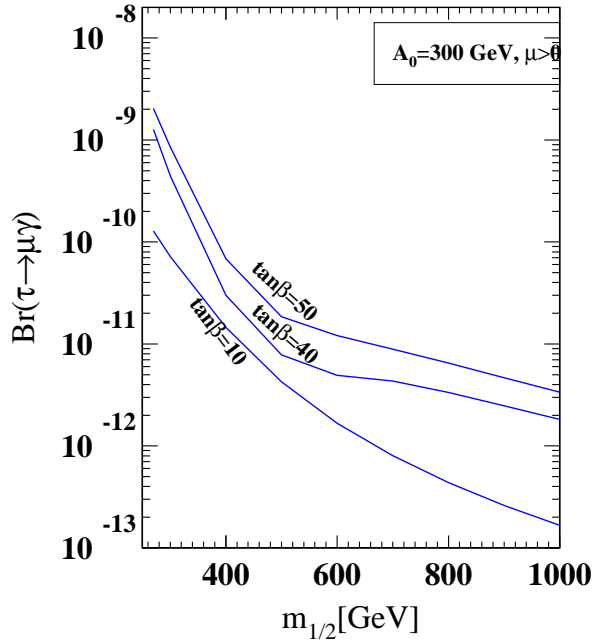}
    \vspace{0.1cm}
    \caption{\label{fig:fig2}  BR[$\tau\rightarrow
\mu\gamma$] is plotted as a function of $m_{1/2}$ for
 $\tan\beta=10$, 40 and 50 in pure type II seesaw.}
 \end{figure}

\begin{figure}
    \centering
    \includegraphics*[angle=0,width=8cm]{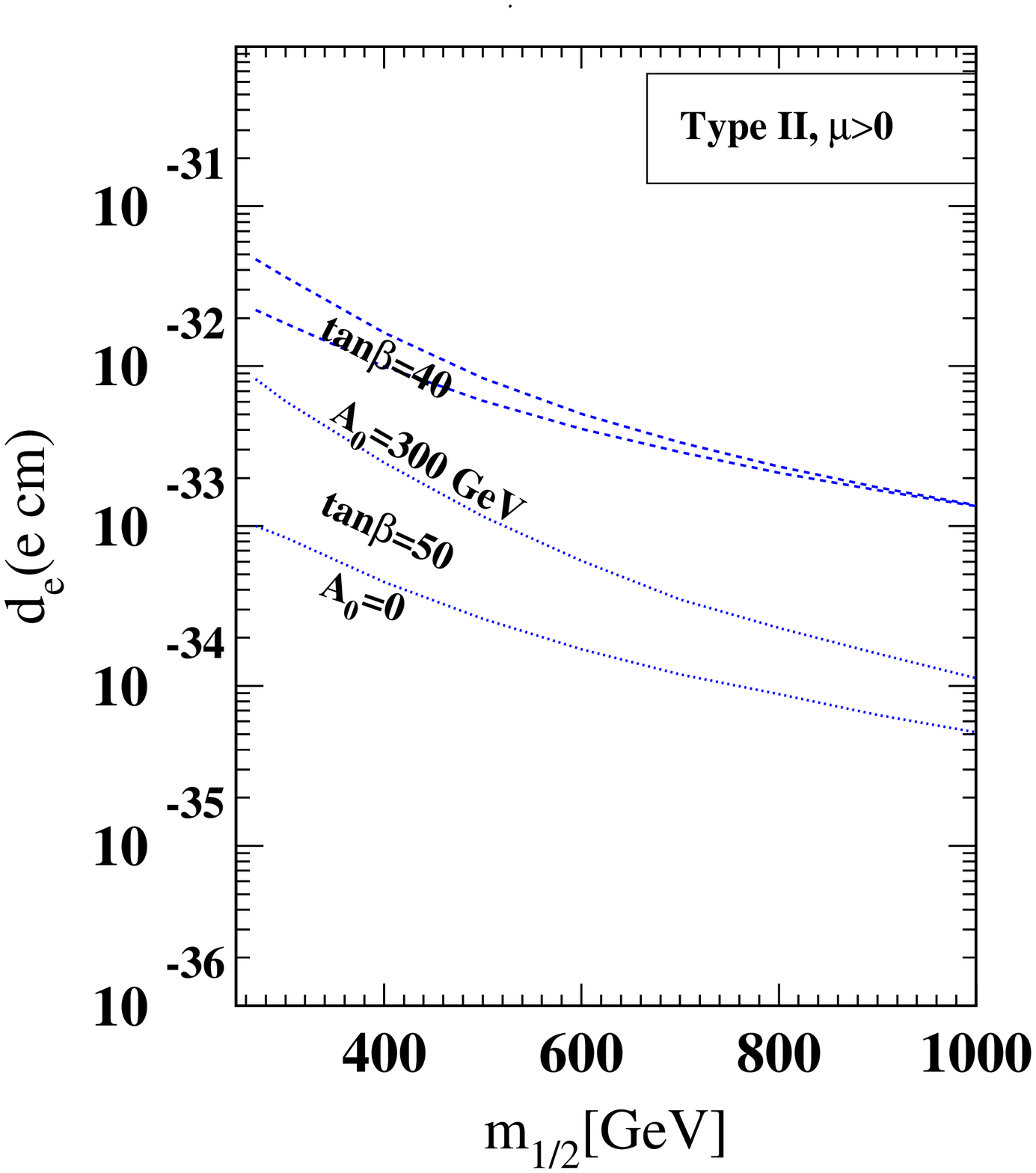}
    \vspace{0.1cm}
    \caption{\label{fig:fig3}  The electron EDM is plotted as a function of
    $m_{1/2}$ for different values $A_0$ and $\tan\beta=40$
 and 50 in pure type II seesaw.}
\end{figure}

\begin{figure}
    \centering
    \includegraphics*[angle=0,width=8cm]{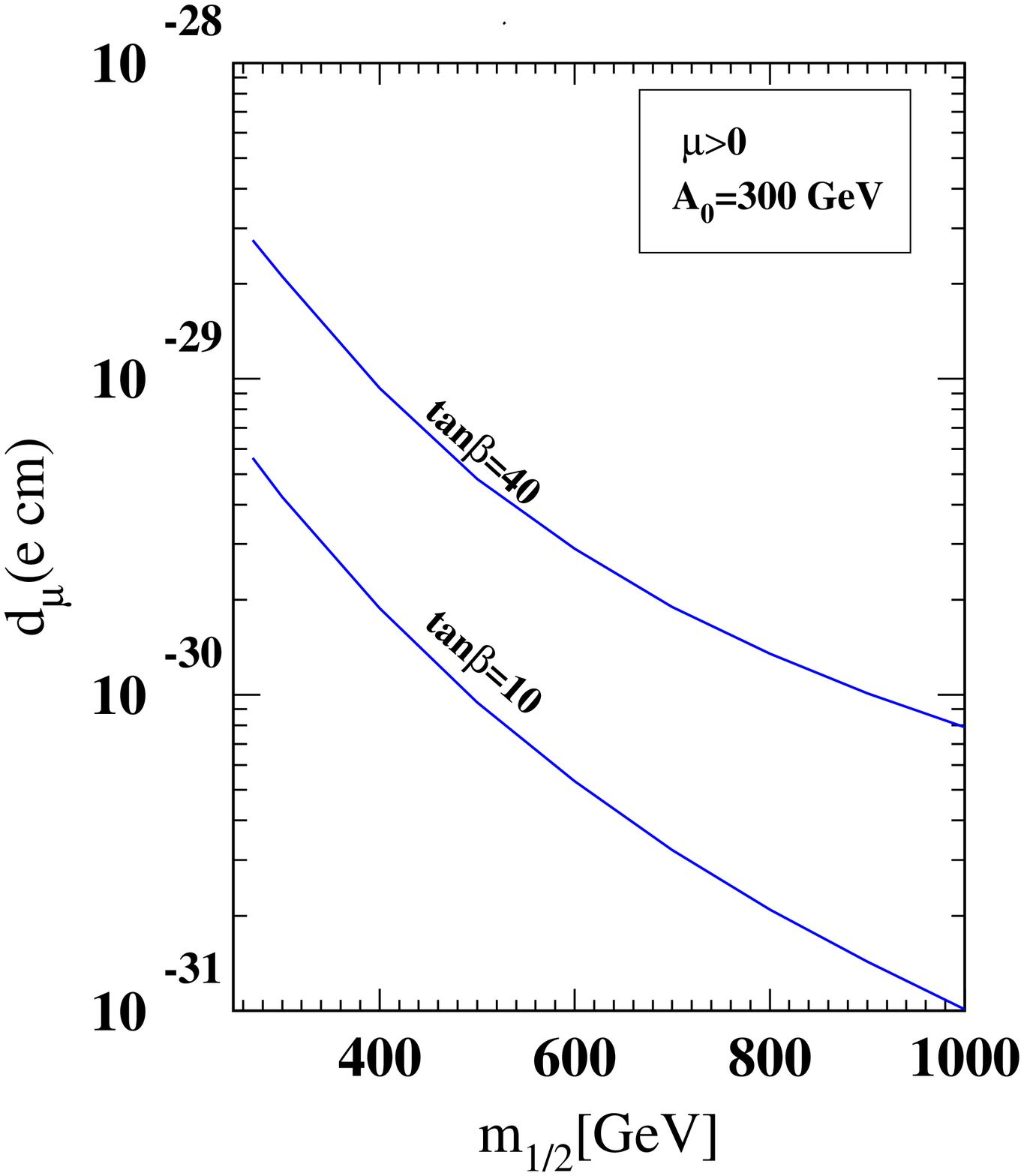}
    \vspace{0.1cm}
    \caption{\label{fig:fig4}  The muon EDM is plotted as a function of
$m_{1/2}$  for
 $\tan\beta=10$, 40 and 50 in pure type II seesaw.}
\end{figure}

The mSUGRA parameter space is constrained by the experimental lower limit on
$m_h$, the measurements of $b\rightarrow s\gamma$ and the recent results on
dark matter relic density~\cite{wmap}. For low $\tan\beta$, the parameter space
has a lower bound on $m_{1/2}$ stemming from the light Higgs mass bound of
$m_h\geq 114$ GeV. For larger $\tan\beta$ the lower bound on $m_{1/2}$ is
obtained by the CLEO constraint on BR($b\rightarrow s\gamma$). The lightest
neutralino is the dark matter candidate in this model and we satisfy the
2$\sigma$ range of the recent relic density constraint $\Omega_{\rm
CDM}=0.1126^{+0.008}_{-0.009}$~\cite{wmap} in the parameter space. The allowed
parameter space of mSUGRA mostly reduces to the neutralino-stau co-annihilation
region for $m_0,\,m_{1/2}\leq 1000$ GeV and when we satisfy the relic density
constraint, $m_0$ gets determined within a very narrow band. For example, $m_0$
varies between 60--100 GeV for $A_0=0$ line in the graph. In
Figs.~\ref{fig:fig1}--\ref{fig:fig4}, we show BR[$\mu\rightarrow e\gamma$] and
BR[$\tau\rightarrow \mu\gamma$] as a function of $m_{1/2}$ for different values
of $A_0$.  We find that the BR is large in most of the parameter space and can
be observable.  In addition, BR[$\tau\rightarrow\mu\gamma$] can also be
observable in the near future.  The figures demonstrate that lepton flavor
violation typically increases with increasing $\tan\beta$.



 The
electron EDM is plotted in Fig.\ \ref{fig:fig3}.
We find that the maximum value of EDM is $\sim 10^{-31}$ e cm.  The muon EDM is
shown in Fig.\ \ref{fig:fig4} and the maximum value shown is about $10^{-29}$ e
cm. The scaling is broken in this model. We do not assume any new CP phases in
SUSY parameters, hence all CP phases arise from the Yukawa and Majorana
couplings.

It is clear that if the seesaw mechanism eventually turns out to be the
explanation of small neutrino mass, the crucial question becomes whether
it is of type I or type II. One may then use lepton flavor violation as a
way to discriminate between these two possibilities.

\newpage
\subsection{Leptogenesis in the Type I Seesaw}
\label{sec:Leptogenesis}
The origin of matter is a fundamental puzzle of
cosmology and  particle
physics. The seesaw provides many
mechanisms to generate this excess;
 we discuss what we can learn about neutrino physics, as well
as the pattern of right-handed neutrino masses, from the observed baryon
asymmetry.

Three ingredients are required to generate the observed Baryon Asymmetry of the
Universe \cite{sakharov}: baryon number violation, C and CP violation and some
out-of-thermal equilibrium dynamics.  The seesaw model
\cite{minkowski,Yanagida:1980,Gell-Mann:1980vs,Glashow:1979vf,Mohapatra:1980ia}, which
was introduced to give small neutrino masses, naturally satisfies these
requirements, producing the baryon asymmetry by the ``leptogenesis'' mechanism~\cite{FY}.  It is interesting to investigate the relation between the requirements of
successful leptogenesis and the observable neutrino mass and mixing matrices.
In particular, does the CP violation that could be observed in neutrino
oscillations bear any relation to leptogenesis?  Do the Majorana phases
that appear in neutrinoless double beta decay experiments do so?

The next subsection reviews the thermal leptogenesis scenario,
focusing on the Type I seesaw,  with hierarchical RH neutrinos
($M_1 \lappeq M_{2,3}/10$).
The relation with light neutrino parameters in type I seesaw models with
three generations is discussed in Subsection \ref{type1}.
The situation in type II seesaw is given in  Section \ref{sec:YBII},
and  the case of quasi-degenerate $N_R$ masses is discussed
in subsection \ref{Resonant}.

\subsubsection{Thermal Leptogenesis}\label{sec:typeIleptogenesis}

The idea of leptogenesis is to use the lepton number violation of the $N_R$
Majorana masses, in conjunction with the $B+L$ violation contained in the
Standard Model~\cite{'tHooft:up}, to generate the baryon asymmetry.  The most
cosmology-independent implementation is ``thermal leptogenesis''
\cite{FY,FPSCRV,AP,BP,gian} which will be reviewed in the following paragraph.
Other leptogenesis scenarios, where the initial number density of (s)neutrinos
is produced non-thermally (by inflaton decay, scalar field dynamics,$\ldots$)
depend on additional parameters of the cosmological model.

 If the temperature $T_{RH}$ of the thermal bath after inflation is $ \gappeq
 M_{R1}$, the lightest $N_R$ will be produced by scattering.  If the $N_R$
 subsequently decay out of equilibrium, a CP asymmetry $\epsilon_1$ in the
 decay produces a net asymmetry of Standard Model leptons. This asymmetry is
 partially transformed into a baryon asymmetry by the non-perturbative $B+L$
 violation~\cite{Kuzmin:1985mm}.  Thermal leptogenesis has been studied in detail
 \cite{BP,FPSCRV,AP,APTU,gian,Buchmuller:2004nz}; the baryon to entropy ratio
 produced is
\begin{equation}
Y_B \simeq C \kappa   \frac{n}{s} \epsilon_1
\label{thlep}  ~~,
\end{equation}
where $\kappa \leq 1$ is an efficiency factor to be discussed in a moment, $n/s
\sim 10^{-3}$ is the ratio of the $N_R$ equilibrium number density to the
entropy density, and $\epsilon_1$ is the CP asymmetry in the $N_{R1}$ decay.
$C \sim 1/3$ tells what fraction of the produced lepton asymmetry is
reprocessed into baryons by the $B+L$ violating processes.
$Y_B$ depends
largely on three parameters: the $N_{R1}$ mass $M_{R1}$, its decay rate
$\Gamma_1$, and the CP asymmetry $\epsilon_1$ in the decay.  The decay rate
$\Gamma_j$ of $N_{Rj}$ can be conveniently parametrized as $ \Gamma_j =
\frac{[Y_\nu^\dagger Y_\nu ]_{jj} M_j} {8 \pi } \equiv \frac{\tilde{m}_j
M_j^2}{8 \pi v_{\rm wk}^2}~~, $ where $\tilde{m}_j$ is often of order of the
elements of the $\nu_L$ mass matrix, although it is a rescaled $N_R$ decay
rate.
The requisite $CP$ violating decay asymmetry $\epsilon_1$
is caused by the interference of
the tree level contribution and
the one-loop corrections in the decay rate
of  the heavy
Majorana neutrinos. For hierarchical neutrinos it is given by:
\begin{equation} \label{eq:eps}
\begin{array}{ll}
 \epsilon_1 & \equiv \frac{\displaystyle \Gamma (N_1 \rightarrow \Phi^- \, \ell^+) -
\Gamma (N_1 \rightarrow \Phi^+ \, \ell^-)}{\displaystyle \Gamma (N_1 \rightarrow \Phi^- \, \ell^+) +
\Gamma (N_1 \rightarrow \Phi^+ \, \ell^-)} \\
&   \simeq -  \frac{\displaystyle 3}{\displaystyle 16 \, \pi}
\sum\limits_{j\neq 1} \frac{\displaystyle{\rm Im} (Y_\nu^\dagger Y_\nu)^2_{1j}}{\displaystyle(Y_\nu^\dagger Y_\nu)_{11}} \,
~ \frac{\displaystyle M_1}{\displaystyle M_j},
\end{array}
\end{equation}
%
where $\Phi$ and $\ell$ indicate the Higgs field and
the charged leptons, respectively.

\begin{figure}
\begin{center}
\includegraphics[height=4cm,width=12cm]{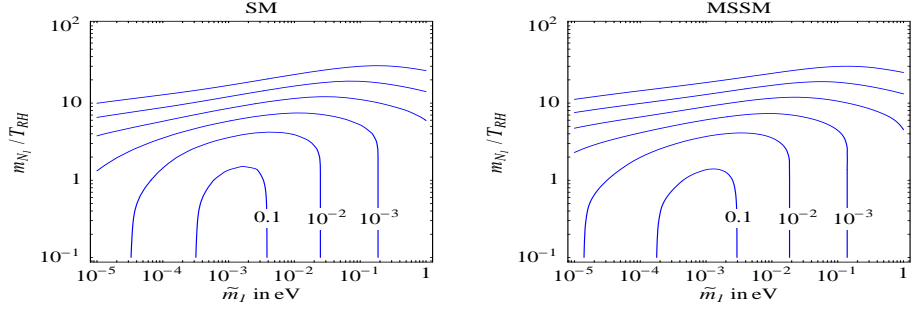}
\end{center}
\caption{
The efficiency parameter
$\kappa$  as function of $(\tilde{m}_1, M_{R1}/T_{\rm RH}$) for
the Type I Seesaw  with hierarchical $N_R$, in the  SM and MSSM. In this plot
$M_{R1} = 10^{10}$ GeV; the plot would  only change slightly
for $M_{R1}\ll 10^{14}$ GeV.
}
\label{rehgian}
\end{figure}

Eq.~(\ref{thlep}) can be of the order of the observed
$Y_B \sim 3 \times 10^{-11} $  when
the following conditions are satisfied:
\begin{enumerate}
\item $M_{R1}$ should be $ \lappeq T_{RH}$\footnote{In the so-called `strong
washout' regime, $T_{RH}$ can be an order of magnitude smaller than $M_1$
\cite{Buchmuller:2004nz}.}.  This temperature is unknown, but bounded above in
certain scenarios ({\it e.g.}  $ T_{RH} \lappeq 10^9\,\mathrm{GeV}$ due to gravitino overproduction in some
supersymmetric models, see \ref{sec:GravitinoProblem}).  This can be seen in Figure
\ref{rehgian}, where the efficiency factor $\kappa$ falls off rapidly for
$M_{R1} > T_{RH}$.
\item The $N_{R1}$ decay rate
$\propto \tilde{m}_1 $  should sit in a narrow
range.  To be precise, $\tilde{m}_1$ must be large enough
to produce an approximately thermal number density
of $N_{R1}$, and small enough that the $N_{R1}$
lifetime is of order the age of the Universe at
$T \sim M_{R1}$ (the out-of-equilibrium  decay condition).
These two constraints are encoded in the efficiency factor $\kappa$,
plotted in Figure  \ref{rehgian}.
\item $\epsilon_1$ must be $\gappeq 10^{-6}$.
\end{enumerate}

\begin{figure}
 \begin{center}
 \includegraphics[height=5cm,width=8cm]{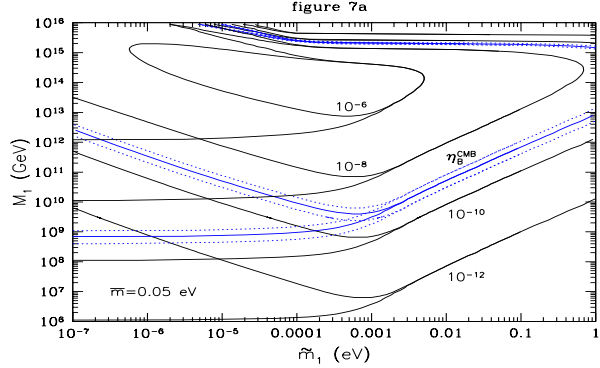}
 \end{center}
 \caption{Contour plot, from \protect\cite{Buchmuller:2002rq}, of the
 baryon to photon
 ratio produced in thermal leptogenesis,
 as a function of  $M_{R1} $ and $  \tilde{m}_1$. The decay asymmetry
 $\epsilon_1$
 was taken to be $10^{-6}$. The three (blue) close-together
 lines are the observed asymmetry. The horizontal contours, for
 small $\tilde{m}_1$ assume a thermal $N_R$ abundance as initial
 condition.}
\label{YBBP}
\end{figure}

In Figure \ref{YBBP} is plotted the baryon asymmetry, produced by thermal
leptogenesis, as a function of $M_{R1}$ and $\tilde{m}_1$, for $T_{RH} \gg
M_{R1}$, and $\epsilon_1 = 10^{-6}$.  To reproduce the observations, $M_{R1}$
and $\tilde{m}_1$ must be inside the three neighboring (blue) lines.

\subsubsection{Parametrizing the type I seesaw}
 \label{type1}

Twenty-one parameters are required to determine the three generation lepton
sector of the type I seesaw model.  This includes the charged lepton masses,
and a mixing matrix (with three complex parameters, {\it e.g.} the MNSP 
matrix)
in the left-handed lepton sector.  The remaining 9 real numbers and 3 phases
can be chosen in various ways:
\begin{enumerate}
\item `` top-down'' --- input the $N_R$ sector:
the eigenvalues of the  mass matrix $M_{R}$ and of $Y_\nu$, and a
matrix transforming between the bases where these matrices are
diagonal \cite{wir} (see also Refs.~\cite{Ellis:2001xt,Davidson:2001zk}).
\item `` bottom-up'' --- input the $\nu_L$ sector:
the eigenvalues of the  mass matrix $M^I_{\nu}$ and of $Y_\nu$, and a
matrix transforming between the bases where these matrices are
diagonal.
\item ``intermediate'' --- the Casas-Ibarra  parametrization
\cite{casasibarra}:  the  ${M_R}$ and  ${M^I_\nu}$ eigenvalues,
and a complex orthogonal matrix ${R}$ which transforms between
these two bases.
\end{enumerate}
To relate the RH parameters relevant for leptogenesis to the
LH ones, many of which are accessible at low energy, it is useful to consider
the first and second parametrization.

\subsubsection{Implications for CP conserving observables}
\label{real}

\begin{figure}[ht]
\centering
\includegraphics[width=12cm]{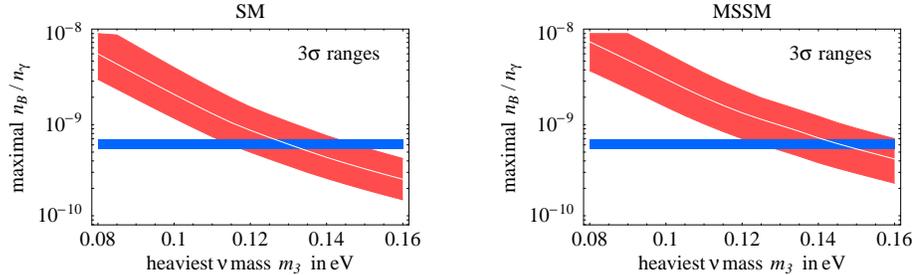}
\caption{
Upper bound on the light neutrino mass scale, assuming hierarchical $N_R$,
taken from \cite{strumia}.  The plot shows the measured baryon asymmetry
(horizontal line) compared with the maximal leptogenesis value as function of
the heaviest neutrino mass $m_3$.
}
\label{m3strumia}
\end{figure}

The second requirement (i.e., the range of $\tilde{m}_1$) sets an upper bound on
the mass scale of light neutrinos.  The scaled decay rate $\tilde{m}_1$ is
usually $\sim m_2, m_3$; for hierarchical light neutrinos, it naturally sits in
the desired range.  One can show \cite{Fujii:2002jw,di2} that $m_1 <
\tilde{m}_1$, so that $m_1 \lappeq 0.15 $ eV
\protect\cite{BdBP,gian,strumia,Buchmuller:2004nz}  is required for thermal
leptogenesis in the type I seesaw, with hierarchical $N_R$.  This is shown in
Figure \ref{m3strumia}.

In type I seesaw models with hierarchical $N_R$, the third condition
($\epsilon_1 \gappeq 10^{-6}$) imposes $M_{R1} \gappeq 10^8$ GeV, because
 $\epsilon_1 \leq 3 M_{R1} (m_3 - m_1)/(8 \pi v_{\rm wk}^2)$ in most of
parameter space \protect\cite{di2,strumia}.
 For three $N_{R}$, the value of $M_{R1}$ has
little implication on low energy neutrino observables.  If $\epsilon_1$ is
maximal --- that is, $M_{R1}$ close to its lower bound, --- this sets one
constraint on the 21 parameters of the type I seesaw.  This has no observable
consequences among Standard Model particles, because at most 12 masses, angles
and phases are measurable, and $\epsilon_1$ can be maximized by choice of the
nine other parameters.  The situation is more promising \cite{dip,wir} in SUSY
models with universal soft terms, where some of the 9 additional parameters can
contribute to slepton RGEs and thereby to lepton flavor violating branching
ratios, as discussed in Section \ref{sec:LFV}.



\subsubsection{Relations between Leptogenesis and leptonic CP violation}
 \label{phases}

The leptogenesis parameter $\epsilon_1$ is a $\CPV$ asymmetry, suggesting a
possible correlation with CP violation in $\nu$ oscillations (the phase
$\delta$),
or to the low
energy Majorana phases ($\phi_{1,2}$).
Let us assume that
$\epsilon_1$ is large enough --- so thermal leptogenesis
works --- and  concentrate on the implications for  low-energy
CP violation.


The first thing that must always be said, in discussing potential connections
between phases in the MNSP lepton mixing matrix and leptogenesis, is that 
there
is no linear relation: leptogenesis can work when there is no $\CPV$ in 
MNSP,
and measuring low energy leptonic phases does not imply that there is CP
violation available for leptogenesis. This was clearly and elegantly shown by
Branco, Morozumi, Nobre and Rebelo in \cite{Branco:2001pq}.  The problem is
that six phases are included in the general three neutrino seesaw scenario ---
it would be astonishing if the $\CPV$ parameter we are interested in
($\epsilon_1$) is proportional to the low energy  phases  ($\delta$,
$\phi_{1,2}$).


Nonetheless, some sort of relation between $\epsilon_1$ and the low energy
phases would be interesting --- so what can we say?  Needing more inputs that
the data provides is a familiar problem for extensions of the SM.  The usual
solutions are to scan over unknowns, or to fix them.  But this is subtle: any
relation depends on
 the choice of ``independent'' phases. For instance, if
 $\epsilon_1$ and $\delta$ are chosen as inputs,
then it follows that they are unrelated \footnote{This
arises in the Casas-Ibarra seesaw parametrization.}.
A choice of parametrization is not obvious:
the $\CPV$ parameter $\epsilon_1$
  is a function of $N_R$ phases, masses, mixing angles,
whereas the observable phases are those of
$U$.

A useful step is to write $\epsilon_1 \propto$ a Jarlskog invariant,
which can be done for thermal leptogenesis with hierarchical
$N_R$ \cite{ryu}:
\begin{equation}
\epsilon_1 \propto
\Im \{ {\rm Tr} [ M_\nu^\dagger  M_\nu  M_\nu^\dagger
( Y_\nu Y_\nu^\dagger)^{-1}
  M_\nu ( Y_\nu^* Y_\nu^T)^{-1} ] \}.
\label{trace}
\end{equation}
The advantage is that Jarlskog invariants
can be evaluated in any basis/parametrization.
It is easy to see, evaluating Eq.\ (\ref{trace}) in the $\nu_L$ mass eigenstate
basis, that the $\CPV$ for leptogenesis is controlled by a matrix $W$ which
transforms between the bases were $Y_\nu$ and $ M_\nu$ are diagonal. This
matrix is unobservable, verifying the no-go theorem of
\cite{Branco:2001pq}. However, in many popular/common Yukawa texture
models, where $Y_\nu$ and $M_{\ell}$ are almost simultaneously diagonalizable,
$W \sim U$ and the MNSP phases are relevant for
thermal leptogenesis.
For $s_{13}$ larger than the mixing angles between
diagonal $Y_\nu$ and $M_{\ell }$, $\epsilon_1 \propto
\sin 2(\phi_1 - \phi_2 + \delta)$.

\subsubsection{Model dependent approaches}

Within specific models
interesting links between the phase relevant for leptogenesis
and the phase $\delta$ measurable in neutrino oscillation experiments
have been made. The precise link depends on how many ``texture''
zeroes are assumed to be present in the neutrino Dirac mass matrix.
For example within the class of two right-handed neutrino models,
if two texture zeroes are assumed then there is a
direct link between $\delta$ and the leptogenesis phase,
with the sign of $\delta$ being predicted from the fact that we are
made of matter rather than antimatter \cite{min,Frampton:2002qc}.
Two right-handed neutrino models can be obtained as a limiting
case of sequential dominance models, and in such models if
only the physically motivated texture zero in the 11 entry of the
Dirac mass matrix is assumed, then the link is more indirect
\cite{King:2002qh}. Other approaches which give rise to
a link between leptogenesis and CP violation include
GUT models \cite{GUTS}, or textures \cite{Ellis:2001xt,RaidalJE,textures,wir}
or left-right symmetric models, to be discussed in the next Subsection. 
On the other hand, if the charged lepton sector contributes significantly to 
the lepton mixing $\theta_{13}$ and therefore also to $\delta$, such links 
may be spoiled \cite{Antusch:2005kw}.

\subsubsection{Leptogenesis in supersymmetric scenarios}
\label{sec:GravitinoProblem}

In supersymmetric scenarios, the history of the early universe is subject to
various constraints. Many of them are associated to the gravitino problem
\cite{Khlopov:1984pf,Moroi:1993mb}. In short, unstable gravitinos are
notoriously in conflict with nucleosynthesis (see \cite{Cyburt:2002uv} for a
recent analysis), while stable ones may `overclose' the universe.  If the
gravitino is very light ($m_{3/2}\lappeq \mathrm{keV}$ \cite{Pagels:1981ke}) or
very heavy  ($m_{3/2}\gappeq 10\,\mathrm{TeV}$), these bounds disappear, and
thermal leptogenesis works (see, e.g., \cite{Ibe:2004tg}). For all other
masses, nucleosynthesis or `overclosure' constraints translate into bounds on
the gravitino abundance at $T\sim 1\,\mathrm{MeV}$ or today, respectively.
Assuming that gravitinos are not produced by inflaton decays (see
\cite{Nilles:2001ry}), this gravitino abundance is linear in the reheat
temperature \cite{Bolz:2000fu}.

Unstable gravitinos with masses below  $\sim10\,\mathrm{TeV}$ lead to severe
constraints on the reheat temperature $T_{RH}$ \cite{Cyburt:2002uv} which are
in conflict with thermal leptogenesis where $T_{RH}\gappeq10^9\,\mathrm{GeV}$.
There are, however, various alternative leptogenesis scenarios such as
non-thermal leptogenesis \cite{Lazarides:1991wu} where the heavy neutrinos are
directly produced by inflaton decays, or mechanisms using the superpartner of
the neutrino, the sneutrino \cite{Murayama:1992ua}, or sneutrino
oscillations (see Subsec.~\ref{sec:SneutrinoOscillation}). These scenarios
can be consistent with unstable gravitinos. 

Stable gravitinos, on the other hand, may evade the constraints from
nucleosynthesis provided that the decays of the next-to-lightest superpartner
into the gravitino are harmless \cite{Feng:2004mt}.
However,  the `overclosure' constraint leads to
$T_{RH}\lappeq(10^9-10^{10})\,\mathrm{GeV}$. Such an upper bound on the reheat
temperature is suggested independently by string-theoretical arguments where
$T_{RH}\lappeq\sqrt{m_{3/2}\,M_\mathrm{P}}$ \cite{Buchmuller:2003is}. Stable
gravitinos are thus (marginally) consistent with thermal leptogenesis, and
provide a natural dark matter candidate \cite{Bolz:1998ek}. It is clear that the
neutrino mass bound, as discussed in \ref{real}, becomes much tighter now since
$\widetilde{m}_1\sim 10^{-3}\,\mathrm{eV}$ (cf.\ Fig.~\ref{YBBP}) and
$m_1\lappeq \widetilde{m}_1$ \cite{Fujii:2002jw,di2}.  This scenario therefore
predicts hierarchical light neutrinos as well as gravitino cold dark matter.
These predictions will be tested in neutrino experiments and at future colliders
\cite{Buchmuller:2004rq}.


\subsection{\label{sec:YBII}Leptogenesis and type II seesaw mechanism}
In type II seesaw scenarios the neutrino mass matrix $M_\nu$
reads
\begin{eqnarray}
M_\nu^{II} = M_L  - M_\nu^D~M_R^{-1}~(M_\nu^D)^T \equiv
 M_L + M_\nu^{I}~,
\end{eqnarray}
where we divided the mass matrix in the conventional type I part
$M_\nu^{I} = - M_\nu^D~M_R^{-1}~(M_\nu^D)^T$ and the part
characteristic for type II, $M_L$.
A type II seesaw term can, e.g., be present in $SO(10)$ models, in
which the $B-L$ symmetry is broken by a {\bf 126} Higgs fields.
Depending on the parameters of the model, either $M_L$ or
$M_\nu^{I}$ can be the dominating source of $M_\nu^{II}$.
As mentioned above, in case of the conventional type I seesaw mechanism
with three families of light and heavy Majorana neutrinos, there are
six phases. As already discussed, in
this case there is in general
no relation between the PMNS phase and leptogenesis
phase.

For the type II case the phase counting gives the result
of 12 independent CP phases and there is no connection between
low and high energy CP violation either.
The number of CP phases can be obtained by going to a basis in which both
$M_L$ and $M_R$ are real and diagonal. Any CP violation will then
stem from the matrices $M_\nu^D$ and $M_\ell M_\ell^\dagger$ (with
$M_\ell$ being the charged lepton mass matrix). Those two matrices
posses in total 9 + 3 = 12 phases.

The term $M_L$ is induced by a $SU(2)_L$ Higgs triplet,
whose neutral component acquires a vev
$v_L \propto v_{\rm wk}^2/M_{\Delta_L}$, where $M_{\Delta_L}$ is the
mass of the triplet and $v_{\rm wk}$ the weak scale.
Consequently, the triplet
contribution to the neutrino mass matrix is
\begin{eqnarray}
M_L = v_L~f_L~,
\end{eqnarray}
with $f_L$ a symmetric $3 \times 3$ coupling matrix.
The magnitude of the contribution of $\Delta_L$ to $M_\nu^{II}$
is thus characterized by its vev $v_L$.
In left-right symmetric theories the left-right symmetry necessarily
implies the presence of a $SU(2)_R$ triplet, whose coupling matrix
is given by $f_R = f_L \equiv f$ and its vev is given by $v_R$, where
$v_L~v_R = \gamma~v_{\rm wk}^2$ with $\gamma$ a model dependent
factor of order one. The right-handed Majorana neutrino mass matrix is
thus given by $M_R = v_R~f = v_R/v_L~M_L$. Before acquiring its vev,
the presence of the doubly charged Higgs and the coupling of the
Higgs triplet with the doublet introduces the possibility of
additional diagrams capable of generating a lepton asymmetry.
First, there is the possibility that in the decay
$N_1 \rightarrow L~H_u$ a virtual Higgs triplet is exchanged in the
one-loop diagram, which
contributes to the decay asymmetry in the decay of the heavy Majorana
neutrinos \cite{O'Donnell:1994am,Lazarides:1998iq,Hambye:2003ka,anki}.
The corresponding term
$\epsilon_1^{\Delta} $ adds up to the conventional term
$\epsilon_1$ whose properties were
discussed in the previous Subsection.

The second new diagram possible is given by the decay of the doubly
charged Higgs triplet in two charged leptons. One-loop exchange of a
heavy Majorana neutrino gives rise to the decay asymmetry
\cite{O'Donnell:1994am}
\begin{eqnarray}
\epsilon_{\Delta} \equiv
\frac{\Gamma(\Delta_L \rightarrow l^c~l^c) -
\Gamma(\Delta_L^\ast \rightarrow l~l) }
{\Gamma(\Delta_L \rightarrow l^c~l^c) +
\Gamma(\Delta_L^\ast \rightarrow l~l)}
~.
\end{eqnarray}
If $M_1 \ll M_{\Delta_L}$ ($M_1 \gg M_{\Delta_L}$) the decay of
the Majorana neutrino (Higgs triplet) will
govern the baryon asymmetry. Thus, depending on which term dominates
$M_\nu^{II}$, four different situations
are possible \cite{Hambye:2003ka}. The discussion of 3 of the
cases has so far not been discussed in as much detail as the
conventional leptogenesis in type I seesaw mechanisms.

If $M_1 \ll M_{\Delta_L}$ and the conventional term $M_\nu^I$
dominates $M_\nu^{II}$, we recover the usual seesaw and leptogenesis
mechanisms and the statements given in
Sec.\ \ref{sec:typeIleptogenesis} apply.
\begin{figure}
\centering
\includegraphics[width=6.2cm,height=5cm]{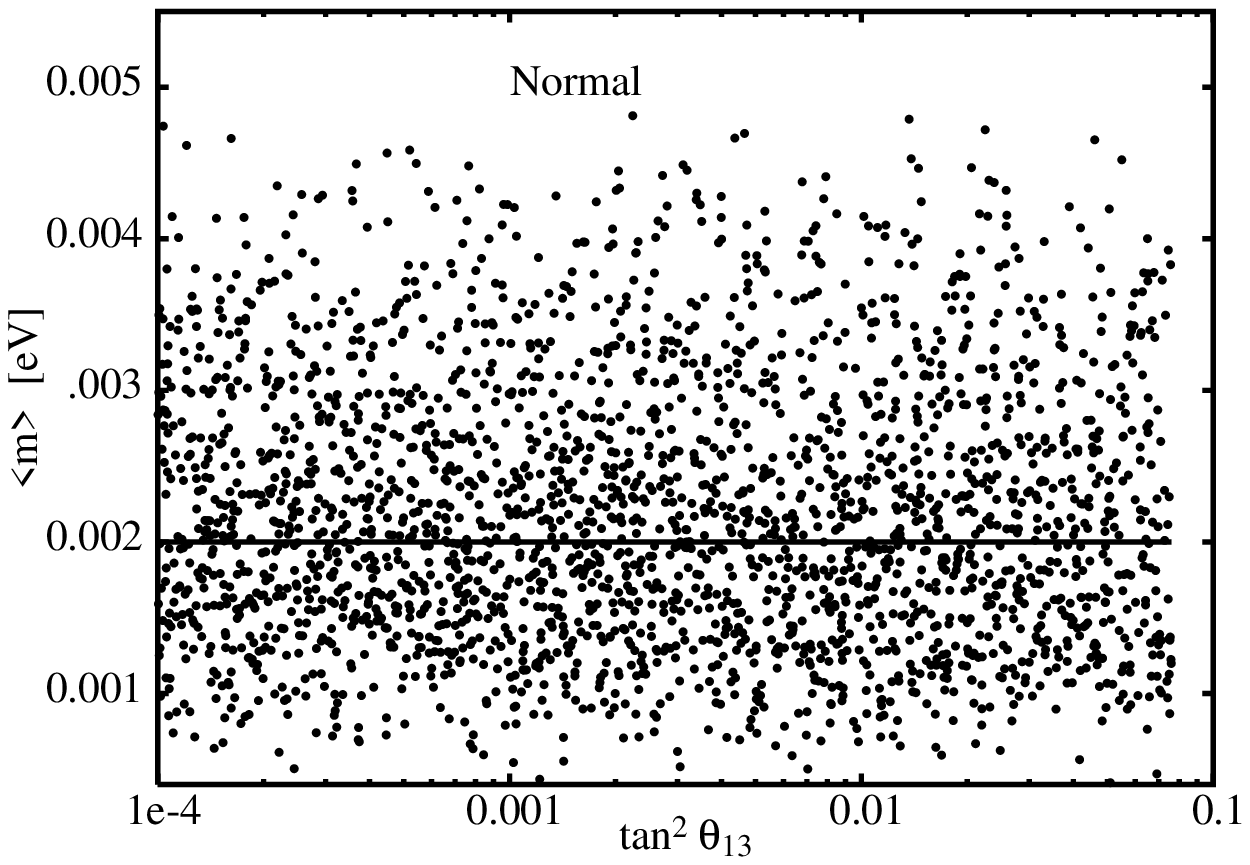}
\includegraphics[width=6.2cm,height=5cm]{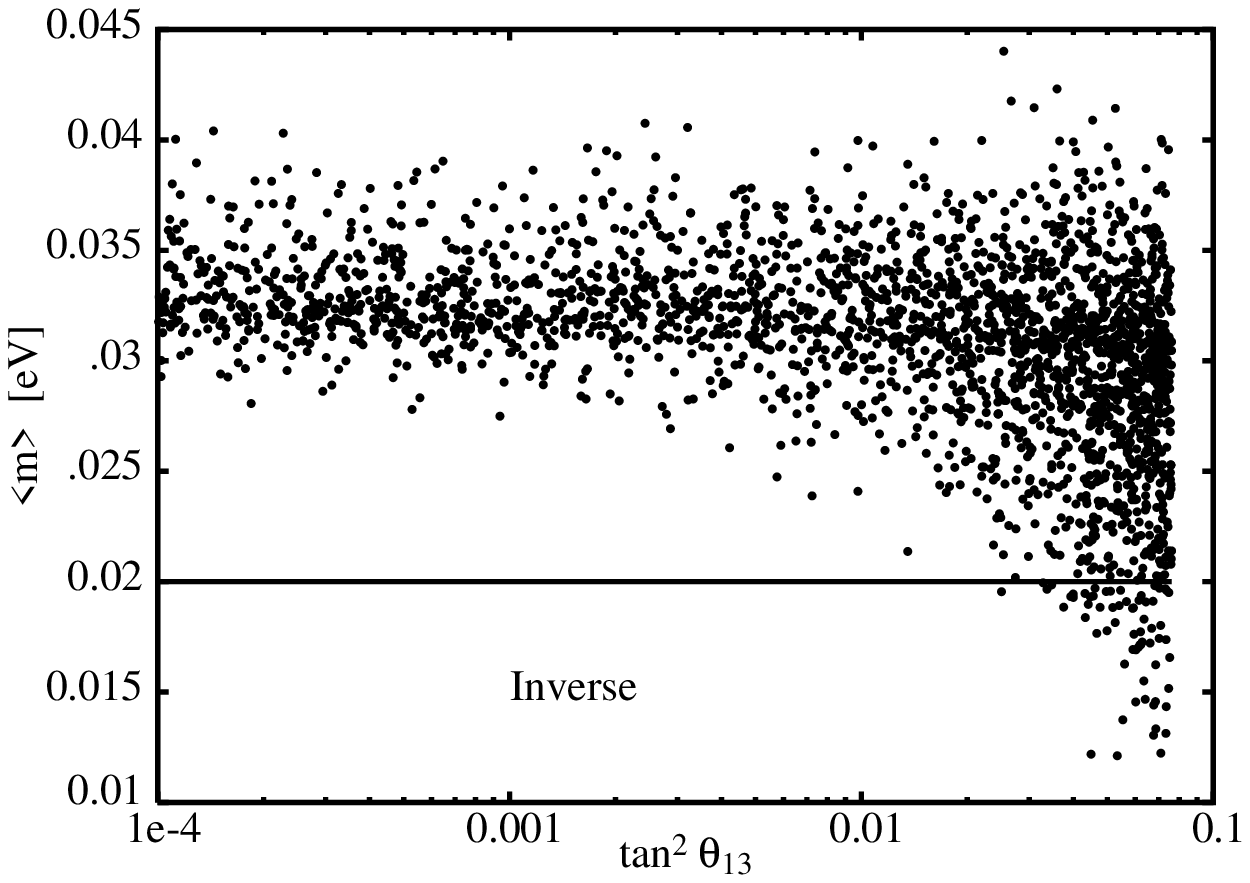}\\
\includegraphics[width=6.2cm,height=5cm]{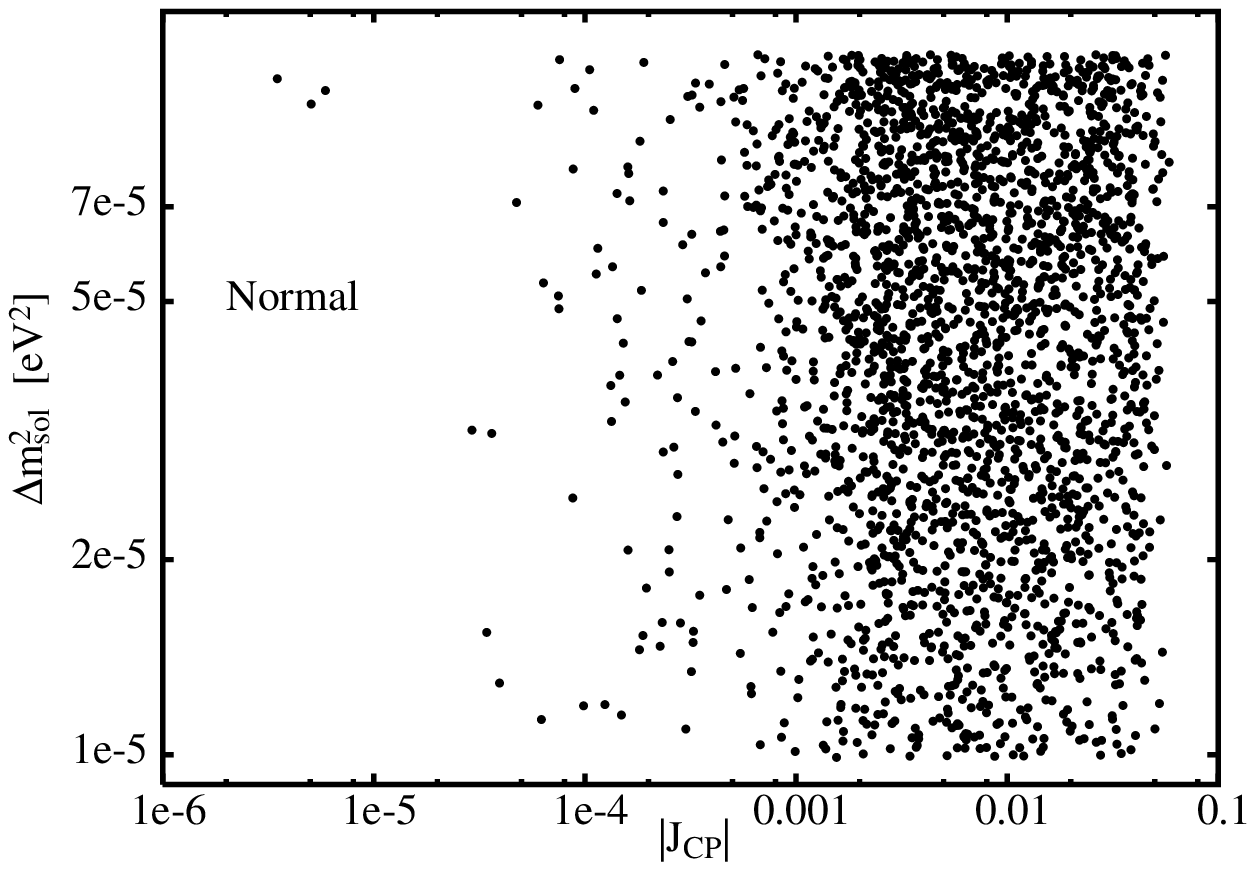}
\includegraphics[width=6.2cm,height=5cm]{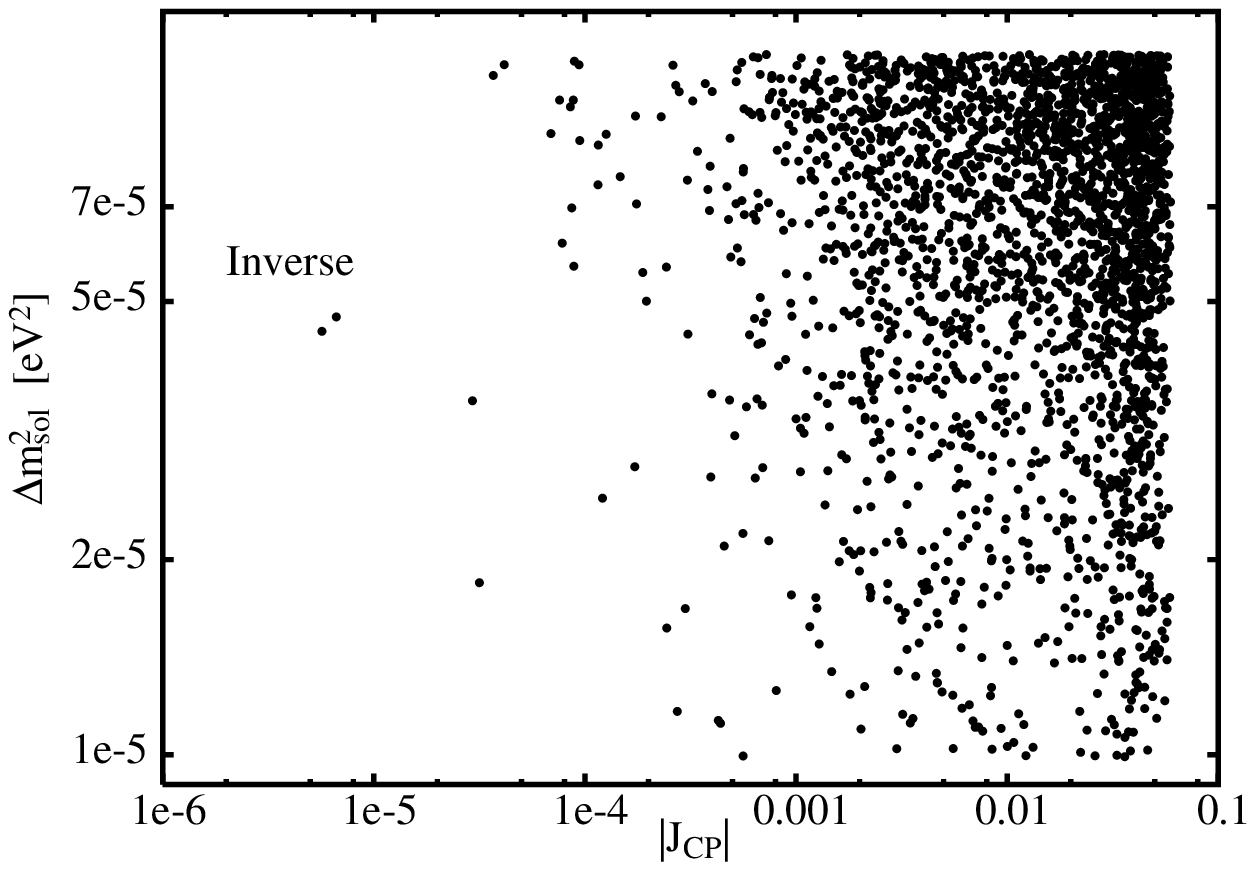}
\caption{Scatter plots of the effective mass against
$\tan \theta_{13}$ and $\Delta m^2_\odot$ against $J_{CP}$ for the normal
(left)
and inverted (right) neutrino mass spectrum.
Taken from \cite{Joshipura:2001ui}.
}
\label{fig:fig}
\end{figure}

In situations in which $M_1 \ll M_{\Delta_L}$, the heavy Majorana neutrinos
display a hierarchical structure and $M_L$ dominates $M_\nu^{II}$, it has been
shown in \cite{Hambye:2003ka,anki} that one can rewrite the decay asymmetries
such that $\epsilon_1^{\Delta} $ depends on $M_L$ and $\epsilon_1$ on
$M_\nu^{I}$. However, since matrices are involved, $\epsilon_1$ can still be
the dominant contribution to the decay asymmetry, a situation which in the
context of left-right symmetry has intensively been investigated in
\cite{Joshipura:1999is,Joshipura:2001ui}, see also \cite{sche}.  Calculating in
this framework the baryon asymmetry in terms of light neutrino parameters (a
bottom-up approach) leads typically to a main dependence on the Majorana phases
in the PMNS matrix.  If $M_\nu^D$ is given by the up-quark mass matrix and the
light neutrinos display a normal hierarchal spectrum, one of the low energy
Majorana phases has to be very close to zero or $\pi/2$
\cite{Joshipura:1999is}.  For $M_\nu^D$ given by the down quark or charged
lepton mass matrix one finds that the in general unknown mass spectrum of the
heavy Majorana neutrinos is {\it exactly} given by the measurable mass spectrum
of the light Majorana neutrinos.  In case of a normal hierarchy, the Majorana
phases should lie around $\pi/4$ or $5\pi/4$. Both values give comparable
results for the rate of neutrinoless double beta decay. Thus, measuring
neutrinoless double beta decay fully determines the neutrino mass matrix in
this scenario.  It is also possible to set limits on the lightest neutrino mass
$m_1$ because the baryon asymmetry is proportional to $m_1$.  It should be
larger than $10^{-5}$ eV in order to produce a sufficient baryon asymmetry
\cite{Joshipura:2001ui}.  For an inverted hierarchy of the neutrinos it turns
out that rather sizable values of $\theta_{13}$ are required.  Thus, sizable
effects of CP violation in future long-baseline neutrino oscillation
experiments are possible. The preferred value of the Majorana phase implies in
addition a rather sizable rate of neutrinoless double beta decay.  Furthermore,
the lightest neutrino mass should be heavier than $10^{-3}$ eV.  Figure
\ref{fig:fig} shows for the normal and inverted hierarchy typical examples for
the expected values of $\theta_{13}$, the effective mass and the CP violating
parameter $J_{CP}$ in neutrino oscillations. A similar example
within a framework incorporating spontaneous CP violating is discussed
in Section \ref{sec:muchun}.

Finally, if neutrinos possess a quasi-degenerate mass spectrum, one of
the Majorana phases is required to lie around $\pi$ or $\pi/2$. A measurement
of neutrinoless double beta decay can resolve this
ambiguity.\\

The other possible scenarios have not been discussed in detail
in the literature so far (see, e.g., \cite{IIothers}).
General statements are however possible.
If, e.g., $M_1 \ll M_{\Delta_L}$ and
the term $\epsilon_1^{\Delta}$
dominates the decay asymmetry, the limits on the light neutrino masses
of order 0.1 eV (see Sec.\ \ref{real}) no longer apply \cite{Hambye:2003ka},
since the couplings responsible for the
neutrino masses do not influence the wash-out processes.
For hierarchical light neutrinos the upper bounds on the
decay asymmetry $\epsilon_1$ and $\epsilon_1^\Delta$ are identical.
In case of quasi-degenerate neutrinos, however, the
limit in case of type II seesaw is weaker by a factor of
$2m_0^2/\Delta m^2_{\rm A}$ \cite{anki}, where $m_0$ is the common
neutrino mass scale.
Along the same lines, the lower limit of order $10^9$ GeV on the
lightest of the heavy Majorana neutrinos can be relaxed by roughly one order
of magnitude \cite{anki}, thereby making thermal leptogenesis less
in conflict with the gravitino problem.

Consider the case when $M_1 \gg M_{\Delta_L}$ and $M_\nu^{I}$
dominates $M_\nu^{II}$. Then $\epsilon_{\Delta}$ will
produce the baryon asymmetry and again the limits on light neutrino masses
do not apply.
The same is true when $M_1 \gg M_{\Delta_L}$ and
$M_L$ is the main contribution to $M_\nu^{II}$.
A smaller range of allowed
parameters is expected in this case \cite{Hambye:2003ka}.

Therefore, given
the fact that quasi-degenerate light neutrinos are hard to reconcile
with standard thermal leptogenesis in type I seesaw models, our discussion
implies that if we learn from
future experiments that neutrinos are indeed quasi-degenerate, triplet
induced leptogenesis represents a valid alternative (this was first noted
in \cite{Chun:2000dr}).

Also possible is --- in inflationary scenarios --- that the decay of the
inflaton into light particles together with interference of one-loop
diagrams with exchanged
$SU(2)$ triplets and heavy Majorana neutrinos generates a lepton
asymmetry. Various slepton decays in future colliders
are expected to be observable \cite{Dent:2003dn}.\\

Up to now the discussion was constrained to the presence of
only one triplet. If only one triplet is present, right-handed
Majorana neutrinos are necessary to produce a decay asymmetry. Introducing
two or more triplets allows self-energy diagrams which can
produce a decay asymmetry without heavy Majorana neutrinos \cite{Ma:1998dx}.
The possibility of lowering the triplet mass scale due to a resonance
effect of close-in-mass triplets is possible \cite{Hambye:2000ui},
giving the prospect of collider phenomenology. The presence of light
and detectable Majorons is also possible.
There are models implementing
this kind of triplet self-energy scenarios with light
left-handed neutrinos \cite{Hambye:2000ui} and with
quasi-degenerate ones \cite{Lazarides:1998jt}. The latter also
predicts a stable proton due to $R$ parity conservation.
Introducing the triplet induced neutrino mass matrix of the type II seesaw
mechanism along the lines of
\cite{Mohapatra:1999zr}, i.e., by a conjunction of flavor and permutation
symmetries will typically include many additional Higgs fields.
Rich phenomenology in form of rare charged lepton decays or
charged lepton EDMs will be among the interesting consequences.\\

To sum up, the type II seesaw mechanism displays the most general but
more complicated framework of neutrino mass and leptogenesis.
Nevertheless, richer phenomenology is expected, most of which remains to
be explored.

\subsection{Dirac Leptogenesis}

Since the seesaw mechanism coupled with existing data on neutrino
masses and mixings does not give complete information about the
RH neutrino sector, one must consider leptogenesis within various
scenarios for RH neutrino masses that correctly explain neutrino
observations.
In this section, we discuss a possibility that the leptogenesis occurs with
Dirac neutrinos.  The conventional leptogenesis \cite{FY}, which we call
Majorana leptogenesis for definiteness, was based on the fact that the standard
model violates $B+L$ \cite{'tHooft:up},  while the Majorana neutrinos violate
$L$, and hence both $B$ and $L$ are violated.  Therefore it is possible to
create $L$ from the decay of right-handed neutrino that is subsequently
converted to $B$ \cite{Kuzmin:1985mm}.  On the other hand, Dirac neutrinos
conserve $L$ and hence $B-L$ is an exact symmetry.  Therefore $B-L$ stays
vanishing throughout the evolution of the universe and it appears impossible to
generate non-vanishing baryon asymmetry\footnote{An obvious exception is
electroweak baryogenesis, where $B-L=0$ while $B=L\neq$ after the electroweak
phase transition.}.

Dirac leptogenesis overcomes this problem by the following simple observation
\cite{Dick:1999je}.  Recall that the Dirac neutrinos have tiny Yukawa couplings,
$M_\nu^D = Y_\nu v_{\rm wk}$,  $Y_\nu \simeq 10^{-13}$.  If this is the only
interaction of the right-handed neutrinos, thermalization is possible only by
processes like $N L \rightarrow H W$  and they do not thermalize for $T \gs g^2
Y_\nu^2 M_{\rm Pl} \sim 10$~eV.\footnote{In contrast, the Yukawa coupling of the
right-handed electron $e_R$ is large enough to equilibrate the $e_R$s before
sphaleronic processes switch off \cite{Campbell:1992jd}.}  At this low
temperature, obviously both $H$ and $W$ cannot be produced and the
thermalization is further delayed until $T_\nu \simeq M_\nu$ when neutrinos
become non-relativistic. Therefore the number of left-handed and right-handed
neutrinos are separately conserved practically up to now.  We call them $L$ and
$N$, respectively, and the total lepton number is $L+N$.  The combination
$L+N-B$ is strictly conserved.

\begin{figure}[tbp]
  \centering \includegraphics[width=0.3\textwidth]{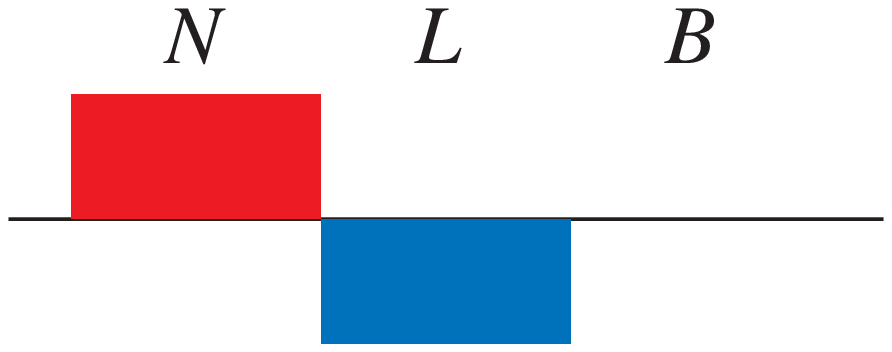}
 \includegraphics[width=0.3\textwidth]{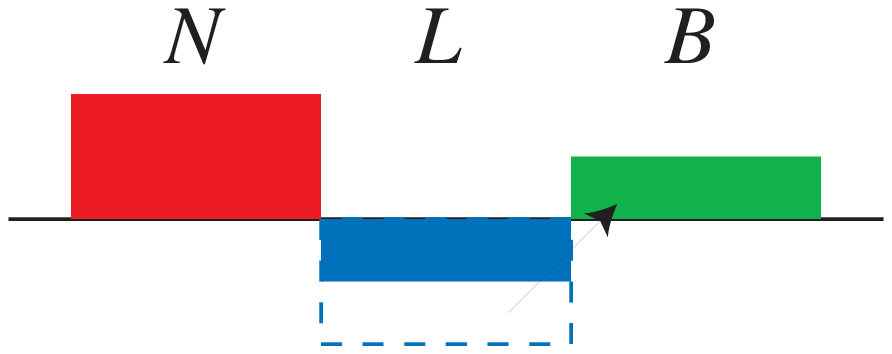}
 \includegraphics[width=0.3\textwidth]{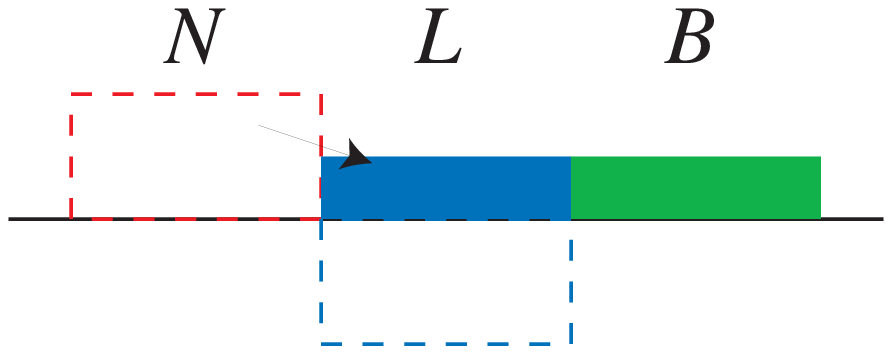}
  \caption{The evolution of the lepton asymmetry in Dirac leptogenesis
    models.  At the first stage, an asymmetry between the ordinary
    leptons and the right-handed neutrinos is created without
    lepton-number violation.  Then the asymmetry in the ordinary
    leptons is partially converted to the baryon asymmetry.  Finally,
    the right-handed neutrinos come in thermal equilibrium with the
    other leptons.  The net baryon and lepton asymmetries remain while
    the overall $B-L$ vanishes.}
  \label{fig:Dirac}
\end{figure}

Suppose the decay of a heavy particle produced an asymmetry $L = -N
\neq 0$.  The overall lepton number is conserved (see
Fig.~\ref{fig:Dirac}).  $N$ is frozen down to $T_\nu$.  On the other
hand, the lepton asymmetry $L$ is partially converted to the baryon
asymmetry via the Standard Model anomaly.  After the electroweak phase
transition $T \ls 250$~GeV, the anomaly is no longer effective.
Finally at $T_\nu$, $L$ and $N$ equilibrate.  In the end there is a
baryon asymmetry $B = -(L+N)$.

The original paper \cite{Dick:1999je} introduced new electroweak doublet scalar
$\Phi$ that has the same quantum numbers as the Higgs doublets and Yukawa
couplings $\Phi L N$ and $\Phi^* L E$.  If there are two sets of them, there is
CP violation and their decays can create the asymmetry $L = -N \neq 0$.
However, these doublets are there just for this purpose and have no other
motivations.

On the other hand, light Dirac neutrinos are natural in models where
the neutrino Yukawa couplings are tied to the small supersymmetry
breaking effects \cite{Arkani-Hamed:2000bq,Borzumati:2000mc}.  The
Dirac neutrino mass is due to the effective operator
\begin{equation}
  \label{eq:2}
  \int d^2 \theta \frac{\chi}{M} L H_u N,
\end{equation}
where $\chi$ is the hidden sector field which acquires a vacuum expectation
value $\langle \chi \rangle \simeq m_{3/2} + \theta^2 m_{3/2} M_{\rm Pl}$ and
$M$ is the heavy mass scale.  The neutrino Yukawa coupling is $Y_\nu \simeq
m_{3/2}/M$, and is naturally small.  The operator can be obtained by integrating
out (two sets of) new doublets $\phi+\phi^c$ that couple as $W = \phi N H_u +
\phi^c L \chi + M \phi \phi^c$.  The asymmetries $L = -N \neq 0$ are created by
the decay of $\phi$ \cite{Murayama:2002je}. Then the origin of small neutrino
mass and the origin of the lepton asymmetry are tied in the same way as the
Majorana leptogenesis. Also concerning the gravitino problem (cf.\
\ref{sec:GravitinoProblem}), Dirac and Majorana leptogenesis are on the
same footing \cite{Murayama:2002je,Boz:2004ga}. Such a scenario may be supported
by the lack of neutrinoless double beta decay as well as the existence or
right-handed sneutrino at LHC and Linear Collider.


\subsection{Resonant Leptogenesis}
\label{Resonant}

The right handed neutrino sector of generic seesaw models is almost
entirely unconstrained by existing data on neutrino masses and
mixings. It is therefore necessary to consider all the various
possibilities for the structure of the RH neutrino sector which are
compatible with current experimental data. In this section, we
consider the case where two or more right-handed neutrinos are nearly
degenerate in mass.

An important, further motivation for this possibility comes from the
severe limits on the right handed neutrino sector that exist in models
of thermal leptogenesis with hierarchical right handed neutrinos. In
particular there exists a bound on the mass of the lightest right
handed neutrino, $M_{R1} \stackrel{<}{{}_\sim} T_{RH}$, discussed in
section 4.2.1. In supersymmetric theories with unstable gravitinos,
this bound can be in conflict with the bound $T_{RH}
\stackrel{<}{{}_\sim} 10^9$~GeV, coming from nucleosynthesis
considerations (see 4.2.6). This motivates us to consider scenarios
where the scale of the right handed neutrino masses can be lowered
whilst still being compatible with thermal leptogenesis
\cite{AP,APIJMPA,boubekeur}. This may be achieved naturally in scenarios with
nearly degenerate right handed neutrinos \cite{AP}, in complete
accordance with current neutrino data, and with the advantage that the
final baryon asymmetry generated is independent of the initial lepton,
baryon or heavy neutrino abundances \cite{APTU}.

If the mass difference between two heavy Majorana neutrinos happens to
be much smaller than their masses, the self-energy ($\epsilon$-type)
contribution to the leptonic asymmetry becomes larger than the
corresponding ($\epsilon'$-type) contribution from vertex
effects~\cite{FPSCRV,AP}.  Resonant leptogenesis can occur when this
mass difference of two heavy Majorana neutrinos is of the order of
their decay widths, in which case the leptonic asymmetry could be even
of order one~\cite{AP,APTU}.  As a result, one can maintain the RH
neutrino masses around the GUT scale \cite{afsmirnov} or one can
contemplate the possibility that the heavy neutrino mass scale
relevant to thermal leptogenesis is significantly lower, for example
in the TeV~range~\cite{AP}. This of course requires a different
realization of the seesaw mechanism \cite{valle2} but it can be in
complete accordance with the current neutrino data~\cite{APTU}.

The magnitude of the $\epsilon$-type CP violation occurring in the
decay of a heavy Majorana neutrino $N_i$ is given by \cite{AP},
\begin{equation}
\epsilon_{N_i} = \frac{\mathrm{Im}
(Y_\nu^\dagger\,Y_{\nu})^2_{ij}}{(Y_\nu^\dagger\,Y_{\nu})_{ii}
(Y_\nu^\dagger\,Y_{\nu})_{jj}}\,\frac{(m_{N_i}^2-m_{N_j}^2) m_{N_i}
\Gamma^{(0)}_{N_{j}}}{(m_{N_i}^2-m_{N_j}^2)^2 + m_{N_i}^2
\Gamma^{(0)\,2}_{N_{j}}}\,,
\label{eps}
\end{equation}
where $\Gamma^{(0)}_{N_i}$ is the tree level total decay width of
$N_i$. It is apparent that the CP asymmetry will be enhanced, possibly
to $\epsilon \sim 1$, provided
\begin{eqnarray}
m_{N_2}-m_{N_1} \sim \frac{1}{2}\,\Gamma^{(0)}_{N_{1,2}}\,,\nonumber\\
\frac{\mathrm{Im}
(Y_\nu^\dagger\,Y_{\nu})^2_{ij}}{(Y_\nu^\dagger\,Y_{\nu})_{ii}
(Y_\nu^\dagger\,Y_{\nu})_{jj}} \sim 1\,.
\label{cond}
\end{eqnarray}
It is  important to note that Eq.~(\ref{eps}) is  only valid for
the mixing of two heavy  Majorana neutrinos. Its generalization to the
three  neutrino  mixing  case  is   more  involved  and  is  given  in
\cite{APTU}.
\begin{figure}[htb]
\begin{center}
\includegraphics{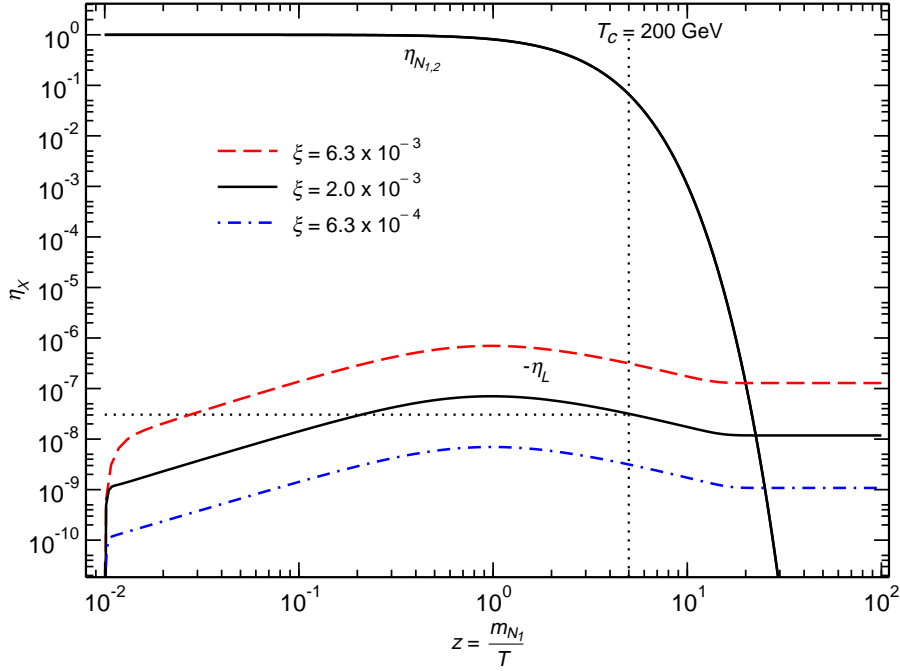}
\end{center}
\vspace{-0.5cm}
\caption{Numerical estimates of the lepton to photon, and neutrino
to photon ratios, $\eta_L$ and $\eta_{N_{1,2}}$ as functions of
$z=m_{N_1}/T$ for scenarios with $m_{N_1}=1\,\mathrm{TeV}$. The model
is based on the Froggatt-Nielsen mechanism and is completely
consistent with all current neutrino data (see \cite{APTU} for
details). It naturally provides a degeneracy of
$\frac{m_{N_2}}{m_{N_1}} - 1 = 9.2 \times 10^{-11}$, with $K$-factors
and CP asymmetries of $K_1 = K_2 = 6570$ and $\delta_{N_1} =
\delta_{N_2} = -0.003$. $\xi$ is a free parameter. The
horizontal dotted line shows the value of $\eta_L$ needed to produce
the observed $\eta_B$. The vertical dotted line corresponds to $T =
T_c = 200\,\mathrm{GeV}$.}
\label{largeK}
\end{figure}

Successful leptogenesis requires conditions out of thermal
equilibrium.  To quantify this, we introduce the parameter, $K^l_{N_i}
= \Gamma^l_{N_i} / H(T=m_{N_i})$ where $H(T)$ is the Hubble parameter
and $\Gamma^l_{N_i}$ is the total decay width of $N_i$ into a lepton
species $l$ ($l=e,\mu,\tau$). In typical hierarchical leptogenesis
scenarios $K_{N_i}^l$ is small, usually $K_{N_i}^l \sim 1$. This
constraint can be translated directly into an upper bound on the
Yukawa couplings of the neutrinos which can be expressed in terms of
effective light neutrino masses, $\widetilde{m}_i$,
\begin{equation}
  \label{meff_bound}
\widetilde{m}_i\ \equiv\ \frac{v^2\, (Y_\nu^\dagger Y_\nu)_{ii}}{2\,
m_{N_i}}\ \simeq\ 10^{-3}\, K_{N_i}~{\rm eV}\,,
\end{equation}
where $K_{N_i} = \sum_l K^l_{N_i}$. However, resonant leptogenesis can
be successful with values of $K_{N_i}$ larger than $1000$ \cite{APTU}
(see Figure~\ref{largeK}).  This has implications for leptogenesis
bounds on the absolute mass scale of the light neutrinos. If a large,
$\stackrel{>}{{}_\sim}0.2\,\mathrm{eV}$, Majorana mass was seen in
neutrinoless double beta decay, this could be naturally accommodated
with resonant leptogenesis, whereas thermal leptogenesis models based
on hierarchical heavy neutrinos would be strongly disfavored, as they
naturally require smaller values of
$K_{N_i}$~\cite{BBP,Buchmuller:2002rq}.

Conditions close to thermal equilibrium (with large $K_{N_i}$) endow
resonant leptogenesis models with another particularly attractive
feature; the final baryon asymmetry generated is almost independent of
the initial baryon, lepton or heavy neutrino abundances
\cite{APTU,APTU2}.

For $K_{N_i} \stackrel{>}{{}_\sim} 1$, and under the assumption that
the neutrino Yukawa couplings are the same for each lepton
flavour, an order of magnitude estimate for the baryon to photon ratio
may be obtained by \cite{APTU}
\begin{equation}
  \label{etaBapprox}
\eta^{\,\mathrm{univ.}}_B\ \sim\ -\, 10^{-2}\,\times\, \sum_{N_{i}}\:
e^{-(m_{N_{i}} - m_{N_1})/m_{N_1}}\,\frac{\delta_{N_{i}}}{K}\,,
\end{equation}
where $\delta_{N_i}$ is the leptonic CP asymmetry in the decay of
$N_i$, $K = \sum_l K_l$ and
\begin{equation}
K_l \ =\ \sum_{N_{i}}\, e^{-(m_{N_{i}} - m_{N_1})/m_{N_1}}\,
K^{l}_{N_{i}}\,.
\end{equation}
It is apparent that if the CP-asymmetry is enhanced, for example
through resonant effects, then $K$ can be increased without an impact
on the final baryon asymmetry.

In resonant leptogenesis scenarios with neutrino Yukawa couplings that
are not universal for each lepton flavour, the effects of individual
lepton flavours on the resultant baryon asymmetry may become very
important \cite{APtau,APTU2}. This also applies to scenarios with
mildly hierarchical RH neutrinos \cite{APTU2}. These effects may
result in an enhancement of the baryon asymmetry predicted using
(\ref{etaBapprox}) by a factor as large as $10^6$ in some models of
resonant leptogenesis. An order of magnitude estimate for the final
baryon asymmetry, when the neutrino Yukawa couplings are not
flavour-universal, may be obtained by \cite{APTU2}
\begin{equation}
\eta_B\ \sim\ -\, 10^{-2}\,\times\, \sum_{l=1}^3\, \sum_{N_{i}}\:
e^{-(m_{N_{i}} - m_{N_1})/m_{N_1}}\, 
\delta^l_{N_{i}}\: \frac{K^{l}_{N_{i}}}{K_l\,K_{N_{i}}}\ ,
\end{equation}
where $\delta_{N_i}^l$ is the CP asymmetry in the decay of $N_i$ to
leptons of flavour $l$. For a more precise computation of the
resultant baryon asymmetry, a network of Boltzmann equations must be
solved, one for each heavy Majorana neutrino species and one for each
lepton flavour \cite{APTU2}. By adding a further Boltzmann equation
for the baryon abundance, and including effects due to the rate of
$B+L$ violating sphaleron transitions, it can be shown that successful
resonant leptogenesis is possible with heavy Majorana neutrinos as
light as the electroweak scale \cite{APTU2}.

In particular, models of resonant $\tau$-leptogenesis \cite{APtau},
where a lepton asymmetry is generated predominantly in the
$\tau$-family, can allow large Yukawa couplings between the $e$ and
$\mu$ lepton families and some of the RH neutrinos. These couplings,
in conjunction with the low RH neutrino scale, lead to a significant
amount of accessible phenomenology, such as potentially observable
neutrinoless double beta decay, $\mu \to e\gamma$, $\mu \to eee$ and
coherent $\mu \to e$ conversion in nuclei, and the possibility of the
collider production of heavy Majorana neutrinos \cite{APtau,APTU2}.

The conditions for resonant leptogenesis can be met in several
ways. Models based on the Froggatt-Nielsen mechanism can naturally
provide nearly degenerate heavy Majorana neutrinos satisfying
Eq.~(\ref{cond}) and provide a light neutrino spectrum fulfilling all
experimental constraints. It can be shown that a model like this can
produce the observed baryon asymmetry by solving the network of
Boltzmann equations -- including gauge mediated scattering effects
\cite{APTU} (see Figure \ref{largeK}). In this model the `heavy'
Majorana neutrinos can be as light as 1 TeV.

$SO(10)$ models with a ``type III seesaw mechanism'' naturally predict
pairs of nearly degenerate heavy Majorana neutrinos suitable for
resonant leptogenesis \cite{AB}. In addition, a model of neutrino mass
from SUSY breaking has been shown to naturally lead to conditions
suitable for resonant leptogenesis \cite{THJMRSW}.

In the radiative leptogenesis mechanism \cite{radlepto}, small mass
differences arise through renormalization group corrections between RH
neutrinos which are exactly degenerate in mass at some high scale. The
leptonic CP asymmetry induced in this scenario is sufficient to
produce the observed baryon asymmetry.

In soft leptogenesis \cite{DGRGKNR,ejchun}, soft SUSY breaking terms
lead to small mass differences between sneutrinos. Resonant effects
allow sneutrino decay to generate the required CP asymmetry.

Several other mechanisms for leptogenesis where the right handed
neutrinos can be at a TeV scale have been suggested
\cite{boubekeur}. Clearly for the seesaw mechanism to operate in such
models, the Dirac mass must be constrained, e.g., by a leptonic global
U(1) symmetry \cite{babumoh}.

\newpage
\section{The Heavy Majorana Mass Matrix}

\subsection{General Considerations}
We have already seen from the Introduction that the type I seesaw
mechanism requires the existence of right-handed neutrinos $N_R$,
and then the light Majorana mass matrix is given as
\begin{eqnarray}
M_{\nu}=-M^D_{\nu}M_R^{-1}{M^D_{\nu}}^T
\label{seesawI}
\end{eqnarray}
where $M^D_{\nu}$ is the Dirac neutrino matrix (to be thought of as perhaps
similar to the quark and charged lepton mass matrices) and $M_R$ is the heavy
Majorana mass matrix. While the elements of $M^D_{\nu}$ must be at or below the
electroweak scale, the characteristic scale of right-handed neutrino masses can
and must be much higher. Having introduced right-handed neutrinos into the
Standard Model for the purpose of accounting for light physical neutrino masses
via the type I seesaw mechanism\footnote{Here we shall assume
$M_{\nu}=M_{\nu}^{\mathrm{I}}$. We shall later comment also on the
type II seesaw mechanism.}  it is clearly an important question to understand
the mass spectrum and couplings of the right-handed neutrinos. Since their only
couplings are their Yukawa couplings to Higgs and left-handed neutrino fields,
it will clearly not be an easy task to answer this question. However there are
three areas where important clues may emerge: the light-neutrino mass matrix
$M_{\nu}$; the baryon asymmetry of the universe; and (assuming supersymmetry)
lepton flavor violation. Taken together with other theoretical ideas, we shall
show that it may be possible to shed light on the right-handed neutrino sector
of the theory.

\subsubsection{The Three Right-Handed Neutrino Paradigm}

It is most common to assume that there are exactly three right-handed
neutrinos. Such an assumption is motivated by unified theories such as $SO(10)$
which predicts that the number of right-handed neutrinos is equal to the number
of quark and lepton families, since a single right-handed neutrino makes up
each 16-plet of the theory. In fact this prediction also follows more generally
from any theory which contains a gauged right-handed group $SU(2)_R$, such as
left-right symmetric theories, Pati-Salam and so on.

Assuming three right-handed neutrinos one can ask whether their mass spectrum
is hierarchical, or contains an approximate two or three-fold degeneracy. From
the point of view of the type I seesaw mechanism in Eq.\ (\ref{seesawI}) it is
clear that the answer to this question depends on the nature of the Dirac
neutrino mass matrix $M^D_{\nu}$.  For example, suppose that the right-handed
neutrinos had a three-fold degeneracy $M_R={\rm diag}(M,M,M)$, then the Eq.\
(\ref{seesawI}) would predict
\begin{eqnarray}
M_{\nu}=-\frac{M^D_{\nu}{M^D_{\nu}}^T}{M}.
\label{seesaw2}
\end{eqnarray}
Then if the Dirac neutrino mass matrix were hierarchical and
approximately proportional to the
up-type quarks, for example, then Eq.\ (\ref{seesaw2}) would imply
\begin{eqnarray}
m_1:m_2:m_3\approx m_u^2:m_c^2:m_t^2
\label{naive}
\end{eqnarray}
which is much too strong a mass hierarchy compared to
the rather mild experimentally measured ratio $0.1\leq m_2/m_3\leq 1$.

The remaining two possibilities are that the three right-handed
neutrinos are either hierarchical or contain an approximate two-fold
degeneracy. In either case it is convenient
to work in a basis where their mass matrix is diagonal:
\begin{equation}
M_R=
\left(\begin{array}{ccc}X' & 0 & 0 \\
0 & X & 0 \\
0 & 0 & Y\end{array}\right)
.
\label{MR1}
\end{equation}
The neutrino Dirac mass matrix $M^D_{\nu}$ in this basis can be written as
\begin{equation}
M^D_{\nu}=
\left(\begin{array}{ccc}a' & a & d \\
b' & b & e \\
c' & c & f\end{array}\right)
\label{MD}
\end{equation}
where in this convention the first column of Eq.~(\ref{MD})
couples to the first right-handed neutrino, the second
column of Eq.~(\ref{MD})
couples to the second right-handed neutrino and so on.
Note that in the hierarchical case in Eq.\ (\ref{MR1})
we do not specify which of the three
right-handed neutrinos $X',X,Y$ is the lightest one,
which is the intermediate one and which is the heaviest one,
since the columns of $M^D_{\nu}$ and the eigenvalues $X',X,Y$ of $M_R$ may
simultaneously be re-ordered without changing $M_{\nu}$.

Having displayed the unknown Yukawa couplings associated with
the Dirac neutrino mass matrix in Eq.\ (\ref{MD}) it is clear that
without further input it is not possible to say anything about the
right-handed neutrino masses or couplings from the experimentally
determined light Majorana mass matrix $M_{\nu}$.
On the other hand, rather natural theoretical assumptions
can lead to a great deal of information about the unknown masses
and couplings of the right-handed neutrinos, as we now discuss.

Regarding the implementation of the type I seesaw mechanism there seem to be
two possible options: either all the right-handed neutrinos contribute equally
(democratically) to each element of $M_{\nu}$, or some right-handed neutrinos
contribute more strongly than others. In the second case, called right-handed
neutrino dominance \cite{King:2002nf}, a rather natural implementation
of the seesaw mechanism is possible.  For example if the right-handed neutrino
of mass $Y$ contributes dominantly to the physical neutrino mass $m_3$, and the
right-handed neutrino of mass $X$ contributes dominantly to the physical
neutrino mass $m_2$, while the right-handed neutrino of mass $X'$ contributes
dominantly to the physical neutrino mass $m_1$, then a sequential dominance of
these three contributions leads to a neutrino mass hierarchy $m_1\ll m_2\ll
m_3$. With such sequential dominance the mixing angles are then given as simple
ratios of Dirac neutrino mass matrix elements: $\tan \theta_{23} \approx e/f$,
$\tan \theta_{12} \approx \sqrt{2}a/(b-c)$, which can be naturally large
independently of the neutrino mass hierarchy.

The physical neutrino masses
are given by $m_3\approx (e^2+f^2)/Y$,
$m_2\approx 4a^2/X$, $m_1\approx (a',b',c')^2/X'$
and the mass ordering of the right-handed neutrino
masses $X',X,Y$ is not determined unless further information
is specified about the Dirac neutrino masses.
In general there are six possible mass orderings of the
three right-handed neutrinos:
\begin{eqnarray}
Y<X<X' \label{1}\\
Y<X'<X \label{2}\\
X<Y<X' \label{3}\\
X'<Y<X \label{4}\\
X'<X<Y \label{5}\\
X<X'<Y \label{6}
\end{eqnarray}
The dominant right-handed neutrino of mass $Y$
(the one mainly responsible for the mass $m_3$) may thus
be the lightest one as in Eqs.\ (\ref{1}, \ref{2}), the intermediate
one as in Eqs.\ (\ref{3}, \ref{4}), or the heaviest one as in
Eqs.\ (\ref{5}, \ref{6}). The neutrino of mass $X'$ is essentially
irrelevant from the point of view of the
light Majorana mass matrix $M_{\nu}$, since the lightest
physical neutrino of mass $m_1$ is approximately zero in the
hierarchical case. Thus $X'$ cannot be determined from
any low energy experiments such as neutrino oscillations
or neutrinoless double beta decay. If $X'$ happens to be the heaviest
right-handed neutrino, as in Eqs.\ (\ref{1}, \ref{3}) then if its mass
is above the GUT scale then it completely
decouples from observable physics. In this case
the three right-handed neutrino model becomes effectively
a two right-handed neutrino model \cite{King:2002nf}.
However even in this
case there is a remaining ambiguity about whether the dominant
right-handed neutrino is the lightest one or the
next-to-lightest one as in Eqs.\ (\ref{1}, \ref{3}).

\subsubsection{Grand Unification}

It is clear that further theoretical input is required in order to elucidate
the nature of the masses and couplings of the right-handed neutrinos. In some
GUT model, one can expect generically that the Dirac neutrino masses are
related to the other quark and charged lepton masses and this additional
information about $M^D_{\nu}$ can then be input into the type I seesaw formula
Eq.\ (\ref{seesawI}) to help to yield information about the right-handed
neutrino mass matrix $M_R$. For example assuming that $M^D_{\nu} \approx M^u$,
the up-type quark mass matrix, and inputting the approximately determined light
Majorana mass matrix, the seesaw formula can then be rearranged to yield
right-handed neutrino masses with a very hierarchical mass spectrum
\cite{afsmirnov}:
\begin{eqnarray}
M_1:M_2:M_3\approx m_u^2:m_c^2:m_t^2~,
\label{naive2}
\end{eqnarray}
which can be compared to the naive expectation for the
physical neutrino masses in Eq.~(\ref{naive}).
Numerically Eq.~(\ref{naive2}) yields the order of magnitude estimates
$M_1\sim 10^5 $ GeV, $M_2\sim 10^{10} $ GeV, $M_1\sim 10^{15} $ GeV,
with an uncertainty of a least one or two orders of magnitude in each
case. In addition there may be special cases
 which completely invalidate these estimates.

In specific GUT models the above expectations can also be very
badly violated. For example in the $SO(10)$  model with
$SU(3)$ family symmetry \cite{King:2001uz},
although the neutrino Dirac mass matrix
is strikingly similar to the up-type quark mass matrix,
a very different pattern of right-handed neutrino masses
emerges:
\begin{eqnarray}
M_1:M_2:M_3\approx \epsilon^6\bar{\epsilon}^3: \epsilon^6\bar{\epsilon}^2:1
\label{naive3}
\end{eqnarray}
where $\epsilon\approx 0.05$, $\bar{\epsilon}\approx 0.15$.
In this model the dominant right-handed neutrino is the lightest
one $Y=M_1$, with $X=M_2$,
while the heaviest right-handed neutrino is decoupled $X'=M_3$
as in Eq.\ (\ref{1}). This model therefore acts effectively as
a two right-handed neutrino model, with the two right-handed
neutrinos being very similar in mass, with interesting implications
for leptogenesis, to which we now turn.

\subsubsection{Leptogenesis}

Leptogenesis and lepton flavor violation are important indicators
which can help to resolve the ambiguity of right-handed neutrino
masses in Eqs.\ (\ref{1}-\ref{6}). In the simplest case of two
right-handed neutrino models, leptogenesis has been well studied
with some interesting results \cite{min,Frampton:2002qc,King:2002qh}.
In general, successful thermal 
leptogenesis for such models requires the mass of the lightest
right-handed neutrino model to be quite high, and generally to exceed the
gravitino constraints, required if supersymmetry is assumed.
Such a strong bound is also at odds with the strong right-handed
neutrino mass hierarchy
expected from GUTs as in Eqs.\ (\ref{naive2}, \ref{naive3}).
In unified theories with type II see-saw, this potential 
problem can be resolved (see e.g.\ \cite{Antusch:2005tu}).

In three right-handed neutrino models with sequential
dominance, if the dominant right-handed neutrino is the lightest
one, then the washout parameter $\tilde{m_1}\sim {\cal{O}}(m_3)$
is rather too large compared to the optimal value
of around $10^{-3}$ eV for thermal leptogenesis.
However, if the dominant right-handed neutrino is either the
intermediate or the heaviest one then one finds
$\tilde{m_1}\sim {\cal{O}}(m_2)$ or arbitrary $\tilde{m_1}$,
which can be closer to the desired value
\cite{Hirsch:2001dg}.

\subsubsection{Sneutrino Inflation}
It has been suggested that a right-handed sneutrino,
the superpartner to a right-handed neutrino, could be
a candidate for the inflaton in theories of cosmological inflation.
This has interesting consequences for the masses and couplings
of the right-handed neutrinos.
For example in the case of chaotic sneutrino inflation,
the mass of the right-handed sneutrino inflaton must be
about $10^{13}$ GeV \cite{Ellis:2003sq},
while in sneutrino hybrid inflation its mass could be considerably lighter
\cite{Antusch:2004hd}.
In both scenarios, the decaying sneutrino inflaton can be responsible
for non-thermal leptogenesis, and can give a reheat temperature
compatible with gravitino constraints providing its Yukawa couplings
are sufficiently small. This typically implies that the
associated right-handed neutrino must be effectively decoupled from the
see-saw mechanism, so that it corresponds to the decoupled
right-handed neutrino of mass $X'$ in sequential dominance discussed above.

\subsubsection{Type II Seesaw Models}

Once the more general type II seesaw framework is permitted
\cite{seesaw2},
then it apparently becomes more problematic to determine the
properties of the right-handed neutrinos which contribute
to $M_\nu$ via the type I part of the seesaw mechanism.
On the other hand, the type II seesaw mechanism provides
the most direct way of raising the neutrino mass scale
to a level that will be observable in neutrinoless double beta
decay. Furthermore, the difficulties of providing consistency of
leptogenesis scenarios with the gravitino bound in supersymmetric
theories, or simply with the strong right-handed neutrino
mass hierarchy expected from GUT models, motivates a more
general type II seesaw framework which can in principle
help resolve some of these difficulties. It has recently been shown
\cite{Antusch:2004re,Antusch:2004xd} how to construct
natural models for partially degenerate neutrinos
by using an $SO(3)$ family symmetry to add a type II contribution
to the light neutrino Majorana mass matrix proportional to the
unit matrix, with large neutrino mixing
originating from sequential dominance.
Compared to the pure type I limit,
the masses of the right-handed neutrinos become
larger if the mass scale of the light neutrinos is increased via the type II
contribution. 
This can also help to resolve the potential conflict between the typical 
predictions for $M_1$  as in Eqs.~(47,48) and thermal leptogenesis 
\cite{Antusch:2005tu}. 
In addition, increasing the neutrino mass scale has
interesting phenomenological consequences, such as a decreasing CP violating phase $\delta$
and a decreasing mixing angle $\theta_{13}$, testable in future experiments.

\subsubsection{Right-Handed Neutrinos in Extended Technicolor}
In the mechanism that has been constructed for producing light neutrinos in
extended Technicolor \cite{nt,lrs,nuf03,ckm}, there are two right-handed
neutrinos.  Interestingly, this mechanism involves a seesaw, but one in which
the relevant Dirac neutrino mass matrix elements are greatly suppressed down to
the level of a few keV, and the Majorana masses are also suppressed, of order
100 MeV to 1 GeV.  The origin of this suppression stems from the fact that the
left- and right-handed chiral components of neutrinos transform differently
under the ETC gauge group. Although the mechanism does involve a seesaw,
it does not involve any GUT-scale masses.  This is clear, since extended
Technicolor models do not contain any such mass scales.  It serves as an
existence proof of how a seesaw mechanism can work with much lower Dirac and
Majorana mass scales than the usual GUT-scale seesaw.


\subsection{Seesaw Neutrino mass and Grand unification}
One of the major ideas for physics beyond the Standard Model is supersymmetric
grand unification (SUSY GUT) \cite{raby}. It is stimulated by a number of
observations that are in accord with the general expectations from SUSY GUTs :
(i) A solution to the gauge hierarchy problem i.e why $v_{\rm wk}\ll M_{\rm
Pl}$; (ii) unification of electroweak, i.e.\
$SU(2)_L\times U(1)_Y$ and strong
$SU(3)_c$ gauge couplings assuming supersymmetry breaking masses are in the TeV
range, as would be required by the solution to the gauge hierarchy; (iii) a
natural way to understand the origin of electroweak symmetry breaking.
Supersymmetric grand unified theories generically predict proton decay via
dimension-five operators, which typically give large branching ratios for modes
like $p \to \bar\nu_\mu K^+$.  The current lower limits on proton decay modes
place significant constraints on these theories and probably rule out a number
of simpler SUSY GUT models \cite{su5out}. Nevertheless, the idea of grand
unification is so attractive that we will proceed on the basis that appropriate
modifications allow supersymmetric GUTs to evade current nucleon decay limits.

Gauge coupling unification leads to a unification scale of about $10^{16}$ GeV
and simple seesaw intuition leads to a seesaw scale of $10^{15}$ GeV to fit
atmospheric neutrino data. This suggests that seesaw scale could be the GUT
scale itself; thus the smallness of neutrino mass could go quite well with the
idea of supersymmetric grand unification. However, in contrast with the items
(i) through (iii) listed above, the abundance of information for neutrinos
makes it a highly nontrivial exercise to see whether the neutrino mixings
indeed fit well into SUSY GUTs.  In turn, the freedom in constructing realistic
GUT models allows many different ways to explain current neutrino
observations. Thus even though neutrino mass is a solid evidence for physics
beyond the Standard Model, the true nature of this physics still remains
obscure. The hope is that the next round of the experiments will help to narrow
the field of candidate theories a great deal.

To see how this is likely to come about, the first point is the choice of
the grand unification group. Even though attempts to implement the
seesaw mechanism using an extension of the
SU(5)  with the addition of a right-handed neutrinos have been made, a
more natural GUT gauge group from the point of view of neutrino mass is
SO(10) since its basic spinor representation contains the right-handed
neutrino automatically along with the other fifteen fermions of the
Standard Model (for each family). Thus in some sense one could argue that
small neutrino masses have already chosen $SO(10)$ GUT as the most natural
way to proceed beyond the Standard Model. $SO(10)$ has therefore rightly
been the focus of many attempts to understand neutrino mixings.
It turns out that within the $SO(10)$ SUSY GUTs there are many
ways to understand large mixings. We outline below only the major
differences among the different ideas.
The hope is that they differ in their predictions
sufficiently so that they can be tested by planned experiments.

One of the features that distinguishes $SO(10)$ from SU(5) is the presence
of local $B-L$ symmetry as a subgroup and the $SO(10)$ models divide into two
classes depending on whether $B-L$ symmetry is broken by a {\bf 16} Higgs
field or an {\bf 126}. In the first case the right-handed neutrino mass
necessarily arises out a nonrenormalizable coupling whereas in the second
case it arises from a renormalizable one. Secondly, the breaking of $B-L$ by
{\bf 16 } Higgs necessarily leads to low energy MSSM with R-parity
breaking so that the model cannot have cold dark matter without additional
assumptions, whereas {\bf 126} breaking of $B-L$ preserves R-parity at low
energies so that the low energy MSSM that derives from such an $SO(10)$ has
a natural dark matter candidate i.e.\ the lightest SUSY particle.

As noted in the Introduction, the $SO(10)$ model has in general the type II
seesaw formula for neutrino masses which can reduce to type I for some
range of parameters. For instance, in the {\bf 16} based models,
the first term in type II seesaw formula is negligible and therefore the
neutrino masses are dictated by type I seesaw formula. In contrast
in {\bf 126} Higgs models, the neutrino mass can be given either by the
first term or the second term in the type II seesaw formula or both.

\subsubsection[A minimal {\bf 126}-based $SO(10)$ model]{A minimal {\bf
 126}-based $\boldsymbol{SO(10)}$ model}

As mentioned, in $SO(10)$ models where a {\bf 126} Higgs breaks $B-L$
symmetry, the right-handed neutrino masses can arise from renormalizable
couplings. A minimal model of this type based on a single {\bf 10} and a
single {\bf 126} field has a number of attractive features \cite{babu,lee}.
Since the {\bf 126} field also contributes to charged fermion masses
through the MSSM doublets in it, this model unifies the flavor structure
in the quark and the neutrino sector thereby increasing the predictivity
of the model in the neutrino sector.  In fact in the absence of CP
violation, the model has only 12 free parameters all of which are
determined by the quark masses and mixings along with the charged lepton
masses. As a result all mixings and masses are predicted by the model
up to an overall scale. It has been shown that if one uses the type I
seesaw formula, the model fails to reproduce the observed solar mixing
angle and also the solar mass difference squared and is therefore ruled
out \cite{fukuyama}. It has however been shown recently that if one uses the
type II seesaw mechanism with the first term dominating the neutrino mass
matrix, the large mixings come out due to b-tau mass convergence in a very
natural manner \cite{bajc}. In particular, an interesting prediction of
this model is that $\theta_{13}\simeq 0.18$ making it quite accessible to
the next generation of experiments such as long-baseline, off-axis and the
reactor experiments.

\subsubsection{{\bf 16}-based models}
The main characteristic of the $SO(10)$ models where a {\bf 16} Higgs breaks
$B-L$ is that right-handed neutrino masses arise from nonrenormalizable
couplings in the superpotential, the implicit assumption being that there
is a high-scale theory (perhaps string theory or a renormalizable high
scale theory with heavier fields) that below the heavy scale leads to this
version of $SO(10)$. This means that without additional symmetry
restriction, there are more parameters than the physical inputs. Often in
these models symmetries that tend to explain quark mixings restrict the
number of couplings somewhat and one can make predictions in the neutrino
sector. There exist several interesting examples of this kind of
models \cite{so1016}. Several of these models tend to give values for
$\theta_{13}$ which are much below the range that can be probed by the
next generation of planned experiments. We give a very small sample of the
different predictions for $\theta_{13}$ in models with both {\bf 16} and
{\bf 126} in  Table~\ref{tab:PredT13}.

\begin{table}
\centering
\begin{tabular}{|c||c|}\hline
Model & $\theta_{13}$ \\ \hline
{\it {\bf 126}} based models & \\
Goh, Mohapatra, Ng & 0.18 \\
Chen, Mahanthappa & 0.15 \\ \hline
{\it {\bf 16}} based models & \\
Babu, Pati, Wilczek & 0.0005 \\
Albright, Barr & 0.014 \\Ross, Velasco-Sevilla & 0.07\\
Blazek, Raby, Tobe & 0.05 \\
\hline
\end{tabular}
\caption{The table lists some typical predictions for
$\theta_{13}$ in different $SO(10)$ models and shows how the next
generation of experiments can narrow the field of possible
SO(10) unification models.}
\label{tab:PredT13}
\end{table}


\subsubsection[Summary of what we can learn about $SO(10)$]{Summary of what we
can learn about $\boldsymbol{SO(10)}$}

A review on different neutrino mass models based on $SO(10)$ can be found in
Ref.~\cite{Chen:2003zv}. From these models, we learn that

\begin{enumerate}

\item First a very generic  prediction of all $SO(10)$ models is that the
neutrino mass hierarchy is normal. The basic reason for this is the quark
lepton symmetry inherent in the model, which tends to make the neutrino
Dirac mass to be of similar hierarchy as the quarks, which via seesaw
mechanism implies normal hierarchy for neutrinos. leading to $\Delta
m^2_{23}\geq 0$. Again this is a result that can be probed in the
long-baseline oscillation or neutrinoless double beta decay
experiments;

\item The second point about the $SO(10)$ models is that they make definite
predictions about the mixing angle $\theta_{13}$ as given in Table~\ref{tab:PredT13}
and often for the other mixing angles.
The planned experiments will therefore considerably narrow the field of
viable $SO(10)$ models through their measurement or upper limit on these
mixing parameters.

\end{enumerate}

\subsubsection{\label{sec:muchun}Implications of Models with Spontaneous CP Violation}

Relations between leptogenesis and CP violation in low energy processes
generally  do not exist due to the presence of un-known mixing angles and
phases in  the heavy neutrino sector, as discussed in
Sec.~\ref{phases}. In models with spontaneous CP violation, all Yukawa
coupling constants are real.
CP violation occurs due to the presence of the phases in the expectation
values of the scalar fields, which break the gauge symmetry spontaneously.
Recently it has been shown that~\cite{Chen:2004ww} in the minimal
left-right $SU(2)_{L} \times SU(2)_{R}$ symmetric model~
\cite{moh}
with spontaneous CP violation, there exist very pronounced relations
between the CP violation in low energy processes, such as neutrino
oscillation and neutrinoless double beta decay, and leptogenesis, which
occurs at a  very high energy scale. The minimal left-right symmetric
model contains a bi-doublet and a pair of triplet Higgses. Using the
gauge degrees of freedom, one can rotate away all but two phases present
in the expectation values of the scalar fields. Thus there are only two
intrinsic phases, the relative phase between the two vevs in the
bi-doublet and that between the left- and right-handed triplets, to
account for {\it all} CP violation in the quark sector and in the lepton
sector.
The relative phase between the two vevs in the bi-doublet is responsible
for CP violation observed in the quark sector,
while CP violation in the lepton sector dominantly comes from the
relative phase between the vevs of the two triplet Higgses. The relative
phase between the two vevs in the bi-doublet appears in the lepton
sector only at the sub-leading order due to the large hierarchy in the
bi-doublet vevs required by a realistic quark sector. As a result, the
relation between CP violation in the quark sector and CP violation in the
lepton sector is rather weak.
Due to the left-right parity, the RH and LH neutrino Majorana mass terms
are proportional to each other, which further reduces the unknown
parameters in the model.  In this model, both leptogenesis and the
leptonic Jarlskog invariant are proportional to the sine of the relative
phase between the vevs of the two triplet Higgses.
Using the experimentally measured neutrino oscillation parameters as
inputs, to obtain sufficient amount of lepton number asymmetry, the
leptonic Jarlskog invariant has to be larger than
$\sim 10^{-5}$. As the Type-II seesaw mechanism is at work, the hierarchy
in the heavy neutrino sector required to obtain the observed neutrino
oscillation parameters is very small, leading to a heavier mass for the
lightest RH neutrino, compared to the case utilizing the Type-I seesaw
mechanism. As a result, the requirement that the decay of the lightest RH
neutrino is out-of-equilibrium in thermal leptogenesis can be easily satisfied.

Similar attempts have been made to induce spontaneous CP violation  from
a single source.
In one such attempt SM is extended by a singlet scalar field which
develops a complex VEV which breaks CP
symmetry~\cite{Branco:2003rt}.
Another attempt assumes that there is one complex VEV of the field which
breaks the $B - L$ symmetry in SO(10)~\cite{Achiman:2004qf}.
Unlike in the minimal left-right symmetry model described
above~\cite{Chen:2004ww},
there is no compelling reason why all other vevs have to be real in these models.

\subsection{\label{sec:RGE}Renormalization group evolution
of neutrino parameters}

Neutrino masses and mixing parameters are subject to the renormalization
group (RG) evolution or running, i.e.\ they depend on energy.
As theoretical predictions for these quantities typically arise from
models at high energy scales such as the GUT scale, this implies
that in general RG corrections have to be included in the testing of model
predictions. In the case of leptonic mixing angles and CP phases the
changes can be large for partially or nearly degenerate neutrino masses.
On the other hand, strongly hierarchical masses bring about very small RG
corrections for the mixing angles, while the running of the mass squared
differences is sizable even in this case.

\subsubsection{Running masses, mixings and CP phases below the seesaw
scale}
\label{sec:AnalyticalApproximations}

At energies below the seesaw scale $M_{\rm R}$, the masses of the light
neutrinos can be described in a rather model-independent way by an
effective dimension 5 operator if they are Majorana particles. The RG
equation of this operator in the SM and MSSM
\cite{Babu:1993qv,Chankowski:1993tx,Antusch:2001ck,Antusch:2001vn,Antusch:2002ek}
leads to differential equations for the energy dependence of the mass
eigenvalues, mixing angles and CP phases
\cite{Chankowski:1999xc,Casas:1999tg,Antusch:2003kp,Mei:2003gn,Luo:2005sq}.
Up to ${\mathcal{O}} (\theta_{13})$ corrections, the evolution of the
mixing angles is given by \cite{Antusch:2003kp}
\begin{eqnarray}
\nonumber
\dot{\theta}_{12} &=&
        -\frac{C y_\tau^2}{32\pi^2} \,
        \sin 2\theta_{12} \, s_{23}^2\,
        \frac{
      | {m_1}\, e^{i \varphi_1} + {m_2}\, e^{i  \varphi_2}|^2
     }{\Delta m^2_{\odot} }\;,
\\ \label{eq:RGEthetaij}
\dot{\theta}_{13} &=&
        \frac{C y_\tau^2}{32\pi^2} \,
        \sin 2\theta_{12} \, \sin 2\theta_{23} \,
        \frac{m_3}{\Delta m^2_{\rm A} \left( 1+\zeta \right)}
        \times I(m_i, \varphi_i, \delta)
\;,
\\ \nonumber
     \dot{\theta}_{23} &=&
        -\frac{C y_\tau^2}{32\pi^2} \, \sin 2\theta_{23} \,
        \frac{1}{\Delta m^2_{\rm A}}
       \left[
         c_{12}^2 \, |m_2\, e^{i \varphi_2} + m_3|^2 +
         s_{12}^2 \, \frac{|m_1\, e^{i \varphi_1} + m_3|^2}{1+\zeta}
        \right]
\;,
\label{eq:Theta23Dot}
\end{eqnarray}
where the dot indicates the differentiation $d/ dt=\mu\,d/d\mu$ ($\mu$ being
the renormalization scale), $s_{ij}:=\sin\theta_{ij}$,
$c_{ij}:=\cos\theta_{ij}$, $\zeta:=\Delta m^2_{\odot}/\Delta m^2_{\rm A}$,
$C=-3/2$ in the SM and $C=1$ in the MSSM, and   $I(m_i, \varphi_i,
\delta):=\left[ m_1 \cos(\varphi_1-\delta) - \left( 1+\zeta \right) m_2 \,
\cos(\varphi_2-\delta) - \zeta m_3 \, \cos\delta \right]$. $y_\tau$ denotes the
$\tau$ Yukawa coupling, and one can safely neglect the contributions coming
from the electron and muon. For the matrix $K$ containing the Majorana phases,
we use the convention $K = \mbox{diag}\, (e^{-i \varphi_1/2},e^{-i
\varphi_2/2},1)$ here.
For a discussion of RG effects in the case of exactly degenerate
neutrino masses, where the above expressions cannot be applied, see e.g.\
\cite{Ellis:1999my,Casas:1999tp,Casas:1999ac,Adhikari:2000as,%
Joshipura:2000ts,Chun:2001kh,Joshipura:2002xa,Joshipura:2002gr,Singh:2004zu}.
>From Eqs.~\eqref{eq:RGEthetaij} one can easily understand the typical
size of RG effects as well as some basic properties. First, in the SM and in
the MSSM with small $\tan\beta$, the RG evolution of the mixing angles is
negligible due to the smallness of the $\tau$ Yukawa coupling. Next, the RG
evolution of the angles is the stronger the more degenerate the mass spectrum
is. For a strong normal mass hierarchy, it is negligible even in the MSSM with
a large $\tan\beta$, but for an inverted hierarchy a significant running is
possible even if the lightest neutrino is massless
\cite{Haba:1999fk,Miura:2000bj}.
Furthermore, non-zero
phases $\delta$, $\varphi_1$ and $\varphi_2$ can either damp or enhance
the running. For instance, the running of $\theta_{12}$ can be damped by non-zero
Majorana phases \cite{Balaji:2000gd,Haba:2000tx,Chankowski:2001mx}.
Typically, $\theta_{12}$ undergoes the
strongest RG evolution because the solar mass squared difference is much
smaller than the atmospheric one. Finally, in the MSSM, $\theta_{12}$ runs from
smaller values at high energies to larger values at low energies
\cite{Miura:2002nz}.

The RG equations for the CP phases
\cite{Casas:1999tg,Antusch:2003kp,Mei:2003gn,Luo:2005sq}
show that their changes are proportional to $1/\Delta m^2_{\odot}$.
Therefore, whenever the mixing angles run sizably, the same happens for
the CP phases. This is very important for the relation between the phases
relevant for high-energy processes like leptogenesis and those appearing
in neutrino oscillations and neutrinoless double beta decay.
The evolution of the Dirac CP phase can be especially drastic for a small
CHOOZ angle, since $\dot\delta$ contains
a term proportional to $1/\theta_{13}$.  It is also possible to generate
a non-zero value of this phase radiatively if at least one of the
Majorana phases is non-zero \cite{Casas:1999tg}.  An exception is the
CP conserving case where all phases are $0$ or $\pi$ and do not change
with energy.

Finally, the neutrino masses always change significantly with energy due
to flavor-blind terms in the RG equations which contain large
quantities like gauge couplings and the top Yukawa coupling.  For
strongly hierarchical masses and small $\tan\beta$, these terms
dominate, so that the masses experience a common rescaling which is
virtually independent of the mixing parameters \cite{Chankowski:2001mx}.

Radiative corrections for Dirac neutrino masses have also been studied
\cite{Chiang:2000um,Lindner:2005as}. Roughly speaking, the RGEs for the
Dirac case are obtained from Eqs.~\eqref{eq:RGEthetaij} by averaging
over the Majorana phases.

\subsubsection{Details of the running in seesaw models}
\label{sec:RunningAbove}

In order to obtain precise results, one has to go beyond the simple
approximations listed above and solve the RG equations numerically.
This involves solving a rather complex system of coupled differential
equations, as all parameters of the theory have to be evolved from
high to low energy.

A further complication arises in seesaw models with
heavy singlet neutrinos which are in general non-degenerate in mass.
The running above their mass thresholds is typically at least as important
as the evolution below both in the SM and in the MSSM unless the
neutrino Yukawa couplings are tiny
\cite{Casas:1999tp,Casas:1999ac,King:2000hk,Antusch:2002rr,Antusch:2002hy,%
Antusch:2002fr,Mei:2004rn,Ellis:2005dr}.
This part of the RG evolution depends on many parameters of the model.
An analytic understanding has been obtained only recently
\cite{Antusch:2005gp,Mei:2005qp}.
If the singlet masses are non-degenerate, one can calculate the
evolution of the neutrino mass parameters by considering
a series of effective theories arising from integrating out the singlets
successively at the respective thresholds \cite{King:2000hk,Antusch:2002rr}.
In general, it is not a good approximation to integrate out
all singlets at the same energy scale, since the threshold corrections
can be very large.

\subsubsection{Implications for model building}

\begin{figure}[htb]
\centering
\includegraphics[width=220pt]{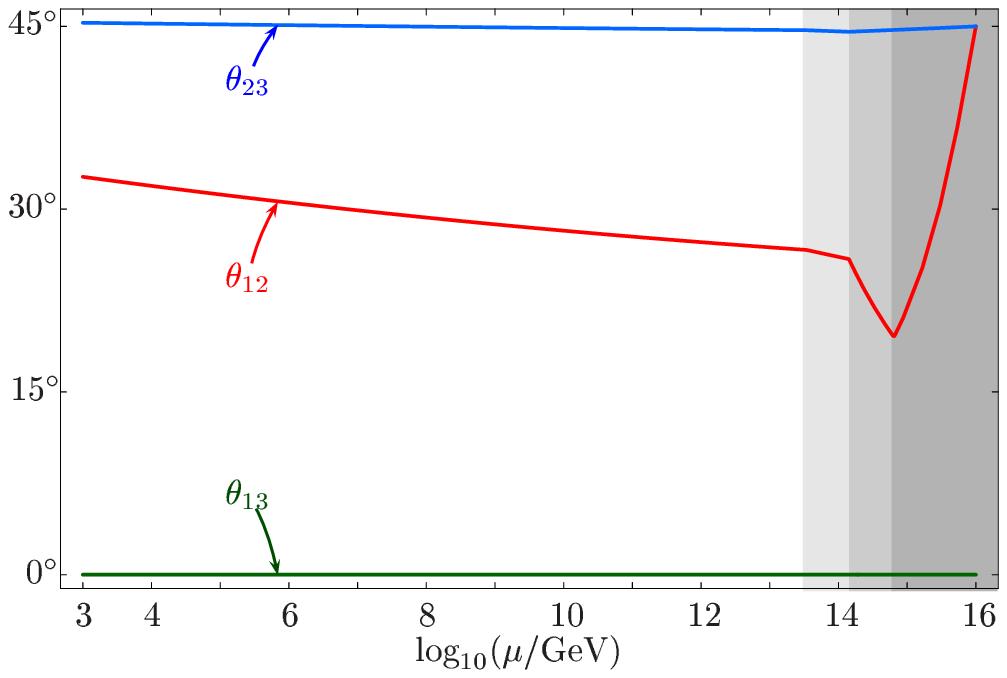}
\hfil
\includegraphics[width=220pt]{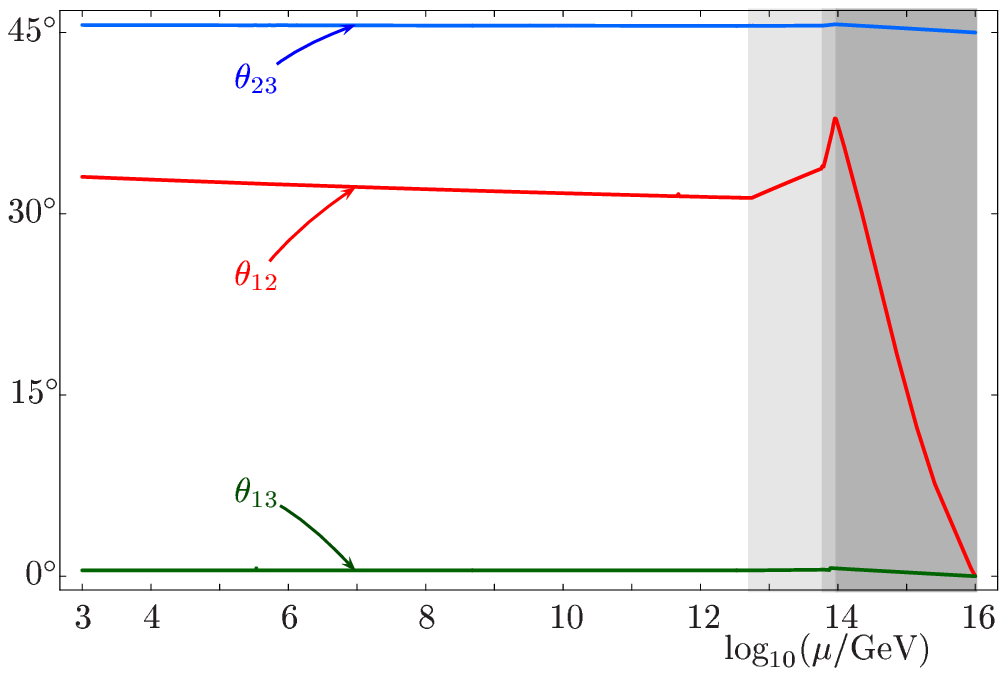}
\vspace*{-4mm}
\caption{Examples for the RG evolution of the lepton mixing angles
from the GUT scale to the SUSY-breaking scale (taken to be
$\approx 1$ TeV)
in the MSSM extended by 3 heavy singlets (right-handed neutrinos)
{\protect \cite{Antusch:2002hy,Antusch:2002fr}}.
The masses of the lightest neutrinos for these examples are around
$0.05\,$eV.
The figures illustrate how the large but non-maximal value of the solar
mixing angle $\theta_{12}$ is reached by RG running if one starts
with bimaximal lepton mixing or with vanishing solar mixing at the GUT
scale.  The kinks in the plots
correspond to the mass thresholds at the seesaw scales,
where the heavy singlets are successively integrated out.
The gray-shaded regions mark the various effective theories between the
seesaw scales.}
\label{fig:lindner3}
\end{figure}

As discussed above, predictions of high-energy mass models can differ
substantially from low-energy experimental results due to the running.
Therefore, RG corrections have
to be included in the analysis. The RG evolution also opens up new
interesting possibilities for model building, like the radiative
magnification of mixing angles. In particular, small or vanishing
solar mixing at high energy can be magnified to the observed large
mixing at low energy (see e.g.\
\cite{Dutta:2002nq,Antusch:2002fr,Bhattacharyya:2002aq,Hagedorn:2004ba}).
Vice versa,
the large but non-maximal solar mixing $\theta_{12}$ can also be reached
starting  from bimaximal mixing at the GUT scale
\cite{Antusch:2002hy,Miura:2003if,Shindou:2004tv}
(for examples, see Fig.~\ref{fig:lindner3}).
It is, however, important to stress that large mixing is no fixed point under
the RGE in the usual see-saw framework.
It has been observed that in SUSY models large mixing can be a fixed point for
different (i.e.\ non-seesaw) types of neutrino mass operators
\cite{Casas:2002sn,Casas:2003kh}.
In addition, the small neutrino mass squared differences can be produced
from exactly degenerate neutrino masses at high energy (see e.g.\
\cite{Chankowski:2000fp,Chen:2001gk,Babu:2002dz,Joshipura:2002xa,%
Joshipura:2002gr,Singh:2004zu}),
if the neutrino masses are nearly degenerate.
For further specific models where the RG evolution is relevant for neutrino
masses and mixings see, for example,
\cite{Haba:2000rf,Kuo:2000ig,GonzalezFelipe:2001kr,Parida:2002gz,%
Mei:2003gn,Antusch:2004xd}.

\subsubsection{High-scale mixing unification and large mixings}

Another question one can ask is whether starting
with small mixing angles at the seesaw scale (as would be naively expected
in models with quark-lepton unification) can one get large mixings at the
weak scale due to RG extrapolation. In a specific model
where neutrino masses are quasi-degenerate, starting with neutrino mixings
that are equal to quark mixings at the GUT scale i.e. $\theta_{12}\simeq
V_{us}$, $\theta_{23}\simeq V_{cb}$ and $\theta_{13}\simeq V_{ub}$, one can
indeed get mixing angles at the weak scale which are consistent with
present observations as shown in Fig.~\ref{mpr1} below. We have chosen
$\tan\beta=55$ in this calculation. An interesting point is that this
mechanism works only if the common mass of the neutrinos is bigger than
$0.1$ eV, a prediction which can easily be tested in the proposed
neutrinoless double beta decay experiments
\cite{Balaji:2000gd,Balaji:2000au,Balaji:2000ma,Babu:2002ex,Mohapatra:2003tw}.
\begin{figure}[htb]
\centering
\epsfxsize=8.5cm
\epsfbox{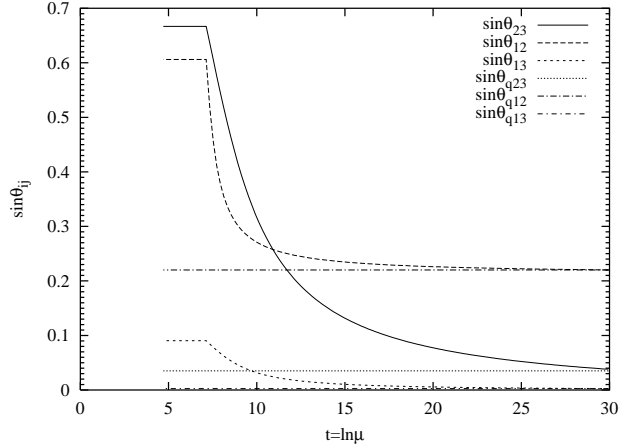}
\caption{Evolution of small quark-like mixings at the seesaw scale to
bi-large
neutrino mixings at low energies for the seesaw scale $M_R=10^{13}$ GeV
with
$\tan\beta=55$, $M_{\rm SUSY}= 1$ TeV. The solid, long-dashed and
short-dashed lines represent
$\sin\theta_{23}$, $\sin\theta_{13}$, and $\sin\theta_{12}$, respectively.
 The evolution of the sines of quark mixing angles,
$\sin\theta_{qij}$($i,j=1,2,3$),
is presented by almost horizontal lines.}
\label{mpr1}
\end{figure}

\subsubsection[Deviations of $\theta_{13}$ from 0 and of
$\theta_{23}$ from $\pi/4$ due to RG effects]{Deviations of
$\boldsymbol{\theta_{13}}$ from 0 and of $\boldsymbol{\theta_{23}}$ from
$\boldsymbol{\pi/4}$ due to RG effects}
\label{sec:RGtheta13andtheta23}

At present, observations are compatible with $\theta_{13}=0$ and
$\theta_{23}=\pi/4$.  New experiments are being planned to lower the
limits on deviations from these values.  Additional motivation for this
kind of measurements is provided by the RG: although it is possible that
a symmetry produces an exactly vanishing $\theta_{13}$ and exactly
maximal atmospheric mixing, this symmetry would typically operate at a
high scale, and therefore its predictions would be subject to the RG
evolution.  Hence, without fine-tuning, one expects non-zero values
of $\theta_{13}$ and $\theta_{23}-\pi/4$ at low energy
\cite{Antusch:2003kp,Antusch:2004yx,Mei:2004rn,Singh:2004zu}.
For example, in the MSSM one finds a shift
$\Delta\!\sin^2 2\theta_{13} > 0.01$ for a considerable parameter range,
i.e.\ one would expect to measure a finite value of $\theta_{13}$
already in the next generation of experiments \cite{Antusch:2003kp}.
On the other hand, there are special configurations of the parameters,
especially the phases, where RG effects are suppressed. Furthermore,
there may be symmetries which stabilize some mixing angle completely
against radiative corrections. For instance, for an inverted hierarchy
with $m_3=0$, $\theta_{13}=0$ is stable under the RG \cite{Grimus:2004cj}
(see also Eqs.~\eqref{eq:RGEthetaij}).
Hence, if future precision measurements do not find $\theta_{13}$ and
$\theta_{23}-\pi/4$ of the size of the generic RG change, one can
restrict parameters,
or even obtain evidence for a new symmetry.

\subsubsection{Implications for leptogenesis}

As has been discussed in Sec.~\ref{sec:Leptogenesis}, the requirement of
successful baryogenesis via leptogenesis places an upper bound on the mass of
the light neutrinos. It is important to note that in order to relate
constraints on the neutrino mass spectrum coming from physics at $M_1$ to
observation, one has to take into account radiative corrections.  It turns out
that there are two effects operating in opposite directions
\cite{Antusch:2003kp,gian,Buchmuller:2004nz,Barbieri:1999ma} which
partially cancel each other:
since the mass scale is increasing, the washout driven
by Yukawa couplings is stronger. On the other hand, larger $\Delta m^2$s allow
for a larger decay asymmetry. Taking into account all these effects, one finds
that the upper bound on the neutrino mass scale becomes more  restrictive
\cite{Antusch:2003kp,gian,Buchmuller:2004nz}. The RG evolution, together with
thermal effects or spectator processes \cite{Buchmuller:2001sr}, gives rise to
the most important corrections to the mass bound \cite{Buchmuller:2004nz}.

\section{Non-Standard Neutrino Interactions}

\subsection{Neutrino magnetic moments}\label{mm}

Once neutrinos are massive, they can have transition magnetic and electric
magnetic dipole moments and, in the Dirac case, also diagonal magnetic and
electric dipole moments \cite{mmom1,mmom2}, \cite{LeeShrock:77},
\cite{leftright}--\cite{mmom2004}.  The lepton-number conserving (diagonal and
transition) magnetic and electric dipole moment operators are given by
$(1/2)\bar \nu_i\sigma^{\mu\nu}\nu_j F_{\mu\nu}$ and $(-i/2)\bar
\nu_i\sigma^{\mu\nu}\gamma_5 \nu_j F_{\mu\nu}$.  Analogous $\Delta L=2$
expressions hold (with $i \ne j$ because of their antisymmetry under $i
\leftrightarrow j$) for Majorana neutrinos.  Therefore, a magnetic or electric
dipole moment always connects one species of neutrino with another.  These
moments are defined for mass eigenstates.  In the case of a Dirac mass
eigenstate, one sees that the operator connects a left-handed
electroweak-doublet neutrino to a right-handed electroweak-singlet (sterile)
neutrino.  In the case of Majorana mass eigenstates, the (transition) dipole
moment operators connect two neutrino fields of the same chirality.  The two
have fundamentally different physical implications.  We have reviewed some
basic properties of these magnetic and electric dipole moments above.

Neutrino magnetic moments can be directly measured in terrestrial experiments
using the neutrino beam from the Sun as in Super-K \cite{superK} or with
neutrinos from close by nuclear reactors as in the MUNU \cite{munu} and in the
Texono \cite{texono} experiments. These experiments have put upper bounds of
the order of $10^{-10} \mu_B$ on the effective neutrino magnetic moment
(defined below). It is interesting that in models involving right-handed
charged currents the diagonal and transition neutrino magnetic moments
\cite{mmom1,mmom2,ms77,LeeShrock:77,leftright,rs82} are not suppressed by
the neutrino mass (as in models with just the Standard Model interactions
\cite{fs}) and could thus be somewhat larger.  However, in general, the same
interactions that can enhance neutrino magnetic moments can give corrections
enhancing neutrino masses, so in a particular model one must be careful to
avoid excessive loop contributions to the latter.  The Borexino prototype
detector has recently been utilized to put a bound of $|\mu_{eff}|_{MSW} < 5.5
\times 10^{-10} \mu_B$ at $90 \%$ C.L. using the elastic scattering of
electrons by the solar $pp$ and $^7Be$ neutrinos \cite{borexino-nu}. At these
sub-MeV energies, the solar neutrino beam contains roughly equal proportion of
$\nu_1$ and $\nu_2$ ($P_1=P_2=0.5$).  As a result, for the same $\nu_{eff}$ the
bounds on $\mu_{11}, \mu_{13}$ would be much better in experiments utilizing
the $pp, ^7Be$ neutrinos compared to the bounds that can be obtained by using
the $^8 B$ neutrinos (which are predominantly $\nu_2$).

Neutrinos with non-zero magnetic moments contribute to the elastic
scattering of electrons in water Cerenkov detectors
\cite{vogel,joshi,grimus}. The effective neutrino magnetic moment
$\mu_{eff}$  (we neglect the contribution
of the electric dipole moment here) responsible for the scattering event
$\nu_i + e^-
\rightarrow \nu_j+e^-$ is proportional to the incoherent sum of
outgoing neutrino states $\nu_j$ as follows:
\begin{equation}
|\mu_{eff}|^2=  \sum_j \big|  \sum_i A_i(L) \mu_{ij} \big|^2
\end{equation}
where $A_i(L)$ is the probability amplitude of a neutrino produced
as a flavor eigenstate (lets say $\nu_e$ or $\bar \nu_e$) to be in
the $i'th$ mass eigenstate on propagating over the source-detector
distance $L$. For vacuum oscillations $A_i(L)=U_{ei}~\exp(-i E_i
L)$ and the effective magnetic moment depends upon the $\Delta m^2$
and the mixing angles as follows
\begin{eqnarray}
|\mu_{eff}|^2_{VO}&=& c_{12}^2 ~(\mu_{11}^2 + \mu_{12}^2 +
\mu_{13}^2) + s_{12}^2 ~(\mu_{21}^2 + \mu_{22}^2 +
\mu_{23}^2)\nonumber \\ &+& 2 c_{12} ~s_{12}~ (\mu_{11} \mu_{21} +
\mu_{12} \mu_{22}+ \mu_{13} \mu_{23}) ~\cos \left( \frac{\Delta
m^2_{12} L}{2 E_\nu}\right) \label{muvo}
\end{eqnarray}
In the expression for $|\mu_{eff}|^2_{VO}$ above  we have assumed that
$U_{e3}$ is negligibly small and the atmospheric mixing angle is
maximal ($s_{23}^2=1/2$). We have also dropped CP violating
phases. In MUNU \cite{munu} and Texono \cite{texono} , $\bar
\nu_{e}$ from nuclear reactors were detected by the elastic
scattering with $e^-$. The source-detector distance is small
($L=18 m $ in MUNU and $L=28 m$ in TEXONO) compared to the $\nu_1
-\nu_2$ oscillation length so that the Cosine term in (\ref{muvo})
is unity. The magnetic moment matrix $\mu_{ij}$ is symmetric for
Dirac neutrinos and anti-symmetric if the neutrinos are Majorana.
It is clear from (\ref{muvo}) that there exists a possibility that
there may be a cancellation between the last (interference) term
which could be negative and the first two terms. So  experimental
upper bounds on $|\mu_{eff}|_{VO}$ which is $0.9 \times
10^{-10}\mu_B$ (MUNU) \cite{munu} and $1.3 \times 10^{-10} \mu_B$
(Texono) (both at $90\%$ C.L.) do not constrain the elements of the
$\mu_{ij}$ matrix without making added assumptions that there is
no cancellation between the different terms in (\ref{muvo}).

This problem does not arise for solar neutrinos as the interference
term averages to zero since $2 E_{\nu}/\Delta m_{12}^2 <<
L_{earth-sun}$. For the solar neutrinos the expression for
$|\mu_{eff}|^2$ reduces to a sum of two positive definite
quantities
\begin{eqnarray}
|\mu_{eff}|^2_{MSW}= P_1 ~(\mu_{11}^2 + \mu_{12}^2 + \mu_{13}^2) +
P_2 ~(\mu_{21}^2 + \mu_{22}^2 + \mu_{23}^2)
 \label{mumsw}
\end{eqnarray}
where $P_1=|A_{e1}(L)|^2$ and $P_2=|A_{e2}(L)|^2=1-P_1$ are the
probabilities of the solar neutrinos to be in the mass eigenstate
$\nu_1$ and $\nu_2$ respectively at the earth. The recent upper
bound $|\mu_{eff}|_{MSW} < 1.1 \times 10^{-10} \mu_B$ at $90 \%$
C.L. established by Super-Kamiokande \cite{superK} can be
translated into bounds on individual elements of  $\mu_{ij}$
without extra assumptions. The $^8B$ neutrinos which are detected
by electron scattering at Super-K are predominantly $\nu_2$ state
($P_2 =0.94$ and $P_1 =0.06$). The Super-K bound on
$|\mu_{eff}|_{MSW}$ therefore implies $|\mu_{12}| < 1.1 \times
10^{-10} \mu_B$ ; $ |\mu_{22}|,\,\, |\mu_{23}| < 1.13 \times 10^{-10}
\mu_B$ and $|\mu_{11}|,\,\, |\mu_{13}| < 4.49 \times 10^{-10} \mu_B$.

It is also possible to put bounds on $\mu_{ij}$ from SNO-NC data
using the fact that neutrinos with non-zero magnetic
moments can dissociate deuterium \cite{grifols} in addition to the
weak neutral currents. The bounds established from SNO-NC data do
not depend upon the oscillation parameters unlike in the case of
Super-K. However the bounds are poorer  due to the large
uncertainty in our theoretical knowledge of the theoretical $^8B $
flux from the sun \cite{bahcall1}.

We will see in a subsequent section (the one on extra dimensions) that the
effective magnetic moment of the neutrinos can get substantially enhanced
in a certain class of extra dimensions models. Searching for $\mu_{\nu}$
can therefore be used to put limits on theories with extra dimensions.


\subsection{Flavor changing and conserving nonstandard neutral
current interactions}

The latest results of neutrino oscillation experiments indicate that the
conversion mechanism between different neutrino flavors is driven by a
non-vanishing mass difference between mass eigenstates together with large
mixing angles between families.  These analyses are done supposing that no
non-standard neutrino interactions (NSNI) are present.  In the
presence of electroweak-doublet and electroweak-singlet neutrinos, the neutral
weak current is, in general, nondiagonal in mass eigenstate neutrino fields
 \cite{LeeShrock:77,valle80}.  This is
the same type of nondiagonality in the neutral weak current that was present in
the original Weinberg electroweak model before the advent of the
Glashow-Iliopoulos-Maiani (GIM) mechanism.  It will be recalled that in this
original Weinberg model the $d$ and $s$ quarks were assigned to a left-handed
SU(2)$_L$ doublet $(u,d \cos\theta_C + s \sin\theta_C)_L$ and a left-handed
SU(2)$_L$ singlet $-d \sin\theta_C + s \sin\theta_C$.  The necessary condition
for the diagonality of the neutral weak current is that all of the fermions of
a given charge and chirality must transform according to the same weak $T$ and
$T_3$.  Alternatively, NSNI in the form of
nondiagonal couplings
between different neutrino flavor eigenstates \cite{LW78}
and/or neutrino flavor-diagonal but flavor
nonsymmetric couplings, can exist \cite{GMP91}.
Including NSNI can modify the
characteristics of neutrino conversion, and in general large values of NSNI
parameters worsen the quality of the fit to data.  We summarize here the
present limits that can be obtained to NSNI parameters, using the result of
neutrino oscillation experiments.

\subsubsection{Atmospheric neutrinos}

As repeatedly mentioned, atmospheric neutrino data are well
described by the oscillation driven
by one mass scale, $\Delta m^2_{32}$, and with maximal mixing between
second and third families.
One important prediction for these numbers is that the high-energy neutrino
events that generate the through-going muon data are well described together
with the low-energy neutrino events, due to the energy dependence of the
Hamiltonian that describes the neutrino evolution.

Assuming a non-vanishing NSNI acting together with mass and mixing, the
solution to the atmospheric neutrino discrepancy can be spoiled if the NSNI
parameters have too large values. This happens because the NSNI entries in the
Hamiltonian that describes the neutrino evolution are energy independent
\cite{GMP91}. Since
a simultaneous explanation of low-energy and high-energy neutrino events
requires a strong energy dependence in the $\nu_\mu,\nu_\tau$ conversion
probability, inclusion of energy independent terms in the Hamiltonian tends to
decrease the quality of the theoretical predictions fit to atmospheric neutrino
data.

The NSNI can be parametrized as a relative strength of such interactions
to the $ \epsilon_{ij}^f=\frac{G_{\nu_i\nu_j}^f}{G_f}$
where $f$ stands for the fermion involved in the new interaction:

\begin{eqnarray}
\epsilon&=&\frac{G_{\nu_\mu\nu_\mu}^d-G_{\nu_e\nu_e}^d}{G_f}
       =\frac{G_{\nu_\tau\nu_\tau}^d-G_{\nu_e\nu_e}^d}{G_f}
       ~~~{\rm (d-quarks)} \nonumber \\
       &=&\frac{G_{\nu_\mu\nu_\mu}^u-G_{\nu_e\nu_e}^u}{G_f}
             =\frac{G_{\nu_\tau\nu_\tau}^u-G_{\nu_e\nu_e}^u}{G_f}
       ~~~{\rm (u-quarks}) \nonumber \\
       &=&\frac{G_{\nu_\mu\nu_\mu}^e-G_{\nu_e\nu_e}^e}{G_f}
      =\frac{G_{\nu_\tau\nu_\tau}^e-G_{\nu_e\nu_e}^e}{G_f}
       ~~~{\rm (electrons)} \nonumber \label{NSNI}
\end{eqnarray}
The solar neutrino data is able to stablish new limits in the
flavor-diagonal NSNI (\ref{NSNI}). With the definitions written above,
these  limits stands for:

In~\cite{fornengo} an analysis of atmospheric neutrinos
and NSNI is performed, and found the following limits, at $3\sigma$:
\begin{eqnarray}
|\epsilon_{\mu\tau}| &<& 0.03 \nonumber \\
|\epsilon'_{\mu\mu}-\epsilon'_{\tau\tau}| &<& 0.05
\end{eqnarray}

These bounds refer to NSNI with d-quarks, and were obtained assuming a two-flavor ($\nu_{\mu}$ and
$\nu_{\tau}$) system. A three family analysis significantly relaxes some of the bounds above \cite{atm_NSNI_3}, such that
order one values for some of the $|\epsilon|$s are not currently ruled out. For many more details, see  \cite{atm_NSNI_3}.

For u-quarks the bounds are
expected to be of the same order, and for NSNI with electrons we expect bounds
looser by a factor of $\sim 3$.

\subsubsection{KamLAND and solar neutrinos}

The excellent agreement between the LMA parameter region that provides a
solution to the solar neutrino problem and the parameter region compatible
with KamLAND data
provides us with an opportunity to use these data sets
to establish a limit on non-standard neutrino interactions (NSNI).


The effect of NSNI is negligible in KamLAND due to the short distance
traveled inside the earth, but due to the high density and long travel distances
in the sun, the presence of a NSNI could displace the best fit
point of solar neutrino analyses, and spoil the agreement between
solar and KamLAND allowed regions.

The oscillation of solar neutrinos is driven by only one mass scale,
$\Delta m_{21}^2$. The higher mass scale $\Delta m_{32}^2$, relevant
for atmospheric neutrino oscillations, decouples, and the mixing
between the first and third family is very small, and will be set to zero
in what follows.

In this approach, after rotating out the third family from the
evolution equation, we can write the 2x2 Hamiltonian that describes the
neutrino evolution as

\begin{eqnarray}
H_{MSW} &=&
\left[
\begin{array}{cc}
+\sqrt2 G_F N_e(r) - \frac{\Delta m^2}{4E} \cos{2\theta} &
\frac{\Delta m^2}{4E} \sin{2\theta}  \\
\frac{\Delta m^2}{4E} \sin{2\theta}    &
\frac{\Delta m^2}{4E} \cos{2\theta}
\end{array}
\right]
\nonumber \\
&+&\left[
\begin{array}{cc}
0 &
\sqrt2 G_F \epsilon_f N_f(r)  \\
\sqrt2 G_F \epsilon_f N_f(r) &
\sqrt2 G_F  \epsilon'_f N_f(r)
\end{array}
\right] ~~,
\end{eqnarray}
where $N_f(r)$ is an effective density felt by the neutrino,
given by $N_f=N_p+2N_n$ for d-quarks,
$N_f=2N_p+N_n$ for u-quarks and $N_f=N_p$ for electrons.

The NSNI parameters can be written as:
\begin{eqnarray}
\epsilon'_f&=&\frac{\epsilon_{\tau\tau}+\epsilon_{\mu\mu}}{2}+
\frac{\epsilon_{\tau\tau}-\epsilon_{\mu\mu}}{2}\cos2\theta_{23}-
\epsilon_{\mu\tau}\sin2\theta_{23}-\epsilon_{ee} \nonumber \\
\epsilon_f&=&\epsilon_{e\mu}\cos\theta_{23}-\epsilon_{e\tau}\sin\theta_{23}~~.
\end{eqnarray}

The atmospheric neutrino analyses have put strong bounds on
$\epsilon_{\mu\tau}$ and $\epsilon_{\mu\mu}-\epsilon_{\tau\tau}$, and
we expect a near to maximal mixing between second and third families
($\cos2\theta_{23}\simeq 0$). 
By further assuming that  $\epsilon_f=0$ (which would be the case if $\epsilon_{e\mu}$ and $\epsilon_{e\tau}$ were negligible) we can write
$\epsilon'_f$ as
\begin{equation}
\epsilon'_f\sim\epsilon_{\mu\mu}-\epsilon_{ee}=
\epsilon_{\tau\tau}-\epsilon_{ee}
\end{equation}

With the present data set and the assumptions listed above, we are able to establish the following limits
to the NSNI parameters, at $1\sigma$ ($2\sigma$):
\begin{eqnarray}
-0.20<&\epsilon'&<0.12~~(\epsilon'<0.30) ~~~\mbox{ d-quarks} \nonumber \\
-0.18<&\epsilon'&<0.10~~(\epsilon'<0.30) ~~~\mbox{ u-quarks} \nonumber \\
-0.55<&\epsilon'&<0.25~~(\epsilon'<0.86) ~~~\mbox{ electrons}~~.
\end{eqnarray}

The limits obtained at $2\sigma$ reflect the weak bounds on $\Delta m^2$
obtained by KamLAND. At present there are a number of possible
``islands'' in the parameter region compatible with KamLAND data, so the
displacement in $\Delta m^2$ could make the best fit point of a combined
analysis jump between consecutive islands. The increase of statistics
at KamLAND will determine in which of these islands the true values
of neutrino parameters lie, avoiding this kind of jump and improving
the limits on the NSNI parameters.

Simulating 1 kton-year of data at KamLAND, the allowed range of
$\Delta m^2$ would be reduced, and in the case of an agreement
between the neutrino parameters coming from KamLAND and solar neutrino
analysis (that would indicate a vanishing NSNI) new limits
could be obtained. Although the $1\sigma$ regions do not change
significantly, at $2\sigma$ we have:
\begin{eqnarray}
-0.42<&\epsilon'&<0.24 ~~~{\rm (d-quarks)} \nonumber \\
-0.40<&\epsilon'&<0.18 ~~~{\rm (u-quarks)} \nonumber \\
&\epsilon'&<0.40 ~~~{\rm (electrons)}~~.
\end{eqnarray}
Further increase of KamLAND statistics will not improve these bounds, which
are now determined by the uncertainty in $\Delta m^2$ in the solar neutrino
analysis.

Details of the analysis presented here can be found in~\cite{holanda2}. Similar
analyses were also done in~\cite{pena1}. It should be pointed out that more
severe constraints on (or a positive hint of) NSNI can be obtained in
next-generation solar neutrino experiments, especially those sensitive to
neutrino energies below $~6$~MeV \cite{pena1}.

\subsubsection{Bounds from non-oscillating phenomena}

Apart from phenomena that involve neutrino oscillations, bounds on
NSNI can also come from the effects of such non-standard interactions
on the charged leptons~\cite{bergmann1,holanda3}.
We should only be careful in translating such
bounds to the neutrino sector, since usually this translation can only
be possible with a few assumptions on the model that generates the
non-standard interactions. Some bounds can also be found using neutrino
scattering experiments~\cite{rossi,pena2}.
We quote here some of the numbers obtained by these analyses. Details of
the calculations can be found in the references. The following tables should be
read as limits on $\epsilon_{i,j}$, where $i$ and $j$ stand for $e$, $\mu$ and
$\tau$, and are the lines and columns of the tables. 

The values quoted here depend on details of the models that generate the
NSNI, such as $SU(2)_L$ breaking effects, absence of fine-tuning
cancellations and the
scale of new physics. Also, since neutrino oscillations are
sensitive only to the vector coupling constant of the NSNI, correlations
between the limits in $\epsilon_L$ and $\epsilon_R$ should be taken into
account in order to compare the numbers presented here with the ones coming
from neutrino oscillation experiments.

\vspace{5mm}
{
{\bf \parindent=0pt Electrons:}
\newcommand{\up}[1]{\raisebox{1.5ex}[0pt]{#1}}
\setlength{\tabcolsep}{1pt}

\nopagebreak
\vspace*{2mm}
\noindent
\begin{tabular}{c@{\hspace{5pt}}|c|c|c}
              & $e$ & $\mu$ & $\tau$   \\
\hline
              & $-0.07<\epsilon_L<0.1$,    &
                $\epsilon< 10^{-6}$~\cite{holanda3}            &
                $\epsilon< 4.2\times 10^{-3}$~\cite{holanda3}   \\
\up{$e$}
              & $-1<\epsilon_R<0.5$~\cite{pena2}             &
                $\,\epsilon_{L,R}<5\times 10^{-4}$~\cite{pena2}  &
                $\,|\epsilon_L|<0.4$, $|\epsilon_R|<0.7$~\cite{pena2}        \\
\hline        &   &  &
                     $\epsilon<3.1\times10^{-3}$~\cite{bergmann1}      \\
\up{$\mu$}    &   &  $|\epsilon_{L,R}|<0.03$~\cite{pena2}         &
                     $\epsilon_{L,R}<0.1$~\cite{pena2}   \\
\hline $\tau$ &   &   & $[-0.05;0.05]_{L,R}$~\cite{pena2}
\end{tabular}
}

\vspace{5mm}
{
{\bf \parindent=0pt d-quarks:}
\newcommand{\up}[1]{\raisebox{1.5ex}[0pt]{#1}}
\setlength{\tabcolsep}{1pt}

\nopagebreak
\vspace*{2mm}
\noindent
\begin{tabular}{c@{\hspace{5pt}}|c|c|c}
              & $e$ & $\mu$ & $\tau$   \\
\hline        & $|\epsilon_L|<0.3$,  &
                $\epsilon< 10^{-5}$~\cite{holanda3}            &
                $\epsilon< 10^{-2}$~\cite{holanda3}   \\
\up{$e$}      & $-0.6<\epsilon_R<0.5$~\cite{pena2}             &
                $\epsilon_{L,R}<7.7\times 10^{-4}$~\cite{pena2}  &
                $\epsilon_{L,R}<0.5$~\cite{pena2}              \\
\hline        &   &  $\epsilon<0.1$~\cite{holanda3}      &
                     $\,\epsilon<1.2\times10^{-2}$~\cite{bergmann1}      \\
\up{$\mu$}    &&$\,|\epsilon_L|<0.003$, $-0.008<\epsilon_R<0.015$~\cite{pena2}&
                     $\epsilon_{L,R}<0.05$~\cite{pena2}   \\
\hline  &   &   & $|\epsilon_L|<1.1$, \\
\up{$\tau$} & & & $|\epsilon_R|<6$~\cite{pena2}
\end{tabular}
}

\vspace{5mm}
{
{\bf \parindent=0pt u-quarks:}
\newcommand{\up}[1]{\raisebox{1.5ex}[0pt]{#1}}
\setlength{\tabcolsep}{1pt}

\nopagebreak
\vspace*{2mm}
\noindent
\begin{tabular}{c@{\hspace{5pt}}|c|c|c}
              & $e$ & $\mu$ & $\tau$   \\
\hline        & $-1<\epsilon_L<0.3$,      &
                $\epsilon< 10^{-5}$~\cite{holanda3}            &
                $\epsilon< 10^{-2}$~\cite{holanda3}   \\
\up{$e$}      & $-0.4<\epsilon<0.7$~\cite{pena2}          &
                $\epsilon_{L,R}<7.7\times 10^{-4}$~\cite{pena2}  &
                $\epsilon_{L,R}<0.5$~\cite{pena2}              \\
\hline        &   &  $\epsilon<0.1$~\cite{holanda3}      &
                     $\epsilon<1.2\times10^{-2}$~\cite{bergmann1}     \\
\up{$\mu$}    &&$|\epsilon_L|<0.003$, $-0.008<\epsilon_R<0.003$~\cite{pena2} &
                     $\epsilon_{L,R}<0.05$~\cite{pena2}   \\
\hline $\tau$ &   &   & $\,|\epsilon_L|<1.4$, $|\epsilon_R|<3$~\cite{pena2}
\end{tabular}
\vspace{2mm}
}

Concluding, using the neutrino
oscillation data we are able to find limits to NSNI parameters,
without assuming any detail about the nature of new physics behind 
these interactions. More work is needed to improve the situation.

\section{Beyond the three neutrino picture}

\subsection{The search for {\em other} light neutrinos}
\label{lightsterile}

A neutrino that does not participate in Standard Model interactions (i.e., is
sterile) might seem of little interest, but this concept includes reasonable
theoretical constructs such as right-handed neutrinos themselves. We note in
passing that in one-family Technicolor theories there are also technineutrinos
that would couple with the usual strength to the $W$ and $Z$ but, because of
their Technicolor interactions, are confined and gain large dynamical masses of
order several hundred GeV; they are therefore not relevant for usual low-energy
neutrino oscillation experiments.  The hypothesis of `sterility' concerns the
weak forces; gravity is expected to be felt anyway, and we cannot exclude that
the `sterile' neutrino participates in new forces, perhaps, mostly coupled to
quarks; or carried by new heavy mediators; or that sterile neutrinos have
preferential couplings with new particles --- say, with Majorons.  Even putting
aside these possibilities, we can probe sterile neutrinos by the search for
observable effects due to their mixing with the ordinary neutrinos. In this
section, we will further restrict our attention on `light' sterile neutrinos
(say, below $10$~eV) and discuss the impact on oscillations.  We make extensive
reference to Ref.~\cite{cmsv}, an updated overview on the phenomenology of one
extra sterile neutrinos.

\subsubsection{Issues of theoretical justification}

Many extensions of the Standard Model incorporate particles behaving as sterile
neutrinos.  The main question is: Why are these light and do they have the couplings
needed to mix with ordinary neutrinos?  A recent discussion is
in Ref. \cite{pl}. Models with
mirror matter \cite{mirr1,mirr2} contain mirror neutrinos \cite{mirrF,mirrB}
and offer a straightforward answer: ordinary
and mirror neutrinos are light for the same reason.  It is easy to arrange a
`communication' term between ordinary and mirror worlds, e.g., due to the
operator $\sim \nu\phi\nu'\phi'/M_{\rm Pl}$.  This leads to long-wavelength
oscillations into sterile neutrinos. 
(see Fig.\ \ref{fig:mirr}, from
\cite{mohan}).  
There are many other possibilities.  
For example, higher-dimensional operators in the superpotential in models
involving a scalar field with an intermediate scale expectation value can
naturally lead to small Dirac and Majorana masses of the same magnitude,
and therefore to light ordinary and sterile neutrinos which can mix \cite{pl}.
Already with mirror matter,
the VEV $\langle \phi'\rangle$ could be different from $\langle
\phi\rangle=174$~GeV, and this has important carryings for the phenomenology
\cite{mirrF,mirrB}.
Alternatively, one could guess on dimensional ground the value
${\rm TeV}^2/M_{\rm Pl}$ as the mass (or mixing) of sterile neutrinos, and
relate the TeV-value, e.g., to supersymmetry breaking \cite{alyos} or high GUT
theories such as $E_6$ \cite{frank}.  In theories with dynamical electroweak
symmetry breaking, the mechanism that has been found for light neutrinos
predicts also that there are sterile neutrinos with masses
of order 100's of MeV to GeV \cite{nt,lrs,nuf03,ckm}.

Right-handed neutrino masses of about 1 or 100~TeV lead to exceedingly large
masses that pose a question for extra dimensions if ``large'' scales should be
absolutely avoided and if the Dirac Yukawa couplings are not small on their
own.  A popular way out is to postulate that the masses are Dirac in character.
As a benefit, one explains why neutrino masses are small when right-handed
neutrinos propagate in the bulk \cite{rmpl}.  But even if neutrinos turn out to
be Dirac particles we would not have an evidence for these theories, since
(1)~Dirac neutrino masses are possible in conventional 4-dimensional theories;
(2)~even in theories with extra dimensions, one can assume that the usual
dim.-5 term is the dominant source of neutrino masses. We just need to fix the
scale of mass to the desired value of $10^{13-15}$~GeV, and neutrinos receive
(presumably large) Majorana masses \cite{mohan}. To summarize, small neutrino
masses are possible, and even more interestingly an infinite number of sterile
neutrinos, but neutrino masses seem not to be a clean signal of low scale
gravity theories.  Oscillations have special character and there are
interesting phenomenological constraints \cite{snsns}.

We believe that there should be increased attention paid not only toward
phenomenology, but also toward theories of (light- or heavy-mass)
sterile neutrinos. The fact that we do not understand usual fermion masses
should be not taken as an excuse to avoid confrontation with theory in our
view.  With these considerations in mind, we will give some emphasis to mirror
neutrinos, and at the same time point out the interest of understanding the
unknown ${\cal O}(1)$ coefficients in these theories, that have an important
impact on the phenomenological implications.

\subsubsection{Phenomenological manifestations}

In the following discussion, we will be concerned mostly with oscillations.
However, the implications can be also elsewhere.  To see that, it is sufficient
to recall that when we add 3 sterile neutrinos we can form Dirac masses, which
means that there is no contribution to the neutrinoless double beta decay
process.

\paragraph{Terrestrial oscillation experiments}

Broadly speaking, there are two types of terrestrial experiments.  The first
one includes several disappearance experiments and LSND; the second one
includes atmospheric neutrinos and long-baseline experiments. The first type is
sensitive mostly to the mixing of $\nu_e$ and a sterile state, the other one
also to $\nu_\mu$ or $\nu_\tau$.  Both types of experiments probe only
relatively large mixing angles, $\theta_s\sim 0.1$.  Sterile neutrinos within
the sensitivity regions are disfavored if standard cosmology (mostly BBN)
applies; further important tests will be done by CMB+LSS or BBN data.  None of
these experiments {\em alone} requires the existence of sterile neutrinos.  A
case for sterile neutrinos can be made interpreting in terms of oscillations
LSND together with solar and atmospheric data \cite{bgg}. The hypothesis
that LSND signal is due to a relatively heavy and mostly sterile neutrino
should be regarded as conservative \cite{ales}, even though it leads to some
problems with disappearance in terrestrial experiments, and interesting
predictions for cosmology (BBN and CMB+LSS spectra). In view of this situation,
the test of the LSND result is of essential importance. At the same time, we
should not forget that sterile neutrinos could manifest themselves in other
manners.

\begin{figure}[htb]
$$\hspace{-5mm}
\includegraphics[width=7cm]{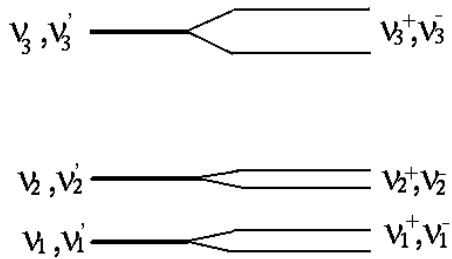}
$$
\caption{The double degeneracy between mass eigenstates of ordinary and mirror
world ($\nu_i$ and $\nu'_i$) is lifted when the small mixing terms are included
in the $6\times 6$ mass matrix.  The new mass eigenstates ($\nu_i^+$ and
$\nu_i^-$) are in good approximation maximal superpositions of $\nu_i$ and
$\nu'_i$.
\label{fig:mirr}}
\end{figure}
\begin{figure}[htb]
$$\hspace{-5mm}\includegraphics[width=18cm]{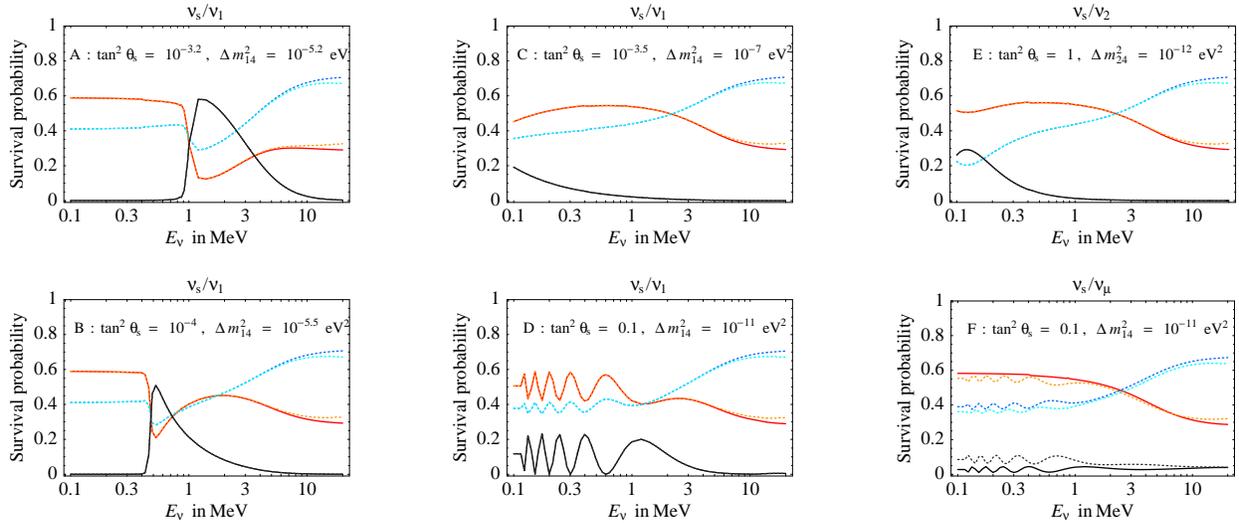}$$
\caption{Examples of oscillation into
sterile neutrinos allowed from present solar neutrino data.
In the plots are shown
the 3 probabilities of oscillations
$P(\nu_e\to\nu_e)$ (decreasing red curve),
$P(\nu_e\to\nu_{\mu,\tau}$) (increasing blue curve) and
$P(\nu_e\to \nu_{\rm s})$ (lower black curve),
as a function of the neutrino energy.
The continuous (dotted) curve are the values
during day (night). \label{fig:nus_sol}}
\end{figure}

\paragraph{Solar and KamLAND neutrinos}
The solar and KamLAND data can be explained well without sterile
neutrinos. Even more, the `LMA' solution received significant confirmations:
the sub-MeV energy regions have been probed by Gallium experiments and the
super-MeV ones by SNO and Kamiokande, and LMA is in agreement with
KamLAND. Thus we are led to consider minor admixtures of sterile neutrinos,
presumably not more than 20~\%.  In many interesting cases sterile neutrinos
are invisible at KamLAND but affect the survival probability of solar
neutrinos.  Quite generally, to test the hypothesis of oscillations into
sterile states it would be important to improve on (or measure precisely) the
fluxes from Beryllium and pp-neutrinos.  A few selected cases are shown in
Fig.~\ref{fig:nus_sol} from~\cite{cmsv}.

\paragraph{Neutrinoless double beta decay}

Massive Majorana sterile neutrinos, mixed with the active ones,
would participate in mediating  neutrinoless double beta decay.
In the case of one light sterile neutrino,
$\langle m \rangle_{eff}$ is given by the sum of
the contributions of four massive neutrinos~\cite{BPP2}:
\begin{equation}
\langle m \rangle_{eff} = \sum_{i=1,4} m_i U_{ei}^2,
\end{equation}
where $m_i$ denotes the mass of the massive eigenstates and
$U_{ei}$ indicates their mixing with the electron neutrino.

\paragraph{Supernovae}

Supernovae are an ideal place to search for manifestations of sterile
neutrinos, either through long wavelength vacuum oscillations of a galactic
supernova (whose distance is in the kpc scale-SN1987A was at 52~kpc) or through
MSW effects (since nuclear densities are reached during the collapse); the main
trouble is that we are still not able to understand supernova explosions
theoretically (the data from SN1987A do not contradict the simplest current
picture for neutrino emission, but they have puzzling features that suggest
caution).  Vacuum oscillations induced by mirror neutrinos 
(as in Fig.\
\ref{fig:mirr}) can led to a disappearance of half of the emitted flux
\cite{mohan}.
  MSW oscillations into sterile neutrinos can produce even 
more
dramatic suppressions; in certain regions of the parameter space this can reach
80~\% \cite{cmsv}. The last type of oscillations are due to the fact that the
SN core is deleptonized, and require that the electron neutrino mixes with the
sterile one. The trouble to verify these predictions is the accuracy of the
expectations on the emitted neutrino energy, that amounts roughly to a factor
of 2; thus it seems that, with present knowledge of supernovae, and using the
data from SN1987A, we can only safely exclude the occurrence of dramatic MSW
effects.  We remark that with a quantitative theory of core collapse
supernovae, this test of sterile neutrinos would become very powerful.  There
are other interesting effects possibly related to heavy sterile neutrinos, such
as r-process nucleosynthesis, re-heating of the shock, rocketing of pulsars,
not discussed here.

\paragraph{Big-bang nucleosynthesis and other cosmological probes}

An often stressed doubt concerning cosmology is about neglected or unknown
effects. This said, we must recall that there has been impressive progress in
the last years. The impact on neutrinos can be summarized as follows: (1)~The
number of neutrinos at the photon decoupling is bounded to be $3\pm 2$.
(2)~The contribution to the energy density of the universe $\Omega_\nu h^2$ is
below or at about 1~\%, or, equivalently, the sum of neutrino masses is below
or at about 1~eV. (It should be recalled that there is an interesting, not
universally accepted claim that the cosmology gives not a bound but a value for
neutrino masses \cite{allen}.  This testifies the high sensitivity reached by
these methods and points to the interest in the value of the bias parameter
$\sigma_8$).  (3)~The effective number of neutrinos at nucleosynthesis time is
$3\pm 2$, when extracted from deuterium abundance, or $2.4\pm 0.7$, when
extracted from helium abundance. These numbers already imply strong bounds on
sterile neutrinos, but do not rule out the sterile hypothesis for interesting
regions of the parameters.  One example is given by mirror neutrinos, when the
new mass splittings are small enough. Another one is given by a new sterile
state, that has only small mixing with the ordinary neutrinos. A second
possibility is to have post-BBN phase transition involving the vacuum
expectation value of a light scalar field which mixes the active and
sterile neutrinos so that at the time of BBN, the active and sterile
neutrinos are unmixed. There are however strong constraints on the nature
and interaction of the scalar field~\cite{chacko,MohapatraNasri} and an interesting
feature of these models is that they leave an imprint on the cosmic
microwave background that can be tested in future experiments such as the
Planck satellite mission.
Yet another possibility is a large electron neutrino asymmetry in the early
universe, which can compensate the effects of a number of extra
neutrino types \cite{Barger:2003rt}.

It has also been pointed out that the production of sterile neutrinos in 
the Early
Universe is strongly suppressed in cosmological scenarios for which
the reheating temperature is
as low as few MeV~\cite{Gelmini:2004ah}. In this case
the bounds on the sterile neutrino parameters
which can be derived from BBN and from the contribution of $\Omega_\nu h^2$ to
the Dark Matter are much weaker than in the standard case.

\paragraph{Ultra-high energy neutrinos}
Although there is a great interest in the search for ultra-high energy
neutrinos, the number of reasonable (or even, less reasonable) mechanisms that
have been discussed to produce them is not large. The reason is that neutrinos
are produced along with electromagnetic radiation, that can be observed in a
variety of ways, even when this is reprocessed.  Following this line of
thought, the astrophysical mechanism that can be conceived to overcome such a
structure is the concept of a `hidden source'.  Another escape way from this
constraint involves sterile neutrinos.  Indeed, if there are ultra-high energy
mirror neutrinos, they inevitably oscillate into neutrinos from our world on
cosmic scales \cite{bv}. This scenario can provide intense fluxes of ultra-high
energy neutrinos, subject only to the observable electromagnetic radiation from
their interaction with the relic neutrino sea.

\subsubsection{Summary of what we can learn on light sterile neutrinos}

In the view of many, there is something embarrassing in the hypothesis of
(light) sterile neutrinos, and some of us believe that this is `a solution
searching for a problem'. However, history tells that the most prominent
characteristic of neutrinos is that they are amazing! Said more seriously, we
should certainly aim to measure neutrino properties, but we should not forget
that we could make discoveries.  And, when we think to new experiments, we
should evaluate their potential for the investigation of sterile neutrinos.

In this section, we recalled that light sterile neutrinos can play a role not
only for LSND but also in terrestrial oscillations experiment, solar neutrinos,
supernovae, astrophysics and cosmology. There are links between the various
observables, but it is not impossible to conceive that sterile neutrinos have
an important role only in astrophysics or cosmology (e.g., in core collapse
supernovae or big-bang nucleosynthesis).  More measurements and theoretical
progresses will lead to important tests of the idea that the neutrinos that we
know are not the full story.

\subsection{What can we learn about four-neutrino mass matrices}

The solar, atmospheric and LSND data require three different (mass)$^2$
splittings. These cannot be accommodated by three neutrino flavors, which
provide only two independent $\Delta m^2$.  Additional degrees of freedom are
necessary in order to understand all this data put together. The easiest option
is to add a sterile neutrino and interpret the data in terms of oscillations of
four neutrino flavors.

 The MiniBooNE experiment at Fermilab is crucial for confirming or refuting the
LSND evidence for neutrino oscillations. If the LSND result is confirmed, a
very exciting epoch in neutrino physics is just about to begin, as the number
of questions that need to be answered becomes even larger than in the standard
three-flavor case.

A general 4-neutrino Majorana mass matrix
 is described by 4 masses, 6 mixing angles and 6 CP violating phases, 3 of
which would affect oscillations.

\begin{figure}[htb]
\begin{center}
\hspace*{-7mm}
\epsfig{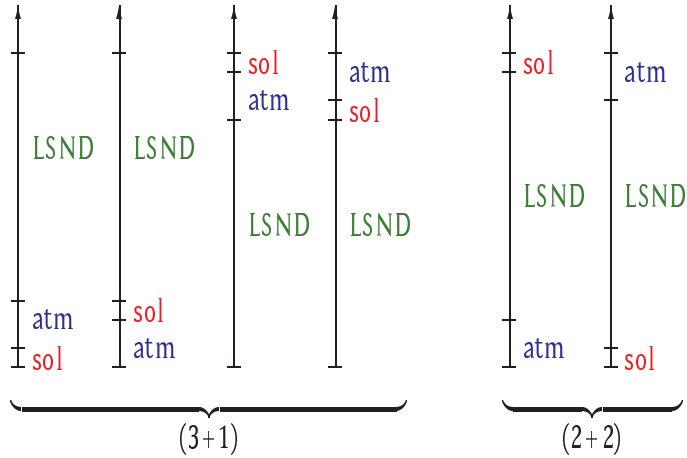}\\
\caption{{\small Four neutrino mass spectra.
}}
\label{4schemes}
\end{center}
\end{figure}

 In this case there are 6 possible mass spectra, as shown in
 Fig.\ \ref{4schemes}. These can be divided in two
 categories: ``3+1'' and ``2+2''. The ``3+1'' mass patterns are comprised
of one sterile neutrino separated by
 $\Delta m^2_{\rm LSND}$ from the other three. The group of three is the usual
group of ``active'' neutrinos, one
 pair  separated by $\Delta m^2_{\odot}$ and the third separated from these
by $\Delta m^2_{\rm A}$.
The ``2+2'' patterns are comprised of
 two pairs of neutrinos, one separated by $\Delta m^2_{\odot}$ and the other
by $\Delta m^2_{\rm A}$. The two
pairs are separated by $\Delta m^2_{\rm LSND}$.

Both categories are already very strongly constrained by experiment
\cite{Maltoni:2003yr, 4nu}.

 In the ``3+1'' scenario the 3 states relevant for solar and atmospheric
oscillations are mostly active and the forth state is almost entirely sterile.
This pattern has the usual three active flavor scenario as a limiting case, so
it agrees very well with all solar and atmospheric data. It is however harder
to accommodate short baseline neutrinos oscillation experiments.  This is
related to the irony of the fact that LSND is an active flavor appearance
experiment, showing $\bar\nu_\mu\to\bar\nu_e$ oscillations, but its solution
has to involve an almost entirely sterile neutrino, while the other experiments
remain unaffected by the presence of the sterile state.  The bounds on
$\sin^22\theta_{\rm LSND}$ coming from KARMEN, CDHS, CHOOZ and atmospheric data
are rather strong, almost conflicting with the value required to explain the
LSND signal.  The fit to all data is not very good, but the ``3+1'' scenario is
not completely excluded at this point.

 In the `2+2'' scenario both solar and atmospheric neutrino oscillations
involve some fraction of conversion into a sterile state. This fractions are
now very strongly constrained by the atmospheric, solar and reactor data,
making the fit to the data to be rather poor in the`2+2'' case. The global
analysis are usually performed by considering three mixing angles and
neglecting the other three, which are known to be small. Including these
additional small angles might improve the quality of the fits, but the ``2+2''
scenario is strongly disfavored.

 The presence of the fourth, sterile neutrino also has implications in
cosmology and cosmological observations
 impose further constraints on the allowed parameter space, as discussed
in the previous section.

 The first question regarding four neutrino mass matrices, namely if they are
indeed necessary to interpret the experimentally observed neutrino oscillation
data will soon be answered by the MiniBooNE experiment.  Assuming the answer is
positive, a whole new set of questions arises. Just as in the three flavor
case, one would like better measurements of all $\Delta m^2$'s and mixing
angles. Given the much larger number of phases involved in the 4 neutrino case,
the possibilities for observing CP violation in the neutrino sector become very
rich and maybe more easily accessible \cite{4nunufact}.  If the LSND signal is
confirmed, than there must be a state with mass higher than $\sqrt{\Delta
m^2_{\rm LSND}}$.  This would be in the range of sensitivity of future tritium
$\beta$ decay experiments like KATRIN, raising the possibility of determining
the absolute scale of neutrino mass. For Majorana neutrinos, neutrinoless
double beta decay might also be accessible \cite{BPP2,4nu0nubeta}.  By
combining data from all types of experiments, the specific mass pattern could
also be determined.

\subsection{Heavy sterile neutrinos}

The general motivation for considering sterile neutrinos has already been
discussed.  Many extensions of the Standard Model imply the existence of more
than one sterile neutrino with couplings to the active ones.  The right-handed
neutrinos participating in the seesaw mechanism, mirror neutrinos, the
neutrinos in extra-dimensional models, and the right-handed technisinglet
neutrinos in the ETC model discussed in other sections are some examples of
such sterile neutrinos.

 ``Light'' (below $\sim$ 10 eV) sterile neutrinos have been discussed in
Sec.~\ref{lightsterile}.  Here we concentrate on ``heavy'' ones, by which we mean
sterile neutrinos with masses above $\sim$ 10 eV, but below $\sim 1 GeV$.  As
noted, the mechanism constructed in Refs.  \cite{nt,lrs,nuf03} for light
neutrinos in Technicolor theories leads to (two) heavy neutrino mass
eigenstates in this range.  We do not discuss here ``very heavy'' neutrinos
(e.g.\ GUT scale), whose properties have been talked about above.  Once sterile
neutrinos are introduced, there is no definitive prediction for either the
number or the masses of these light neutral fermions.  Answering the question
of the total number of neutrinos (active and sterile) and their masses is
fundamental, as it would lead to much progress in understanding physics beyond
the Standard Model. It is thus very important to address these issues from the
experimental/observational point of view.

Heavy sterile neutrinos with couplings to the active ones have profound
implications in cosmology and astrophysics. They can also be constrained
by several types of laboratory experiments. We discuss here the present status
and future prospects for determining the properties of these heavy sterile
neutrinos.

\subsubsection{Laboratory Experiments}

The existence and mixing of heavy sterile neutrinos has many effects on
particle and nuclear decays.  These include contributions to $\mu^+ \to e^+
\gamma$, $\mu^+ \to e^+ e^+ e^-$, $K_L \to \mu^\pm e^\mp$, and $K^+ \to \pi^+
\mu^\pm e^\mp$, among others.  However, given limits on mixing, these
contributions are expected to be quite small (see further below on $\mu^+ \to
e^+ \gamma$).  

The two-body leptonic decays of charged pions and kaons, and also measurement
of the differential decay distribution in $\mu$ decay, can be used to search
for, and set bounds on, the emission of massive neutrinos via lepton mixing
\cite{Shrockhs,Shrocklep}.  The experimental signature for the emission of a
massive neutrino via lepton mixing in $\pi^+_{\ell 2}$ or $K^+_{\ell 2}$ decay
would be the appearance of an additional peak in the momentum spectrum of the
outgoing charged lepton $\ell^+ = \mu^+$ or $e^+$.  The position of the extra
peak is determined by the mass of the heavy neutrino and the size of the extra
peak is proportional to the mixing $|U_{\ell h}|^2$, where $\ell=e$ or $\mu$,
between the extra state and the neutrino $\nu_\ell$.  Initial bounds from
retroactive data analyses were given in \cite{Shrockhs}, and dedicated searches
were carried out in $\pi^+_{\mu 2}$ \cite{pimu2}, $K^+_{\mu 2}$ \cite{kmu2},
$\pi^+_{e2}$ \cite{pie2}, and $K^+_{e2}$ \cite{ke2} decays. Because of renewed
interest \cite{armbruster}, some recent searches in $\pi^+_{\mu 2}$ are
reported in \cite{pimu2recent}.  Resultant upper bounds on $|U_{\ell h}|^2$
range down to $10^{-5}$ -- $10^{-7}$ in the mass range from several MeV to
$\sim 300$ MeV.  Admixed massive neutrinos also affect the observed ratio of
branching ratios $BR(\pi^+_{e2})/BR(\pi^+_{\mu 2})$ and this has been used to
set limits (e.g., \cite{britton92b}).

The nuclear beta decay spectrum is also affected by the presence of heavy
neutrinos mixed with $\nu_e$. The spectrum would have a ``kink'' at the
endpoint energy $E_{max} (m_h)$ \cite{maki_nuclear,Shrockhs,mckellar,kobzarev}.
The position of the kink determines the mass of the heavy state, $m_h$, and the
change in the slope of the spectrum determines the mixing $|U_{eh}|^2$.  Many
experiments searching for such kinks in the Kurie plots of nuclear beta decays
were carried out in the 1980's and beginning of the 1990's; although a few
claimed positive results, these were refuted \cite{pdg}.  Resultant upper
limits were of the order $|U_{eh}|^2\approx 10^{-3}$ for masses between $10$
keV and $ \sim 1$ MeV.  

The nuclear transition involving electron capture, $e^- + (Z,A) \to (Z-1,A) +
\nu_e$ \cite{derujula} and muon capture, $\mu^- + (Z,A) \to (Z-1,A) + \nu_\mu$
\cite{deutsch} can also be used to search for, and put limits on, massive
neutrino emission via mixing. 

Assuming that the additional sterile neutrinos are Majorana, there are several
$|\Delta L|=2$ transitions and meson and hyperon decays (processes analogous to
neutrinoless nuclear double beta decay) that can occur.  One of these is the
nuclear transition $\mu^- + (A,Z) -> \mu^+ + (A, Z-2)$ \cite{mmm}. Meson decays
include $K^+ \to \pi^- \mu^+ \mu^+$.  A first upper limit on the branching
ratio for this decay was set in \cite{ls1}, and this has been greatly improved
by a dedicated search in a recent BNL experiment \cite{e865} (see also
\cite{ls2,dib,cernk,krev}).  Analogous $|\Delta L|=2$ decays of heavy-quark
mesons are also of interest. Early searches include one by the Mark II detector
at PEP for the decays $D^+ \to K^- \mu^+\mu^+$ and $D^+ \to \pi^- \mu^+\mu^+$
\cite{pep} and one by the CLEO experiment at CESR for $B^+ \to K^-\mu^+\mu^+$
\cite{cleo}; current limits are given in Ref. \cite{pdg}.  One can also
consider $|\Delta L|=2$ hyperon decays such as $\Xi^- \to p \mu^-\mu^-$ and
$\Sigma^- \to p \mu^- \mu^-$.  A first upper limit, on $BR(\Xi^- \to p
\mu^-\mu^-)$, was set in Ref. \cite{lsh}; a recent dedicated search 
reporting a
much improved limit is by the HyperCP experiment at Fermilab \cite{hypercp}.

Mixing between heavy and light neutrinos also leads to neutrino decay
\cite{Shrockdecay,gronau,gnr84}.  A number of searches for neutrino decays in
accelerator experiments have been carried out \cite{nudecay}.  Depending 
on the
assumed mass of the neutrino mass eigenstate, various decays are possible,
including $\nu_h \to \nu' e^+ e^-$, $\nu_h \to \nu' \mu^\pm e^\mp$, $\nu_h \to
\nu' \mu^+ \mu^-$.  Bounds on various combinations of mixing angle factors from
these experiments are reported in \cite{nudecay}; published limits range down
to $|U_{\ell h}|^2 < 10^{-9}$, $\ell=e,\mu$, for heavy neutrino masses of
several hundred MeV \cite{nudecay,pdg}. In addition to weak charged current
contributions to these neutrino decays, there are also contributions from the
weak neutral current, since, in the presence of sterile neutrinos, the weak
neutral current is not, in general, diagonal in mass eigenstates
\cite{LeeShrock:77,valle80,Kusenko:2004qc}. Present and future experiments as
MiniBooNE and MINOS might have the possibility to improve on some of the
present bounds on heavy neutrino decays~\cite{Kusenko:2004qc}.  Searches for
heavy neutrino production and decay have also been carried out at $e^+e^-$
colliders; see \cite{pdg} for limits.  For masses between a few GeV and $m_Z$
the best bounds come from measurements at the Z pole, where the Z boson could
decay into a standard neutrino and a sterile one, that would further decay
\cite{pdg}.

\subsubsection{Astrophysics and cosmology}

The existence of sterile neutrinos with even very small mixing to the active
ones can have dramatic consequences in astrophysics and cosmology. These are
discussed by a different working group \cite{APSastro}. Because astrophysical
and cosmological observations provide the strongest constraints and prospects
for future answers regarding heavy sterile neutrinos, we do, however, include
here an overview of this subject.

{\it Cosmology}

Massive sterile neutrinos could be produced in the early universe and can
provide some or even all the required dark matter. Heavy sterile neutrinos can
be produced by scattering-induced conversion of active neutrinos
\cite{DodelsonWidrow}. These neutrinos are produced non-resonantly and they can
be a warm dark matter candidate. A different mechanism of production of heavy
sterile neutrinos appears if there is a non-vanishing initial lepton number in
the Universe \cite{ShiFuller}. In this case sterile neutrinos can be produced
resonantly and the energy spectrum is in this case highly non-thermal. The
sterile neutrino can act then as a warm, cool or even cold dark matter
\cite{Fullerdm}.

 Cosmological observations impose strong constraints on massive sterile
neutrinos \cite{Fullerdm, steriledm}.  The radiative decay of such neutrinos to
a light neutrino and a photon would affect the diffuse extragalactic background
radiation, by producing a large number of photons of energy of the order
$m_H$. The DEBRA experiment is now constraining the parameter space of sterile
neutrinos based on this. The Chandra X-ray observatory has a great potential of
resolving a considerable fraction of the observed X-ray background and
consequently imposing much stronger constraints or potentially detecting X-ray
fluxes from dark matter sterile neutrinos in the gravitational potential wells
of clusters of galaxies.

 Heavy sterile neutrino decay prior to cosmic microwave background (CMB)
decoupling increases the energy density in relativistic particles, leading to
further constraints on the allowed parameter space.

 Big bang nucleosynthesis (BBN) is one of the big successes of the Standard
Model of cosmology, successfully predicting the primordial abundance of light
elements. The energy density in the sterile neutrino sea prior to BBN must not
be too high in order not to spoil the successful predictions of
BBN. Photoproduction of deuterium and $^6{\rm Li}$ from decay of sterile
neutrinos after BBN also imposes additional constraints on the sterile neutrino
parameter space.

 Large scale structure observations are also essential, as they can constrain
the nature of the dark matter (hot, warm or cold), consequently setting the
scale for the mass of the sterile state.

\begin{figure}[htb]
\begin{center}
\hspace*{-7mm}
\epsfig{file=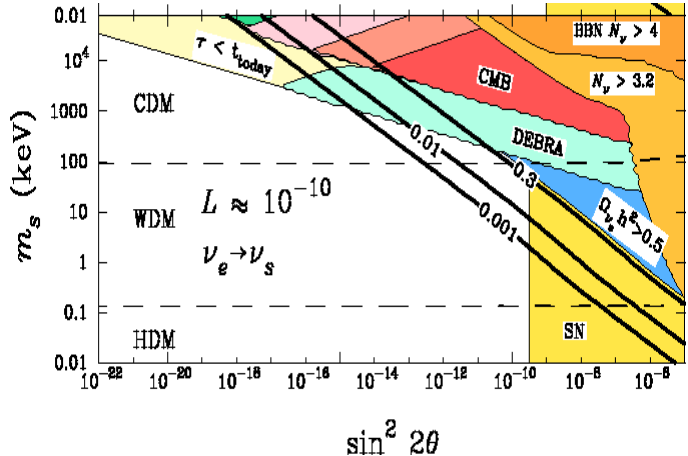,width=9cm,height=6cm}\\
\caption{{\small Cosmological/astrophysical constraints
 on heavy sterile neutrinos.}}
\label{cosmconstraint}
\end{center}
\end{figure}

Cosmological constraints are illustrated in Fig.\ \ref{cosmconstraint}
(from \cite{Fullerdm}).

{\it Supernovae}

 Neutrinos play a dominant role in core collapse supernovae. Even small
admixtures of heavy sterile neutrinos can have profound implications for
supernova physics.  Too much active neutrino conversion into sterile states can
lead to too much energy loss to sterile neutrinos, contradicting observations
from supernova SN1987a \cite{Fullerdm, steriledm}. The energy emitted in
sterile neutrinos depends on the mixing angle between the sterile and active
states in matter. For a long time it has been thought that most of the emission
occurs in the resonant region, where the effective mixing angle becomes
$\pi/4$.  This is not necessarily true because of a non-linear effect that
limits the resonant emission \cite{Fullerdm}.  The effective matter potential
for the $\nu_e\leftrightarrow \nu_s$ is given by
\begin{equation}
V=G_F\rho/\sqrt{2}/m_N (3 Y_e-1+4Y_{\nu_e}+2Y_{\nu_\mu}+2Y_{\nu_\tau})
\end{equation}
where $Y_i=(n_i-n_{\bar i})/n_B$.  For antineutrinos the matter potential
changes sign.  Due to the presence of the $Y_\nu$ terms, coming from neutral
current neutrino-neutrino scattering, this effective potential has zero as a
fixed point. Once a resonance is reached for, let's say, neutrinos, $\nu_e$'s
start converting with maximal effective mixing angle into $\nu_s$. This
decreases the $Y_\nu$ term, driving the parameters off resonance and thus
limiting the emission. The matter potential is effectively driven to zero on a
time scale of less than a second and the emission continues with vacuum mixing
angle. The parameter space where the sterile neutrino emission from supernovae
could be relevant is also marked on Fig.\ \ref{cosmconstraint}.

 It is interesting to note that a sterile neutrino in the few keV region could
also account for pulsar kicks \cite{pulsarkick}.  In the presence of a strong
magnetic field, neutrinos are emitted asymmetrically in a supernova core. This
asymmetry is washed out for active neutrinos which are trapped. If some
conversion to sterile neutrinos occurs, these can escape the core of the star
preserving the initial asymmetry. Only a few percent asymmetry is sufficient to
account for the initial kick of the star that would explain the unusually high
velocities of pulsars.

 To summarize, the presence of heavy sterile neutrinos with some (even very
small) mixing to active neutrinos has numerous implications in astrophysics and
cosmology.  At present, a sterile neutrino with mass of the order of a few keV
and very small mixing ($\sin^22\theta\approx 10^{-8}$) with an active one seems
to be allowed by constraints, could account for all or some fraction of the
dark matter in the Universe, would affect emission of supernovae neutrinos,
could explain the pulsar kicks and might lead to observable contributions to
X-ray measurements. In the future, more and more precise cosmological
observations can constrain very strongly the parameter space for such sterile
neutrinos, potentially closing the window or leading to detection of some
signal.

\newpage
\section{Supersymmetry and neutrinos}

Neutrino masses are not the only motivation to extend the standard model. One
also likes to extend it in order to solve the gauge hierarchy problem. Models
of low-energy supersymmetry are attractive candidates for the theory of TeV
scale physics.  In the minimal supersymmetric extension of the Standard Model
(MSSM) neutrinos are massless. Thus, we need to consider supersymmetric
extensions of the Standard Model that allow for neutrino masses.

There are basically three questions we like to answer when we talk
about the relations between supersymmetry and neutrinos
\begin{enumerate}
\item
Can successful predictions for neutrino masses of non-supersymmetric
extensions of the Standard Model be retained once these models are
supersymmetrized? In particular, can supersymmetry help in making
such models more motivated?
\item
Are there models where neutrino masses arise only due to
supersymmetry?
\item
Are there interesting phenomena in the slepton sector that can shed
light on the issue of neutrino masses, lepton number violation and
lepton flavor violation?
\end{enumerate}

The questions in the first item were already discussed in previous
sections of this review. Generically, supersymmetry does not upset
model predictions regarding neutrinos, and in some cases it in fact
helps. For example, in the case of GUT, making the model
supersymmetric helps to achieve coupling unification and thus to make
the model realistic.

In the following we concentrate on the last two items.
We briefly
describe two frameworks where neutrino masses are tightly connected to
supersymmetry, or more precisely, to supersymmetry breaking,
neutrino masses from R-parity violation and from supersymmetry
breaking.  We then discuss two effects, that of sneutrino oscillation
and sneutrino flavor oscillation, that can help us disentangle the
origin of neutrino masses using supersymmetric probes.
The aspects of supersymmetric seesaw models, i.e.\ the connection to
decays such as $\mu \rightarrow e \gamma$ have been discussed in previous
Sections.

\subsection{Neutrino masses from R-parity violation}

Neutrino masses from R-parity violation have been extensively
studied. Here we briefly summarize the main results. For a more
complete reference list where more details can be found, see, for
example, \cite{Grossman:2003gq}.

Once R-parity is violated there is no conserved quantum number that
would distinguish between the down-type Higgs doublet and the lepton
doublets. (For definitions and notation see, for example,
\cite{Grossman:1998py}.) Thus, these fields in general mix. Such
mixings generate neutrino masses; in fact, they generically
produce too large masses. One neutrino gets a tree level mass
which depends on the mixings between the
Higgs and the sneutrinos \cite{JN}. The other two
neutrinos get their masses at the one loop level, and thus their
masses are smaller by, roughly, a loop factor.

There are many different one loop contributions to the neutrino
masses.  (For a complete list see
\cite{Davidson:2000uc,Davidson:2000ne}.) The ``standard'' diagrams
are those that arise from the R-parity violating trilinear couplings
in the superpotential, the so-called $\lambda$ and $\lambda'$
couplings. These are the only contributions that are present in the
supersymmetric limit. Once supersymmetry breaking is included, there
are many more contributions (in fact, also the tree level
contribution is present only due to supersymmetry breaking). These
contribution are likely to be much larger than the $\lambda$ and
$\lambda'$ loop induced masses. The dominant diagrams are likely to be
those that arise due to bilinear couplings
\cite{Grossman:2003gq}. These are the couplings that mix the scalar
components of the Higgs and the neutrino superfields. The basic reason
that the $\lambda$ and $\lambda'$ loop induced masses are very
small is that they are proportional to the small down-type quark or
charged lepton Yukawa couplings.  This suppression factor is absent in
the bilinear induced masses.

The most attractive feature of R-parity violation models of neutrino masses is
that they naturally generate hierarchical neutrino masses with large mixing
angles. This is due to the fact that only one neutrino gets a mass at tree
level, while the other neutrinos only acquire loop induced masses. Numerically,
however, the predicted mass hierarchy is in general somewhat too strong.
Models with R-parity violation also predict observable lepton-number violating
processes at possibly observable rates (e.g., \cite{rabirev,lsh,dib}).

The biggest puzzle posed by R-parity violation models is to understand the
smallness of the neutrino masses. There must be a mechanism that generates very
small R-parity violating couplings. There are several ideas of how to do
it. For example, the small R-parity violation couplings can be a result of an
Abelian horizontal symmetry \cite{Banks:1995by}.

\subsection{Neutrino masses from supersymmetry breaking}

The smallness of neutrino masses can be directly related to the
mechanism of supersymmetry breaking, in particular, to the mechanism
that ensures a weak scale $\mu$ parameter
\cite{Arkani-Hamed:2000bq,Borzumati:2000mc,Arkani-Hamed:2000kj,%
Borzumati:2000ya,Abel:2004tt,March-Russell:2004uf}.
In general, there is no reason why the MSSM $\mu$ parameter is of the
order of the weak scale. Generically, it is expected to be at the
cut-off scale of the theory, say the Planck or the GUT
scale. Phenomenologically, however, $\mu$ is required to be at the
weak scale. One explanation, which is known as the Giudice-Masiero
mechanism, is that a $\mu$ term in the superpotential is not allowed
by a global symmetry. The required effective weak scale $\mu$ is
generated due to supersymmetry breaking effects.

The Giudice-Masiero mechanism can be generalized to generate small
neutrino masses. It might be that the large Majorana mass term that
drives the seesaw mechanism is forbidden by a global
symmetry. Effective Majorana mass terms for the right-handed
neutrinos, of the order of the weak scale, are generated due to
supersymmetry breaking.  The same global symmetry can also suppress
the Dirac mass between the right and left-handed neutrinos. Then, the
left-handed neutrinos have very small Majorana or Dirac masses as
desired.

The emerging neutrino spectrum depends on the exact form of the global
symmetry that is used to implement the Giudice-Masiero
mechanism. Nevertheless, the feature that the left-handed neutrino
masses are very small is generic.

\subsection{Sneutrino oscillation}
\label{sec:SneutrinoOscillation}

As already discussed in this report, it is interesting to find out
whether neutrinos have Majorana masses. In other words, we would
like to find out if total lepton number is violated in nature. As
already discussed, the most promising way to do it, is to look
for neutrinoless double beta decay. If supersymmetry is realized in
nature we may have other probes. Here we describe one example, that
of sneutrino-anti-sneutrino mixing and oscillation
\cite{Grossman:1997is}.

Consider a supersymmetric extension of an extended Standard Model that
contains Majorana neutrino masses.  For simplicity we consider only
one neutrino generation. In such models, the effect of $\Delta L=2$
operators is to introduce a mass splitting and mixing into the
sneutrino-anti-sneutrino system.  This phenomena is analogous to the
effect of a small $\Delta B=2$ perturbation to the leading $\Delta
B=0$ mass term in the $B$-system which results in a mass splitting
between the heavy and light neutral $B$ mesons. The very small mass
splitting can be measured by observing flavor oscillations.  The
flavor is tagged in $B$-decays by the final state lepton charge.
Since $\Delta m_B \sim
\Gamma_B$, there is time for the flavor to oscillate before the meson
decays.

The sneutrino system can exhibit similar behavior. The lepton number
is tagged in sneutrino decay using the charge of the outgoing lepton.
The relevant scale is the sneutrino width. If the sneutrino mass
splitting is large, namely when
\begin{equation} \label{xsnudef}
x_{\tilde \nu} \equiv {\Delta m_{\tilde \nu} \over \Gamma_{\tilde \nu}}
\gs 1,
\end{equation}
and the sneutrino branching ratio into final states with a charged
lepton is significant, then a measurable same sign dilepton signal is
expected.  Any observation of such oscillation will be an evidence for
total lepton number violation, namely for Majorana neutrino masses.

The size of the same sign lepton signal is model dependent. It is
generically expected that the sneutrino mass splittings are of the
order of the neutrino masses, as both are related to $\Delta L=2$
operators. The sneutrino width can vary a lot. When two body decay
channels are dominant the width is roughly of the order of
$\Gamma_{\tilde \nu} \sim \alpha M_{\tilde \nu}\sim {\cal O}(1\;{\rm GeV})$,
and thus $x_{\tilde \nu} \ll 1$ and a very small signal is expected. In
models where the two body decay channels are forbidden the sneutrino
width can be much smaller. For example, in models where the stau is
the LSP there can be a situation where the sneutrino has only three body
decay channels allowed. Then $x_{\tilde \nu}$ may be large enough for
the oscillation signal to be observed.

\subsection{Sneutrino flavor oscillation}

As we know there are large mixing angles in the lepton sector. An
independent probe of lepton flavor violation can be provided in
supersymmetric models via the slepton sector
\cite{Krasnikov:1995qq,Arkani-Hamed:1996au,Arkani-Hamed:1997km,Dine:2001cf}.
While generally the slepton mixings are not directly related to the
lepton mixings, both are lepton flavor violating effects.

Slepton flavor oscillations arise if the sleptons mass eigenstates are
not flavor eigenstates. Experimentally, the signal consists of a
lepton flavor imbalance.  For example, in a linear collider one can
search for a signal like $e^+ e^- \to \tilde\nu \tilde \nu^* \to X e^+
\mu^-$, where $X$ is the non leptonic part of the final state. In
hadron colliders, one can look for signals like $\chi^- \to \tilde\nu
\mu^- \to X e^+ \mu^-$, where $\chi^-$ is a chargino. The oscillation
probabilities depend on the slepton mass splittings and their mixing
angles. In principle, oscillation signals can be used to measure these
parameters. Even if this cannot be achieved in practice, just the
observation of an oscillation signal will provide an independent
confirmation for lepton flavor non-conservation.

In many proposals for supersymmetry breaking, a high degree of
degeneracy among sleptons is predicted.  As a result, there is the
potential for large flavor mixing among the sleptons. This can lead to
substantial and observable flavor violating signals at colliders.  To
be readily observable, it is necessary that mass splittings between
the states are not much smaller than the decay widths, and that the
mixing angles are not very small.  In a large class of supersymmetry
breaking models, the splittings can be comparable or even larger than
the widths, and the mixing angles may be large. In such cases, dramatic
collider signatures are possible.

\newpage

\section{Expectations in Superstring Constructions}

There has been relatively little work on the implications of
superstring theories for neutrino masses. However, it is known that
some of the ingredients employed in grand unified theories and other
four-dimensional models may be difficult to implement in known types of
constructions. For example, the chiral supermultiplets that survive in
the effective four-dimensional field theory are generally bi-fundamental
in two of the  gauge group factors (including the case
of fundamental under one factor and charged under a $U(1)$) for lowest
level heterotic constructions,  or either bi-fundamental or adjoint
for intersecting brane constructions.  This makes it difficult to
break the GUT symmetry, and even more so to find the high-dimensional
Higgs representations (such as the {\bf 126} of $SO(10)$) usually
employed in GUT models for neutrino and other fermion masses. Thus,
it may be difficult to directly embed many of the models, especially
GUT models involving high-dimensional representations rather than
higher-dimensional operators, in a string framework. Perhaps more likely
is that the underlying string theory breaks directly to an effective
four-dimensional theory including the Standard Model and perhaps other
group factors~\cite{Langacker:2003xa}. Some of the aspects of grand
unification, especially in the gauge sector, may be maintained in such
constructions. However, the GUT relations for Yukawa couplings are often
not retained~\cite{Dine:1985vv,Breit:1985ud,Witten:1985bz} because the
matter multiplets of the effective theory may have a complicated origin in
terms of the underlying string states.  Another difference is that Yukawa
couplings in string derived models may be absent due to symmetries in
the underlying string construction, even though they are not forbidden
by any obvious symmetries of the four-dimensional theory, contrary to
the assumptions in many non-string models. Finally, higher-dimensional
operators, suppressed by inverse powers of the Planck scale, are common.

Much activity on neutrino masses in string theory occurred following
the first superstring revolution. In particular, a number of authors
considered the implications of an $E_6$ subgroup of the heterotic $E_8
\times E_8$ construction~\cite{Dine:1985vv}-\cite{Nandi:1985uh}.  Assuming
that the matter content of the effective theory involves three {\bf 27}'s,
one can avoid neutrino masses altogether by fine-tuned assumptions
concerning the Yukawa couplings~\cite{Dine:1985vv}. However, it is
difficult to implement a canonical type I seesaw.  Each {\bf 27} contains
two Standard Model singlets, which are candidates for right-handed
neutrinos, and for a field which could generate a large Majorana mass for
the right-handed neutrinos if it acquires a large vacuum expectation value
and has an appropriate trilinear coupling to the neutrinos. However, there
are no such allowed trilinear couplings involving three {\bf 27}'s (this
is a reflection of the fact that the {\bf 27} does not contain a {\bf 126}
of the $SO(10)$ subgroup). $E_6$ string-inspired models were constructed
to get around this problem by invoking additional fields not in the {\bf
27}~\cite{valle1,Witten:1985bz}  or higher-dimensional
operators~\cite{Nandi:1985uh}, typically leading to extended versions
of the seesaw model involving fields with masses/vevs at the TeV scale.

Similarly, more recent heterotic and intersecting brane constructions, e.g., involving
orbifolds and twisted sectors, may well have the necessary fields for a type I seesaw, 
but it is again required that the necessary Dirac Yukawa couplings and Majorana masses for
the right-handed neutrinos be present simultaneously. 
Dirac couplings need not emerge at the renormalizable level, but can be of the
form
\begin{equation}
 \langle S'_1 \cdots S'_{d-3}\rangle N LH_u/M_{\rm PL}^{d-3},
\end{equation}
where the $S'_i$ are standard model singlets which acquire large expectation
values ($d=3$ corresponds to a renormalizable operator). Similarly,
Majorana masses can be generated by the operators
\begin{equation}
\langle S_1 \cdots S_{n-2}\rangle N N/M_{\rm  PL}^{n-3}.
\end{equation}
Whether such couplings are present at the appropriate orders depends on the
underlying string symmetries and selection rules, which are often very restrictive.
It is also necessary for the relevant $S$ and $S'$ fields to acquire the needed
large expectation values, presumably without breaking supersymmetry at a large
scale.   Possible mechanisms involve
approximately flat directions of the potential, e.g.,\ associated with
an additional $U(1)'$ gauge symmetry~\cite{Cleaver:1997nj,pl}, stringy
threshold corrections~\cite{Cvetic:1992ct,Mochinaga:1993td}, or
 hidden sector condensates~\cite{Faraggi:1993zh}.

There have been surprisingly few investigations of neutrino masses in
explicit semi-realistic string constructions. 
It is difficult to obtain canonical  Majorana masses in intersecting brane constructions 
\cite{Blumenhagen:2005mu}, because there are no interactions involving the
same intersection twice. 
Two detailed studies \cite{Ibanez:2001nd,Antoniadis:2002qm}
 of nonsupersymmetric
models with a low string scale concluded that lepton number was conserved,
though a small Dirac mass might emerge from a large internal dimension. Large
enough internal dimensions for the supersymmetric case may be difficult to
achieve, at least for simple toroidal orbifolds. 

There are
also difficulties for heterotic models.
An early study of $Z_3$ orbifolds
yielded no canonical Dirac neutrino Yukawas~\cite{Font:1989aj} at low order.
Detailed analyses of free fermionic models and their flat directions
were carried out in \cite{Faraggi:1993zh,Coriano:2003ui} 
and \cite{Ellis:1997ni,Ellis:1998nk}. Both studies concluded that small Majorana
masses could be generated if one made some assumptions about dynamics
in the hidden sector. In  \cite{Faraggi:1993zh,Coriano:2003ui} the masses
were associated with an extended seesaw involving a low mass scale.
The seesaw found in  \cite{Ellis:1997ni,Ellis:1998nk} was of the canonical type
I type, but in detail it was rather different from GUT-type models. 
A seesaw was also claimed in a heterotic $Z_3$ orbifold model with $E_6$ breaking
to $SU(3)\times SU(3)\times SU(3)$ \cite{Kim:2004pe}. A recent
study of $Z_6$ orbifold constructions found Majorana-type operators \cite{Kobayashi:2004ya}, 
but (to the order studied) the $S_i$ fields did not have the required expectation values.

In \cite{Giedt:2005vx} a large class of vacua of the bosonic $Z_3$ orbifold were
analyzed with emphasis on the neutrino sector to determine whether the minimal type I
seesaw is common, or if not to find possible
guidance to model building, and possibly to get clues concerning textures and mixing
if examples were found.
 Several examples from each of 20 patterns of vacua were studied, and the nonzero
superpotential terms through degree 9 determined.
There were huge number of $D$ flat directions, with the number reduced
greatly by the $F$-flatness condition.
Only two of the patterns had Majorana mass operators, while {\em none} 
had simultaneous Dirac operators of low enough degree to allow neutrino
masses larger than $10^{-5}$ eV. (One apparently successful model was
ruined by off-diagonal Majorana mass terms.) It is not clear whether this failure
to obtain a minimal seesaw is a feature of the particular class of construction,
or whether it is suggesting that stringy constraints and selection rules might
make string vacua with minimal seesaws rare. Systematic analyses of
the neutrino sector of other classes of constructions would be very useful.

There are other possibilities for obtaining small neutrino masses in string 
constructions, such as
extended seesaws and
small Dirac masses from higher dimension operators. 
Small Dirac neutrino masses from type I anisotropic compactifications have been 
discussed recently in \cite{Antusch:2005kf}.  
The possibility of embedding type II
seesaw ideas (involving Higgs triplets) in string constructions
was considered in \cite{Langacker:2005pf}. It is
possible to obtain a Higgs triplet of $SU(2)$ with non-zero hypercharge
in a higher level construction (in which $SU(2) \times SU(2)$ is broken
to a diagonal subgroup). In this case, because of the underlying $SU(2)
\times SU(2)$ symmetry the Majorana mass matrix for the light neutrinos
should involve only off-diagonal elements (often with one of the three
off-diagonal elements small or vanishing), with profound phenomenological
consequences, including an inverted hierarchy
and two large mixings.  This is a top-down motivation for the texture
C in Table~\ref{texturetable}.

\newpage

\section{Theories with a TeV-scale $\boldsymbol{U(1)'}$}

Many extensions of the Standard Model and MSSM include the existence
of additional non-anomalous $U(1)'$ gauge symmetries. These include
many superstring constructions \cite{Cvetic:1995rj}, grand unified
theories, Little Higgs models, and models of dynamical symmetry
breaking (DSB)~\cite{Hill:2002ap}. In a regular grand unified theory
the $U(1)'$ breaking needs to be at a large scale, because scalars that
can mediate proton decay can have masses no larger than the $U(1)'$
breaking scale. In string theories, the $U(1)'$ symmetry breaking
is usually induced by soft supersymmetry breaking effects at the TeV
scale~\cite{Cvetic:1995rj,Cvetic:1997wu,Erler:2002pr}, although in some
cases there is the possibility of breaking along an $F$ and $D$ flat
direction at an intermediate scale~\cite{Cleaver:1997nj} (depending on
the sign of a combination of soft mass-squares).  The Little Higgs and
DSB models are at the TeV scale.

A TeV scale $Z'$ has many interesting phenomenological
consequences~\cite{Langacker:2003bc}, but here we are concerned
with neutrino masses. In the intermediate $Z'$-scale case,
higher-dimensional operators, involving one or more powers of the fields
with intermediate-scale vevs but suppressed by powers of the Planck mass,
can yield naturally small Dirac neutrino masses~\cite{Cleaver:1997nj,pl}.
Variants can also lead to mixing between light ordinary and sterile
neutrinos, as suggested by the LSND results, or even to a type I seesaw.

Models in which the $U(1)'$ breaking is at the TeV scale generally
do not allow a canonical type I seesaw model, because the Majorana
mass for the right-handed neutrino $N_R$ requires $U(1)'$ breaking
(unless the $N_R$ carries no $U(1)'$ charge). However, a number of
other possibilities are allowed for the neutrino masses~\cite{KLL},
including small Dirac masses (e.g., associated with a second $U(1)'$
broken at an intermediate scale), and Majorana masses associated with
a TeV-scale seesaw~\cite{valle1} or a heavy Higgs triplet (Type
II seesaw)~\cite{Hambye:2000ui}. The small Dirac mass case involves
a strong constraint from Big Bang Nucleosynthesis, because the $Z'$
interactions could efficiently produce the right-handed components prior
to nucleosynthesis, leading to too much $^4$He~\cite{Olive:wz}. For
generic couplings of the $Z'$ to the $N_R$ the observed abundance
implies a $Z'$ mass larger than around 4 TeV, stronger than indirect or
collider constraints~\cite{Barger:2003zh}. This can be evaded or weakened
if the $N_R$ carries no $U(1)'$ charge (as can occur naturally in some
models involving two $U(1)'$ factors~\cite{Langacker:2003bc,KLL}) or if the
mass is Majorana.

\newpage

\section{Neutrino Masses in Theories with Dynamical Symmetry Breaking}

The source of electroweak symmetry breaking (EWSB) remains unknown, and a
dynamical origin of this breaking is an appealing possibility.  This can be
realized via the formation of a bilinear condensate involving fermions with a
new strong gauge interaction, generically called Technicolor (TC) 
\cite{tc,dsb}.  Indeed, one
may recall that in both of the well-known two cases in which scalar fields have
been used in models of spontaneous symmetry breaking, namely the
Ginzburg-Landau effective Hamiltonian for superconductivity and the Gell-Mann
Levy sigma model for hadronic chiral symmetry breaking, the scalar fields were
not fundamental, and the true underlying physics responsible for these
respective phenomena involved the formation of bilinear fermion condensates
(Cooper pairs in superconductivity and the $\langle \bar q q \rangle$
condensate in QCD).  In order to communicate this symmetry breaking in the
Technicolor sector to the standard-model (technisinglet) fermions, one embeds
the Technicolor model in a larger, extended Technicolor (ETC) theory
\cite{etc,tcrev}. To satisfy constraints from flavor-changing neutral-current
processes, the ETC vector bosons that mediate generation-changing transitions
must have large masses. These masses arise from the sequential breaking of the
ETC gauge symmetry on mass scales ranging from $10^3$ TeV down to the TeV
level.  Precision measurements place tight constraints on these models,
suggesting that there are a small number of new degrees of freedom at the TeV
scale and that the Technicolor theory has a slowly running (``walking'') gauge
coupling with large anomalous dimensions \cite{wtc}.

Since ETC models do not involve any superheavy GUT-scale mass, there was for a
long time a puzzle of how one could explain light neutrino masses in these
models.  A possible solution to this puzzle was given in Ref.\ \cite{nt} and
studied further in Refs. \cite{lrs,nuf03,ckm}.  This does involve a seesaw,
but one of a new type not involving any superheavy GUT scale.  The resultant
formula $M_\nu \simeq (M_\nu^D)^2/m_R$ holds, with the largest Dirac neutrino
masses of order a few keV and the relevant Majorana neutrino mass of order
O(0.1) GeV to O(100) GeV.  These Dirac and Majorana neutrino masses are
greatly suppressed relative to conventional values.  This suppression is a
natural consequence of the representations of the ETC gauge group for the
various neutrino fields.  These ETC models are not yet developed sufficiently
to make detailed predictions for leptonic mixing angles, but it seems possible
to get substantial neutrino mixing.  One interesting feature of this mechanism
for neutrino masses is that there are only two, rather than three right-handed
electroweak-singlet neutrinos, in contrast, e.g., to $SO(10)$ GUT models.

The ETC gauge group SU($N_{ETC}$) commutes with the Standard Model (SM) group
$G_{SM}$.  The ETC group gauges the three generations of technisinglet fermions
and connects them with the technicolored fermions.  The ETC gauge symmetry is
chiral, so that when it becomes strong, sequential breaking occurs naturally.
The ETC symmetry breaking takes place in stages, leaving the residual exact
technicolor gauge symmetry SU($N_{TC}$).  This entails the relation
$N_{ETC}=N_{gen}+N_{TC}=3+N_{TC}$, where $N_{gen}$ is the number of
standard-model fermion generations.  The choice of $N_{TC}=2$ is required for
the mechanism of Ref.\ \cite{nt} to work.  This thus implies $N_{ETC}=5$; i.e.,
one uses an $SU(5)_{ETC}$ gauge theory.  A related $SU(5)_{ETC}$ theory had
previously been studied in Ref.\ \cite{at94}. The choice $N_{TC}=2$ has two
other motivations: (a) it minimizes the TC contributions to the electroweak $S$
parameter, which is a stringent constraint on TC theories, (b) with a
standard-model family of technifermions, $Q_L = {U \choose D}_L$, $L_L = {N
\choose E}_L$, $U_R$, $D_R$, $N_R$, $E_R$ transforming according to the
fundamental representation of SU(2)$_{TC}$, it can yield an approximate
infrared fixed point and the associated walking behavior.  This sequential
breaking of the $SU(5)_{ETC}$ is driven by the condensation of SM-singlet
fermions.

One can explore whether this dynamical neutrino mass mechanism could take place
in the context of ETC theories in which the strong-electroweak gauge group is
extended beyond that of the Standard Model. Theories with the left-right
symmetry group $G_{LR} = {\rm SU}(3)_c \times {\rm SU}(2)_L \times {\rm
SU}(2)_R \times {\rm U}(1)_{B-L}$ \cite{moh} and the group $G_{422}={\rm
SU}(4)_{PS} \times {\rm SU}(2)_L \times {\rm SU}(2)_R$ \cite{ps} are of
particular interest here, where $B$ and $L$ denote baryon and lepton number and
$SU(4)_{PS}$ combines color and lepton number.  Ref.\ \cite{lrs} presented a
full ETC model in which these extended strong-electroweak gauge symmetries can
be broken dynamically and showed that the mechanism of Ref.\ \cite{nt} can also
hold here.  Dynamical symmetry breaking of $G_{LR}$ has also been studied in
Ref.\ \cite{lrslindner}. Further, dynamical symmetry breaking of the
electroweak symmetry can be triggered by a neutrino condensate
\cite{Martin:1991xw}.  ETC theories have many other testable implications. Some
recent work has focused on ETC contributions to dimension-5 dipole moment
operators \cite{dml,qdml} and dimension-6 four-fermion operators and their
effects \cite{ckm}.

\newpage
\section{Neutrinos in extra dimensions}

The pioneering idea by  Kaluza and Klein~(KK)~\cite{KK} that our world
may  have more than four dimensions  has attracted renewed interest
over the last ten years~\cite{IA,EW,ADD,DDG1}.  The possible existence
of extra dimensions   has  enriched dramatically our   perspectives in
searching for  physics  beyond the  Standard Model.   Obviously, extra
dimensions have  to be sufficiently  compact to  explain why they have
escaped   detection so  far, although their   allowed  size is  highly
model-dependent~\cite{pheno}.  This means that the derived constraints
not  only  depend  on the   number  of the fields   sensitive to extra
dimensions and their transformation  properties with respect to those,
but also on the geometry  and/or the shape of  the new dimensions.  In
the  latter  case, higher-dimensional   theories may  be distinguished
between  those  formulated on a flat  space  and those  that utilize a
warped geometry.

Higher-dimensional theories  may also provide interesting alternatives
for explaining the smallness  of  the observed light neutrino  masses.
Their predictions  for the light-neutrino   spectrum can be confronted
with  recent neutrino oscillation data.  In  the following, we discuss
a generic higher-dimensional neutrino scenario 
in which a
flat
geometry is realized.


The        original  formulation   of     higher-dimensional  neutrino
models~\cite{DDG2,ADDM} relies  on the  possible existence of  singlet
neutrinos that propagate  in  a higher $[1 +  (3+\delta)]$-dimensional
space which is usually  termed the bulk, where  $\delta$ is the number
of the additional  spatial compact  dimensions.  In this  formulation,
the  ordinary SM particles  reside in  a $(1+3)$-dimensional Minkowski
subspace, which is called the wall.   Hence, the left-handed neutrinos
and  the Higgs  bosons   live on the    wall.   The overlap  of  their
wave-functions with the bulk neutrinos  is suppressed by the volume of
the   extra-dimensional  space  $(R\,  M_F)^{\delta/2} \approx  M_{\rm
P}/M_{\rm  F}$,   where $R$  is  the   common compactification radius,
$M_{\rm F}$ is  the fundamental gravity  scale and  $M_{\rm P} \approx
10^{16}$~TeV is the usual Planck mass.  This volume-suppression factor
gives  rise to effective  neutrino Yukawa couplings that are naturally
very small, i.e.~of  order $M_{\rm  F}/M_{\rm  P} \sim  10^{-15}$, for
$M_{\rm F} = 10$~TeV, although  the original higher-dimensional Yukawa
couplings  of the theory could   be of order  unity.  This suppression
mechanism~\cite{DDG2,ADDM}     is  a   generic     feature  of   these
higher-dimensional neutrino models  realized on  an toroidal bulk;  it
has some dependence  on  the compactification  properties of the  bulk
neutrinos and    the    structure  of  the   Higgs    sector    of the
theory~\cite{DS,AP1}.

To  illuminate the  discussion  that follows,  we  consider a  minimal
extension of  the SM where  one 5-dimensional (bulk)  sterile neutrino
$N$ has  been added.   Furthermore, the extra  dimension, say  $y$, is
compactified on a $S^1/Z_2$ orbifold.  The  SM fields are localized on a
4-dimensional Minkowski  subspace at  the orbifold fixed  point $y=0$.
Different minimal models may be  constructed depending on the way that
lepton number is broken:

\begin{itemize}

\item[(i)] One  may   add  lepton-number violating  bilinears   of the
Majorana type  in  the  Lagrangian~\cite{DDG2},  e.g.~operators of the
form $N^T  C^{(5)-1} N$, where $C^{(5)} =  - \gamma_1 \gamma_3$ is the
charge conjugation operator.

\item[(ii)] Lepton-number-violating mass terms can be generated through
the Scherk-Schwartz mechanism~\cite{SS}. This mechanism turns out to
be equivalent to (i), after KK reduction.

\item[(iii)]  Lepton  number  can  be  broken if  the  $Z_2$-even  and
$Z_2$-odd   two-component  spinors   of  the   bulk   neutrino  couple
simultaneously to the same  left-handed charged lepton state.  This is
only  possible  if the  3-brane  describing  our  observable world  is
shifted from the $S^1/Z_2$ orbifold fixed point~\cite{DDG2,BKPP}.

\item[(iv)] Violation  of   the  lepton  number  can be  achieved   by
introducing  operators  of higher dimension  in the  number of fields,
e.g.~$(L\Phi)^2/M_{\rm F}$. Such   operators may be generated  through
gravity effects~\cite{MNP}.

\end{itemize}

The current neutrino oscillation  data  provide an important test  for
singling out a good  candidate model that  includes higher-dimensional
neutrinos.       For  example,   orbifold     models   with   one bulk
neutrino~\cite{DDG2,DS,MNP,MP,BCS,LRRR,DLP}, based on models of type
(i) and/or (ii) mentioned above, seem to prefer the Small Mixing Angle
(SMA)  solution  which is highly   disfavored by recent neutrino data
analyses.  Alternatively, if all neutrino data  are to be explained by
oscillations of active neutrinos with  a small admixture of a sterile KK
component, then the compactification scale has  to be much higher than
the brane-Dirac mass terms.   After integrating out the bulk  neutrino
of the model, the resulting effective light-neutrino mass matrix is of
rank~1.  Because of this restricted form  of the neutrino mass matrix,
two out  of the three  active neutrinos are  massless.  Since only one
neutrino-mass difference  can  be  formed  in   this case,  it  proves
difficult to    accommodate all neutrino   oscillation  data  in these
models~\cite{BCS,LRRR,DLP}.

\subsection{Three bulk neutrinos}
One way to avoid this problem is to add three bulk
neutrinos, one for each generation \cite{MP,DLP}. This model, in the absence
of CP
phases, is characterized by seven parameters: three neutrino masses
$m_{1,2,3}$ and three mixing angles for left handed neutrinos as defined
earlier and the radius of the large extra dimension. Since the three
mixing angles are arbitrary, the model can easily accommodate the
bilarge mixing solution preferred by oscillation data. In the diagonal
mass basis, the bulk neutrinos are associated with mass eigenstates. The
mixing of the
$i$th active neutrino with the nth KK mode of the corresponding bulk
neutrino is given by $\xi_{i,n}\simeq m_i R/n$. It is interesting that
all mixings are intimately connected with the masses. There are limits on
$\xi_i$ from laboratory data such as CHOOZ-Palo-Verde as well as from big
bang nucleosynthesis \cite{GM}. BBN constraints for one extra dimension
give $\xi^2_3 \leq 1.7\times 10^{-4}{(eV.R)}^{0.92}$. For a hierarchical
pattern, using $\xi= m_3 R$, one gets $R\leq 0.03$ eV$^{-1}$. The bounds
from neutrino data such as solar and atmospheric etc.\ are less
stringent \cite{DLP} and are roughly given by: $R\leq 4$ eV$^{-1}$.

Among the consequences of this model, two are especially interesting. Both
of these concern the KK tower of sterile neutrinos.

\begin{figure}[htb]
\begin{center}
\epsfxsize15cm\epsffile{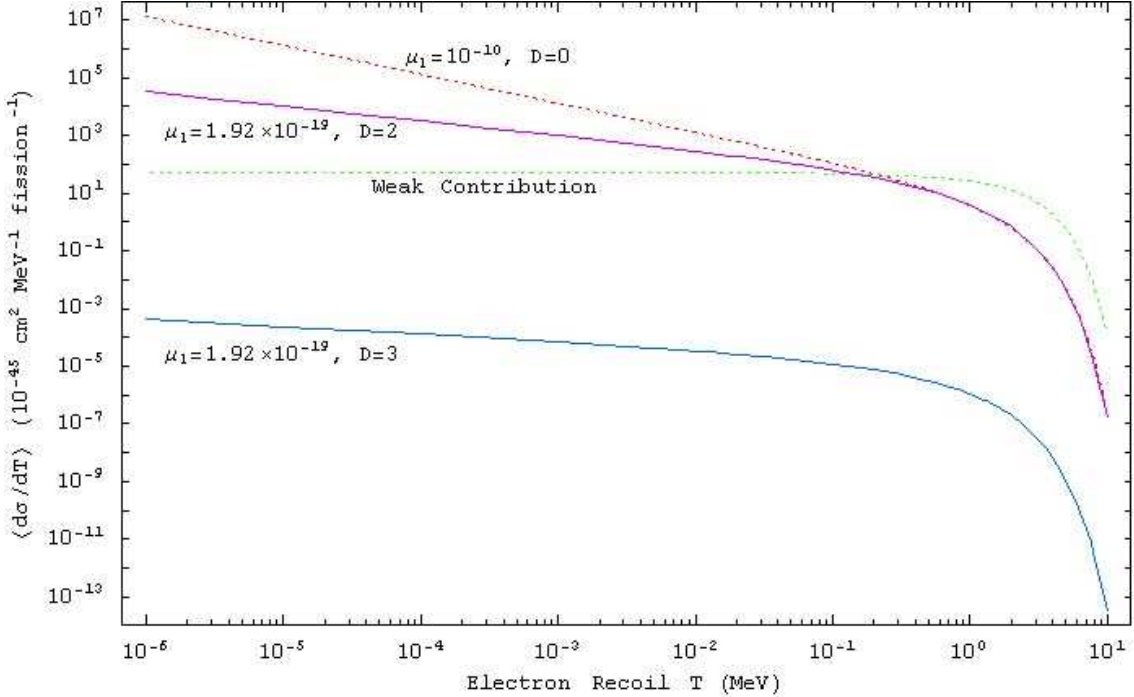}
\caption{
The figures give the contribution of neutrino magnetic moment for the
case of single Dirac neutrino, for two large extra dimensions (and a
comparison between the two) to differential cross section
$\frac{d\sigma}{dT}$ (where $T$ is the electron recoil energy) for neutrino
electron scattering and compares it with the case of one right-handed
neutrino (``Standard Model with one right-handed neutrino").
\label{fig:cstr1}}
\end{center}
\label{haibo}
\end{figure}
(i) In the presence of the infinite tower of states, the magnetic moment
of the neutrino gets contribution from all the states \cite{ng}. For
instance in
the scattering of a neutrino of energy $E\approx 10 $ MeV (corresponding
to a reactor neutrino beam) the number of
states contributing to the magnetic moment is given by $(ER)^2\sim
10^{18}$. Since
all the states add incoherently, the effective magnetic moment is
increased from $10^{-20}\mu_B$ to $10^{-11}\mu_B$ ($\mu_B$ is the Bohr
magneton). The effect on the differential cross section $d\sigma/dT$,
where $T$ is the electron recoil energy has recently been calculated and
given in Fig.~\ref{haibo} \cite{haibo}.

(ii) A second consequence of the existence of the KK tower of sterile
neutrinos is the possibility that when neutrinos travel through dense
matter there can be MSW resonances \cite{DS} and give rise to a dip
pattern \cite{DS,cmy} in the neutrino survival probability corresponding to
energies spaced by $E\approx \Delta m^2_{\nu_F\nu_{KK}}/2\sqrt{2}G_F
N_e$ (i.e. $E, 4E, 9E,\ldots$). The dip arises because typically the survival
probability goes like $e^{-c\frac{\Delta m^2}{E}}$. Therefore at lower
energies, there is more suppression which with increase in energy becomes
less and less effective. Also the resonance condition is not satisfied as
energy increases, if it was satisfied at lower energies. For the solar
neutrinos, such dip structure is quite pronounced \cite{cmy}.
In the hierarchical pattern for neutrino masses, this would correspond to
$E\approx $ 10 MeV for densities comparable to solar core. The value of
the energy clearly depends on the size of the extra dimensions, growing
with $R^{-1}$. This is a very interesting phenomenon which could be used
to probe this class of extra dimension models.

\subsection{Lepton number breaking in the bulk}
In the three bulk neutrino picture, all the neutrinos are Dirac neutrinos
since the model has an additional global $B-L$ symmetry.
An interesting possibility is the scenario (iii) described above,
where  sizable lepton-number  violation  is induced  by shifting  the
$y=0$  brane by  an  amount $a  \sim  1/M_W$.  In  this scenario,  the
tree-level rank-1  form of the  effective neutrino mass matrix  can be
significantly modified  through lepton-number violating  Yukawa terms,
thus offering sufficient freedom  to describe the neutrino oscillation
data~\cite{BKPP}.

In  addition to  constraints   from neutrino  oscillation  data, other
experiments can  also    play an   important role in    constraining
higher-dimensional neutrino models.   Specifically,  strong  limits on
$M_F$ and the Yukawa couplings of the  theory may be obtained from the
non-observation  of lepton-flavor-violating  decays  in  muon and tau
decays  and    from   the    absence   of   $\mu$-$e$   conversion  in
nuclei~\cite{IP,FP}.  Table~\ref{Tabextra} gives  a  brief  summary of
these limits. These  phenomenological constraints are complementary to
those obtained   from   pure  theoretical  considerations,    such  as
perturbative unitarity~\cite{CGY}.

\begin{table}
\centering
\begin{tabular}{|cccc|}
\hline
 & $h_e\ =\ h_\mu$ &\hspace{-0.7cm}$=\
h_\tau\ =\ h\ \stackrel{>}{{}_\sim}\ 1$ & \\
Observable & Lower ~limit &
\hspace{-0.2cm}on $M_F/h^2\ [\,{\rm TeV}\,]$& \\
& $\delta =2$ & $\delta = 3$ & $\delta =6 $  \\
\hline
BR$(\mu \to e \gamma)$ & $75$ & $43$ & $33$\\
\hline
BR$(\mu \to e e e)   $ & $250$ & $230$ & $200$\\
\hline
BR$(\mu \ ^{48}_{22}{\rm Ti} \to e \ ^{48}_{22}{\rm Ti})$ & $380$ & $320$
& 300\\
\hline
\end{tabular}
\caption{ \label{Tabextra}
One-loop-level limits on $M_F/h^2$ from~\cite{IP}.
}
\end{table}

Another low-energy experiment of  great importance is the neutrinoless
double beta    decay  of     a  nucleus.    The   recently  claimed
experimental evidence ~\cite{klapdor} of
an effective  neutrino mass of  order  0.4~eV (see
however \cite{antiklapdor}),
combined  with information from   solar and atmospheric neutrino data,
restricts the admissible  forms of the light-neutrino mass hierarchies
in 4-dimensional models with   3 left-handed (active)  neutrinos.  The
allowed scenarios  contain  either  degenerate neutrinos or  neutrinos
that   have   an  inverse  mass    hierarchy~\cite{KS}.   A   positive
interpretation  of     the   recently    reported     $0\nu\beta\beta$
signal~\cite{klapdor} imposes additional constraints on  model-building.
For  example,  higher-dimensional  models   that utilize   the shining
mechanism from a distant brane~\cite{MLP} may accommodate an effective
neutrino mass of 0.4~eV but also predict the emission of Majorons.  On
the other hand, 5-dimensional models  formulated on a warped geometric
space~\cite{Huber} face difficulties to   account for  the  observable
excess in~\cite{klapdor}.

In   the context of $S^1/Z_2$  orbifold  models, one  has  to solve an
additional theoretical problem.   The   resulting KK  neutrinos  group
themselves   into  approximately  degenerate  pairs   of  opposite  CP
parities.  Because  of  this, the  lepton-number-violating KK-neutrino
effects   cancel each other  leading to   unobservably small predicted
values for  the $0\nu\beta\beta$  decay.  These   disastrous CP parity
cancellation  effects  can  be  avoided   by  arranging the   opposite
CP parity  KK neutrinos to  couple to the  $W^\pm$ bosons with unequal
strength.   This feature can naturally  be implemented if  the $y = 0$
wall displaced  from one of the  $S^1/Z_2$ orbifold fixed points by an
amount  of    order  $1/M_W$.    A unique     prediction   of such   a
model~\cite{BKPP} is that the effective neutrino mass and the scale of
the light neutrino masses can be completely decorrelated.


\section{Other new physics and neutrinos}

\subsection{New long range forces}

Long range forces in the context of particle physics originated
with the ideas of Yang and Lee \cite{yang} and Okun \cite{okun}
who proposed that gauging the baryon number or lepton number would
give rise to a composition dependent long range force which could
be tested in the classic Eotovos type experiments \cite{adelberger}.
A special class of long range forces which distinguish between
leptonic flavors have far reaching implications for the neutrino
oscillations \cite{am,grifols-masso} which may be used as a
probe of such forces.

The standard
model Lagrangian is invariant under four global symmetries corresponding
to the Baryon and three Lepton numbers $L_\alpha$ $(\alpha=e,\mu,\tau)$.
Of these, only three
combinations \cite{foot} of lepton numbers (i) $L_e-L_\mu$,
(ii) $L_e-L_\mu$ or
(iii) $L_\mu-L_\tau$, can be gauged in an anomaly free way without
extending the matter content. The existence of neutrino oscillations
implies that these symmetries have to be broken but the relevant gauge
bosons can still be light if the corresponding couplings are very weak.
It
is possible in this case to obtain light gauge boson induced forces having
terrestrial range
(e.g.\ the Sun-Earth
distance) without invoking extremely low mass scales \cite{am}.
The exchange of such boson would induce matter effects in
terrestrial, solar and atmospheric neutrino oscillations. For example,
the electrons inside the sun generate a potential $V_{LR}$ at the earth
surface given by
\begin{equation} \label{vlr}
V_{LR}=\alpha {N_e\over R_{es}}\approx (1.04 \times 10^{-11}
\,\mathrm{eV})\left({\alpha\over 10^{-50}}\right) ~, \end{equation} where
$\alpha\equiv {g^2\over 4 \pi}$ corresponds to the gauge coupling
of the $L_{e}-L_{\mu,\tau}$ symmetry , $N_e$ is the number of
electrons inside the sun and $R_{es}$ is the earth-sun distance
$\approx 7.6 \times 10^{26} \,\mathrm{GeV}^{-1}$. The present bound on the
$Z$-dependent force with range $\lambda\sim 10^{13}$ cm is given
\cite{adelberger} by $\alpha< 3.3\times 10^{-50}$. Eq.\ (\ref{vlr})
then shows that the potential $V_{LR}$ can introduce very
significant matter-dependent effects in spite of the very strong bound
on $\alpha$. One can define a parameter $$ \xi\equiv {2 E_\nu
V_{LR}\over \Delta m^2}$$ which measures the effect of the long
range force in any given neutrino oscillation experiment. Given
the terrestrial bound on $\alpha$, one sees that $\xi$ is given by
$\xi_{atm}\sim 27.4$ in atmospheric or typical long-baseline
experiments while it is given by $\xi_{solar}\sim 7.6$ in the case of
the solar or KamLAND type of experiments. In either case, the long
range force would change the conventional oscillation analysis.
Relatively large value of $\alpha$ suppresses the oscillations of
the atmospheric neutrinos. The observed oscillations then can be
used to put stronger constraints on $\alpha$ which were analyzed
in \cite{am}. One finds the improved 90\% CL bound .
\begin{equation} \label{atmbound}
\alpha_{e\mu}\leq 5.5\times  10^{-52}~~~~;~~~~\alpha_{e\tau}\leq 6.4
\times
10^{-52} ~,
\end{equation}
in case of the $L_{e}-L_{\mu,\tau}$ symmetries, respectively.

Although these bounds represent considerable improvement over the
conventional fifth force bound, they still allow interesting effects which
can be used as a probe of such long range forces in future long-baseline
experiments with super beam or at neutrino factories. As a concrete
example,
let us consider the influence of the $L_e-L_\mu$ gauge interactions on the
long-baseline oscillations of muon neutrinos of ${\cal O}$(GeV) energy.
The oscillations of these neutrinos are governed by the following $3\times
3$ (mass)$^2$ matrix in the flavor basis:
\begin{equation} \label{3by3}
M_{\nu}^2=U^*
\mathrm{Diag}(m_1^2,m_2^2,m_3^2) U^\dagger+\mathrm{Diag}(A_{CC}+A_{LR},-A_{LR},0) ~.
\end{equation}
$U$ denotes the (vacuum) mixing matrix for which we adopt the conventional
parametrization. $A_{CC}=2E_\nu \sqrt{2} G_F n_e\approx (1.04 \times
10^{-13}\,\mathrm{eV}) 2 E_\nu$ describes the conventional MSW matter contribution
generated
by the earth matter (density $\rho\sim 2.8~ \rm gm/cm^3$; electron
fraction $Y_e\sim 0.49$). The
$A_{LR}\approx (1.04 \times 10^{-13}\,\mathrm{eV}) 2 E_\nu \alpha_{52}$
with $\alpha_{52}$ denoting the coupling of the long range
force measured in units of $10^{-52}$. The $A_{LR}$ term dominates over
$A_{CC}$ if $\alpha$ saturates the bound in Eq.\ (\ref{atmbound}).
\begin{figure}[htb]
\begin{center}
\hspace*{-8mm}
\epsfig{file=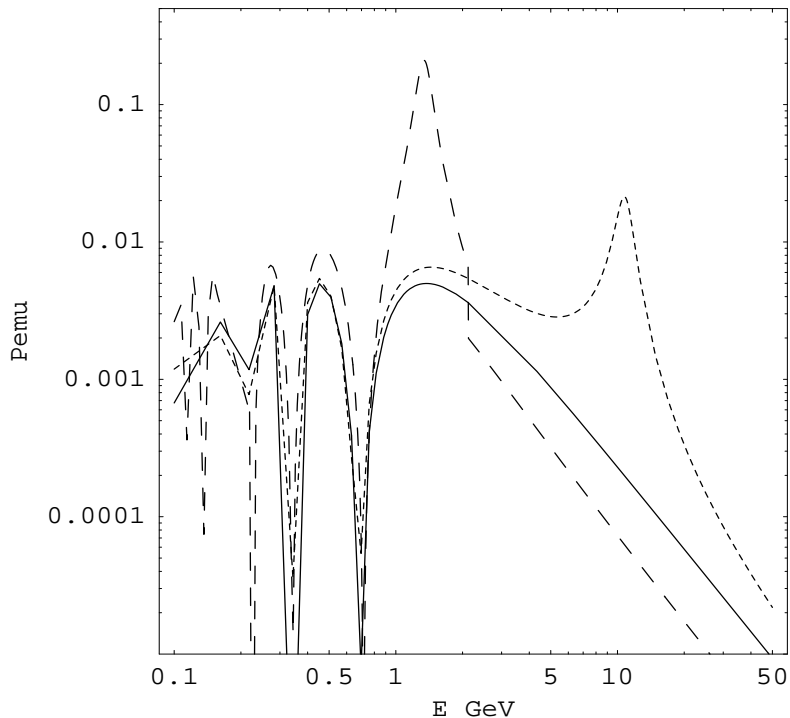,width=11cm,height=12cm}\\
\caption{{\small The long-baseline neutrino oscillation probability
$P_{e\mu}$ in case
of vacuum (solid), earth matter effects (dotted) and with inclusion
of the long range potential $V_{LR}$ (dashed). The plotted curves
correspond to a baseline of 740 km, $\Delta m_{32}^2=2.5 \times
10^{-5}$ eV$^2$, $\Delta m_{21}^2=7.0 \times 10^{-5}$ eV$^2$,
($\theta_{12},\theta_{23})=(32^\circ,45^\circ)$ , $\alpha_{e\mu}=5.5
\times 10^{-52}$ and $\sin \theta_{13}=0.05$.
 }}
\label{amfig1}
\end{center}
\end{figure}
The matter induced terms in Eq.\ (\ref{3by3}) modify the neutrino
oscillation in a non-trivial manner. This effect is analyzed in
the limit of the vanishing solar scale and $A_{LR}=0$ in
\cite{matter}. The $23$ mixing angle remains unaffected by the
matter induced contribution but the $13$ angle can get resonantly
enhanced for a neutrino energy given by
\begin{equation} E_\nu\approx
{\cos 2 \theta_{13}\Delta m_{32}^2\over 2 \sqrt{2} G_F n_e}\approx
11.8 \,\mathrm{GeV} ~.
\end{equation}
This leads to a rise in the oscillation
probability $P_{e\mu}$, Figure(\ref{amfig1}). The additional long
range contribution results in a noticeable shift in the resonance
energy as seen from Fig.\ (\ref{amfig1}). We assumed normal
hierarchy in this figure. The resonance behavior would be absent
in case of the inverted hierarchy or in case of the anti-neutrino
beam. While more detailed study is required to distinguish these
cases, it is clear that future observations of matter effects in
the long-baseline neutrino experiments provide a good probe of
additional long range forces.

\subsection{Lorentz noninvariance, CPT violation and decoherence}

\subsubsection{CPT Violation}

In this section, we discuss neutrino oscillation phenomenology in the
presence of CPT violation.
CPT is a symmetry in any theory that satisfies the three assumptions
that are normally taken for granted: (1) locality, (2) Lorentz
invariance, and (3) hermiticity of the Hamiltonian.  In particular,
it predicts that the mass is common for a particle and its
anti-particle.  Any violation of CPT therefore would have a profound
consequence on fundamental physics.

The best limit on CPT violation is in the neutral kaon system,
$|m(K^0) - m(\overline{K}^0)| < 10^{-18} m_K = 0.50 \times
10^{-18}$~GeV \cite{Hagiwara:fs}.  Having such a stringent bound does
not seem a sizable CPT violation in neutrino at the first sight.
However, the kinematic parameter is mass-squared instead of mass, and
the constraint may naturally be considered on the CPT-violating
difference in mass-squared $|m^2(K^0) - m^2(\overline{K}^0)| <
0.25$~eV$^2$.  In comparison, the combination of SNO and KamLAND data
leads to the constraint $|\Delta m^2_\nu - \Delta m^2_{\bar{\nu}}| <
1.3 \times 10^{-3}$~eV$^2$ (90\% CL) and hence currently the best
limit on CPT violation \cite{Murayama:2003zw}.

Having seen that the CPT violation in neutrino masses may be of size
relevant to neutrino oscillation, it is useful to discuss how it may
affect the phenomenology.  In fact, the primary motivation for recent
discussions on CPT violation in neutrino oscillation has been to
reconcile LSND data with other data \cite{Murayama:2000hm}.  It is
well known that the LSND data is not quite consistent with the other
oscillation evidence and limits even if a sterile neutrino state is
introduced, both for $2+2$ and $3+1$ spectra (see
\cite{Maltoni:2003yr} for a recent analysis; adding more sterile states helps
\cite{sorel}).  The main point is that the LSND
oscillation is primarily an anti-neutrino oscillation $\bar{\nu}_\mu
\rightarrow \bar{\nu}_e$, while the solar neutrino oscillation is
purely neutrinos $\nu_e \rightarrow \nu_{\mu,\tau}$.  It was shown to
fit the LSND, solar, and atmospheric neutrino data simultaneously
without invoking a sterile neutrino at that time
\cite{Murayama:2000hm,Barenboim:2001ac,Strumia:2002fw}.  Phenomenology
had been further developed in
Refs.~\cite{Barenboim:2002rv,Barenboim:2002ah}.

\begin{figure}[htb]
  \centering
  \includegraphics[width=0.4\textwidth]{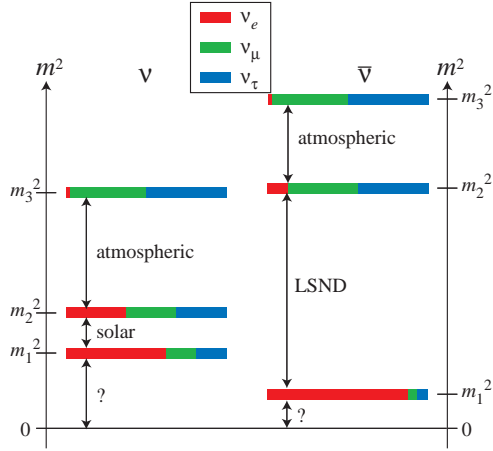}
  \caption{The proposed spectra of neutrinos and anti-neutrinos in
    \cite{Murayama:2000hm} and \cite{Barenboim:2001ac}.  Excluded by
    KamLAND.}
  \label{fig:oldCPTviolation}
\end{figure}

However, KamLAND data shows $\bar{\nu}_e \rightarrow
\bar{\nu}_{\mu,\tau}$ oscillation with parameters consistent with the
solar neutrino oscillation, and the CPT-violation alone cannot explain
LSND.  A new proposal tried to explain LSND and atmospheric
anti-neutrino oscillations with a single $\Delta m^2$
\cite{Barenboim:2002ah}, which was excluded by a global fit in
\cite{Gonzalez-Garcia:2003jq}.  Currently the best fit to the data is
obtained by allowing for one sterile neutrino {\it and}\/ CPT
violation \cite{Barger:2003xm}.  Because the short-baseline
experiments that are constraining the interpretation of LSND data with
sterile neutrino are mostly in neutrinos but not in anti-neutrinos,
the $3+1$ spectrum is allowed if there is little mixing of the sterile
state with others in neutrinos.

\begin{figure}[htb]
  \centering
  \includegraphics[width=0.4\textwidth]{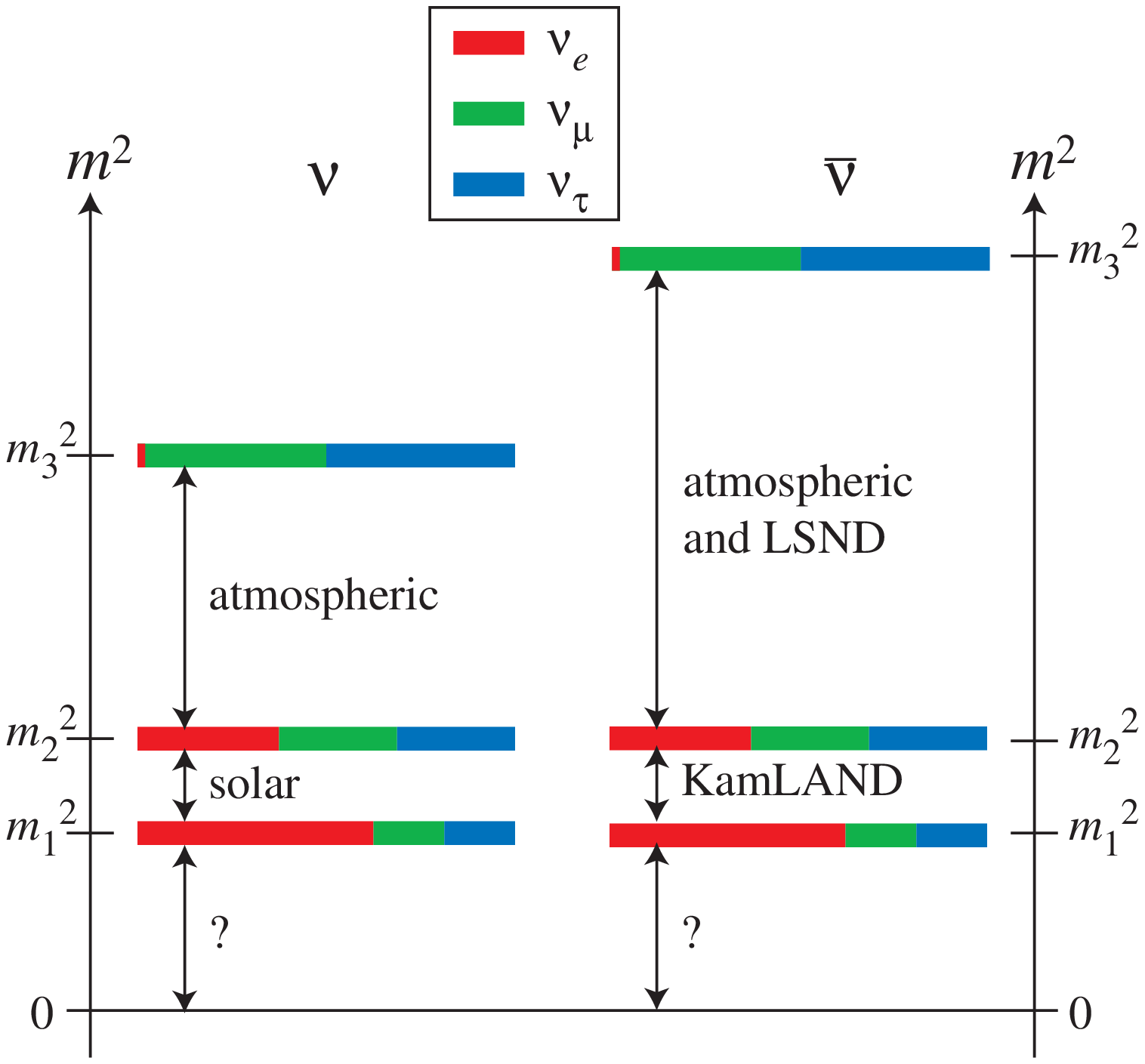}
  \includegraphics[width=0.4\textwidth]{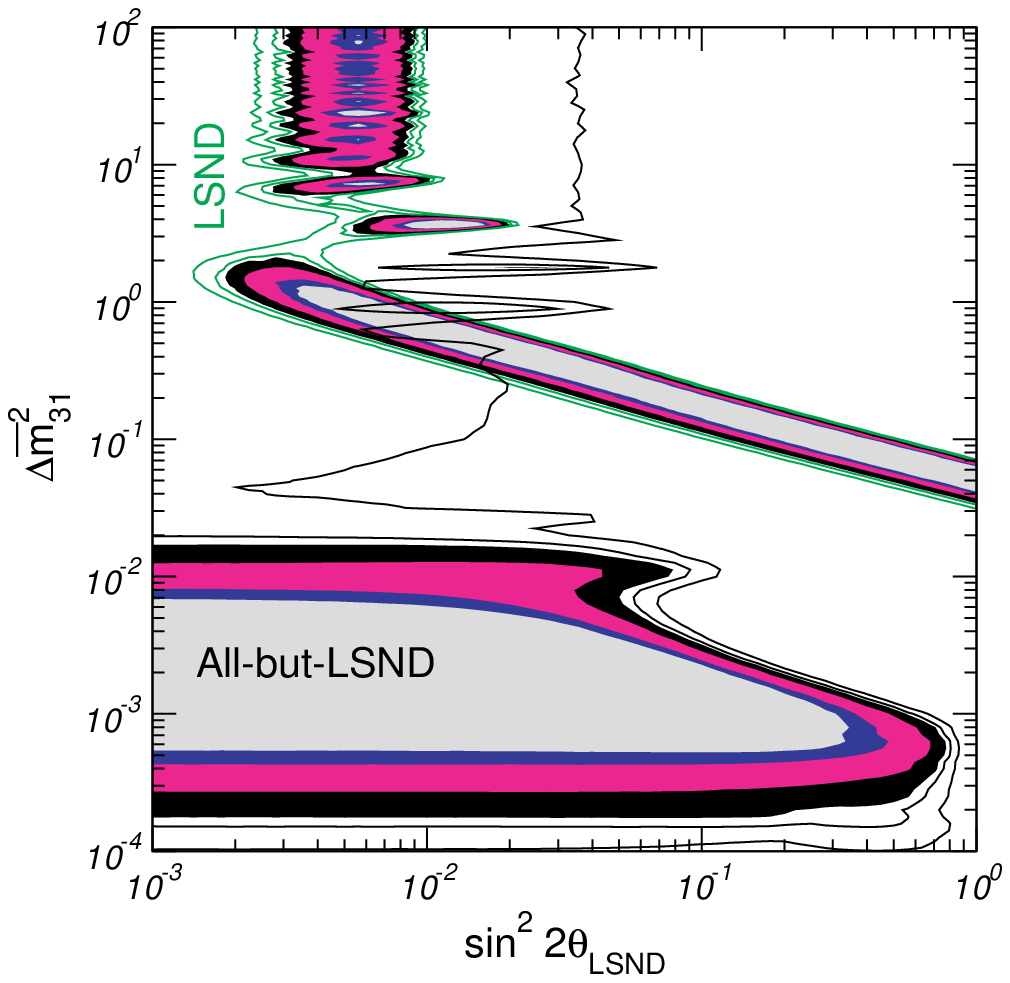}
  \caption{The revised proposal in \cite{Barenboim:2002ah}.  Excluded
    by the analysis taken from \cite{Gonzalez-Garcia:2003jq}.}
  \label{fig:newCPTviolation}
\end{figure}

\begin{figure}[htb]
  \centering
  \includegraphics[width=0.4\textwidth]{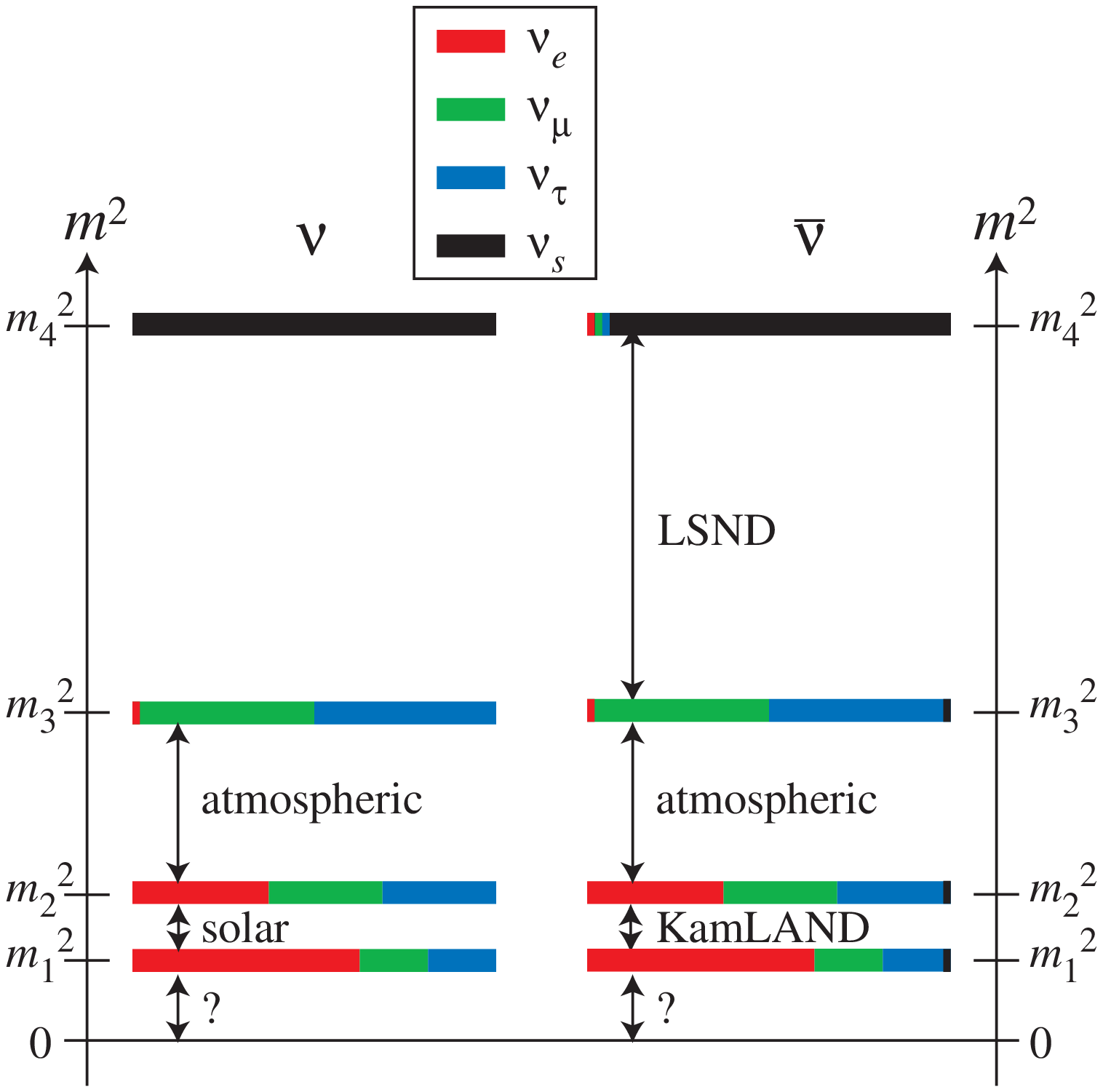}
  \includegraphics[width=0.4\textwidth]{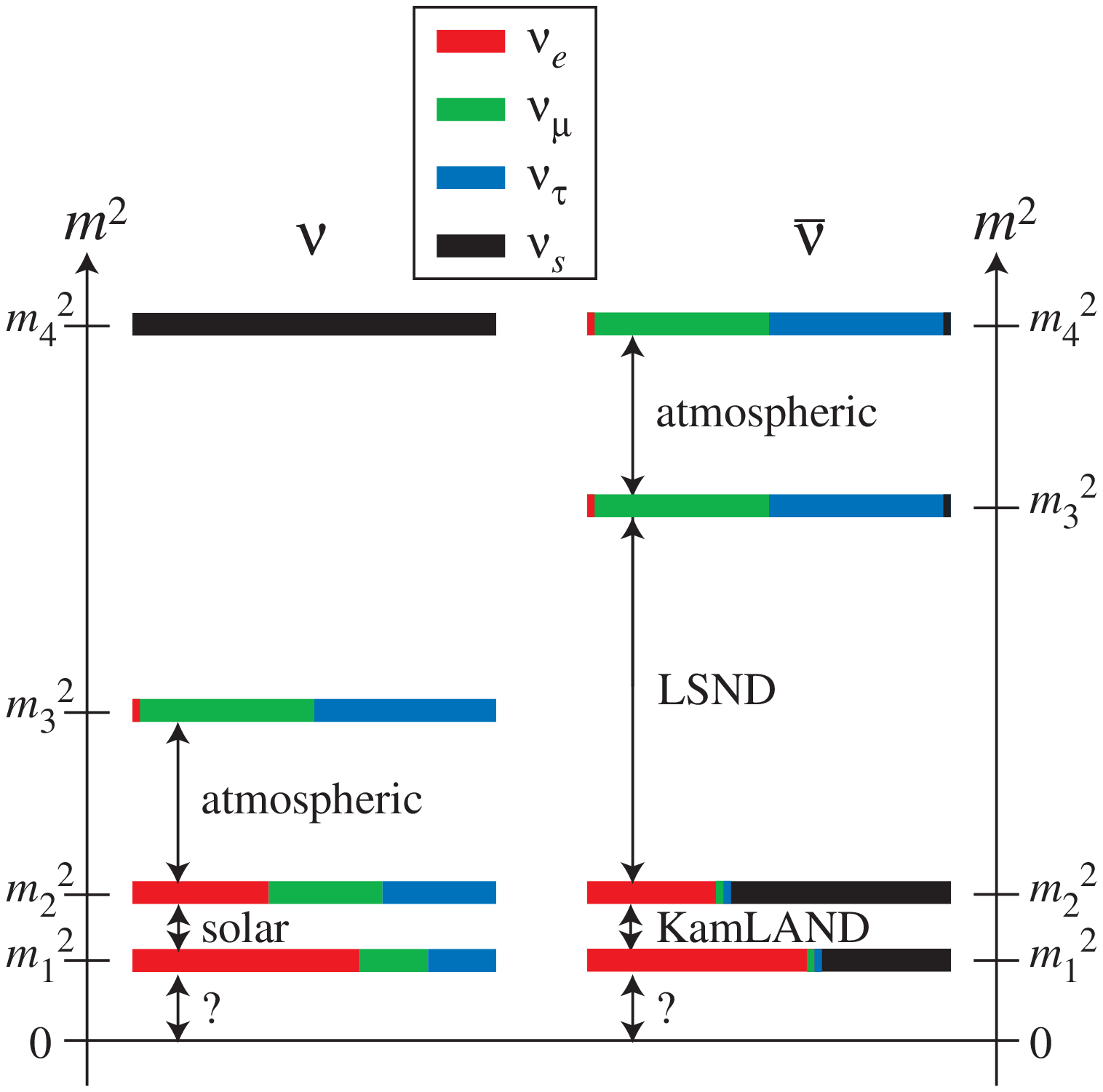}
  \caption{The revised proposal in \cite{Barger:2003xm} that combines
    CPT violation and a sterile neutrino.  The neutrinos always have
    $2+2$ spectrum, while the anti-neutrinos may have either $3+1$ or
    $2+2$ spectrum.}
  \label{fig:newCPTviolation2}
\end{figure}

Even though arbitrarily changing the neutrino and anti-neutrino masses
seems to preserve Lorentz invariance, interacting theory also
violates Lorentz invariance \cite{Greenberg:2002uu}.  All discussions
above assumed Lorentz invariance and hence should be regarded as
purely phenomenological exercise.  One theoretically well-defined way
to break CPT is to introduce a cosmological ``matter effect,'' namely
a background number density coupled to neutrinos.  However, such
framework does not explain data consistently \cite{DeGouvea:2002xp}.
See also \cite{Kostelecky:2003xn} for a different framework of CPT
violation and a recent discussion on the use of decoherence and
CPT violation \cite{Barenboim:2004wu}. 
Lorentz invariance violation in the neutrino sector can arise via
the see-saw mechanism. As discussed in \cite{Choubey:2003ke}
this could explain why it would not be seen in the charged lepton sector.

\subsubsection{Decoherence}

So far, CPT violation as an inequality of masses between
particles and antiparticles was the only way we understood
CPT violation in high energy physics.
However, is this CPT violation the only
way a violation of this symmetry can
manifest itself in nature? Such a question
becomes extremely relevant for the case of LSND, because it is
possible that other mechanisms leading to CPT violation exist,
unrelated, in principle, to mass differences between
particles and antiparticles. Such additional mechanisms for
CPT violation may well be capable of explaining the LSND results
within a three generation scenario without invoking a sterile neutrino
(a scenario which, on the other hand, is getting totally excluded as
new experimental data become available).
It is therefore necessary to explore whether alternative ways
exist to account for the LSND result without invoking extra (sterile)
neutrino states.

Quantum decoherence is the key to answer this question. Indeed, quantum
decoherence in matter propagation occurs when the matter subsystem interacts
with an `environment', according to the rules of open-system quantum mechanics.
At a fundamental level, such a decoherence may be the result of propagation of
matter in quantum gravity space-time backgrounds with `fuzzy' properties, which
may be responsible for violation of CPT in a way not necessarily related to
mass differences between particles and antiparticles.

A characteristic example of such a violation occurs in quantum gravity models
that involve singular space-time configurations, integrated over in a path
integral formalism, which are such that the axioms of quantum field theory, as
well as conventional quantum mechanical behavior, cannot be maintained. Such
configurations consist of wormholes, microscopic (Planck size) black holes, and
other topologically non-trivial solitonic objects, such as {\it geons}
etc. Collectively, we may call such configurations {\it space time foam}.

It has been argued that, as result, a mixed state description must be used
({\it QG-induced decoherence})~\cite{ehns}, given that such objects cannot be
accessible to low-energy observers, and as such must be traced over in an
effective field theory context. As a consequence of that CPT invariance in its
strong form must be abandoned in a foamy quantum gravity theory.  Such a
breakdown of CPT symmetry is a fundamental one, and, in particular, implies
that a proper CPT operator may be {\it ill defined} in such QG decoherence
cases.

Some caution should be paid regarding CPT violation through decoherence.  From
a formal view point, the non-invertibility of the S-matrix, which implies a
strong violation of CPT, does not preclude a softer form of CPT invariance, in
the sense that any strong form of CPT violation does not necessarily have to
show up in any single experimental measurement. This implies that, despite the
general evolution of pure to mixed states, it may still be possible in the
laboratory to ensure that the system evolves from an initial pure state to a
single final state, and that the weak form of CPT invariance is manifested
through the equality of probabilities between these states.  If this is the
case, then the decoherence-induced CPT violation will not show up in any
experimental measurement.

In the parametrization of \cite{ehns} for the decoherence effects, one uses
three decoherence parameters with dimensions of energy, $\alpha,\beta,\gamma$,
where the positivity of $\rho$, required by the fact that its diagonal elements
express probability densities, implies $\alpha, \gamma \ge 0$, and $\alpha
\gamma \ge \beta^2$.  If the requirement of a completely positive map $\rho
(t)$ is imposed in the two generation case, then ${\cal L}$ becomes diagonal,
with only one non vanishing entry occupied by the decoherence parameter $\gamma
> 0$~\cite{benatti}.  Following this approach, for a three generation scenario,
we will assume for the $9 \times 9$ decoherence matrix ${\cal L}$: $[{\cal
L}_{\mu\nu}]= {\rm Diag}\left(0, -\gamma_1,-\gamma_2,
-\gamma_3,-\gamma_4,-\gamma_5,-\gamma_6,-\gamma_7,-\gamma_8\right)$ in direct
analogy with the two-level case of complete positivity~\cite{lisi,benatti},
although there is no strong physical motivation behind such restricted forms of
decoherence. This assumption, however, leads to the simplest possible
decoherence models, and, for our {\it phenomenological} purposes in this work,
we will assume the above form, which we will use to fit all the available
neutrino data. It must be clear to the reader though, that such a
simplification, if proved to be successful (which, as we shall argue below, is
the case here), just adds more in favor of decoherence models, given the
restricted number of available parameters for the fit in this case.

In order to check these models, we have performed a $\chi^2 $ comparison (as
opposed to a $\chi^2 $ fit) to SuperKamiokande sub-GeV and multi GeV data (40
data points), CHOOZ data (15 data points) and LSND (1 datum), for a sample
point in the vast parameter space of our extremely simplified version of
decoherence models.  Let us emphasize that we have {\bf not} performed a
$\chi^2$-fit and therefore the point we are selecting (by ``eye'' and not by
$\chi$) is not optimized to give the best fit to the existing data. Instead, it
must be regarded as one among the many equally good sons in this family of
solutions, being extremely possible to find a better fitting one through a
complete (and highly time consuming) scan over the whole parameter space.

Cutting the long story short, and to make the analysis easier, we have set
all the $\gamma_i$ in the neutrino sector to zero, restricting this way,
all the decoherence effects to the antineutrino one where we have assumed
for the sake of simplicity, $
\overline{\gamma_{1}}= \overline{\gamma_{2}} = 2 \cdot 10^{-18} \cdot E ~~{\rm and}~~
\overline{\gamma_{3}} = \overline{\gamma_8}=
 1 \cdot 10^{-24}/E~ $,
where
$E$ is  the  neutrino energy,
and barred quantities
refer to the antineutrinos, given that decoherence takes place only
in this
sector in our model.
All the other parameters
are assumed to be zero.  All in all, we have introduced only
two new parameters, two new degrees of freedom, $ \overline{\gamma_{1}}$
and $ \overline{\gamma_{3}}$,
and we shall try
to explain with them all the available experimental data.
(For further details we refer the reader to~\cite{Barenboim:2004wu}).

At this point it is important to stress that the inclusion of two new degrees
of freedom is not sufficient to guarantee that one will indeed be able to
account for all the experimental observations.  We have to keep in mind that,
in no-decoherence situations, the addition of a sterile neutrino (which comes
along with four new degrees of freedom -excluding the possibility of CP
violating phases) did not seem to be sufficient for matching all the available
experimental data, at least in CPT conserving situations.

In order to test our model we have calculated the $\chi^2$ of the 56 data
points mentioned above for different scenarios, The results with which we hope
all our claims become crystal clear are summarized in the table, were we
present the $\chi^2$ comparison for the following cases: (a) pure decoherence
in the antineutrino sector, (b) pure decoherence in both sectors, (c) mixing
plus decoherence in the antineutrino sector, (d) mixing plus decoherence in
both sectors, and (e) mixing only - the standard scenario:

\begin{table}[h]
\centering
\begin{tabular}{|c|c|c|}
\hline
Model & $\chi^2$ without LSND  & $\chi^2$ including LSND  \\[0.5ex]
\hline\hline
(a) & 980.7 & 980.8\\ \hline
(b) & 979.8 & 980.0\\ \hline
(c) & 52.2 & 52.3\\\hline
(d) & 54.4 & 54.6\\\hline
(e) & 53.9 & 60.7\\[0.5ex]
\hline
\end{tabular}
\end{table}

{}From the table it becomes clear that the mixing plus decoherence scenario in
the antineutrino sector can easily account for all the available experimental
information, including LSND. It is important to stress once more that our
sample point was not obtained through a scan over all the parameter space, but
by an educated guess, and therefore plenty of room is left for improvements. On
the other hand, for the mixing-only/no-decoherence scenario, we have taken the
best fit values of the state of the art analysis and therefore no significant
improvements are expected.  At this point a word of warning is in order:
although superficially it seems that scenario (d), decoherence plus mixing in
both sectors, provides an equally good fit, one should remember that including
decoherence effects in the neutrino sector can have undesirable effects in
solar neutrinos, especially due to the fact that decoherence effects are
weighted by the distance traveled by the neutrino, something that may lead to
seizable (not observed!) effects in the solar case.

One might wonder then, whether decohering effects, which affect the
antineutrino sector sufficiently to account for the LSND result, have any
impact on the solar-neutrino related parameters, measured through antineutrinos
in the KamLAND experiment.  In order to answer this question, it will be
sufficient to calculate the electron survival probability for KamLAND in our
model, which turns out to be $ P_{\bar\nu_{e}\rightarrow \bar\nu_{e}}
\mid_{\mbox{\tiny KamLAND}} \simeq .63$, in perfect agreement with
observations. It is also interesting to notice that in our model, the LSND
effect is not given by the phase inside the oscillation term (which is
proportional to the solar mass difference) but rather by the decoherence factor
multiplying the oscillation term.  Therefore the tension between LSND and
KARMEN data is naturally eliminated, because the difference in length leads to
an exponential suppression.

Having said that, it is now clear that decoherence models (once neutrino mixing
is taken into account) are the best (and arguably the only) way to explain all
the observations including the LSND result. This scenario, which makes dramatic
predictions for the upcoming neutrino experiments, expresses a strong
observable form of CPT violation in the laboratory, and in this sense, our fit
gives a clear answer to the question asked in the introduction as to whether
the weak form of CPT invariance is violated in Nature. It seems that, in order
to account for the LSND results, we should invoke such a decoherence-induced
CPT violation, which however is independent of any mass differences between
particles and antiparticles.

This CPT violating pattern, with equal mass spectra for neutrinos and
antineutrinos, will have dramatic signatures in future neutrino oscillation
experiments. The most striking consequence will be seen in MiniBooNE, according
to our picture, MiniBooNE will be able to confirm LSND only when running in the
antineutrino mode and not in the neutrino one, as decoherence effects live only
in the former. Smaller but experimentally accessible signatures will be seen
also in MINOS, by comparing conjugated channels (most noticeably, the muon
survival probability).  Higher energy neutrino beams or long-baseline
experiments, will have significant deviations from the non-decoherence models,
as our effects scale with energy and distance traveled, being therefore the
best tool to explore decoherence models.

If the neutrino masses are actually related to decoherence as a result of
quantum gravity, this may have far reaching consequences for our understanding
of the Early stages of our Universe, and even the issue of Dark Energy that
came up recently as a result of astrophysical observations on a current
acceleration of the Universe from either distant supernovae data or
measurements on Cosmic Microwave Background temperature fluctuations from the
WMAP satellite.  Indeed, decoherence implies an absence of a well-defined
scattering S-matrix, which in turn would imply CPT violation in the strong
form.  A positive cosmological {\it constant} $\Lambda > 0$ will also lead to
an ill definition of an S-matrix, precisely due to the existence, in such a
case, of an asymptotic-future de Sitter (inflationary) phase of the universe,
with Hubble parameter $\sim\sqrt{\Lambda}$, implying the existence of a cosmic
(Hubble) horizon. This in turn will prevent a proper definition of pure
asymptotic states.

We would like to point out at this stage that the claimed value of the dark
energy density component of the (four-dimensional) Universe today, $\Lambda
\sim 10^{-122}M_P^4$, with $M_P \sim 10^{19}$~GeV (the Planck mass scale), can
actually be accounted for (in an amusing coincidence?)  by the scale of the
neutrino mass differences used in order to explain the oscillation experiments.
Indeed, $\Lambda \sim [(\Delta m^2)^2/M_P^4]M_P^4 \sim 10^{-122} M_P^4$ for
$\Delta m^2 \sim 10^{-5}$~eV$^2$, the order of magnitude of the solar neutrino
mass difference assumed in oscillation experiments (which is the one that
encompasses the decoherence effects).  The quantum decoherence origin of this
mass then would be in perfect agreement with the decoherence properties of the
cosmological constant vacuum, mentioned previously.

\newpage

\section{NuTeV Physics}

The NuTeV experiment \cite{Zeller:2001hh}
at Fermilab has measured the ratios of neutral to charged current events in
muon (anti)neutrino-nucleon scattering:
\begin{eqnarray}
R_\nu
& = &
\frac{ \sigma(\nu_\mu N \rightarrow \nu_\mu X) }{\sigma(\nu_\mu N
\rightarrow \mu^-   X) }
\;=\; g_L^2 + r g_R^2\;, \cr
R_{\bar{\nu}}
& = & \frac{ \sigma(\bar{\nu}_\mu N \rightarrow \bar{\nu}_\mu X) }{ \sigma
(\bar{\nu}_\mu N \rightarrow \mu^+         X) }
\;=\; g_L^2 + \frac{g_R^2}{r}\;,
\end{eqnarray}
where
\begin{equation}
r = \frac{ \sigma( \bar{\nu}_\mu N \rightarrow \mu^+ X) }{ \sigma( \nu_\mu
 N \rightarrow \mu^- X) }
\sim \frac{1}{2}\;,
\end{equation}
and has determined the parameters $g_L^2$ and $g_R^2$
\cite{LlewellynSmith:ie} to be
\begin{eqnarray}
g_L^2 & = & 0.30005 \pm 0.00137\;, \cr
g_R^2 & = & 0.03076 \pm 0.00110\;.
\label{nutev}
\end{eqnarray}
The Standard Model (SM) predictions of these parameters
based on a global fit to non-NuTeV data, cited as
$[g_L^2]_\mathrm{SM}=0.3042$ and
$[g_R^2]_\mathrm{SM}=0.0301$ in Ref.~\cite{Zeller:2001hh},
differ from the NuTeV result by $3\sigma$ in $g_L^2$.
Alternatively, if the SM is fit to the NuTeV result, the preferred range of
the Higgs mass is $660~\mathrm{GeV}< m_H$ (90\% C.L.) \cite{Chanowitz:2002cd},
well above the value of $m_H \sim 90~\mathrm{GeV}$
preferred by the non-NuTeV global fit \cite{LEP/SLD:2003}.

 The significance of the NuTeV result remains controversial
\cite{Davidson:2001ji},
and a critical examination of the initial analysis is ongoing.
Several groups are evaluating potential theoretical uncertainties arising
from purely
Standard Model physics which might be comparable to or larger than the
quoted experimental uncertainty of the NuTeV result.
Candidate sources of large theoretical uncertainty include
next-to-leading-order (NLO) QCD corrections \cite{Dobrescu:2003},
NLO electroweak corrections \cite{Diener:2003}, and parton distribution
functions
(especially as involves assumptions about sea-quark
asymmetries) \cite{Gambino:2003}.
The effect of the former has been estimated to be comparable in size to
the NuTeV experimental uncertainty,
while the latter two might give rise to effects comparable in size
to the full NuTeV discrepancy with the Standard Model.
Elucidation of the actual impact of these effects on the NuTeV result awaits a
reanalysis of the NuTeV data.
However, it remains a distinct possibility that the discrepancy with the
Standard Model prediction is genuine and that its resolution lies in physics
beyond the Standard Model.
Indeed, as Chanowitz has emphasized \cite{Chanowitz:2002cd},
the precision electroweak data indicate new physics
whether anomalous data are excluded from global fits
(since the preferred Higgs mass is then well below the direct search limit) or
included in the fits (in which case anomalous data themselves
demand a new physics explanation).


Note that the NuTeV value for $g_L^2$ in Eq.~(\ref{nutev}) is
\textit{smaller} than its SM prediction.
This is a reflection of the fact that the ratios $R_\nu$ and
$R_{\bar{\nu}}$ were
smaller than expected by the SM. (The $g_R^2$ term is smaller than the
$g_L^2$ term
by an order of magnitude and is insignificant.)
Thus, possible \textit{new} physics explanations of the NuTeV anomaly
would be those
that suppress the neutral current cross sections over the charged current
cross sections,
or enhance the charged current cross sections over the neutral current
cross sections.
Two classes of models have been proposed which accomplish this task.


The first class comprises models which suppress $R_\nu$ and
$R_{\bar{\nu}}$ with the
introduction of new neutrino-quark interactions, mediated by leptoquarks or
extra $U(1)$ gauge bosons ($Z'$'s),
which interfere either destructively with the $Z$-exchange amplitude,
or constructively with the $W$-exchange amplitude \cite{Davidson:2001ji}.
In order to preserve the excellent agreement between the SM and non-NuTeV data,
the new interactions must selectively interfere with the
$\nu_\mu N$ ($\bar{\nu}_\mu N$) scattering process, but little
else.  This severely restricts the types of interactions that may be introduced.

Ref.~\cite{Davidson:2001ji} proposes a model in which the $Z'$ couples to
$B-3L_\mu$.
This model must be fine-tuned to avoid $Z$-$Z'$ mixing
\cite{Loinaz:1999qh} which
would disrupt, among other things, lepton universality at the $Z$-pole.
Fitting the NuTeV anomaly requires
\begin{equation}
\frac{M_{Z'}}{g_{Z'}} \approx 3\;\mathrm{TeV}\;.
\end{equation}
Bounds from direct $Z'$ searches at the Tevatron and LEP limit the
possible range of $M_{Z'}$ to
$M_{Z'} > 600\,\mathrm{GeV}$ for $g_{Z'}\sim 1$, or
$2\;\mathrm{GeV} < M_{Z'} < 10\,\mathrm{GeV}$ for $g_{Z'}\sim 10^{-3}$.

The $Z'$ in the model proposed in Ref.~\cite{Ma:2001md} does not
couple the neutrinos and quarks directly, since the gauged charge is
$L_\mu - L_\tau$.
Rather, it is a tunable $Z$-$Z'$ mixing in the model which is responsible for
suppressing the neutral channel cross section. The same mixing violates lepton
universality on the $Z$-pole and prevents the mechanism from completely
mitigating the NuTeV anomaly. $Z'$ masses in the range
$60\,\mathrm{GeV}<M_{Z'} <72\,\mathrm{GeV}$, or $M_{Z'}> 178\,\mathrm{GeV}$
brings the theoretical value of $g_L^2$ within $1.6\sigma$ of the NuTeV
value while
keeping lepton universality violation within $2\sigma$.

In general, models in this class are constrained strongly by lepton
universality, because $\nu_\ell$ is the $SU(2)_L$ partner of
$\ell_L^-$. New interactions which respect
the $SU(2)_L$ gauge symmetry cannot affect neutrino couplings alone: they
necessarily
affect couplings of the charged leptons.  Nevertheless, they provide
possible explanations
of the NuTeV anomaly, and predict a flavor-selective gauge boson in the several
100~GeV to TeV range, well within reach of the LHC.


Models of the second class suppress the $Z\nu\nu$ coupling by mixing the
neutrino with heavy gauge singlet states (neutrissimos,
i.e.\ right-handed neutrinos) \cite{numix,Chang:1994hz,LOTW1,LORTW2}.
For instance, if the $SU(2)_L$ active $\nu_\mu$ is a linear combination
of two mass eigenstates with
mixing angle $\theta$,
\begin{equation}
\nu_\mu = (\cos\theta) \nu_\mathrm{light} + (\sin\theta) \nu_\mathrm{heavy}\;,
\end{equation}
then the $Z\nu_\mu \nu_\mu$ coupling is suppressed by a factor of
$\cos^2\theta$ (assuming the heavy states are too massive to be created
on-shell).
Likewise, the $W\mu\nu_\mu$ coupling is suppressed by $\cos\theta$.
Although both the numerators and denominators of $R_{\nu}$ and
$R_{\bar{\nu}}$ are suppressed in such a model,
the suppression of the numerators exceeds that of the denominators, and
the ratios are therefore diminished.
More generally, if the $Z\nu_\ell \nu_\ell$ coupling
($\ell=e,\mu,\tau$) is suppressed by
a factor of $(1-\varepsilon_\ell)$,  then the $W\ell\nu_\ell$ coupling is
suppressed by
$(1-\varepsilon_\ell/2)$, and $R_{\nu}$ and $R_{\bar{\nu}}$ are suppressed
by $(1-\varepsilon_\mu)$.

The effect of such suppressions of the neutrino-gauge couplings is not
limited to NuTeV
observables alone. In addition to the obvious suppression of the $Z$
invisible width by a factor of
$[1-(2/3)(\varepsilon_e+\varepsilon_\mu+\varepsilon_\tau)]$, all SM
observables will be affected
through the Fermi constant $G_F$ which is no longer equal to the muon
decay constant $G_\mu$:
\begin{equation}
G_F = G_\mu \left(1+\frac{\varepsilon_e + \varepsilon_\mu}{2} \right)\;.
\end{equation}
This shift in $G_F$ will destroy the excellent agreement between the SM
and $Z$-pole observables.
However, since $G_F$ always appears in the combination $\rho G_F$ in
neutral current amplitudes,
the agreement can be recovered by absorbing the shift in $G_F$ into a
shift in $\rho$, or equivalently, in the oblique correction parameter $T$
\cite{Peskin:1990zt}.
Indeed, it was shown in Ref.~\cite{LOTW1}, that
the $Z$-pole, NuTeV, and $W$ mass data can all be fit with the oblique
correction
parameters $S$, $T$, $U$, and a flavor universal suppression parameter
$\varepsilon = \varepsilon_e = \varepsilon_\mu = \varepsilon_\tau$, the
best fit values given by
\begin{eqnarray}
S & = & -0.03 \pm 0.10 \;,\cr
T & = & -0.44\pm 0.15 \;,\cr
U & = & \phantom{-}0.62\pm 0.16  \;,\cr
\varepsilon & = & 0.0030\pm 0.0010\;,
\end{eqnarray}
for a reference SM with $m_H=115\;\mathrm{GeV}$.
Therefore, for this class of models to work, neutrino mixing with heavy gauge
singlet states must be accompanied
by new physics contributions to $S$, $T$, and $U$.
The values of $S$ and $T$ can be accommodated within the SM by simply
increasing the
Higgs mass to hundreds of GeV, but the $W$ mass requires a large and
positive $U$ parameter
which cannot be generated within the SM. Thus, the models are not
complete until
some mechanism is found which explains the $W$ mass.  But then, if the SM
is fit to the $W$ mass
alone, the preferred Higgs mass is far below direct search limits
\cite{Chanowitz:2002cd}, which could be an
indication that the $W$ mass requires new physics regardless of NuTeV.

At first blush, the preferred value of $\varepsilon$ above is also
problematic.
This implies a large mixing angle, $\theta = 0.055 \pm 0.010$, if
interpreted as due to
mixing with a single heavy state.
The commonly accepted seesaw mechanism
\cite{Yanagida:1980,Gell-Mann:1980vs,Glashow:1979vf,Mohapatra:1980ia}
relates
the mixing angle to the ratio of the neutrino masses:
\begin{equation}
\frac{m_\mathrm{light}}{m_\mathrm{heavy}}\approx \theta^2\;.
\end{equation}
Choosing $m_\mathrm{light} \sim 0.1\,\mathrm{eV}$ and $m_\mathrm{heavy}
\sim 100\,\mathrm{GeV}$
($m_\mathrm{heavy}>M_Z$ is needed to suppress $\Gamma_\mathrm{inv}$)
we find the mixing angle orders of magnitude too small: $\theta \sim 10^{-6}$.
However, this result does not mean that it is impossible to have a large
enough mixing angle
between the light and heavy states.   As pointed out in
Ref.~\cite{Chang:1994hz},
in models with more than one generation, the generic mass matrix includes
enough
degrees of freedom to allow us to adjust all the masses and mixings
independently.
Concrete examples of models with large mass hierarchies AND large mixing
angles can be found in
Refs.~\cite{Glashow:2003wt,LORTW2}.
What is sacrificed, however, is the traditional seesaw explanation of the
small neutrino mass:
i.e.\ since the Majorana mass $M$ in the neutrino mass matrix
should be of the order
of the GUT scale, the neutrino mass $m_\mathrm{light}\sim m^2/M$ is
naturally suppressed
if the Dirac mass $m$ is comparable to that of the other fermions.
An alternative mechanism is used in Ref.~\cite{LORTW2}.
There, an intergenerational symmetry is imposed on the neutrino mass
texture which reduces its rank,  generating naturally light
(massless) mass eigenstates.

Abandoning the seesaw mechanism also frees the masses of the heavy states
from being fixed at the GUT scale.  Indeed, in the model discussed in
Ref.~\cite{LORTW2},
the assumption that neutrinos and up-type quarks have a common Dirac mass implies that the masses of the heavy state could be a few TeV, well within the reach of the LHC.  Without quark-lepton unification $m_\mathrm{heavy}$ could
be even lighter, rendering
them accessible to Tevatron Run II.

Because of the large mixing angles between the light and heavy states in
this class of models,
flavor changing processes mediated by the heavy states may be
greatly enhanced \cite{Glashow:2003wt,LORTW2,Shrock:1977,LeeShrock:77}.
As a result, stringent constraints can be placed on the models from the
experimental limits on
$\mu\rightarrow e\gamma$, $\tau\rightarrow \mu\gamma$, \cite{present}
$\mu$-$e$ conversion in nuclei \cite{Ahmad:1988ur,Simkovic:2001fy},
muonium-antimuonium oscillation \cite{Willmann:1998gd,Halprin:wm}, etc.
For instance,  the MEGA limit on $\mu\rightarrow e\gamma$
leads to the
constraint \cite{LORTW2}
\begin{equation}
\varepsilon_e\varepsilon_\mu \approx 0\;.
\end{equation}
Therefore, lepton universality among the $\varepsilon_\ell$ must be
broken maximally.
Ref.~\cite{LORTW3} shows that it is possible to fit the $Z$-pole, NuTeV,
and lepton
universality data while satisfying this condition.

The MEG (Mu-E-Gamma) experiment at PSI \cite{meg} plans to improve upon the
MEGA limit by about two orders of magnitude.  The MECO (Muon on Electron
COnversion) experiment at Brookhaven \cite{meco} aims to improve the limits on
$\mu-e$ conversion in nuclei by three orders of magnitude.  Further constraints
can be obtained from muon $g-2$ \cite{Sichtermann:2003cc,Ma:2002df}, and the
violation of CKM unitarity \cite{Langacker:2003tv,Abele:2002wc,Tipton:2000qi}.

The NuTeV anomaly, even if it does not ultimately endure sustained scrutiny,
stirs us to look past orthodoxies in our model-building (seesaw, SUSY,
GUTs,...) and to ask broadly what is permitted by the data. The neutrino mixing
solution is relatively conservative in its use of the neutrino sector to
address the NuTeV question.  Nonetheless, it makes interesting predictions
about new particles at LHC, can be probed by a wide range of neutrino
oscillation experiments, precision measurements and rare decay searches, and
introduces an alternative to the seesaw paradigm.  Whether this or another
solution resolves the NuTeV anomaly, the NuTeV result serves to focus the
imagination of the theorist on the opportunities presented by the experiments.

\newpage

\section{Conclusions}
 In this report, we have presented a brief review of the present\footnote{As of the end of summer, 2005.} knowledge
of neutrino physics and what we can learn from the planned experiments
in the next decade. Three very important measurements that are guaranteed to have a
significant impact on the search for physics beyond the Standard Model
are: (i) the rate of $\beta\beta_{0\nu}$, which
will inform us not only whether the neutrino is a Majorana or
Dirac particle but may also provide information about the neutrino masses;
(ii) the value of $\theta_{13}$, which will considerably narrow the
field of flavor models and (iii) the sign of the $\Delta m^2_{13}$, which determines
the neutrino mass hierarchy and will also help guide our understanding of flavor physics. Within the three
neutrino picture, more precise measurements of the solar and atmospheric
mixing angles will be helpful in discriminating among various new physics
possibilities.

Important though somewhat model-dependent constraints can be
drawn from experimental searches for charged lepton flavor violating processes, such as
$\mu \rightarrow e \gamma$ or $\mu\to e$ conversion in nuclei, and from
searches for nonzero electric dipole moments of leptons.
Keep in mind that the matter-antimatter symmetry of the Universe may have its
explanation in the very same mechanism that generates the small
neutrino masses, and that we may be able to test this hypothesis with enough low-energy information.

Beyond the three neutrino picture, a very important issue is the status
of the LSND result and whether the existence of light sterile neutrinos can be inferred from terrestrial
neutrino oscillations experiments. The results of MiniBooNE, assuming they
confirms those from LSND, have the potential to revolutionize our current understanding of neutrinos.
Even if MiniBooNE does not confirm the LSND result, sterile neutrino effects can still be present
in different channels at a more subdominant level, as has been suggested in several theoretical models.

Another important issue in neutrino physics is the magnetic moment of the
neutrino, which is
expected at to be nonzero but very small within the standard picture of eV sized
neutrino masses and in the absence of new physics at the TeV scale. Thus, evidence for
a nonzero neutrino magnetic moment close to the current astrophysical limit of
$~10^{-11}\mu_B$ would have to be interpreted as evidence of TeV scale
new physics such as TeV scale left-right models, horizontal
models, or large extra dimensions. Other unique probes of TeV scale physics are provided by
neutrino oscillation experiments, thanks to their sensitivity to non-standard neutrino interactions.

Finally, one can use results in neutrino physics to test the limits of the
assumptions on which the Standard Model is based, including Lorentz and CPT
invariance.

\vspace{0.5cm}
\begin{center}
{\bf Acknowledgments}
\end{center}
The work of R.N.M.\ is supported by the National Science Foundation grant
No. PHY-0099544 and PHY-0354401.
The work of W.R. and M.L. is supported by the ``Deutsche Forschungsgemeinschaft''
in the ``Sonderforschungsbereich 375 f\"ur Astroteilchenphysik'' and under project number RO-2516/3-1 (W.R.).
The work of M.R. is partially supported by the EU 6th Framework Program
MRTN-CT-2004-503369 ``Quest for Unification'' and MRTN-CT-2004-005104
``ForcesUniverse''.
The work of R.S. is supported in part by the NSF
grant NSF-PHY-00-98527.
The work of J.K. is supported by the ``Impuls- und Vernetzungsfonds'' of
the Helmholtz Assciation, contract number VH-NG-006.
The work of S.A. is supported by the PPARC grant PPA/G/O/2002/00468.
The work of A.d.G. is sponsored in part by the US Department of Energy 
Contract DE-FG02-91ER40684. We thank F. Vissani for participating in the 
shorter version of this report.

\end{document}